\documentclass[11pt,a4paper,openany]{scrbook}
\usepackage{mathpazo}
\usepackage{dcolumn}
\usepackage{bm}
\usepackage[fleqn]{amsmath}
\usepackage{amssymb}
\usepackage{booktabs,longtable,subfigure}
\usepackage{graphicx,psfrag}
\usepackage{epsfig}
\usepackage{color} 
\usepackage{rotating}
\usepackage{axodraw}
\usepackage{slashed}
\usepackage{rotating}
\usepackage{cite}
\usepackage{geometry}
\allowdisplaybreaks[3]
\newcommand{\ie}{\ensuremath{\mathrm{i}}}
\newcommand{\eu}{\ensuremath{\mathrm{e}}}
\newcommand{\cw}[1][{}]{\ensuremath{\cos^{#1} \theta_{w}}}
\newcommand{\sw}[1][{}]{\ensuremath{\sin^{#1} \theta_{w}}}
\newcommand{\tw}[1][{}]{\ensuremath{\tan^{#1} \theta_{w}}}

\newcommand{\vv}{\ensuremath{\bar{v}}}
\newcommand{\uu}{\ensuremath{\bar{u}}}
\newcommand{\real}{\ensuremath{\mathfrak{Re}}}
\newcommand{\imag}{\ensuremath{\mathfrak{Im}}}
\newcommand{\mink}[2]{\ensuremath{\,#1\! \cdot \! #2 \,}}

\newcommand{\dif}{\ensuremath{\mathrm{d}}}
\newcommand{\dd}[1][{}]{\ensuremath{\mathrm{d}^{#1}}}

\newcommand{\M}{\ensuremath{\mathcal{M}}}

\newcommand{\ssl}[1]{\ensuremath{\slashed{#1}}}
\newcommand{\eps}{\ensuremath{\epsilon_{\kappa\lambda\mu\nu}k_1^\kappa p_1^\lambda p_2^\mu q^\nu}}
\newcommand{\half}{\ensuremath{\frac{1}{2}}}

\newcommand{\Backgroundnu}[1][{}]{\ensuremath{e^+e^-  \to \nu \bar\nu \gamma}#1}

\newcommand{\signal}[1][{}]{\ensuremath{e^+e^- \rightarrow \tilde\chi^0_1 \tilde\chi^0_1 \gamma}#1}

\newcommand{\x}[1]{\ensuremath{\tilde{\chi}^0_{#1}}}
\newcommand{\GeV}{\ensuremath{\,\mathrm{GeV}}}
\newcommand{\fb}{\ensuremath{\,\mathrm{fb}}}
\newcommand{\pb}{\ensuremath{\,\mathrm{pb}}}

\DeclareMathOperator*{\diag}{diag}

\def\lsim{\raise0.3ex\hbox{$\;<$\kern-0.75em\raise-1.1ex\hbox{$\sim\;$}}}
\def\gsim{\raise0.3ex\hbox{$\;>$\kern-0.75em\raise-1.1ex\hbox{$\sim\;$}}}

\begin{document}
\begin{titlepage}		
\begin{center}
{\Large \bf
{\LARGE
Constraints on Neutralino masses and mixings 
\medskip

from Cosmology and Collider Physics

} 
}
\vspace{20mm}
{\Large Dissertation\\
	zur\\
	Erlangung des Doktorgrades (Dr. rer. nat.) \\
	der\\
	Mathematisch-Naturwissenschaftlichen Fakult\"at\\
	der\\
	Rheinischen Friedrich-Wilhelms-Universit\"at\\
	zu Bonn \\[50mm]
}
{\Large 
	vorgelegt von\\[3mm]
	{\sc Ulrich Langenfeld}\\[3mm]
	geb. in \\
	Neuwied\\
\vfill 
Bonn 2007
}
\end{center}
\end{titlepage}
\begin{titlepage}
\begin{large}
Angefertigt mit Genehmigung der Mathematisch-Naturwissenschaftlichen Fakult\"at
der Universit\"at Bonn.

\vfill
\begin{tabbing}
xxxxxxxxxxxxxxxxxx \=\kill
Referent:	\> Prof. Herbert Dreiner\\[1mm]
Korreferent: \> Prof. Manuel Drees\\[1mm]
Tag der Promotion:\> \phantom{10. Juli 2007}\\ 
\end{tabbing}
\end{large}
\end{titlepage}
\begin{titlepage}
\begin{large}
\vspace*{19cm}
\parindent 0pt
\hspace*{\fill}	\\[10mm]
Ich versichere, da\ss ~ich diese Arbeit selbst\"andig verfa\ss t und keine
anderen als die angegebenen Quellen und Hilfsmittel benutzt sowie die
Zitate kenntlich gemacht habe.
\begin{tabbing}
xxxxxxxxxxxxxxxxxx \=			\kill
Referent:	\> Prof. Herbert Dreiner\\
Korreferent: 	\> Prof. Manuel Drees\\
\end{tabbing}
\end{large}
\end{titlepage}
\pagenumbering{arabic}

\chapter*{}
\vspace{20mm}
\hspace{100mm}{\LARGE \it To my parents}

\mainmatter
\chapter*{\centering \it Acknowledgements}
\noindent
First I would like to express my gratitude to my supervisor Herbi Dreiner 
for his help and support with this work. 

Furthermore I thank Manuel Drees for enlightening discussions on
dark matter and acting as second referee for this thesis.

I also thank I.~Brock, K.~Desch, G.~Moortgat-Pick, M.~Schumacher, X.~Tata and 
G.~Weiglein for helpful discussions. 

I am also grateful to my collaborator as well as room mate
Olaf Kittel for the effective and also fun research, he 
provided the Fortran code for the neutralino pair production
to me and read parts of the manuscript.

I have always enjoyed the friendly atmosphere in our group and the enspiring 
discussions with Olaf Kittel, Federico von der Pahlen, Markus Bernhardt, 
Jong Soo Kim, Sebastian Grab, Anja Marold and Chung-Lin Shan. 

In addition I would like to thank our always helpful secretaries of the 
Bonn theory group, namely D.~Fassbender, P.~Z\"undorf, and S.~Heidbrink and
our computer and allround specialist A.~Wisskirchen.

My parents supported me during the time this work was being done. I am very grateful
to them.

\newpage
\phantom{bla}
\tableofcontents
\setcounter{chapter}{-1}
\chapter{Abstract}

Bounds on cross section measurements of chargino pair production at LEP yield
a bound on the chargino mass. If the GUT relation $M_1 = 5/3 \tw[2] M_2$
is assumed, then the lightest neutralino must be heavier than 
$\approx 45 -50\GeV$. If $M_1$ is considered as a free parameter
independent of $M_2$ there is no bound on the mass of the lightest neutralino. 
In this thesis, I examine consequences of light, even massless neutralinos
in cosmology and particle physics.

\medskip
In Chapter~\ref{ch:cosmo}, I discuss mass bounds on the lightest neutralino
from relic density measurements. The relic density can be calculated by solving
the Boltzmann equation.
If the relic density is considered as a function 
of the particle mass then there are two mass regions where the relic density
takes on realistic values.  
In the first region the neutralino is relativistic
and its mass must be lower than $0.7\,\mathrm{eV}$, in the second region
the neutralino is nonrelativistic and must be heavier than $\approx 13\GeV$. 
I compare the Cowsik-McClelland bound, the approximate solution of a
relativistic particle for the Boltzmann equation, and the Lee-Weinberg bound,
the non-relativistic approximation, with the full solution and I find
that the approximation and the full solution agree quite well.

\medskip
In Chapter~\ref{ch:chi12}, I derive bounds on the selectron mass from the
observed limits on the cross section of the reaction
$e^+e^-\rightarrow \x{1}\x{2}$ at LEP, if the lightest neutralino
is massless. If \mbox{$M_2,\mu < 200\GeV$}, the selectron must be heavier than
$350\GeV$.

\medskip
In Chapter~\ref{ch:radiative},
I study radiative neutralino production $e^+e^- \to \tilde\chi^0_1
\tilde\chi^0_1\gamma$ at the linear collider with longitudinally
polarised beams.  I consider the Standard Model background from
radiative neutrino production $e^+e^- \to \nu \bar\nu \gamma$, and the
supersymmetric radiative production of sneutrinos $e^+e^- \to
\tilde\nu \tilde\nu^\ast \gamma$, which can be a background for
invisible sneutrino decays. I give the complete tree-level formulas
for the amplitudes and matrix elements squared. In the Minimal
Supersymmetric Standard Model, I study the dependence of the cross
sections on the beam polarisations, on the parameters of the
neutralino sector, and on the selectron masses.  I show that
for bino-like neutralinos
longitudinal polarised beams enhance the signal and simultaneously
reduce the background, such that search sensitivity is significantly enhanced.
I point out that there are parameter regions where
radiative neutralino production is
the {\it only} channel to study SUSY particles at the ILC, since heavier 
neutralinos, charginos and sleptons are too heavy to be pair-produced 
in the first stage of the linear collider with $\sqrt{s} =500\GeV$.

In Section~\ref{sec:beampol},
I focus on three different mSUGRA scenarios in turn at the
Higgs strahlung threshold, the top pair production threshold, and at
$\sqrt{s} =500\GeV$. In these scenarios at the corresponding
$\sqrt{s}$, radiative neutralino production is the only supersymmetric
production mechanism which is kinematically allowed. The heavier
neutralinos, and charginos as well as the sleptons, squarks and
gluinos are too heavy to be pair produced. I calculate the signal
cross section and also the Standard Model background from radiative
neutrino production. For my scenarios, I obtain significances larger
than $10$ and signal to background ratios between $2\%$ and $5\%$, if
I have electron beam polarization $P_{e^-} = 0.0- 0.8$ and positron
beam polarization $P_{e^+} = 0.0 - 0.3$. If I have electron beam
polarization of $P_{e^-} = 0.9$, then the signal is observable with
$P_{e^+} = 0.0$ but both the significance and the signal to background
ratio are significantly improved for $P_{e^+} = 0.3$.

\medskip
In Chapter~\ref{ch:magic}, I present a method to determine neutralino couplings to
right and left handed selectrons and $Z$ bosons from cross section measurements 
of radiative neutralino production and neutralino pair production 
$e^+ e^- \rightarrow \x{1}\x{2/3/4}$, $e^+ e^- \rightarrow \x{2}\x{2}$ at the ILC. 
The error on the couplings is of order
$\mathcal{O}(0.001 -0.01)$. From the neutralino couplings the 
neutralino diagonalisation matrix can be calculated. If all neutralino 
masses are known, $M_1$, $M_2$, and $\mu$ can be calculated with an error
of the order
$\mathcal{O}(1\GeV)$. If also the cross sections of the reactions
$e^+ e^- \rightarrow \x{2}\x{3/4}$ can be measured the error
of $M_1$, $M_2$, and $\mu$ reduces to $\mathcal{O}(1\GeV)$.

\chapter{The Gaugino sector in the MSSM}
The Standard Model (SM) has been tested to high precision. 
But many problems remain unsolved. The SM does not include gravity.
The electro-weak couplings and the strong coupling do not unify in a point
at the GUT scale $\Lambda_{\mathrm{GUT}}$. The SM model cannot explain 
why there is so much more matter than antimatter in the universe and it does 
not provide a dark matter candidate.

One solution to these problems is 
supersymmetry~\cite{Haber:1984rc,aitchison,drees,dreesbook,Nilles:1983ge}. 
In supersymmetric theories,
each fermion is mapped onto a boson and vice versa. 
The spin of the fermion and its partner boson differ by half a unit,
the other quantum numbers are unchanged.

The superpartners of leptons, quarks, gauge bosons, and Higgs bosons
are called sleptons, squarks, gauginos, and higgsinos, respectively. 
The two neutral gauginos $\lambda_0$, $\lambda_3$
and the two neutral higgsinos $\widetilde{h}_1^1$, $\widetilde{h}_2^2$
have the same quantum numbers and mix. 
The physical mass eigenstates are obtained by diagonalisation
of the mass matrix. These neutral particles are called neutralinos. 
They are Majorana fermions. 
The two charged gauginos and two charged higgsinos mix to charginos. 

At low energies no SUSY particles have been observed, so SUSY must be broken.
The most common way is introducing explicitly soft SUSY breaking terms.  

\medskip 
The part of the Lagrangian which describes the neutralino mixing is given by~\cite{dreesbook}
 \begin{eqnarray}
\label{eq:chimix}
\mathcal{L} &=& -\half \lambda_0 \lambda_0 M_1 -\half \lambda_3 \lambda_3 M_2 
		+ \mu \widetilde{h}_1^1\widetilde{h}_2^2 
		- \frac{g_2}{2}\lambda_3(v_1 \widetilde{h}_1^1 - v_2 \widetilde{h}_2^2)
		+ \frac{g_1}{2}\lambda_0(v_1 \widetilde{h}_1^1 - v_2 \widetilde{h}_2^2)\\[2mm]
	     &\equiv& - \half \psi^T M \psi \notag\\[2mm]
\text{with}&&\notag\\[2mm] 
\label{eq:neutralinomatrix}    
M &=&
\begin{pmatrix}
M_1 & 0   & - m_Z \sw \cos\beta & \phantom{-}m_Z\sw \sin\beta \\
0   & M_2 & \phantom{-} m_Z \cw \cos\beta  & -m_Z \cw\sin\beta \\
 - m_Z \sw \cos\beta &\phantom{-} m_Z \cw \cos\beta  & 0 & -\mu\\
\phantom{-}m_Z\sw \sin\beta& -m_Z \cw\sin\beta & -\mu & 0
\end{pmatrix},\,\,\\[2mm]
\psi^T &=& 
\begin{pmatrix}\lambda_0,& \lambda_3,&\widetilde{h}_1^1,&\widetilde{h}_2^2\end{pmatrix}\qquad
(\text{the}\enspace \psi_i\enspace\text{are\enspace Weyl\enspace spinors}).
\end{eqnarray}
$M_1$ and $M_2$ are the $U(1)_Y$ and the $SU(2)_w$ gaugino mass parameters, respectively. 
They break SUSY explicitly. $\mu$ is the higgsino mass parameter and
$\tan\beta = \frac{v_2}{v_1}$ is the ratio of the two vacuum expectation
values of the Higgs fields, $m_Z$ the $Z$ boson mass, and $\tw$ the
weak mixing angle. 
 
$M_1$, $M_2$, and $\mu$ are real parameters, if $CP$ is conserved, in general they are
complex:
\begin{eqnarray}
\label{eq:complex}
M_1 = |M_1|\eu^{\ie \phi_1},\qquad \mu = |\mu|\eu^{\ie \phi_\mu}\enspace . 
\end{eqnarray}

The matrix $M$ is symmetric, even for $M$ complex. The reason for this fact is
that in Eq.~\ref{eq:chimix} there appear no hermitian conjugated fields.
The matrix $M$ can be diagonalised
by an unitary matrix $N$ using Takagi's factorization theorem~\cite{Horn:1985ma}   
\begin{eqnarray}
\label{eq:chidiag}
\diag(m_{\x{1}},m_{\x{2}},m_{\x{3}},m_{\x{4}}) = N^\ast M N^{-1}\,.
\end{eqnarray} 
The diagonal elements $m_{\x{i}}$ are non-negative and are the square roots of the eigenvalues of
$M M^{+}$. The transformation Eq.~(\ref{eq:chidiag}) is not a similarity transformation,
if $N$ is complex.

If $M$ is a real matrix it can also be diagonalised by an orthogonal matrix. 
From the lower right $2\times2$ submatrix one can see that at least one eigenvalue is negative.
This sign is interpreted as the $CP$ eigenvalue of the neutralino. The masses of the neutralinos
are $|m_i|$, $i = 1\ldots 4$. 
The sign can be absorbed in a phase of the corresponding eigenvector, leading back to   
Eq.~(\ref{eq:chidiag}).

The eigenvalues of $M M^+$ and the diagonalisation Matrix $N$ can be obtained 
algebraically, see Ref.~\cite{ElKheishen:1992yv}
or numerically. The algebraic method is problematic because it is numerically unstable.
Gunion and Haber present in~\cite{Gunion:1987yh} approximate solutions to
the eigenvalues of $M$ and the diagonalisation matrix $N$, if $|M_{1,2}\pm \mu| \gg m_Z$.

\medskip
Without loss of generality $M_2$ can be chosen positive. Proof:
Let $M_2 = |M_2|\eu^{\ie \phi_2}$. The phase $\phi_2$ of $M_2$ can be removed
by the transformations:
\begin{eqnarray}
\psi &=& 
\begin{pmatrix}
\lambda_0\\[2mm] 
\lambda_3\\[2mm]
\widetilde{h}_1^1\\[2mm]
\widetilde{h}_2^2\\[2mm]
\end{pmatrix}
\mapsto
\psi^\prime =
\begin{pmatrix}
\lambda_0\,\eu^{-\ie \phi_2/2}\\[2mm] 
\lambda_3\,\eu^{-\ie \phi_2/2}\\[2mm]
\widetilde{h}_1^1\,\eu^{\ie \phi_2/2}\\[2mm]
\widetilde{h}_2^2\,\eu^{\ie \phi_2/2}\\[2mm]
\end{pmatrix}\, .
\end{eqnarray}
The parameters $M_1$ and $\mu$ transform as
\begin{eqnarray}
M_1 &\mapsto& M_1' = M_1 \,\eu^{\ie \phi_2}\,,\\[2mm]
\mu &\mapsto& \mu' = \mu \,\eu^{-\ie \phi_2}\,.
\end{eqnarray}  
It is not necessary to transform the higgsino fields. Alternatively, the vacuum expectation 
values $v_{1/2}$ can be transformed as $v_{1/2}^\prime = v_{1/2}\eu^{-\ie \phi_2/2}$ leaving 
$\tan\beta$ invariant. 

If $\phi_2 = \pi$ then the signs of $M_1$ and $\mu$ are interchanged. 
This transformation reverses also the sign of the eigenvalues of $M$. 

\medskip
In GUT theories, $M_1$ and $M_2$ are related by
\begin{eqnarray}
\label{eq:gut}
M_1 = \frac{5}{3}\tw[2] M_2 \approx \half M_2\,.
\end{eqnarray}
It follows that $M_1$ and $M_2$ can be chosen positive.

$M$ can have zero eigenvalues.
From $\det(M) = 0$ it follows in the $CP$ conserving case, see Ref.~\cite{Gogoladze:2002xp},
\begin{eqnarray}
0 &=& \det(M) = 
\mu\big[M_2 m_Z^2\sw[2]\sin(2\beta) + M_1\big(-M_2 \mu + m_Z^2\cw[2]\sin(2\beta)\big)\big]\notag\\[2mm]
&&\Rightarrow \mu = 0 \quad\vee\quad M_1 = 
	\frac{M_2 m_Z^2\sw[2]\sin(2\beta)}{M_2 \mu - m_Z^2\cw[2]\sin(2\beta)}.
\end{eqnarray}
The solution $\mu = 0$ is excluded due to experimental constraints from the 
$Z^0$-widths measured at LEP~\cite{Choudhury:1999tn}.

In the $CP$ violating case, substitute $M_1 \mapsto M_1 \eu^{\ie \phi_1}$, 
$\mu \mapsto \mu \eu^{\ie \phi_\mu}$ with $M_1$, $\mu \ge 0$. 
This yields two equations, which must be separately fulfilled:
\begin{eqnarray}
\imag \det(M) = 0 &\Rightarrow& 
		\mu = \frac{m_Z^2 \cw[2] \sin(2\beta) \sin\phi_1}{M_2 \sin(\phi_1+\phi_\mu)},\\[2mm] 
\real \det(M) = 0 &\Rightarrow& M_1 =  -M_2 \tw[2]\frac{\sin(\phi_1 + \phi_\mu)}{\sin\phi_\mu},
\end{eqnarray}
or
\begin{eqnarray}
M_2 & = & \frac{m_Z^2 \cw[2] \sin(2\beta) \sin\phi_1}{\mu \sin(\phi_1+\phi_\mu)}\qquad\text{and}\quad
M_1  =   -\frac{m_Z^2 \sw[2] \sin(2\beta)\sin\phi_1}{\mu \sin\phi_\mu},
\end{eqnarray}
or
\begin{eqnarray}
\sin(2\beta) &=& \frac{\mu M_2 \sin(\phi_1+\phi_\mu)}{m_Z^2 \cw[2] \sin\phi_1}\quad\text{and}\quad
M_1  =   -M_2 \tw[2]\frac{\sin(\phi_1 + \phi_\mu)}{\sin\phi_\mu}.
\end{eqnarray}
It follows immediately that $\sin\phi_1/\sin\phi_\mu < 0$ and 
$\sin(\phi_1+\phi_\mu)/\sin\phi_\mu < 0$ must hold. Also in the $CP$ violating case 
one can always find parameters $|M_1|, \phi_1, M_2, |\mu|, \phi_\mu$, and $\tan\beta$ to get
$m_{\x{1}} = 0$.

\medskip
The chargino mixing is described by the following matrix:
\begin{eqnarray}
\mathcal{L} &=& - (\psi^-)^T X \psi^+ ,\\[2mm]
X & \equiv &
\begin{pmatrix}
M_2 & \sqrt{2} m_W \sin\beta\\[2mm]
 \sqrt{2} m_W \cos\beta & \mu
\end{pmatrix},\\[2mm]
\psi^+ &\equiv& (\lambda^+,\widetilde{h}_2^1)^T,\quad \psi^- \equiv (\lambda^-,\widetilde{h}^2_1)^T,\quad
\end{eqnarray}
$X$ is not symmetric, so it must be diagonalised by a biunitary transformation:
\begin{eqnarray}
\diag(m_1^\pm , m_2^\pm) = U^\ast X V^{-1}, 
\end{eqnarray}
with $U$, $V$ unitary $2\times 2$ matrices. The matrices $U$ and $V$ are obtained by
solving
\begin{eqnarray}
\diag\big((m_1^\pm)^2 , (m_2^\pm)^2\big) = V X^+X V^{-1} = U^\ast X X^+ U^T\enspace .
\end{eqnarray}
The eigenvalues can be obtained analytically, see Ref.~\cite{Haber:1984rc, aitchison}.
In practical use it is easier to diagonalize the matrix $X$ numerically but using the 
analytical formulae. 

\medskip
The lower experimental bound on the lightest chargino mass is 
$m_{\widetilde{\chi}_1^\pm} > 104\GeV$~\cite{Yao:2006px}. This bound leads to lower bounds
on $\mu$ and $M_2$: $\mu, M_2 > 100\GeV$. If Eq.~(\ref{eq:gut}) is assumed, then $M_1$ 
depends on $M_2$ and from this fact follows a lower bound on the mass of the
lightest neutralino: $m_{\widetilde{\chi}_1^0} \gsim 49\GeV$~\cite{Abdallah:2003xe}. 
But up to now there is no
evidence that Eq.~(\ref{eq:gut}) holds. So I consider $M_1$ as a free parameter. 
In the following I study implications of massless and light neutralinos. 
In Chapter~\ref{ch:cosmo}, I discuss bounds on the neutralino mass from dark matter density measurements.
In Chapter~\ref{ch:chi12}, I derive bounds on the selectron mass from the observed cross section limits
from \x{1}\x{2} production at LEP, if \x{1} is massless.  In Chapter~\ref{ch:radiative}, I calculate
the cross section for radiative neutralino production and its neutrino and sneutrino background 
at a future $e^+e^-$ linear collider. 
I discuss the influence of beam polarisation on radiative neutralino production
and consequences of SUSY searches at a future linear collider.
Finally, in Chapter~\ref{ch:magic}, I present a method how to determine neutralino couplings to the
right and left handed selectron and the $Z$ boson.

\chapter{Cosmological bounds on neutralino masses}
\label{ch:cosmo}
\section{The Cowsik-McClelland-bound}
I derive bounds on the mass of the lightest neutralino through cosmological
considerations. Neutralinos are neutral and interact only weakly. 
If they are (pseudo-)stabile, they are dark - matter (DM) candidates.
The dark matter density $\Omega_{\mathrm{DM}}h^2$ has been measured by the WMAP 
collaboration~\cite{Spergel:2003cb}. This constrains the mass(es) of the particle(s) which
constitute the dark matter.

In Ref.~\cite{Kolb:1990eu}, Kolb and Turner describe the thermal evolution of the Universe 
and its impact on particle physics. 
I give a short summary in order to clarify the subsequent section.
\subsection{The Expansion of the Universe}
The expansion of the Universe is described by the Einstein field equations with 
the Robertson-Walker (RW) metric. In the RW metric, the Universe is assumed to be homogeneous 
and isotropic.
\begin{eqnarray}
\label{eq:friedmann}
\left(\frac{\dot{R}}{R}\right)^2 +\frac{k}{R^2} &=& \frac{8\pi G}{3}\rho,\\[2mm]
\label{eq:bla}
2\frac{\ddot{R}}{R} + \left(\frac{\dot{R}}{R}\right)^2 +\frac{k}{R^2} &=& - {8\pi G}p,\\[2mm]
\label{eq:firstlaw}
\dif(\rho R^3) &=& - p \dif(R^3)\enspace.
\end{eqnarray}
Here $R$ is the cosmic scale factor, $p$ and $\rho$ denote the pressure and the density,
respectively, and $G$ is Newton's constant.
Eq.~(\ref{eq:friedmann}) is called the Friedmann equation, Eq.~(\ref{eq:firstlaw}) is the 1st law
of thermodynamics. 
The parameter $k$ can be chosen as $\pm 1$ or $0$ to describe spaces with constant positive
or negative curvature, or flat geometry, respectively.
Eq.~(\ref{eq:friedmann}) and Eq.~(\ref{eq:bla}) can be subtracted to yield an equation for 
the acceleration of the scale factor
\begin{eqnarray}
\frac{\ddot{R}}{R} &=& - \frac{4\pi G}{3}(\rho + 3p)\enspace.
\end{eqnarray}
The Hubble parameter $H(t)$ determines the expansion of the Universe. 
It is defined as $H \equiv \frac{\dot{R}}{R}$.
The present day value $H(0) = H_0$ is called the Hubble constant.
With this definition the critical density $\rho_C$ ---the density, where the geometry 
of the Universe is flat--- follows as $\rho_c = \frac{3 H^2_0}{8\pi G}$.  
To solve Eqs.~(\ref{eq:friedmann}) - (\ref{eq:firstlaw}) we need an additional ingredient:
an equation of state, that describes the connection between density and pressure of the matter content of 
the Universe (i.e. radiation, baryonic matter or dark energy).
At the beginning, the Universe was dominated by radiation, after recombination the photons decoupled 
and the Universe was matter dominated. Today the dark energy contributes most of the density of 
the Universe. The equations of state are
\begin{eqnarray}
\label{eq:radiation}
p &=& \phantom{-}\frac{1}{3}\rho \quad\text{for}\,\,\text{radiation},\\[2mm]
\label{eq:matter}
p &=& \phantom{-}0 \phantom{\rho}\quad\text{for}\,\,\text{matter},\\[2mm]
\label{eq:vacuum}
p &=& -\rho \quad\text{for}\,\,\text{dark energy.}
\end{eqnarray}    
The Eqs~(\ref{eq:radiation})-(\ref{eq:vacuum}) can be summarized to
\begin{eqnarray}
p = w \rho, \enspace\text{with}\enspace w = \{\frac{1}{3},\enspace 0,\enspace -1\}, 
\end{eqnarray}
for radiation, matter, and dark energy, respectively. The dark energy is connected to the 
cosmological constant in the Einstein field equation.
\subsection{Basic Thermodynamics}
The particle density, the energy density and the pressure of a particle species 
in the Universe are given by
\begin{eqnarray}
\label{eq:numberdensity}
n &=& \frac{g}{(2\pi)^3}\int \dd[3] p f({\bf{p}}), \\[2mm]
\label{eq:density}
\rho &=& \frac{g}{(2\pi)^3}\int \dd[3] p E({\bf{p}})f({\bf{p}}),\\[2mm]
\label{eq:pressure}
p &=& \frac{g}{(2\pi)^3}\int \dd[3] p \frac{|{\bf{p}}|^2}{3E}f({\bf{p}}),
\end{eqnarray} 
where the phase space distribution (or occupancy) is given by
\begin{eqnarray}
f({\bf{p}}) &=& \frac{1}{\eu^{(E-\mu)/T} \pm 1}.  
\end{eqnarray}
The $+$ sign holds for fermions, the $-$ for bosons, and
$\mu$ is the chemical potential of the particles species. The energy
of a relativistic particle is given by $E({\bf p}) = \sqrt{{\bf p}^2 + m^2}$.
The entropy $S$ follows from
\begin{eqnarray}
\label{eq:entropy}
T\dif S = \dif{\rho V} + p \dif V = d [(\rho + p)V] - V\dif p\enspace.
\end{eqnarray} 

\subsection{Particles in the Universe}
I consider the behaviour of a class of particles $\psi_i$, $i=1\ldots n$ 
(f. e. sparticles in the MSSM) in the thermal bath of the early Universe.
Griest and Seckel discuss in \cite{Griest:1990kh} the mechanisms that are important
in order to determine the number density of these new particles. They assume that the $\psi_i$
have a multiplicatively conserved quantum number which distinguish them from Standard Model (SM) 
particles. In the MSSM, $R$ parity~\cite{Dreiner:1997uz} is such a quantum number. 
The subsequent reactions appear: 
\begin{subequations}
\label{eq:reactiontypes}
\begin{align}
\hspace{50mm}\psi_i \psi_j  &\rightleftharpoons X X',\\[2mm]
\psi_i X &\rightleftharpoons \psi_j X',\\[2mm]
\psi_j &\rightleftharpoons \psi_i X X'.
\end{align}
\end{subequations}
where $X$, $X'$ denote SM particles.
Examples in the MSSM for these three reaction types are:
\begin{subequations}
\begin{align}
\hspace{50mm}\chi_1^0 \chi_2^0 &\rightleftharpoons e^- e^+,\\[2mm]
\chi_1^0 e^- &\rightleftharpoons \nu_e\chi_1^-, \\[2mm]
\chi_2^0 &\rightleftharpoons \chi_1^0 e^+ e^-,
\end{align}
\end{subequations}
respectively. One of these particles is stabile due to the conserved quantum number.
In the MSSM with conserved R-parity, it is the $\chi_1^0$.
Griest and Seckel classify the reaction types, see Eq.~(\ref{eq:reactiontypes}), further.
If the lightest $\psi_i\equiv \psi_1$ is nearly mass degenerate to the next to lightest
particle $\psi_2$, then the number density of $\psi_1$ is also determined by annihilations of
$\psi_2$ which decays later into $\psi_1$. This is called coannihilation.
The masses of annihilation products can be heavier than the masses of the ingoing particles,
if the energy of the ingoing particles is large enough. Griest and Seckel call this
"forbidden" channels.  If annihilation occurs at a pole in the cross section it is called
annihilation near a pole or resonant annihilation.

For the further discussion, I exclude coannihilation and resonant annihilation for simplicity. 

\medskip   
The time evolution of a particle $\psi$ with total cross section $\sigma$ is described 
by the Boltzmann equation:
\begin{eqnarray}
\label{eq:boltzmann}
\frac{\dif n_\psi}{\dif t} + 3 H n_\psi + \langle \sigma |v|\rangle[n^2_\psi - (n^2_\psi)_{\mathrm{Eq}}] = 0,
\end{eqnarray}
with the the particle velocity $v$.
The second term accounts for the dilution of the species due to the expansion of the Universe, 
the third term for the decrease  by annihilation into other particles or coannihilation with other particles.
If we define
\begin{subequations}
\begin{align} 
\label{eq:entropydensity}
s &\equiv \frac{S}{V} = \frac{p + \rho}{T}\enspace (V:\text{volume}), \\[2mm]
\label{eq:y}
Y &= \frac{n_\psi}{s}\enspace ,\\[2mm]
\label{eq:xT} 
x &\equiv \frac{m}{T} \enspace (m\!:\text{particle\,\,mass}),\\[2mm] 
H(m) &= 1.67 g_\ast^{1/2}\frac{m^2}{m_\mathrm{Pl}}\enspace(m_\mathrm{Pl}\!:\text{Planck\,\,mass}),\\[2mm]
g_{\ast} &= \sum_{bosons}g_i\left(\frac{T_i}{T}\right)^4 +
	\frac{7}{8}\sum_{fermions}g_i\left(\frac{T_i}{T}\right)^4\enspace
(T_i:\enspace\text{temperature\,\,of\,\,particle\,\,species}\enspace i),\notag\\[-2mm]
\label{eq:gstar}&&\\
\label{eq:time} 
t &= 0.301 g^{-1/2}_\ast \frac{m_\mathrm{Pl}}{m^2}x^2,
\end{align}
\end{subequations} 
then Eq.~(\ref{eq:boltzmann}) can be cast into 
\begin{eqnarray}
\label{eq:suggestiv}
\frac{\dif Y}{\dif x} &=& -0.167 \frac{x s}{H(m)} \langle \sigma |v|\rangle\left(Y^2 - Y^2_{\mathrm{Eq}}\right)
\end{eqnarray}
or
\begin{eqnarray}
\frac{x}{Y_{\mathrm{Eq}}}\frac{\dif Y}{\dif x} &=& 
     -\frac{\Gamma_A}{H}\left[\left(\frac{Y}{Y_{\mathrm{Eq}}}\right)^2 -1\right],\quad 
     \Gamma_A \equiv n_\mathrm{Eq}\langle\sigma_A|v|\rangle\enspace.
\end{eqnarray}
$g_\ast$ is the number of massless degrees of freedom at $T_i$, where the particle temperature $T_i$ accounts for
the possibility that it is different from the photon temperature $T$. 
The thermal averaged cross section $\langle \sigma |v|\rangle$ is defined as
\begin{eqnarray}
\label{eq:thermalavxsec}
\langle \sigma |v|\rangle = \frac{1}{(n^2_\psi)_{\mathrm{Eq}}}\int \prod_{i=1}^4\frac{\dif^3 p_i}{(2\pi)^3 E_i}
	|\M|^2(2\pi)^4\delta^{(4)}(p_1+p_2 - p_3-p_4) \eu^{-(E_3+E_4)/T}\,.
\end{eqnarray}
If $Y = Y_{\mathrm{Eq}}$, then $Y$ does not change with time, so it is constant as expected,
c. f. Eq.~(\ref{eq:thermalavxsec}). 
If $\Gamma_A/H  < 1$,
then the relative change of $n_\psi$ is small and the annihilation processes stop, which means 
that the number of that particle species remains constant within a comoving volume.

\subsection{Application to Massless Neutralinos}
In the MSSM with $R$-parity conservation, the lightest neutralino is stabile and can be the lightest
supersymmetric particle. Therefore it is a dark matter candidate. 
I discuss the case when the neutralino is (nearly) massless, 
$m_{\widetilde{\chi}} \lesssim \mathcal{O}(1\,\mathrm{eV})$.  
The $Z$ width allows a higgsino contribution of about  
$\sqrt{N_{13}^2+N_{14}^2} < 0.5 \approx (0.08)^{1/4}$, see Choudhury et al.~\cite{Choudhury:1999tn}. 
$M_1$, the bino-mass, is normally chosen smaller than $M_2$ and $\mu$, 
and so the lightest neutralino is almost $100\%$ bino. For simplicity, I assume that it is purely bino.     
The neutralino freezes out at $x_f = m/T_f \ll 3$, and at freeze out it is still relativistic. 
From that it follows $Y(t\rightarrow \infty) = Y_{\mathrm{Eq}}(x_f)$.

\begin{eqnarray}
Y = \frac{n_{\mathrm{Eq}}}{s_0} = \frac{45}{2\pi^2}\zeta(3)\frac{g_{\mathrm{eff}}}{g_{\ast S}},
\end{eqnarray}
where $n_{\mathrm{Eq}}$ and $s$ are given by Eq.~(\ref{eq:numberdensity}) and (\ref{eq:entropydensity}),
$s_0$ is the present entropy density, and $\zeta$ denotes the Riemannian Zeta function. It is assumed
that the entropy per comoving volume is conserved.
$g_{\mathrm{eff}}$ counts the degrees of freedom of the neutralino field multiplied with $3/4$ to correct 
for the fermionic nature of the field, $g_{\ast S}$ counts the number of relativistic fields at freeze out, 
whereby fermionic fields are corrected with $7/8$:
\begin{eqnarray}
g_{\ast S} &=& \sum_{bosons}g_i\left(\frac{T_i}{T}\right)^3 +
	\frac{7}{8}\sum_{fermions}g_i\left(\frac{T_i}{T}\right)^3\enspace ,\\[2mm]
g_{\mathrm{eff}} &=& \left\{
		\begin{array}{ll}
		g,&\psi = \text{boson}\\[2mm]
		\frac{3}{4}g,&\psi = \text{fermion}	
		\end{array}
		\right. \enspace .
\end{eqnarray}
The neutralino density is obtained by
\begin{eqnarray}
\label{eq:relicdensity}
\rho_\chi  &=& m_{\widetilde{\chi}} n_\chi =  m_{\widetilde{\chi}} s_0 Y(t=\infty)=  
m_{\widetilde{\chi}} \frac{45}{2}\frac{\zeta(3)}{\pi^2}\frac{g_{\mathrm{eff}}}{g_{\ast S}(T)},\\[2mm]
\label{eq:relic}
\Omega_\chi h^2 &\equiv& \frac{\rho_\chi}{\rho_c} 
=  \frac{43}{11}\,\frac{\zeta(3)}{\pi^2}\,\frac{8\pi G}{3 H_0^2}\,\frac{g_{\mathrm{eff}}}{g_{\ast S}(T)}\,
			T_\gamma^3 \,m_{\widetilde{\chi}} .
\end{eqnarray}
In Eq.~(\ref{eq:relic}) I relate the relic density $\Omega h^2$ to the photon temperature by using
$s_0 = \frac{86 \pi^2}{11\cdot 45}T_\gamma^3$  and to the critical density.
The constraint on the density is chosen such that the lightest neutralino does not disturb structure
formation, so they cannot form the dominant component of the dark matter. 

Light neutralinos decouple at $T \approx \mathcal{O}(1 - 10\enspace\mathrm{MeV})$.
This temperature is somewhat higher than the temperature, when the neutrinos decouple. 
This is due to the selectron mass which can be larger than the $Z$ mass, leading to smaller cross sections.
But the temperature is below the muon mass so that it is not necessary to know the exact value.  
Nevertheless we have 2 bosonic and 12 fermionic relativistic degrees of freedom (one Dirac electron, 
three left handed neutrino species, one photon, one light Majorana neutralino)
leading to $g_{\ast S} = 12.5$ and $g_{\mathrm{eff}}=1.5$. 
If I demand (value of $\Omega_{\nu} h^2$ taken from WMAP~\cite{Spergel:2003cb})
\begin{eqnarray}
\Omega_\chi h^2 & \le & \Omega_{\nu} h^2 = 0.0067,
\end{eqnarray}
then it follows
\begin{eqnarray}
m_{\widetilde{\chi}} \le 0.7/h^2 \enspace\mathrm{eV}\enspace.
\end{eqnarray}
This idea is due to Gershtein and Zel'dovich~\cite{Gershtein:1966gg} and 
Cowsik and McClelland~\cite{Cowsik:1972gh} to derive neutrino mass bounds.
\section{The Lee - Weinberg bound}
In this section, I discuss mass bounds for heavy nonrelativistic neutralinos with 
$m_{\widetilde{\chi}} \geq \mathcal{O}(10\GeV)$. I use the same method as proposed 
by various authors independently 
in~\cite{Lee:1977ua,Hut:1977zn,Sato:1977ye,Vysotsky:1977pe} to constrain
heavy neutrinos. This bound is now referred as Lee - Weinberg bound.

This case is not as easy, the thermal averaged cross section and the freeze out temperature
have to be calculated to yield an approximate solution of the Boltzmann equation. 
 
For simplicity, I consider only the neutralino annihilation into leptons
\begin{eqnarray}
\tilde{\chi}^0_1 \tilde{\chi}^0_1 \rightarrow \ell \overline{\ell},
	\quad \ell = e,\mu,\tau,\nu_e,\nu_\mu,\nu_\tau.
\end{eqnarray}
The $\tau$ is considered as massless
\footnote{For a lower mass bound of about $\approx 15 \GeV$, this is a good approximation, 
but not for neutralino masses of the order $\mathcal{O}(1\GeV)$~\cite{Bottino:2003iu}.}, 
all sleptons have common mass $M_{\tilde{\ell}}$ 
(not to be confused with the common scalar mass parameter $M_0$), 
so the cross sections are related by 
\begin{eqnarray}
\sigma(\tilde{\chi}^0_1 \tilde{\chi}^0_1 \rightarrow \ell^-_R \overline{\ell}^+_L)
= 16 \sigma(\tilde{\chi}^0_1 \tilde{\chi}^0_1 \rightarrow \ell^-_L \overline{\ell}^+_R)
= 16 \sigma(\tilde{\chi}^0_1 \tilde{\chi}^0_1 \rightarrow \nu_\ell \overline{\nu}_\ell).
\end{eqnarray}
The thermal averaged cross section Eq.~(\ref{eq:thermalavxsec}) can be calculated using the techniques
described in \cite{Gondolo:1990dk}. This leads to a parametrisation of the form 
$\langle\sigma|v|\rangle \approx \sigma_0 x^{-n}$. In the case of a bino the thermal averaged cross 
section reads
\begin{eqnarray}
\langle\sigma(\tilde{\chi}^0_1 \tilde{\chi}^0_1 \rightarrow \ell 
	\overline{\ell})|v|\rangle \approx \sigma_0 x^{-n}
= 54 \pi\frac{\alpha^2}{\cw[4]}\frac{m_{\widetilde{\chi}}^2}{M^4_{\tilde{\ell}}}\,x^{-1},\quad
\end{eqnarray} 
with $x$ defined in Eq.~(\ref{eq:xT}).
The Boltzmann Equation can be cast into the form
\begin{eqnarray}
\frac{\dif Y}{\dif x} &=& -\left(\frac{x \langle\sigma|v|\rangle s}{H(m)}\right)_{x=1} x^{-n-2} 
                            \left(Y^2(x) - Y_{\mathrm{Eq}}^2(x)\right)\enspace .
\end{eqnarray}
Let the difference $\Delta(x)$ denote the deviation $Y(x) - Y_{\mathrm{Eq}}(x)$ of the particle 
density of the bino from equilibrium density $Y_{\mathrm{Eq}}(x) = 0.145(g/g_{* S})x^{3/2}\eu^{-x}$. 
Shortly after the Big Bang, the deviation and its derivative are small. Therefore, a good 
approximation is setting $|\frac{\mathrm{d}}{\mathrm{d}x}\Delta(x)| \equiv|\Delta '(x)| \approx 0$, 
and one gets:
\begin{eqnarray}
\label{eq:early}
\Delta(x) \approx& 
	-\frac{x^{n+2}Y_{\mathrm{Eq}}^\prime(x)}{\left(\frac{x \langle\sigma|v|\rangle s}{H(m)}\right)_{x=1}
	(2 Y_{\mathrm{Eq}}(x) + \Delta)}, &\enspace 1 \le x \ll x_f.
\end{eqnarray}
Later after decoupling, the neutralinos are no longer in thermal equilibrium, and the terms involving  
$Y_{\mathrm{Eq}}(x)$ can be neglected. So one gets the following differential equation:
\begin{eqnarray}
\label{eq:late}
\Delta'(x) \approx& 
	- \left(\frac{x \langle\sigma|v|\rangle s}{H(m)}\right)_{x=1} x^{-n-2}\Delta^2\hspace*{\fill},
	 &\enspace x_f \ll x\enspace.
\end{eqnarray}
To solve Eq.~(\ref{eq:late}), we have to integrate from $x = x_f$ to $x= \infty$. Recall, that we 
transformed the time dependence of the Boltzmann equation into an $x$-dependence by the 
transformations Eq.~(\ref{eq:xT}-\ref{eq:time}). The solutions for Eqs~(\ref{eq:early}) 
and~(\ref{eq:late}) are
\begin{eqnarray}
\label{eq:Delta}
\Delta &\approx & \left\{ \begin{array}{ll}
\frac{1}{2 \left(\frac{x \langle\sigma|v|\rangle s}{H(m)}\right)_{x=1}}x^{n+2} ,&
	  \enspace 1 \le x \ll x_f,\\[6mm]
\frac{n+1}{\left(\frac{x \langle\sigma|v|\rangle s}{H(m)}\right)_{x=1}}x_f^{n+1},&
	  \enspace x_f \ll x\enspace.
	  \end{array}\right.
\end{eqnarray}  
Eq.~(\ref{eq:Delta}) requires the knowledge of the freeze out temperature $T_f$ or, equivalently,
$x_f = m/T_f$. The decoupling temperature is the temperature, when the deviation $\Delta$ 
has grown to order $Y_{\mathrm{Eq}}(x)$. One sets $\Delta(x_f) = c Y_{\mathrm{Eq}}(x)$, $c = \mathcal{O}(1)$, 
and solves Eq.~(\ref{eq:early}) for $x_f$, yielding
\begin{eqnarray}
x_f &\approx & \ln[0.145 (g/g_{*}^{1/2})(n+1)\big({x \langle\sigma|v|\rangle s}/{H(m)}\big)_{x=1}] -
		\notag\\[2mm]
	       &&\left(n + \half\right) \ln\left[\ln\left(
		    0.145 (g/g_{*}^{1/2})(n+1)\big({x \langle\sigma|v|\rangle s}/{H(m)}\big)_{x=1}
	       \right)\right],\hspace*{10mm}\\[2mm]
Y(x=\infty) &=&\Delta(x=\infty) \approx 
	\frac{3.79(n+1)x_f^{n+1}}{(g_{*S}/g_{*}^{1/2})m_{\mathrm{Pl}}m_{\widetilde{\chi}}\sigma_0},\\[2mm]
n_\chi &=& s_0 \Delta(x=\infty) \approx 
     \frac{1.13\times10^{4}(n+1)x_f^{n+1}}{(g_{*S}/g_{*}^{1/2})m_{\mathrm{Pl}}m_{\widetilde{\chi}}\sigma_0}
	\enspace\mathrm{cm}^{-3},\\[2mm] 
\label{eq:reliclate}
\Omega_\chi h^2 &=& m_{\widetilde{\chi}} n_\chi \approx 
     \frac{1.07\times10^{9}(n+1)x_f^{n+1}}{(g_{*S}/g_{*}^{1/2})m_{\mathrm{Pl}}\sigma_0}
	\enspace\GeV^{-1}\enspace.
\end{eqnarray}
The choice $c(c+2) = n+1$~\cite{Kolb:1990eu} has been implemented and yields the best fit 
to the relic density.
There is a remarkable feature of Eq.~(\ref{eq:reliclate}): 
The lower the cross section the larger the relic density. 
This can be understood: The particle density distribution is a Boltzmann distribution. 
If the cross section is large the particles stay longer in thermal
equilibrium, and the particle density decreases stronger with falling temperature.   

\begin{figure}[ht!]
\setlength{\unitlength}{1cm}
\begin{center}
\subfigure[Contour lines of equal relic density of a bino type lsp in the 
	$m_{\widetilde{\chi}}$ - $M_{\widetilde{\ell}}$-plane\label{fig:relic0}]{\scalebox{0.53}{
	\psfrag{mx}{{}}
	\psfrag{Me}{{}}
	\includegraphics{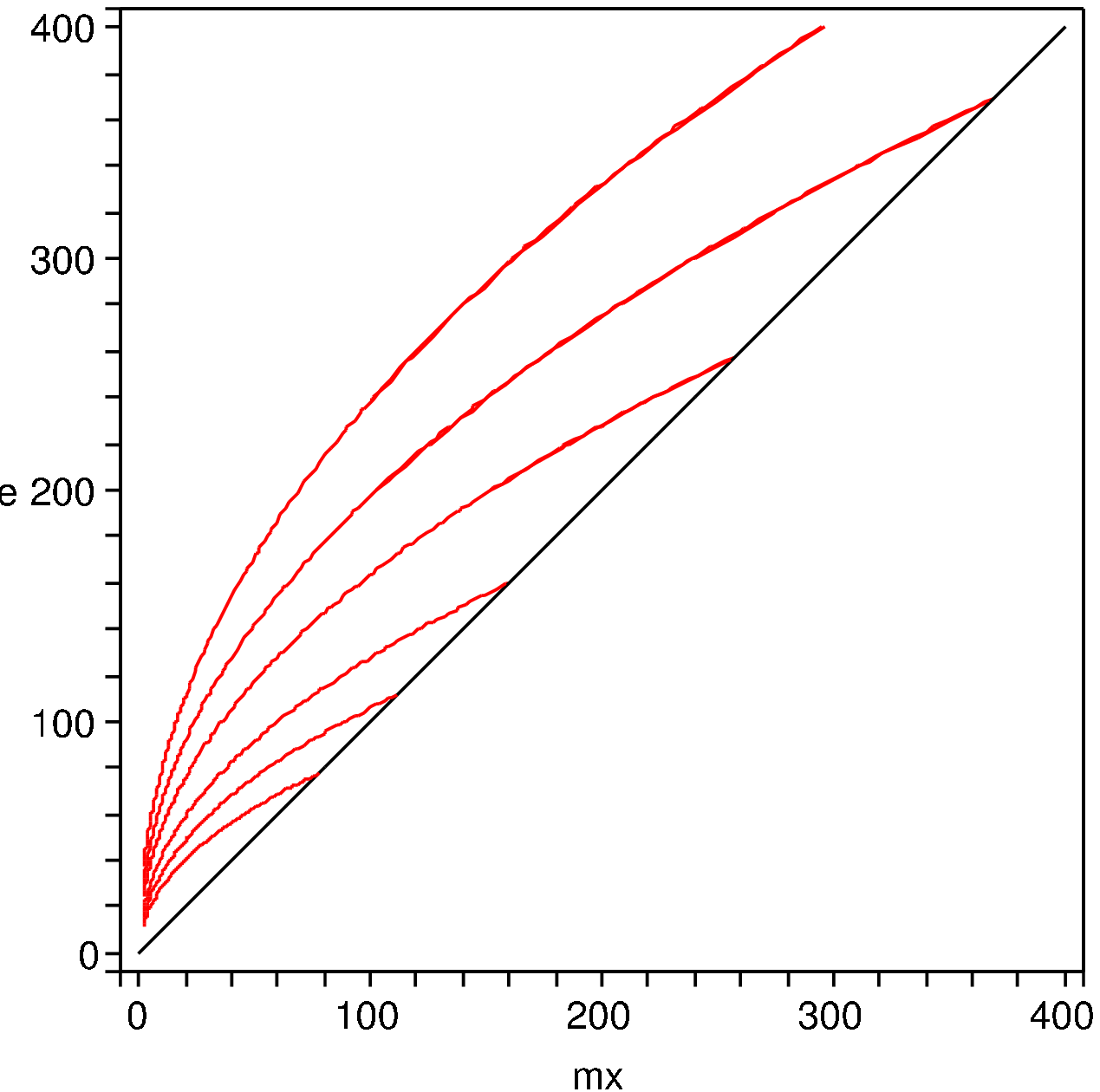}}
	\put(-4,0){$m_{\widetilde{\chi}}\, [\!\GeV]$}
	\put(-7.6,3){\rotatebox{90}{$M_{\tilde{\ell}}\, [\!\GeV]$}}
	\put(-5.5,1.8){\tiny 0.005}
	\put(-5.1,2.2){\tiny 0.01}
	\put(-4.2,3.0){\tiny 0.02}
	\put(-2.9,4.3){\tiny 0.05}
	\put(-1.3,6.0){\tiny 0.1}
	\put(-2.8,6.1){\tiny 0.2}
	\put(-3.0,2.9){$m_{\chi} > m_{\widetilde{\ell}}$}	
}
\hspace*{2mm}
\subfigure[$\Omega_{\mathrm{DM}}h^2\pm 3\sigma$ area of the relic density\label{fig:relic1}]{
	\psfrag{mx}{{}}
	\psfrag{Me}{{}}
	\scalebox{0.53}{\includegraphics{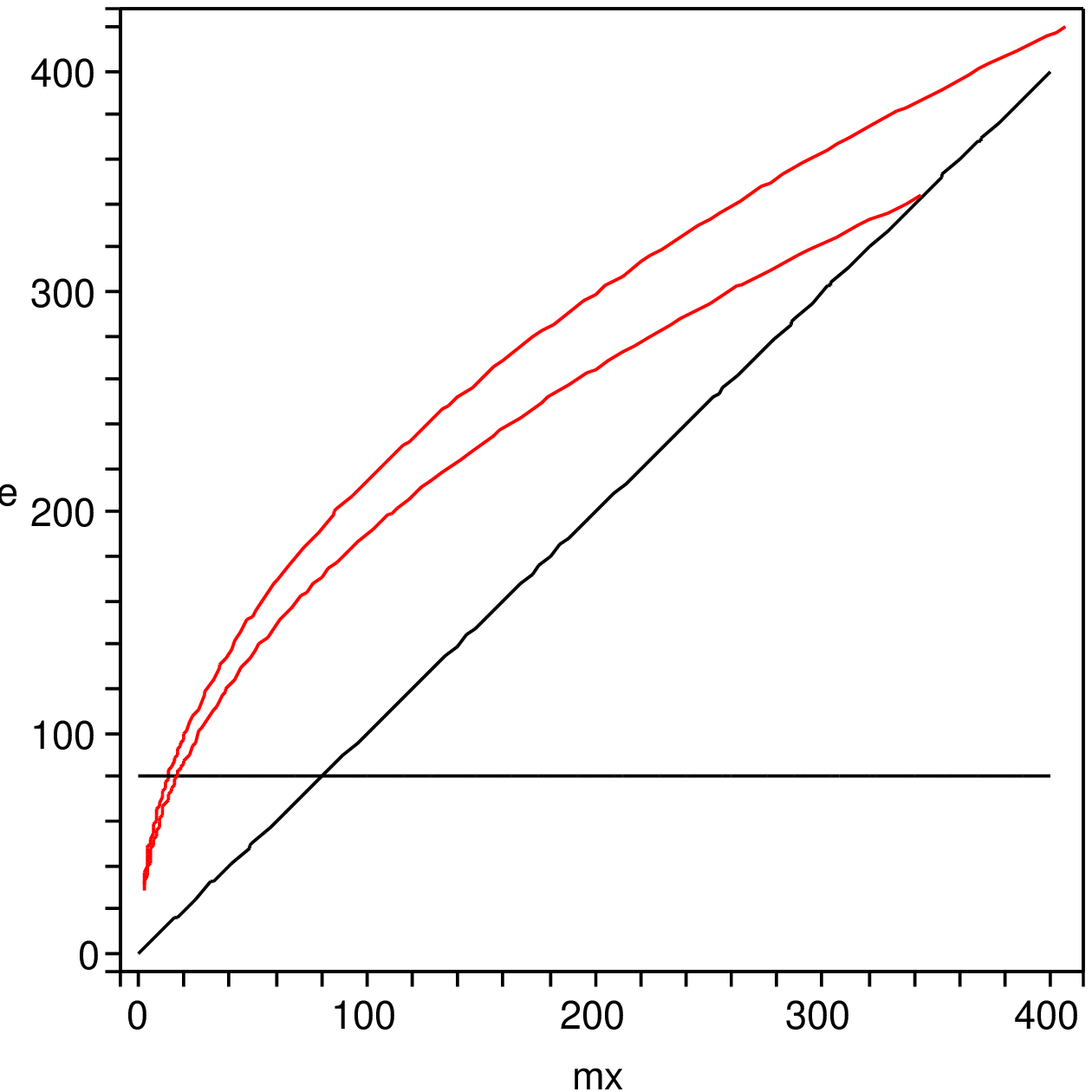}}
	\put(-4,0){$m_{\widetilde{\chi}}\, [\!\GeV]$}
	\put(-7.7,3){\rotatebox{90}{$M_{\tilde{\ell}}\, [\!\GeV]$}}
	\put(-3.0,3.5){$m_{\chi} > m_{\widetilde{\ell}}$}	
	\put(-3.0,1.6){$\scriptstyle{m_{\widetilde{\ell}} = 80\GeV}$}	
	\put(-4.0,5.2){\tiny 0.136}
	\put(-4.0,4.1){\tiny 0.087}	
}
\label{fig:relic}
\caption{relic density of a bino type lsp}
\end{center}
\end{figure}

In Fig.~\ref{fig:relic0}, I show contours of constant relic density in the 
$M$ - $m_{\widetilde{\chi}}$-plane. 
The lower right hand triangle of the figure is excluded since the sleptons are lighter than the neutralino. 
In Fig.~\ref{fig:relic1}, I show the contours limiting the 
$\Omega_{\mathrm{DM}}h^2 \pm 3\sigma_\Omega = 0.113 \pm 3\times 0.008$ area (\cite{Bennett:2003bz}),
$\sigma_\Omega$ denotes the absolute error on $\Omega_{\mathrm{DM}}h^2$. 
The horizontal line indicates the approximate  lower bound on the slepton masses of about $80\GeV$.  
If we demand that the neutralinos constitute the whole dark matter and that the sleptons are heavier 
than $80\GeV$, we find a lower mass bound of the neutralinos of about $13\GeV$. The masses of the slepton 
cannot exceed $\approx 400\GeV$. 
If the next to lightest supersymmetric particle is heavier than $400\GeV$, the neutralino mass bounds
are
\begin{eqnarray}
\label{eq:massbound}
 13\GeV \le m_{\widetilde{\chi}} \le 400 \GeV\enspace.
\end{eqnarray}

The result shows the advantage of estimating the neutralino mass from the dark matter density 
form an approximation rather than doing the full calculation:
There are only two (or three) parameters ($m_{\widetilde{\chi}}$, $M_{\widetilde{\ell}}$, 
or $M_{\tilde{q}}$), which can be plotted in a two dimensional figure.

\medskip
I summarize the assumptions which enter the above mass bounds~(\ref{eq:massbound}):
\begin{itemize}
\item The neutralino is a nonrelativistic bino.
\item The annihilation products are charged leptons, which are considered as massless.
\item Coannihilation and resonant annihilation is unimportant.
\item $R$-parity ($P_6$ - hexality~\cite{Dreiner:2005rd}) is conserved. 
\end{itemize}  

\section{Numerical solution of the full Boltzmann equation}
In the previous section, I derived from an approximate solution of the Boltzmann equation
an upper and lower bound on the neutralino mass and -- with caution -- for the slepton mass. 
Now I compare these results with the exact solution. For this purpose I use the program micrOMEGAs
~\cite{Belanger:2004yn}.

The estimation does not take into account coannihilation and resonant annihilation. Near the threshold
where the neutralino is almost mass degenerate with the sleptons there is coannihilation.
And even for a small Higgsino component, there is large resonant annihilation 
if the neutralino mass is half of the $Z^0$-mass or half of the $h^0$-mass. 

\begin{figure}
\setlength{\unitlength}{1cm}
\subfigure[$\Omega_{\mathrm{DM}}h^2\pm 3\sigma$ area of the relic density. Numerical solution
	of the full Boltzmann equation for a neutralino lsp with input data: 
	$M_2 = 193\GeV$, $\mu = 350\GeV$, $M_3$ = $800\GeV$, $\tan\beta = 10$, 
	$M_{H_3} = 450 \GeV$, $A_\tau = \mu \tan\beta$, $M_{\widetilde{q}}= 1000\GeV$.
	\label{fig:relic2}]{
	\psfrag{x}{}
	\psfrag{y}{}
	\scalebox{0.55}{\includegraphics{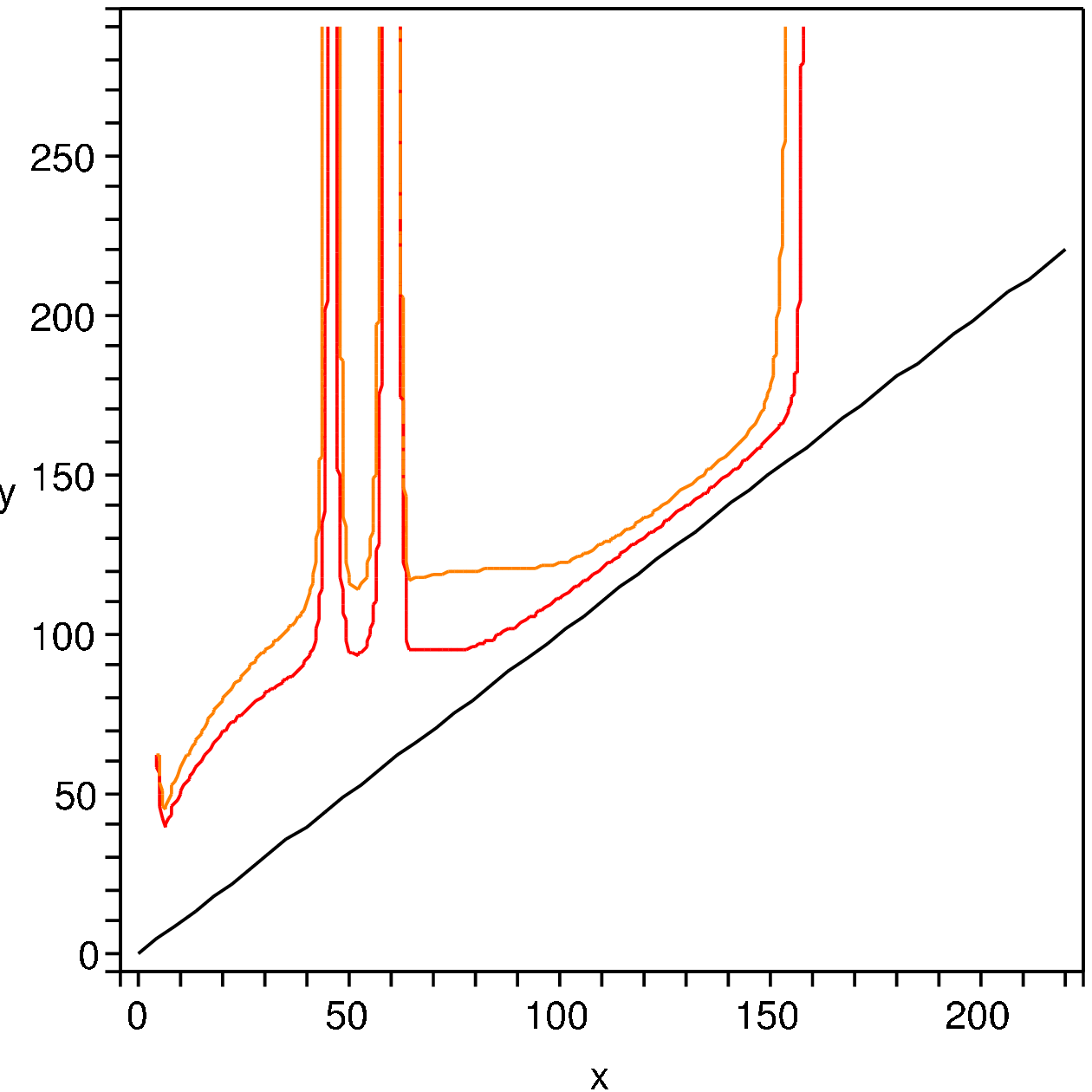}}
	\put(-4,0){$m_{\widetilde{\chi}}\, [\!\GeV]$}
	\put(-8.2,3){\rotatebox{90}{$M_{\tilde{e}}\, [\!\GeV]$}}
	\put(-3.0,2.9){$m_{\chi} > m_{\widetilde{e}}$}
	\put(-5.0,2.65){\tiny 0.087}
	\put(-4.9,3.45){\tiny 0.136}
	}
\hspace{5mm}
\subfigure[Fig.~\ref{fig:relic2} overlayed with Fig.~\ref{fig:relic1} to compare approximate and exact solution.
	\label{fig:relic3}]{
	\psfrag{mx}{}
	\psfrag{Me}{}	
	\scalebox{0.55}{\includegraphics{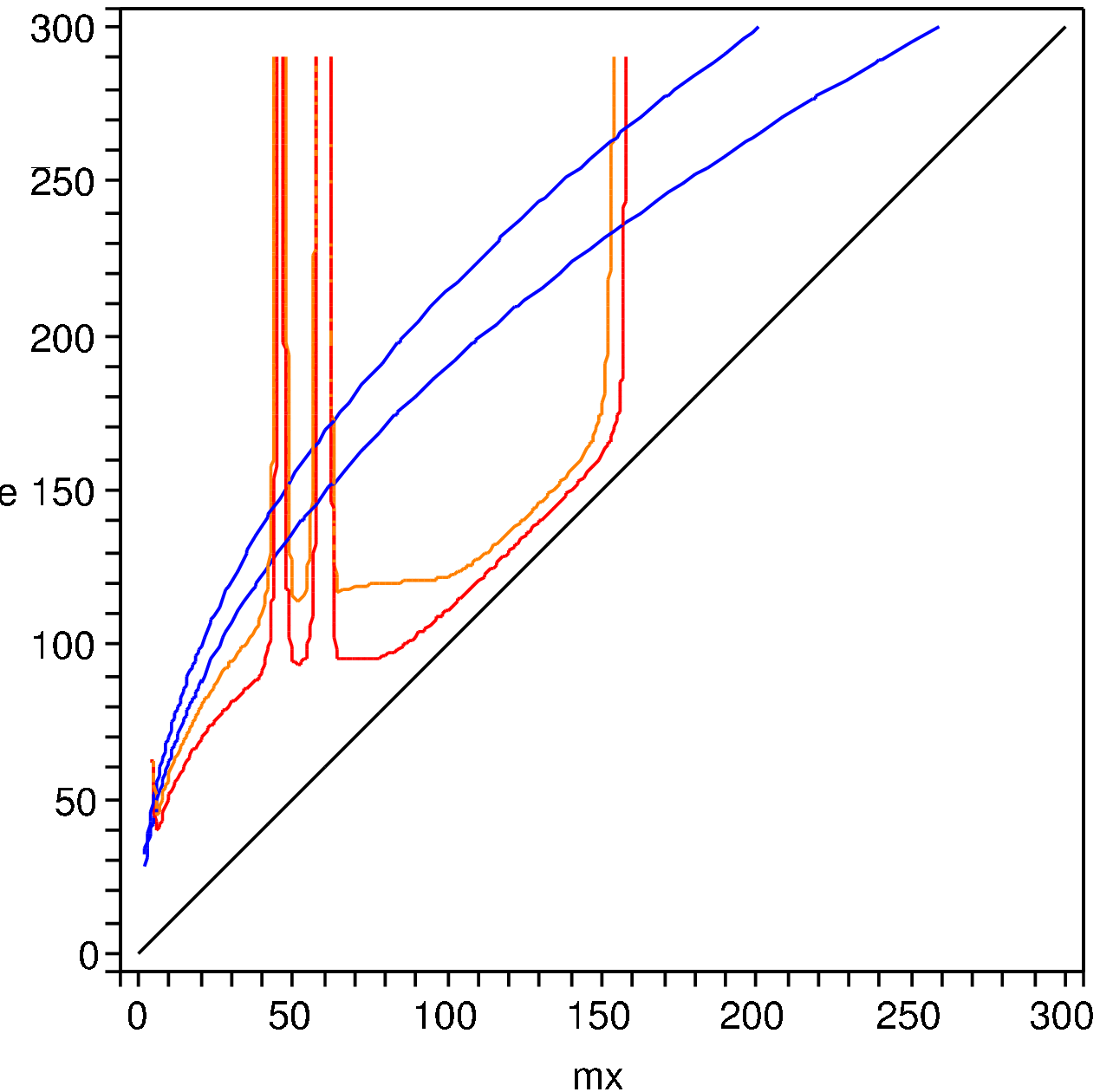}}
	\put(-4,0){$m_{\widetilde{\chi}}\, [\!\GeV]$}
	\put(-8.2,3){\rotatebox{90}{$M_{\tilde{e}}\, [\!\GeV]$}}
	\put(-3.0,2.9){$m_{\chi} > m_{\widetilde{e}}$}
	}
	
\vspace*{5mm}
\begin{center}
\subfigure[Neutralino density as a function of its mass for
	$M_2= 193\GeV$, $\mu = 350\GeV$, $\tan\beta = 10$, 
	common slepton mass $M_{\tilde{\ell}} = 150\GeV$,
	common squark mass $M_{\tilde{q}} = 1000\GeV$, $M_3 = 800\GeV$, $M_{H_3} = 450\GeV$.
	\label{fig:relicdensity}]{\scalebox{0.49}{\includegraphics{relicdensity.eps}}}
\end{center}		
\caption{Comparison of approximate and exact calculation of the relic density for a neutralino lsp.}
\label{fig:relicexact}
\end{figure}
In Fig.~\ref{fig:relicexact}, I show contour lines of the relic density for the following scenario:
$M_2 = 200\GeV$, $\mu = 300\GeV$, $M_3$ = $800\GeV$,
$\tan\beta = 10$, $M_{H_3} = 450 \GeV$, $A_\tau = \mu \tan\beta$, $M_{\widetilde{q}}= 1000\GeV$.
The masses of the sleptons and of the lightest neutralino are varied.
The masses of the particles other than sleptons are kept constant, so one can directly
see the influence of the particle masses on the relic density.
Electron and muon are considered as massless, and the choice of $A_\tau$ leads to equal
slepton masses. 
The corner at the bottom right is excluded since the neutralino is heavier than the sleptons.
Contrary to the Lee-Weinberg-approximation, the exact solution
includes coannihilation near the line $m_{\chi_1^0} = m_{\tilde{\ell}}$.
Fig.~\ref{fig:relic2} shows also the influence of a (small) Higgsino component, 
leading to resonant annihilation due to $Z^0$ and $h$ bosons. 
The resonance increases the cross section dramatically. This allows for larger slepton masses.
The two resonances appear in Fig.~\ref{fig:relic2} as two valleys in the 
$m_{\widetilde{\chi}}$-$M_{\widetilde{\ell}}$ plane at 
$m_{\widetilde{\chi}} = m_Z/2$ and $m_{\widetilde{\chi}} = m_h/2$. 
From this I conclude that in realistic models no bound can be set on the 
slepton mass by relic density calculations. As lower bound on the neutralino mass I get
\begin{eqnarray}
10 -15 \GeV \leq m_{\widetilde{\chi}}\,.
\end{eqnarray}
This agrees with the lower bound from the approximation. 
The upper bound is given by the mass of the next to lightest supersymmetric particle (nlsp).  
For comparison Fig.~\ref{fig:relic3} shows the approximate and the exact solution overlayed
in one plot. Apart from the valleys both plots agree quite well. 

If non-relativistic neutralinos constitute the whole dark matter, they cannot be completely
annihilated due to resonant annihilation. This means that the mass of the neutralino
is sufficiently far away from the relations $m_{\widetilde{\chi}} = m_Z/2$ or 
$m_{\widetilde{\chi}} = m_h/2$.  

Fig. \ref{fig:relicdensity} shows for one parameter set ($M_2= 193\GeV$, $\mu = 350\GeV$, 
$\tan\beta = 10$, $M_{\tilde{\ell}} = 150\GeV$ as common slepton mass, 
$M_{\tilde{q}} = 1000\GeV$ as common squark mass, $M_3 = 800\GeV$, and $M_{H_3} = 450\GeV$)
the relic density $\Omega h^2$ as a function of the neutralino mass.
$M_1$ has been increased from $M_1 = 1.3\GeV$ to $130\GeV$ to vary the neutralino mass.
The two spikes at the end of the curve stem from resonant annihilation due to the 
$Z^0$ and the $h$ resonance.  The nlsp has a mass of about $135\GeV$.
The qualitative shape of the curve is similar to the curve
published in~\cite{Kolb:1990eu}.

The horizontal red dashed lines are lines with 
$\Omega h^2 = \Omega_{\mathrm{DM}} h^2 \pm 3\sigma_\Omega$
with $\Omega_{\mathrm{DM}} h^2 = 0.113$, $\sigma_\Omega = 0.008$. The black curve crosses 
the allowed ribbon twice: at very light neutralinos with mass $\mathcal{O}(10^{-9} \GeV)$ 
and at massive neutralinos with mass $\mathcal{O}(10\GeV)$. In the first case the 
particles which constitute the dark matter cannot only be neutralinos because too many 
relativistic particles disturb structure formation in the early Universe.
To avoid this constraint the neutralinos are only allowed to contribute as much as 
the neutrinos. This lowers the neutralino mass bound a little bit.
The bound for relativistic neutralinos agrees very well with the predictions of the 
Cowsik-McClelland-bound.

The exact value of the lower mass bound in the nonrelativistic case depends on the parameters 
of the model (slepton and squark masses, mass difference to the nlsp, resonant annihilation). 
The upper bound is rather trivial, it is the next to lightest supersymmetric particle. 
Such searches need a lot of CPU time and have recently been done by 
Hooper and Plehn~\cite{Hooper:2002nq}, Bottino and al.~\cite{Bottino:2003iu} and
Belanger et al.~\cite{Belanger:2003wb}.
Hooper and Plehn found a lower bound of about $18\GeV$ for a nonrelativistic neutralino,
Bottino et al. found a lower bound of about $6\GeV$, and Belanger et al. found 
a lower neutralino mass bound of about $6\GeV$ in models with a light pseudoscalar Higgs $A$
with mass $M_A < 200\GeV$.

\chapter{$\x{1}$-$\x{2}$-production at LEP}
\label{ch:chi12}
In this chapter, I derive mass bounds on the selectron mass from upper limits
on the cross section 
$\sigma(e^+e^-\rightarrow \x{1}\x{2})$ measured by the OPAL 
collaboration at LEP~\cite{Abbiendi:2003sc},
if the lightest neutralino \x{1} is assumed as massless.
These bounds on the cross section
translate into bounds on the selectron mass. 
I assume equal right and left handed selectron masses.  

The Delphi~\cite{Abdallah:2003xe} and the Opal collaboration~\cite{Abbiendi:2003sc} 
have searched for SUSY particles.
For neutralino pair production
\begin{eqnarray} 
e^+e^-\rightarrow \x{1}\x{2}
\end{eqnarray}
they present   
upper bounds on the cross sections in the $m_{\x{1}}$-$m_{\x{2}}$ plane.
Their analysis assumes that the hadronic channels 
$\x{2} \rightarrow Z^\ast \x{1},\enspace Z^\ast\rightarrow q \overline{q}$ have a 
$\text{BR}(Z^\ast\rightarrow q \overline{q}) = 100\%$.
The selectron mass is assumed to be $500\GeV$. So the two body decays into 
selectrons is not possible.
The production of $\x{1}\x{2}$ in $e^+e^-$ collision occurs either by $s$ channel exchange
of a $Z$ boson or via $t$ and $u$ channel selectron exchange. 
For massless neutralinos, the \x{1} is nearly pure bino ($N_{11}\ge 0.98$), 
so it couples preferably to $\tilde{e}_{R}$, 
the $\x{2}$ is mostly wino and couples stronger to $\tilde{e}_L$. 
Due to the large selectron mass the $t$ and $u$ channel contributions 
$\sigma_{\tilde{e}}$ to the cross section are
suppressed, so the dominant contribution $\sigma_Z$ comes from the $s$ channel. 
The interference between $Z$ and selectron exchange $\sigma_{Z\tilde{e}}$ is positive. 
If I denote the total cross section as 
$\sigma_Z = \sigma_Z + \sigma_{\tilde{e}} + \sigma_{Z\tilde{e}}$
then the $\tilde{e}_{R/L}$ contribution becomes larger,
if the selectron is lighter, $200\GeV \leq m_{\tilde{e}} \leq 500\GeV$. 
This ensures that the experimental limits on the cross section
are also applicable for selectrons with mass $< 500\GeV$.
%
Therefore, the reported bounds on the cross section are absolute upper bounds.

\begin{figure}[ht!]
\setlength{\unitlength}{1cm}
\begin{center}
\subfigure[Contour lines for the cross section $\sigma(e^+e^-\rightarrow \x{1}\x{2})$ in fb
           for $m_{\tilde{e}}=200\GeV$ and $\sqrt{s} = 208\GeV$ 
           in the $\mu$ - $M_2$ plane\label{fig:cont1}]
	  {\scalebox{0.5}{\includegraphics{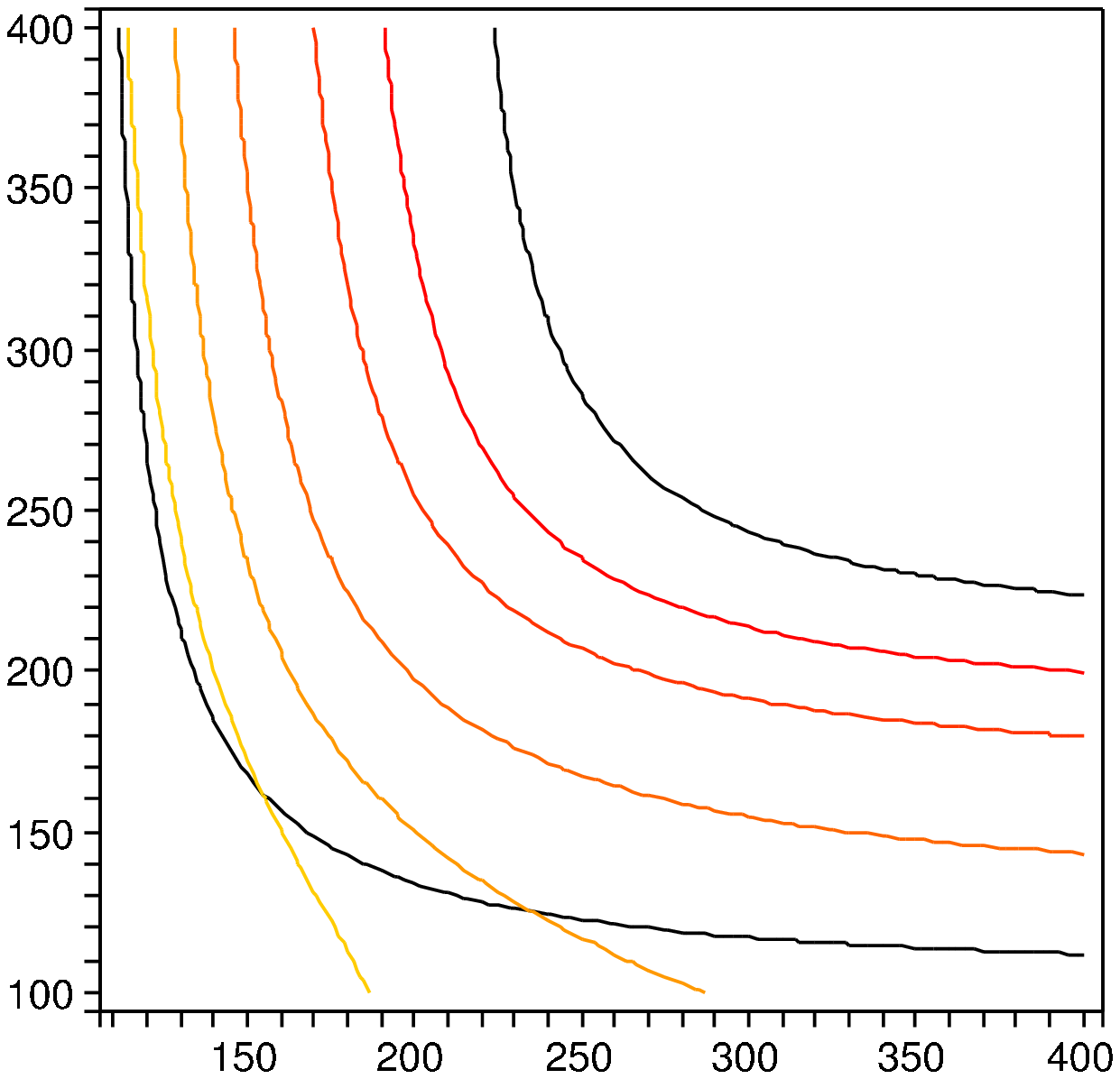}}
\put(-4,0){$\mu\, [\!\GeV]$}
\put(-7.5,3){\rotatebox{90}{$M_2\, [\!\GeV]$}}
\put(-5.4,1){\tiny 200}
\put(-4.2,1){\tiny 150}
\put(-3.2,2){\tiny 100}
\put(-2.7,2.5){\tiny 50}
\put(-2.2,2.8){\tiny 20}
\put(-2.2,3.3){\tiny 0}
\put(-2.7,5){$m_{\x{2}} > \sqrt{s}$}
\put(-2.7,0.9){\tiny $m_{\x{2}} = 104\GeV$}
}
\hspace*{5mm}
\subfigure[Contour lines for the minimal allowed $m_{\tilde{e}}$-mass in the 
           $\mu$ - $M_2$ plane. Below the blue line there is $m_{\widetilde{e}} > m_{\x{2}}$, and above
           $m_{\widetilde{e}} < m_{\x{2}}$.\label{fig:cont2}]
	  {\scalebox{0.5}{\includegraphics{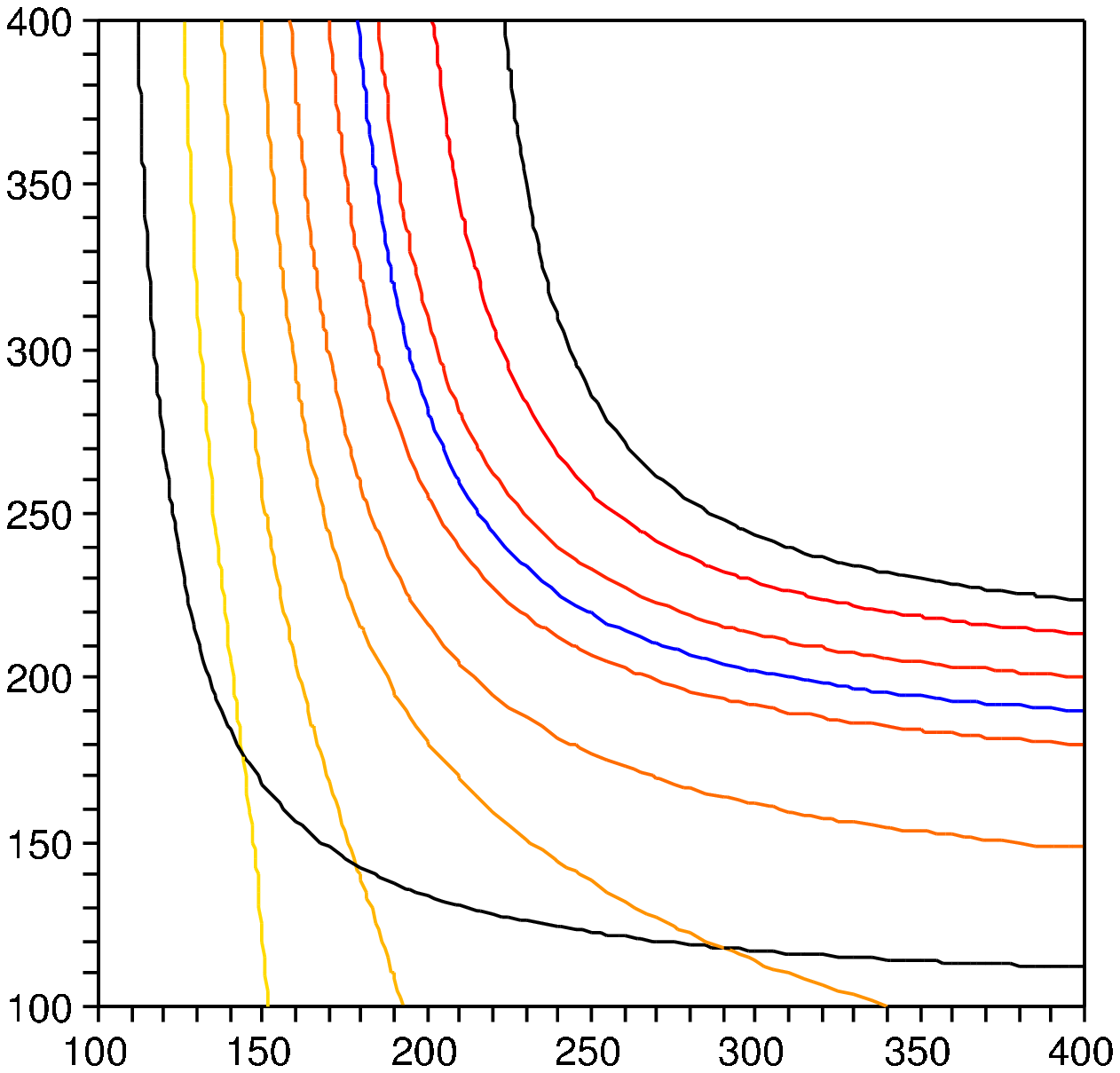}}
\put(-4,0){$\mu\, [\!\GeV]$}
\put(-7.5,3){\rotatebox{90}{$M_2\, [\!\GeV]$}}
\put(-5.8,1){\tiny 600}
\put(-5.1,1){\tiny 500}
\put(-4.4,1.6){\tiny 400}
\put(-3.8,2){\tiny 300}
\put(-3.4,2.4){\tiny 200}
\put(-3,2.7){\tiny 150}
\put(-2.7,3){\tiny 100}
\put(-2.7,5){$m_{\x{2}} > \sqrt{s}$}
\put(-2.7,0.9){\tiny $m_{\x{2}} = 104\GeV$}
}

\subfigure[Contour lines for the ${\x{2}}$ mass in GeV in the $\mu$ - $M_2$ plane  
     \label{fig:cont3}]{\scalebox{0.5}{\includegraphics{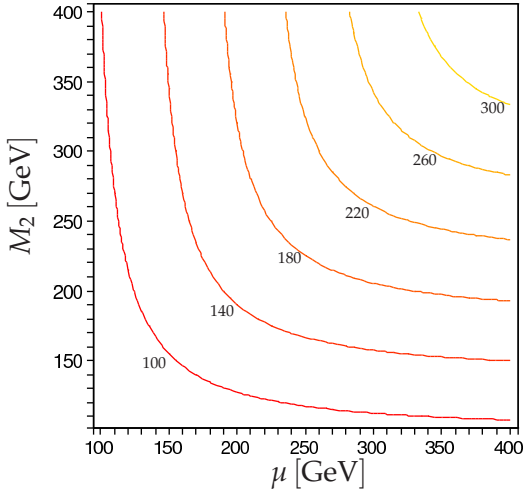}}
\put(-4,0){$\mu\, [\!\GeV]$}
\put(-7.5,3){\rotatebox{90}{$M_2\, [\!\GeV]$}}
\put(-5.7,1.5){\tiny 100}
\put(-4.8,2.2){\tiny 140}
\put(-3.9,2.9){\tiny 180}
\put(-3.0,3.5){\tiny 220}
\put(-2.1,4.2){\tiny 260}
\put(-1.2,4.9){\tiny 300}
}
\subfigure[Observed $95\%$ confidence level upper bound on the cross section 
	$\sigma(e^+e^-\rightarrow \x{1}\x{2})$, based on decays to hadronic final states
	and assuming $100\%$ branching ratio for decay into $Z^{0\ast}$.
	This figure is taken from~\cite{Abbiendi:2003sc}.	 
	\label{fig:cont4}]{\scalebox{0.35}{\includegraphics{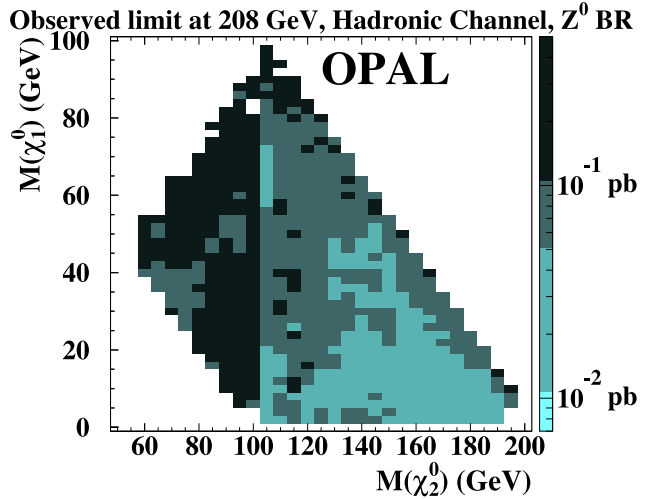}}
}
\label{fig:cont}
\caption{Deriving a lower mass bound on the selectron mass if \x{1} is massless.
	}
\end{center}
\end{figure}

\medskip
In Fig.~\ref{fig:cont1}, I show contour lines for the cross section 
$\sigma(e^+e^-\rightarrow \x{1}\x{2})$
with $m_{\tilde{e}}=200\GeV$ and $\tan\beta = 10$ in the $\mu$-$M_2$ 
plane for $M_1$ chosen such that $\x{1}$ is massless. 
The cross section reaches values up to $200\fb$.
There is a large parameter region where the cross section exceeds 
$50\fb$. From Fig.~\ref{fig:cont4}, taken from~\cite{Abbiendi:2003sc}, 
one reads off that for a massless \x{1} the maximally allowed cross section 
is about $50\fb$ at $\sqrt{s}=208\GeV$ 
(At $m_{\x{2}} = 115\GeV, 125\GeV, 135-145\GeV$, there are dark grey spots,
indicating that the allowed cross section is $100\fb$.
They are most likely due to fluctuations in the data,
I ignore them for simplicity). 
Within the $m_{\chi^+} = 104\GeV$ contour line and the $50\fb$ contour line 
the cross section is larger than $50\fb$ and so this part of the parameter space
is ruled out (note that \x{2} and $\widetilde{\chi}_1^+$ are nearly mass degenerate).

In  Fig.~\ref{fig:cont2}, I show contour lines
of the minimal selectron mass so that the limits from  Fig.~\ref{fig:cont4}, 
$\sigma(e^+e^-\rightarrow \x{1}\x{2}) < 50\fb$, are fulfilled.
The upper black line indicates the kinematical limit.
Below the lower black line, the $\chi_1^+$ is lighter than $104\GeV$,
which is experimentally excluded~\cite{Yao:2006px}.
Along the blue contour the \x{2} and the selectrons have equal masses at about $175\GeV$. 
Above the blue line the selectrons are lighter than \x{2} and the two body decay
$\x{2}\rightarrow \widetilde{e}_{R/L} e$ is allowed. 
For $m_{\x{2}} > 175\GeV$ no part of the parameter space can be excluded. 
In Fig.~\ref{fig:cont3}, I show
contour lines for the mass of \x{2} in the $\mu$-$M_2$ plane.  

For $\mu,\enspace M_2 < 200\GeV$ the OPAL bound is only fulfilled if the selectrons are heavier than
$\approx 350\GeV$. For $\mu = 352\GeV$, $M_2 = 193\GeV$ as in the  SPS1a scenario,
the right handed selectron must be heavier than $180\GeV$. 
 
\medskip
{\bf Conclusion:}    
The experimental limits on the cross section $\sigma(e^+e^-\rightarrow \x{1}\x{2})$ by OPAL set severe
bounds on the selectron mass if $m_{\x{1}} = 0\GeV$ and $m_{\x{2}} < 175\GeV$. 

\chapter{Radiative Neutralino Production}
\label{ch:radiative}
\section{Introduction}
\label{sec:intro}
Supersymmetry (SUSY) is an attractive concept for theories beyond the
Standard Model (SM) of particle physics. SUSY models like the Minimal
Supersymmetric Standard Model 
(MSSM)~\cite{Haber:1984rc,Gunion:1984yn,Nilles:1983ge}
predict SUSY partners of the SM particles with masses of the order of
a few hundred GeV.  Their discovery is one of the main goals of
present and future colliders in the TeV range.  In particular, the
international $e^+e^-$ linear collider (ILC) will be an {excellent}
tool to determine the parameters of the SUSY model with high
precision~\cite{Aguilar-Saavedra:2001rg,Abe:2001nn,Abe:2001gc,
Weiglein:2004hn,Aguilar-Saavedra:2005pw}.  Such a machine provides
high luminosity ${\mathcal L}=500\fb^{-1}$, a center-of-mass 
energy of $\sqrt s = 500\GeV$ in the first stage, and a polarised
electron beam with the option of a polarised positron
beam~\cite{Moortgat-Pick:2005cw}.

The neutralinos are the fermionic SUSY partners of the neutral gauge
and CP-even Higgs bosons.  Since they are among the lightest particles
in many SUSY models, they are expected to be among the first states to
be observed.  At the ILC, they can be directly produced in pairs
\begin{equation}
e^+ + e^-\to\tilde\chi_i^0 + \tilde\chi_j^0\,, 
\label{neut-pairs}
\end{equation}
which proceeds via $Z$ boson and selectron exchange~\cite{Bartl:1986hp,
  neutralino-pair}. At tree level, the neutralino sector depends only
on the four parameters $M_1$, $M_2$, $\mu$, and $\tan\beta$, which are the
$U(1)_Y$ and $SU(2)_L$ gaugino masses, the higgsino mass parameter,
and the ratio of the vacuum expectation values of the two Higgs
fields, respectively. These parameters can be determined by measuring
the neutralino production cross sections and decay
distributions~\cite{Weiglein:2004hn,Choi:2001ww,Choi:2005gt,
  Barger:1999tn,Kneur:1999nx}.  In the MSSM with R-parity (or proton
hexality, $P_6$, \cite{Dreiner:2005rd}) conservation, the lightest
neutralino $\tilde\chi_1^0$ is typically the lightest SUSY particle
(LSP) and as such is stable and a good dark matter 
candidate~\cite{Goldberg:1983nd, Ellis:1983ew}.
In collider experiments the LSP
escapes detection such that the direct production of the lightest
neutralino pair
\begin{eqnarray} 
e^+e^-\to\tilde\chi_1^0\tilde\chi_1^0 
\end{eqnarray}
is invisible.
Their pair production can only be observed indirectly via radiative production 
\begin{eqnarray}
e^+e^-\to\tilde\chi_1^0\tilde\chi_1^0\gamma, 
\end{eqnarray}
where the photon is radiated off the incoming beams or off the exchanged
selectrons.  Although this higher order process is suppressed by the
square of the additional photon-electron coupling, it might be the
lightest state of SUSY particles to be observed at colliders.  The
signal is a single high energetic photon and missing energy, carried
by the neutralinos.

As a unique process to search for, the first SUSY signatures at
$e^+e^-$ colliders, the radiative production of neutralinos has been
intensively studied in the literature~\cite{Fayet:1982ky,Ellis:1982zz,
  Grassie:1983kq,Kobayashi:1984wu,Ware:1984kq,Bento:1985in,
  Chen:1987ux,Kon:1987gi,Bayer,Choi:1999bs,Baer:2001ia,Weidner,
  Fraas:1991ky,Datta:1994ac,Datta:1996ur,Datta:2002jh,
  Ambrosanio:1995it}.\footnote{In addition I found two 
  references~\cite{Ahmadov:2006xr, Ahmadov:2005ci}, which are however 
  almost identical in wording and layout to Ref.~\cite{Fraas:1991ky}.}
Early investigations focus on LEP energies and discuss special
neutralino mixing scenarios only, in particular the pure photino
case~\cite{Fayet:1982ky,Ellis:1982zz,Grassie:1983kq,Kobayashi:1984wu,
  Ware:1984kq,Bento:1985in,Chen:1987ux,Kon:1987gi}.  More recent
studies assume general neutralino mixing~\cite{Bayer,Choi:1999bs,
  Baer:2001ia,Weidner,Fraas:1991ky,Datta:1994ac,Datta:1996ur,
  Datta:2002jh,Ambrosanio:1995it} and some of them underline the
importance of longitudinal~\cite{Bayer,Choi:1999bs,Baer:2001ia,
  Weidner} and even transverse beam polarisations~\cite{Bayer,
  Weidner}.  The transition amplitudes are given in a generic
factorised form~\cite{Choi:1999bs}, which allows the inclusion of
anomalous $WW\gamma$ couplings.  Cross sections are calculated with
the program {\tt CompHEP}~\cite{Baer:2001ia}, or in the helicity
formalism~\cite{Weidner}.
Some of the studies~\cite{Fraas:1991ky,Datta:1994ac,Datta:1996ur,
  Datta:2002jh,Ambrosanio:1995it} however do not include longitudinal
beam polarisations, which might be {essential} for measuring radiative
neutralino production at the ILC.  Special scenarios are considered,
where besides the sneutrinos also the heavier neutralinos~\cite{
  Datta:1994ac,Datta:1996ur,Datta:2002jh}, and even charginos~\cite{
  Chen:1995yu,Kane:1997yr,Datta:1998yw} decay invisibly or almost
invisibly.  However, a part of such unconventional signatures are by
now ruled out by LEP2 data~\cite{Datta:1994ac,Kane:1997yr,
  Abbiendi:2002vz}.  For the ILC, such ``effective'' LSP scenarios
have been analysed~\cite{Datta:1996ur}, and strategies for detecting
invisible decays of neutralinos and charginos have been
proposed~\cite{Chen:1995yu,Datta:1998yw}.
Moreover, the radiative production of neutralinos can serve as a direct test to see, 
whether neutralinos are dark matter candidates. See for example
Ref.~\cite{Birkedal:2004xn}, which presents a model independent calculation for 
the cross section of radiatively produced dark matter candidates at 
high-energy colliders, including polarised beams for the ILC.

The signature ``photon plus missing energy'' has been studied
intensively by the LEP collaborations ALEPH~\cite{Heister:2002ut},
DELPHI~\cite{Abdallah:2003np}, L3~\cite{Achard:2003tx}, and
OPAL~\cite{Abbiendi:2002vz,Abbiendi:2000hh}.  In the SM, 
\begin{eqnarray}
e^+e^- \to \nu \bar\nu \gamma
\end{eqnarray}
is the leading process with this signature.  Since
the cross section depends on the number $N_\nu$ of light neutrino
generations~\cite{Gaemers:1978fe}, it has been used to measure $N_\nu
$ consistent with three.  In addition, the LEP collaborations have
tested physics beyond the SM, like non-standard neutrino interactions,
extra dimensions, and SUSY particle productions.  However, no
deviations from SM predictions have been found, and only bounds on
SUSY particle masses have been set, e.g.  on the gravitino 
mass~\cite{Heister:2002ut,Abdallah:2003np,Achard:2003tx,Abbiendi:2000hh}. 
This process is also important in determining collider
bounds on a very light neutralino~\cite{lightneutralino}.  For a
combined short review, see for example Ref.~\cite{Gataullin:2003sy}.

Although there are so many theoretical studies on radiative neutralino
production in the literature, a thorough analysis of this process is
still missing in the light of the ILC with a high center-of-mass
energy, high luminosity, and longitudinally polarised beams. As noted
above, most of the existing analyses discuss SUSY scenarios with
parameters which are ruled out by LEP2 already, or without taking beam
polarisations into account. In particular, the question of the role of
the positron beam polarisation has to be addressed. If both beams are
polarised, the discovery potential of the ILC might be significantly
extended, especially if other SUSY states like heavier neutralino, chargino or
even slepton pairs are too heavy to be produced at the first stage of
the ILC at $\sqrt s = 500$~GeV.  Moreover, the SM background photons
from radiative neutrino production $e^+e^- \to\nu\bar\nu\gamma $ have
to be included in an analysis with beam polarisations.  Proper beam
polarisations could enhance the signal photons and reduce those from
the SM background at the same time, which enhances the statistics
considerably. In this respect also the MSSM background photons from
radiative sneutrino production 
\begin{eqnarray}
e^+e^-\to\tilde\nu\tilde\nu^\ast\gamma 
\end{eqnarray}
have to be discussed, if sneutrino production is kinematically accessible and 
if the sneutrino decay is invisible.

Finally, the studies which analyse beam polarisations do not give
explicit formulas for the squared matrix elements, but only for the
transition amplitudes~\cite{Bayer,Choi:1999bs,Weidner}.  Other authors
admit sign errors~\cite{Datta:2002jh} in some interfering amplitudes
in precedent works~\cite{Datta:1996ur}, however do not provide the
corrected formulas for radiative neutrino and sneutrino production.
Additionally, I found inconsistencies and sign errors in the $Z$
exchange terms in some works~\cite{Bayer,Weidner}, which yield wrong
results for scenarios with dominating $Z$ exchange.  Thus I will give
the complete tree-level amplitudes and the squared matrix elements
including longitudinal beam polarisations, such that the formulas can
be used for further studies on radiative production of neutralinos,
neutrinos and sneutrinos.

In Sec.~\ref{sec:xsection}, I discuss my signal process, radiative
neutralino pair production, as well as the major SM and MSSM
background processes.  In Sec.~\ref{sec:results}, I define cuts on the
photon angle and energy, and define a statistical significance for
measuring an excess of photons from radiative neutralino production
over the backgrounds.  I analyse numerically the dependence of cross
sections and significances on the electron and positron beam
polarisations, on the parameters of the neutralino sector, and on the
selectron masses.  I summarise and conclude in
Sec.~\ref{sec:conclusion}.  In the Appendix, I define neutralino
mixing and couplings, and give the tree-level amplitudes as well as
the squared matrix elements with longitudinal beam polarisations for
radiative production of neutralinos, neutrinos and sneutrinos.  In
addition, I give details on the parametrisation of the phase space.

\section{Radiative Neutralino Production and Backgrounds}
  \label{sec:xsection}	

\subsection{Signal Process}
\begin{figure}[t!]
{%
\unitlength=1.0 pt
\SetScale{1.0}
\SetWidth{0.7}      
\scriptsize    
\allowbreak
\begin{picture}(95,79)(0,0)
\Text(15.0,70.0)[r]{$e^-$}
\ArrowLine(16.0,70.0)(58.0,70.0) 
\Text(80.0,70.0)[l]{$\gamma$}
\Photon(58.0,70.0)(79.0,70.0){1.0}{5} 
\Text(54.0,60.0)[r]{$e^-$}
\ArrowLine(58.0,70.0)(58.0,50.0) 
\Text(80.0,50.0)[l]{$\widetilde\chi^0_1$}
\Line(58.0,50.0)(79.0,50.0) 
\Text(54.0,40.0)[r]{$\widetilde{e}_R$}
\DashArrowLine(58.0,50.0)(58.0,30.0){1.0} 
\Text(15.0,30.0)[r]{$e^+$}
\ArrowLine(58.0,30.0)(16.0,30.0) 
\Text(80.0,30.0)[l]{$\widetilde\chi^0_1$}
\Line(58.0,30.0)(79.0,30.0) 
\Text(47,0)[b] {diagr. 1/4}
\end{picture} \ 
{} \qquad\allowbreak
\begin{picture}(95,79)(0,0)
\Text(15.0,70.0)[r]{$e^-$}
\ArrowLine(16.0,70.0)(58.0,70.0) 
\Text(80.0,70.0)[l]{$\widetilde\chi^0_1$}
\Line(58.0,70.0)(79.0,70.0) 
\Text(54.0,60.0)[r]{$\widetilde{e}_R$}
\DashArrowLine(58.0,70.0)(58.0,50.0){1.0} 
\Text(80.0,50.0)[l]{$\widetilde\chi^0_1$}
\Line(58.0,50.0)(79.0,50.0) 
\Text(54.0,40.0)[r]{$e^-$}
\ArrowLine(58.0,50.0)(58.0,30.0) 
\Text(15.0,30.0)[r]{$e^+$}
\ArrowLine(58.0,30.0)(16.0,30.0) 
\Text(80.0,30.0)[l]{$\gamma$}
\Photon(58.0,30.0)(79.0,30.0){1.0}{5} 
\Text(47,0)[b] {diagr. 2/5}
\end{picture} \ 
{} \qquad\allowbreak
\begin{picture}(95,79)(0,0)
\Text(15.0,70.0)[r]{$e^-$}
\ArrowLine(16.0,70.0)(58.0,70.0) 
\Text(80.0,70.0)[l]{$\widetilde\chi^0_1$}
\Line(58.0,70.0)(79.0,70.0) 
\Text(54.0,60.0)[r]{$\widetilde{e}_R$}
\DashArrowLine(58.0,70.0)(58.0,50.0){1.0} 
\Text(80.0,50.0)[l]{$\gamma$}
\Photon(58.0,50.0)(79.0,50.0){1.0}{5} 
\Text(54.0,40.0)[r]{$\widetilde{e}_R$}
\DashArrowLine(58.0,50.0)(58.0,30.0){1.0} 
\Text(15.0,30.0)[r]{$e^+$}
\ArrowLine(58.0,30.0)(16.0,30.0) 
\Text(80.0,30.0)[l]{$\widetilde\chi^0_1$}
\Line(58.0,30.0)(79.0,30.0) 
\Text(47,0)[b] {diagr. 3/6}
\end{picture} \ 
{} \qquad\allowbreak
\begin{picture}(95,79)(0,0)
\Text(15.0,60.0)[r]{$e^-$}
\ArrowLine(16.0,60.0)(37.0,60.0) 
\Photon(37.0,60.0)(58.0,60.0){1.0}{5} 
\Text(80.0,70.0)[l]{$\gamma$}
\Photon(58.0,60.0)(79.0,70.0){1.0}{5} 
\Text(33.0,50.0)[r]{$e^-$}
\ArrowLine(37.0,60.0)(37.0,40.0) 
\Text(15.0,40.0)[r]{$e^+$}
\ArrowLine(37.0,40.0)(16.0,40.0) 
\Text(47.0,41.0)[b]{$Z$}
\DashLine(37.0,40.0)(58.0,40.0){3.0} 
\Text(80.0,50.0)[l]{$\widetilde\chi^0_1$}
\Line(58.0,40.0)(79.0,50.0) 
\Text(80.0,30.0)[l]{$\widetilde\chi^0_1$}
\Line(58.0,40.0)(79.0,30.0) 
\Text(47,0)[b] {diagr. 7}
\end{picture} \ 
{} \qquad\allowbreak
\begin{picture}(95,79)(0,0)
\Text(15.0,60.0)[r]{$e^-$}
\ArrowLine(16.0,60.0)(37.0,60.0) 
\Text(47.0,61.0)[b]{$Z$}
\DashLine(37.0,60.0)(58.0,60.0){3.0} 
\Text(80.0,70.0)[l]{$\widetilde\chi^0_1$}
\Line(58.0,60.0)(79.0,70.0) 
\Text(80.0,50.0)[l]{$\widetilde\chi^0_1$}
\Line(58.0,60.0)(79.0,50.0) 
\Text(33.0,50.0)[r]{$e^-$}
\ArrowLine(37.0,60.0)(37.0,40.0) 
\Text(15.0,40.0)[r]{$e^+$}
\ArrowLine(37.0,40.0)(16.0,40.0) 
\Photon(37.0,40.0)(58.0,40.0){1.0}{5} 
\Text(80.0,30.0)[l]{$\gamma$}
\Photon(58.0,40.0)(79.0,30.0){1.0}{5} 
\Text(47,0)[b] {diagr. 8}
\end{picture} \ 
{} \qquad\allowbreak
\begin{picture}(95,79)(0,0)
\Text(15.0,70.0)[r]{$e^-$}
\ArrowLine(16.0,70.0)(58.0,70.0) 
\Text(80.0,70.0)[l]{$\gamma$}
\Photon(58.0,70.0)(79.0,70.0){1.0}{5} 
\Text(54.0,60.0)[r]{$e^-$}
\ArrowLine(58.0,70.0)(58.0,50.0) 
\Text(80.0,50.0)[l]{$\widetilde\chi^0_1$}
\Line(58.0,50.0)(79.0,50.0) 
\Text(54.0,40.0)[r]{$\widetilde{e}_L$}
\DashArrowLine(58.0,50.0)(58.0,30.0){1.0} 
\Text(15.0,30.0)[r]{$e^+$}
\ArrowLine(58.0,30.0)(16.0,30.0) 
\Text(80.0,30.0)[l]{$\widetilde\chi^0_1$}
\Line(58.0,30.0)(79.0,30.0) 
\Text(47,0)[b] {diagr. 9/12}
\end{picture} \ 
{} \qquad\allowbreak
\begin{picture}(95,79)(0,0)
\Text(15.0,70.0)[r]{$e^-$}
\ArrowLine(16.0,70.0)(58.0,70.0) 
\Text(80.0,70.0)[l]{$\widetilde\chi^0_1$}
\Line(58.0,70.0)(79.0,70.0) 
\Text(54.0,60.0)[r]{$\widetilde{e}_L$}
\DashArrowLine(58.0,70.0)(58.0,50.0){1.0} 
\Text(80.0,50.0)[l]{$\widetilde\chi^0_1$}
\Line(58.0,50.0)(79.0,50.0) 
\Text(54.0,40.0)[r]{$e^-$}
\ArrowLine(58.0,50.0)(58.0,30.0) 
\Text(15.0,30.0)[r]{$e^+$}
\ArrowLine(58.0,30.0)(16.0,30.0) 
\Text(80.0,30.0)[l]{$\gamma$}
\Photon(58.0,30.0)(79.0,30.0){1.0}{5} 
\Text(47,0)[b] {diagr. 10/13}
\end{picture} \ 
{} \qquad\allowbreak
\begin{picture}(95,79)(0,0)
\Text(15.0,70.0)[r]{$e^-$}
\ArrowLine(16.0,70.0)(58.0,70.0) 
\Text(80.0,70.0)[l]{$\widetilde\chi^0_1$}
\Line(58.0,70.0)(79.0,70.0) 
\Text(54.0,60.0)[r]{$\widetilde{e}_L$}
\DashArrowLine(58.0,70.0)(58.0,50.0){1.0} 
\Text(80.0,50.0)[l]{$\gamma$}
\Photon(58.0,50.0)(79.0,50.0){1.0}{5} 
\Text(54.0,40.0)[r]{$\widetilde{e}_L$}
\DashArrowLine(58.0,50.0)(58.0,30.0){1.0} 
\Text(15.0,30.0)[r]{$e^+$}
\ArrowLine(58.0,30.0)(16.0,30.0) 
\Text(80.0,30.0)[l]{$\widetilde\chi^0_1$}
\Line(58.0,30.0)(79.0,30.0) 
\Text(47,0)[b] {diagr. 11/14}
\end{picture} \ 
}
\caption{Diagrams for radiative neutralino production $e^+e^- \to
  \tilde\chi_1^0\tilde\chi_1^0\gamma$~\cite{Boos:2004kh}. For the
  calculation in Appendix~\ref{sec:app:chifore}, the first number of the
  diagrams labels $t$-channel, the second one $u$-channel exchange of
  selectrons, where the neutralinos are crossed.}
\label{fig:diagrams}
\end{figure}
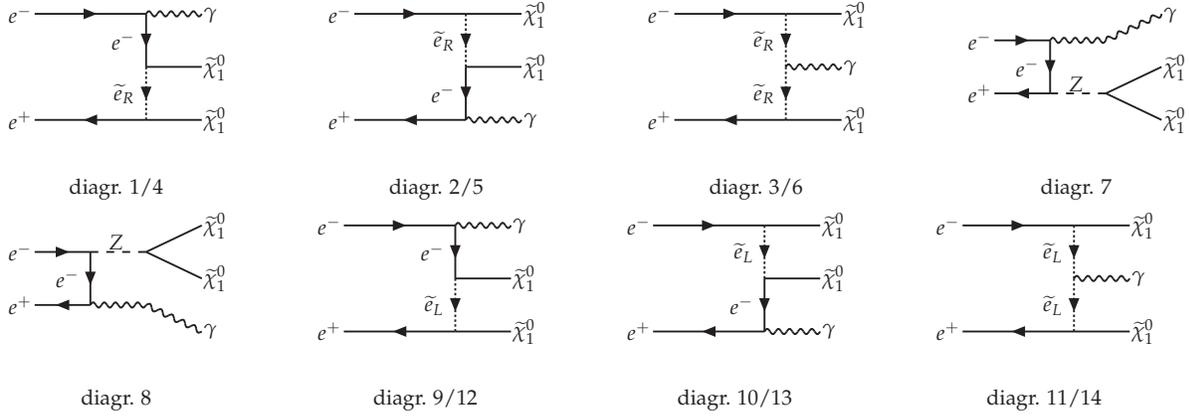

Within the MSSM, radiative neutralino production~\cite{Fayet:1982ky,
Ellis:1982zz,Grassie:1983kq,Kobayashi:1984wu,Ware:1984kq,Bento:1985in,
Chen:1987ux,Kon:1987gi,Bayer,Choi:1999bs,Baer:2001ia,Weidner,
Fraas:1991ky,Datta:1994ac,Datta:1996ur,Datta:2002jh,Ambrosanio:1995it}
\begin{equation}
e^++e^- \to \tilde\chi_1^0+\tilde\chi_1^0+\gamma
\label{productionChi}
\end{equation}
proceeds at tree-level via $t$- and $u$-channel exchange of right and
left selectrons $\tilde e_{R,L}$, as well as $Z$ boson exchange in the
$s$-channel. The photon is radiated off the incoming beams or the
exchanged selectrons; see the contributing diagrams in
Fig.~\ref{fig:diagrams}. I give the relevant Feynman rules for
general neutralino mixing, the tree-level amplitudes, and the complete
analytical formulas for the amplitude squared, including longitudinal
electron and positron beam polarisations, in
Appendix~\ref{sec:app:chifore}. I also summarise the details of the
neutralino mixing matrix there. For the calculation of cross sections
and distributions I use cuts, as defined in Eq.~(\ref{cuts}). An
example of the photon energy distribution and the $\sqrt s$ dependence
of the cross section is shown in Fig.~\ref{plotEdist}.

\subsection{Neutrino Background}
\noindent
Radiative neutrino production~\cite{Datta:1996ur,Gaemers:1978fe,Berends:1987zz,
Boudjema:1996qg,Montagna:1998ce}
\begin{equation}
e^+ +e^- \to \nu_\ell+\bar\nu_\ell+\gamma\,,\;\;\qquad \ell=e,\mu,\tau
\label{productionNu}
\end{equation}
is a major SM background. Electron neutrinos $\nu_e$ are produced via
$t$-channel $W$ boson exchange, and $\nu_{e,\mu,\tau}$ via $s$-channel
$Z$ boson exchange.  I show the corresponding diagrams in
Appendix~\ref{sec:app:nuback}, where I also give the tree-level
amplitudes and matrix elements squared including longitudinal beam
polarisations.

\subsection{MSSM Backgrounds}
Next I consider radiative sneutrino
production~\cite{Datta:1996ur,Franke:thesis, Franke:1994ph}
\begin{equation}
e^+ +e^- \to \tilde\nu_\ell+\tilde\nu^\ast_\ell+\gamma\,, 
\;\qquad \ell=e,\mu,\tau\,.
\label{productionSneut}
\end{equation}
I present the tree-level Feynman graphs as well as the amplitudes and
amplitudes squared, including beam polarisations, 
in Appendix~\ref{sec:app:snuback}.  The process has $t$-channel contributions via
virtual charginos for $\tilde\nu_e\tilde\nu_e^\ast $-production, as
well as $s$-channel contributions from $Z$ boson exchange for
$\tilde\nu_{e, \mu,\tau}\tilde\nu_{e, \mu,\tau}^\ast $-production, see
Fig.~\ref{fig:sneutrino}.  Radiative sneutrino production,
Eq.~(\ref{productionSneut}), can be a major MSSM background to
neutralino production, Eq.~(\ref{productionChi}), if the sneutrinos
decay mainly invisibly, e.g., via $\tilde\nu\to\tilde \chi^0_1\nu$.
This leads to so called ``virtual LSP'' scenarios~\cite{Datta:1994ac,
  Datta:1996ur, Datta:2002jh}.  However, if kinematically allowed,
other visible decay channels like $\tilde\nu\to\tilde\chi^\pm_1\ell^\mp$ 
reduce the background rate from radiative sneutrino production.
For example in the SPS~1a scenario~\cite{Ghodbane:2002kg,Allanach:2002nj}, 
I have ${\rm  BR}(\tilde\nu_e\to\tilde
\chi_1^0\nu_e)=85\%$, see Table~\ref{scenarioSPS1}.

In principle, also neutralino production $e^+e^- \to \tilde\chi_1^0
\tilde\chi^0_2$ followed by the subsequent radiative neutralino
decay~\cite{Haber:1988px} $\tilde\chi^0_2 \to \tilde\chi^0_1 \gamma$
is a potential background.  However, significant branching ratios
${\rm BR}(\tilde\chi^0_2 \to \tilde\chi^0_1 \gamma)>10\%$ are only
obtained for small values of $\tan\beta<5$ and/or $M_1\sim
M_2$~\cite{Ambrosanio:1995it,Ambrosanio:1995az,Ambrosanio:1996gz}.
Thus I neglect this background in the following.  For details see
Refs.~\cite{Ambrosanio:1995az,Ambrosanio:1996gz,Baer:2002kv}.

\section{Numerical Results}
\label{sec:results}

I present numerical results for the tree-level cross section for
radiative neutralino production, Eq.~(\ref{productionChi}), and the
background from radiative neutrino and sneutrino production,
Eqs.~(\ref{productionNu}) and (\ref{productionSneut}), respectively.
I define the cuts on the photon energy and angle, and define the statistical
significance. 
I study the dependence of the cross sections and the significance
on the beam polarisations $P_{e^-}$ and $P_{e^+}$, the supersymmetric
parameters $\mu$ and $M_2$, and on the selectron masses. In order to
reduce the number of parameters, I assume the SUSY GUT relation
\begin{eqnarray}
\label{eq:gutrelation}
M_1 = \frac{5}{3}\tw[2]M_2\,.
\end{eqnarray}
Therefore the mass of the lightest neutralino is $m_{\chi^0_1}\gsim50\,
\mathrm{GeV}$~\cite{aleph}.  I also use the approximate renormalisation
group equations (RGE) for the slepton
masses~\cite{Ibanez:1983di,Ibanez:1984vq,Hall:1985zn},
\begin{eqnarray}
\label{sleptonR}
m_{\tilde e_R}^2 &=& m_0^2 +0.23 M_2^2-m_Z^2\cos 2 \beta \sw[2],\\
\label{sleptonL}
m_{\tilde e_L  }^2 &=& m_0^2 +0.79 M_2^2+
              m_Z^2\cos 2 \beta\Big(-\frac{1}{2}+ \sw[2]\Big),\\
m_{\tilde\nu_e  }^2 &=& m_0^2 +0.79 M_2^2+ \frac{1}{2}m_Z^2\cos 2 \beta,
\label{sneutrino}
\end{eqnarray}
with $m_0$ the common scalar mass parameter. 
Since in my scenarios the dependence on $\tan\beta$ is rather mild,
I fix $\tan\beta = 10$.

\begin{table}
\renewcommand{\arraystretch}{1.2}
\caption{Parameters and masses for  SPS~1a 
scenario~\cite{Ghodbane:2002kg,Allanach:2002nj}.}
\begin{center}
        \begin{tabular}{|c|c|c|c|}
\hline
$\tan\beta=10$ & $\mu= 352$~GeV &$M_2= 193$~GeV &$m_0=100$~GeV \\
\hline\hline
$m_{\chi^0_{1}}=94$~GeV & 
$m_{\chi^\pm_{1}}=178$~GeV & 
$m_{\tilde e_{R}}=143$~GeV & 
$m_{\tilde\nu_e}=188$ GeV \\
\hline
$m_{\chi^0_{2}}=178$~GeV & 
$m_{\chi^\pm_{2}}=376$~GeV & 
$m_{\tilde e_{L}}=204$~GeV &
${\rm BR}(\tilde\nu_e\to\tilde\chi_1^0\nu_e)=85\%$ \\
\hline
\end{tabular}
\end{center}
\renewcommand{\arraystretch}{1.0}
\label{scenarioSPS1}
\end{table}

\subsection{Cuts on Photon Angle and Energy}
\label{subsec:cuts}
To regularise the infrared and collinear divergencies of the
tree-level cross sections, I apply cuts on the photon scattering
angle $\theta_\gamma$ and on the photon energy $E_\gamma$
\begin{equation}
 -0.99 \le \cos\theta_\gamma \le 0.99,\quad \quad
0.02 \le x\le \;1-\frac{m_{\chi_1^0}^2}{E_{\rm beam}^2}, \quad \quad
x = \frac{E_\gamma}{E_{\rm beam}}, 
\label{cuts}
\end{equation}
with the beam energy $E_{\rm beam}=\sqrt{s}/2$.  The cut on the
scattering angle corresponds to $\theta_\gamma \in [8^\circ,172^\circ]
$, and reduces much of the background from radiative Bhabha
scattering, $e^+e^-\to e^+e^-\gamma$, where both leptons escape close
to the beam pipe~\cite{Abdallah:2003np,Heister:2002ut}.  The lower cut
on the photon energy is $E_\gamma= 5 \GeV$ for $\sqrt{s}=500\GeV$. The
upper cut on the photon energy $x^{\mathrm{max}}=1-m_{\chi_1^0}^2/E_
{\rm beam}^2$ is the kinematical limit of radiative neutralino
production. At $\sqrt{s}=500\GeV$ and for $m_{\chi_1^0}\gsim70\GeV$,
this cut reduces much of the on-shell $Z$ boson contribution to
radiative neutrino production, see Refs.~\cite{Baer:2001ia,
  Datta:1994ac,Franke:1994ph,Vest:2000ad} and Fig.~\ref{plotEdist}(a).
I assume that the neutralino mass $m_{\chi_1^0}$ is known from LHC or
ILC measurements~\cite{Weiglein:2004hn}.  If $m_{\chi_1^0}$ is
unknown, a fixed cut, e.g., $E_\gamma^{\rm max}=175$~GeV at $\sqrt{s}=500
\GeV$, could be used instead~\cite{Vest:2000ad}.

\subsection{Theoretical Significance}
In order to quantify whether an excess of signal photons from 
neutralino production, $N_{\mathrm{S}}=\sigma {\mathcal L}$, 
for a given integrated luminosity $\mathcal{L}$, can be measured
over the SM background photons, $N_{\rm B}=\sigma_{\rm B}{\mathcal L}$,
from radiative neutrino production,
I define the theoretical significance  $S$ and the signal to background ratio 
$r$ (or reliability)
\begin{eqnarray}
\label{significance}
S  &=&  \frac{N_{\rm S}}{\sqrt{N_{\rm S} + N_{\rm B}}}=
\frac{\sigma}{\sqrt{\sigma + \sigma_{\rm B}}} \sqrt{\mathcal L},\\[2mm]
\label{sigbck}
r &= & \frac{\sigma_{\mathrm{Signal}}}{\sigma_{\mathrm{Background}}}\enspace .
\end{eqnarray}
A theoretical significance of, e.g., $S = 1$ implies that the signal
can be measured at the statistical 68\% confidence level.  Also the
the signal to background ratio $N_{\mathrm{S}}/N_{\mathrm{B}}$ should
be considered to judge the reliability of the analysis.  For example,
if the background cross section is known experimentally to 1\%
accuracy, I should have $N_{\mathrm{S}}/N_{\mathrm{B}}>1/100$.

I will not include additional cuts on the missing mass or on the
transverse momentum distributions of the photons~\cite{Baer:2001ia,
  Vest:2000ad}. Detailed Monte Carlo analyses, including detector
simulations and particle identification and reconstruction
efficiencies, would be required to predict the significance more
accurately, which is however beyond the scope of the present work.
Also the effect of beamstrahlung should be included in such an
experimental analysis~\cite{Vest:2000ad,Ohl:1996fi,Hinze:2005xt}.
Beamstrahlung distorts the peak of the beam energy spectrum to lower
values of $E_{\rm beam}=\sqrt{s}/2$, and is more significant at
colliders with high luminosity.  In the processes I consider, the
cross sections for $e^+e^- \to \tilde\chi^0_1 \tilde\chi^0_1\gamma$
and $e^+e^- \to \nu \bar\nu \gamma$ depend significantly on the beam
energy only near threshold. In most of the parameter space we
consider, for $\sqrt{s}= 500\GeV$ the cross sections are nearly
constant, see for example Fig.~\ref{plotEdist}(b), so I expect that
the effect of beamstrahlung will be rather small. However, for $M_2,\,
\mu\gsim 300\,\mathrm{GeV}$, $\,\signal$ is the only SUSY production process,
which is kinematically accessible, see Fig.~\ref{CrossSectionMuM2}.
In order to exactly determine the kinematic reach, the ILC
beamstrahlung must be taken into account.

\subsection{Energy Distribution and $\sqrt{s}$ Dependence}

\begin{figure}[t!]
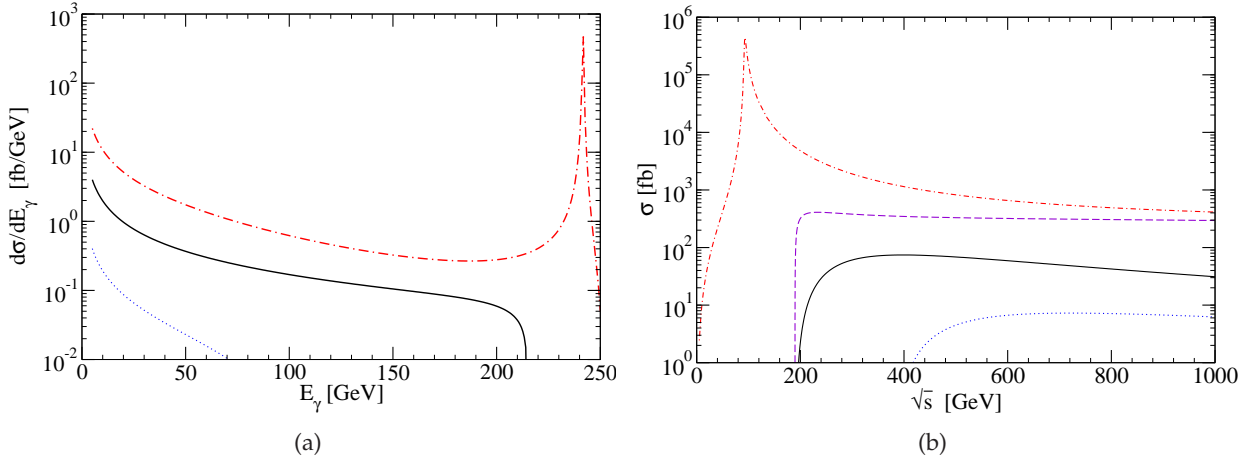

\setlength{\unitlength}{1cm}
\subfigure[
\label{fig:sps1adiff}]%
{\scalebox{0.3}{\includegraphics{sps1adiffnewcut.eps}}}
\hspace{1mm}
\subfigure[
\label{fig:sps1a}]{\scalebox{0.3}{\includegraphics{sps1anewcut.eps}}}
\caption{ (a) Photon energy distributions for $\sqrt s = 500\GeV$, 
        and (b) $\sqrt s$ dependence of the cross sections $\sigma$
        for radiative neutralino production $e^+e^- \to \tilde\chi^0_1
        \tilde\chi^0_1\gamma$ (black, solid), neutrino  production 
        $e^+e^- \to \nu\bar\nu\gamma$ (violet, dashed) and sneutrino  
        production $e^+ e^- \to \tilde\nu\tilde\nu^\ast\gamma$ (blue, dotted)
        for scenario  SPS~1a~\cite{Ghodbane:2002kg,Allanach:2002nj}, 
        see Table~\ref{scenarioSPS1}, with $(P_{e^-},P_{e^+})=(0.8,-0.6)$.
        The red dot-dashed line is in~(a) the photon energy distribution
        for radiative neutrino production $e^+e^- \to \nu\bar\nu\gamma$,
        and in~(b) the cross section without the upper cut
        on the photon energy $E_\gamma$, see Eq.~(\ref{cuts}).        
        \label{plotEdist}}
\end{figure}
In Fig.~\ref{plotEdist}(a), I show the energy distributions of the
photon from radiative neutralino production, neutrino production, and
sneutrino production for scenario SPS~1a~\cite{Ghodbane:2002kg,
  Allanach:2002nj}, see Table~\ref{scenarioSPS1}, with $\sqrt s =
500\GeV$, beam polarisations $(P_{e^-},P_{e^+})=(0.8,-0.6)$, and cuts
as defined in Eq.~(\ref{cuts}). The energy distribution of the photon
from neutrino production peaks at $E_\gamma= (s -m_Z^2)/(2\sqrt{s})
\approx242$~GeV due to radiative $Z$ return, which is possible for
$\sqrt s > m_Z$.  Much of this photon background from radiative
neutrino production can be reduced by the upper cut on the photon
energy $x^{\rm max}=E_\gamma^{\rm max}/E_{\rm beam}=1-m_{\chi_1^0}^2
/E_{\rm beam}^2$, see Eq.~(\ref{cuts}), which is the kinematical
endpoint $E_\gamma^\mathrm{max}\approx215$~GeV of the energy
distribution of the photon from radiative neutralino production, see
the solid line in Fig.~\ref{plotEdist}(a). Note that in principle the
neutralino mass could be determined by a measurement of this 
endpoint $E_\gamma^\mathrm{max} = E_\gamma^\mathrm{max}(m_{\chi^0_1})$
\begin{eqnarray}
 m_{\chi_1^0}^2 = \frac{1}{4}\left(s - 2\sqrt{s}E_\gamma^\mathrm{max}
\right).
\end{eqnarray}
For this one would need to be able to very well separate the signal
and background processes. This might be possible if the neutralino is
heavy enough, such that the endpoint is sufficiently removed from the
$Z^0$-peak of the background distribution.

In Fig.~\ref{plotEdist}(b) I show the $\sqrt s$ dependence of the
cross sections.  Without the upper cut on the photon energy $x^{\rm
  max}$, see Eq.~(\ref{cuts}), the background cross section from
radiative neutrino production $e^+e^- \to \nu\bar\nu\gamma$, see the
dot-dashed line in Fig.~\ref{plotEdist}(b), is much larger than the
corresponding cross section with the cut, see the dashed line.
However with the cut, the signal cross section from radiative
neutralino production, see the solid line, is then only about one
order of magnitude smaller than the background.

\subsection{Beam Polarisation Dependence
\label{beampoldep}}

\begin{figure}
\setlength{\unitlength}{1cm}
 \begin{picture}(20,20)(0,-2)
   \put(2.0,16.5){\fbox{$\sigma(e^+e^- \to \tilde\chi^0_1\tilde\chi^
    0_1\gamma)$ in fb}}
     \put(-4.5,-4.5){\includegraphics{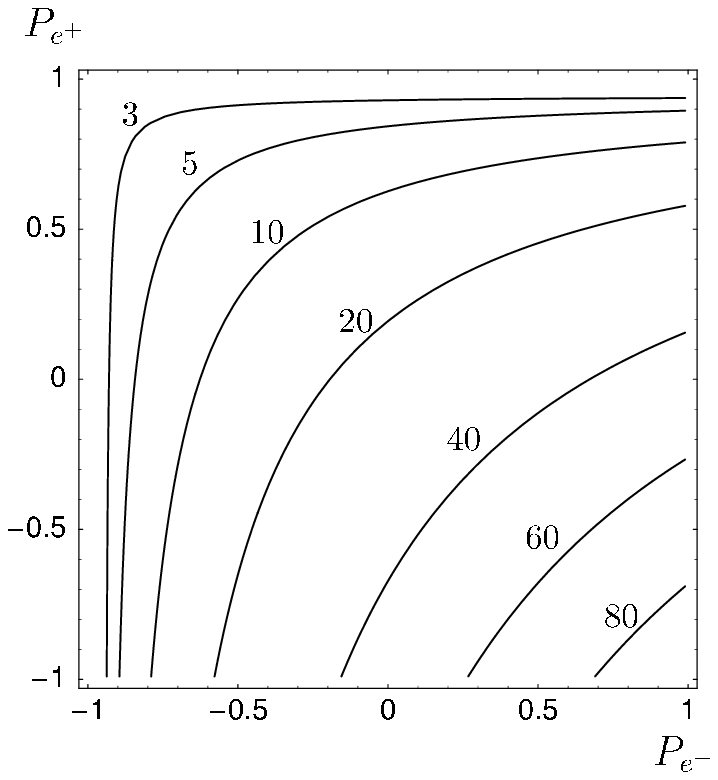}}
         \put(1.,8.8){(a)} 
   
        \put(3.5,-4.5){\includegraphics{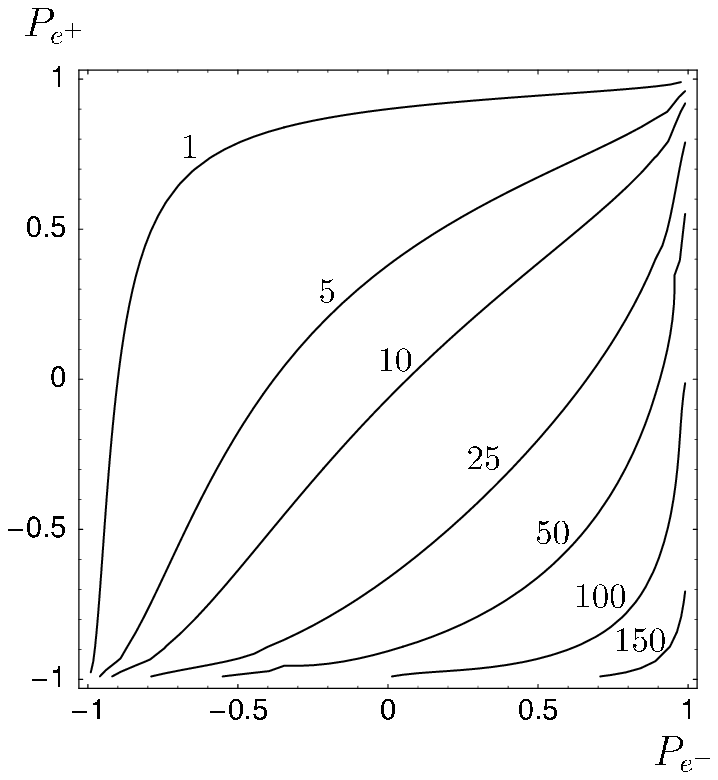}}
        \put(11.,16.5){\fbox{$S=\frac{\sigma}{\sqrt{\sigma+
       \sigma_{\rm B}}}\sqrt{\mathcal{L}} $} }
        \put(9.,8.8){(b)} 
        \put(-4.5,-13.5){\includegraphics{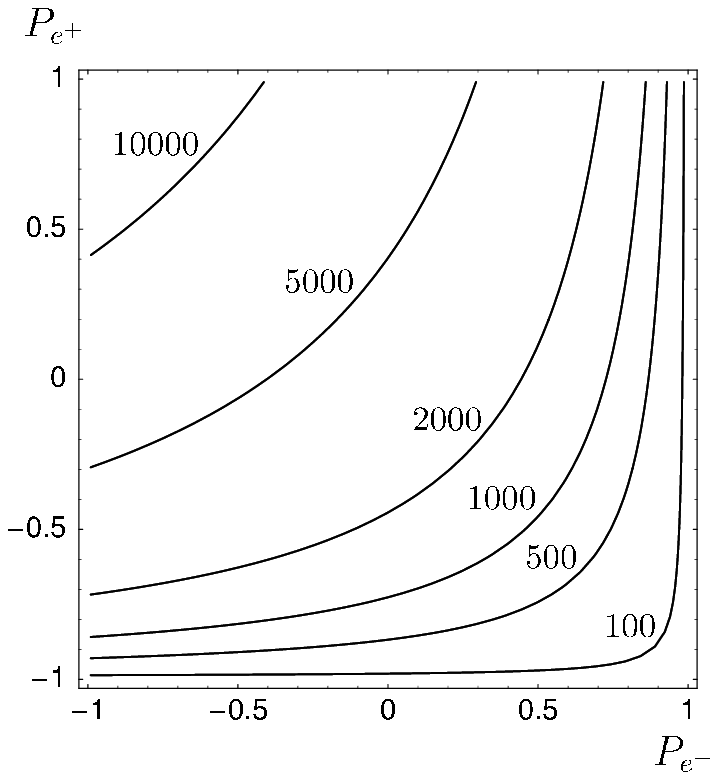}}
        \put(2.0,7.5){\fbox{$\sigma_{\rm B}(e^+e^-\to\nu\bar\nu
          \gamma)$ in fb }}
        \put(1.0,-0.3){(c)} 
        \put(3.5,-13.5){\includegraphics{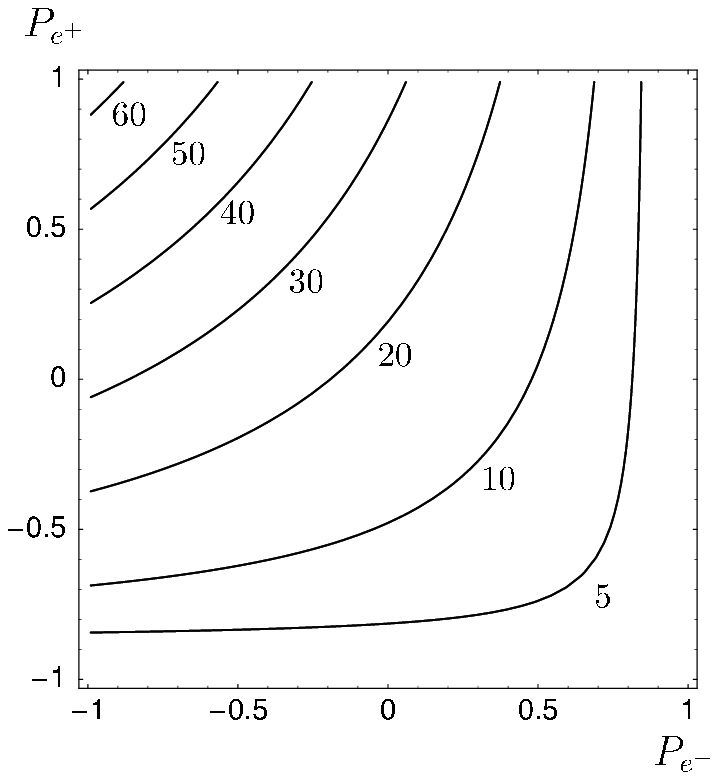}}
        \put(10.,7.5){\fbox{$\sigma(e^+e^-\to\tilde\nu\tilde
          \nu^\ast\gamma)$ in fb}}
          \put(9.0,-0.3){(d)} 
\end{picture}
\vspace*{-1.5cm}
\caption{%
        (a) Contour lines of the cross section and (b) the significance $S$
        for  $e^+e^- \to \tilde\chi^0_1\tilde\chi^0_1\gamma$ at $\sqrt{s}
        =500$~GeV and ${\mathcal{L}}=500~{\rm fb}^{-1}$ for scenario 
        SPS~1a~\cite{Ghodbane:2002kg,Allanach:2002nj}, see 
        Table~\ref{scenarioSPS1}. The beam polarisation dependence of 
        the cross section for radiative neutrino and sneutrino production are shown in~(c) 
        and~(d), respectively. 
        \label{varBeamPol}}
\end{figure}

%
In Fig.~\ref{varBeamPol}(a) I show the beam polarisation dependence
of the cross section $\sigma(e^+e^-\to\tilde\chi^0_1\tilde\chi^0_1
\gamma)$ for the SPS~1a scenario~\cite{Ghodbane:2002kg,
  Allanach:2002nj}, where radiative neutralino production proceeds
mainly via right selectron $\tilde e_R$ exchange.  Since the
neutralino is mostly bino, the coupling to the right selectron is more
than twice as large as to the left selectron.
Thus the contributions from right selectron exchange to the cross section are about
a factor 16 larger than the $\tilde{e}_L$ contributions.
In addition the $\tilde{e}_L$ contributions are suppressed compared 
to  the $\tilde{e}_R$ contributions by a factor of about 2 
since $m_{\tilde{e}_R }<m_{\tilde{e}_L} $, see 
Eqs.~(\ref{sleptonR})-(\ref{sleptonL}). The $Z$ boson exchange is
negligible. The background process, radiative neutrino production,
mainly proceeds via $W$ boson exchange, see the corresponding diagram
in Fig.~\ref{fig:neutrino}. Thus
positive electron beam polarisation $P_{e^-}$ and negative positron
beam polarisation $P_{e^+}$ enhance the signal cross section and
reduce the background at the same time, see Figs.~\ref{varBeamPol}(a)
and \ref{varBeamPol}(c), which was also observed in 
Refs.~\cite{Choi:1999bs,Vest:2000ad}. 
The positive electron beam polarisation
and negative positron beam polarisation 
enhance $\tilde e_R$ exchange and suppress $\tilde e_L$
exchange, such that it becomes negligible. Opposite
polarisations would lead to comparable contributions from both selectrons.
In going from unpolarised beams $(P_{e^-},P_{e^+})=(0,0)$ to polarised
beams, e.g., $(P_{e^-},P_{e^+})=(0.8,-0.6)$, the signal cross section
is enhanced by a factor $\approx 3$, and the background cross section
is reduced by a factor $\approx 10$.  The signal to background ratio
increases from $N_{\rm S}/N_{\rm B}\approx 0.007$ to $N_{\rm S}/N_{\rm
  B}\approx 0.2$, such that the statistical significance $S$, shown in
Fig.~\ref{varBeamPol}(b), is increased by a factor $\approx 8.5$ to
$S\approx 77$.  If only the electron beam is polarised,
$(P_{e^-},P_{e^+})=(0.8,0)$, I still have $N_{\rm S}/N_{\rm B}\approx
0.06$ and $S\approx 34$, thus the option of a polarised positron beam
at the ILC doubles the significance for radiative neutralino
production, but is not needed or essential to observe this process at
$\sqrt{s}=500\GeV$ and ${\mathcal{L}}=500\fb^{-1}$ for the SPS~1a
scenario.

In contrast, the conclusion of Ref.~\cite{Baer:2001ia} is, that an almost
pure level of beam polarisations is needed at the ILC to observe this
process at all.  The authors have used a scenario with $M_1 = M_2$,
leading to a lightest neutralino, which is mostly a wino. Thus larger
couplings to the left selectron than to the right selectron are
obtained.  In such a scenario, one cannot simultaneously enhance the
signal and reduce the background.  Moreover their large selectron
masses $m_{\tilde e_{L,R}}=500\GeV$ lead to an additional suppression
of the signal, see also Sec.~\ref{selectronmassdependence}.

Finally I note that positive electron beam polarisation and negative
positron beam polarisation also suppress the cross section of
radiative sneutrino production, see Fig.~\ref{varBeamPol}(d).  Since
it is the corresponding SUSY process to radiative neutrino production,
I expect a similar quantitative behaviour.

\subsection{$\mu$ \& $M_2$ Dependence}

%
\begin{figure}
\setlength{\unitlength}{1cm}
 \begin{picture}(20,20)(0,-2)
        \put(2.2,16.5){\fbox{$\sigma(e^+e^- \to \tilde\chi^0_1\tilde
 \chi^0_1\gamma)$ in fb}}
\put(-4.5,-4.5){\includegraphics{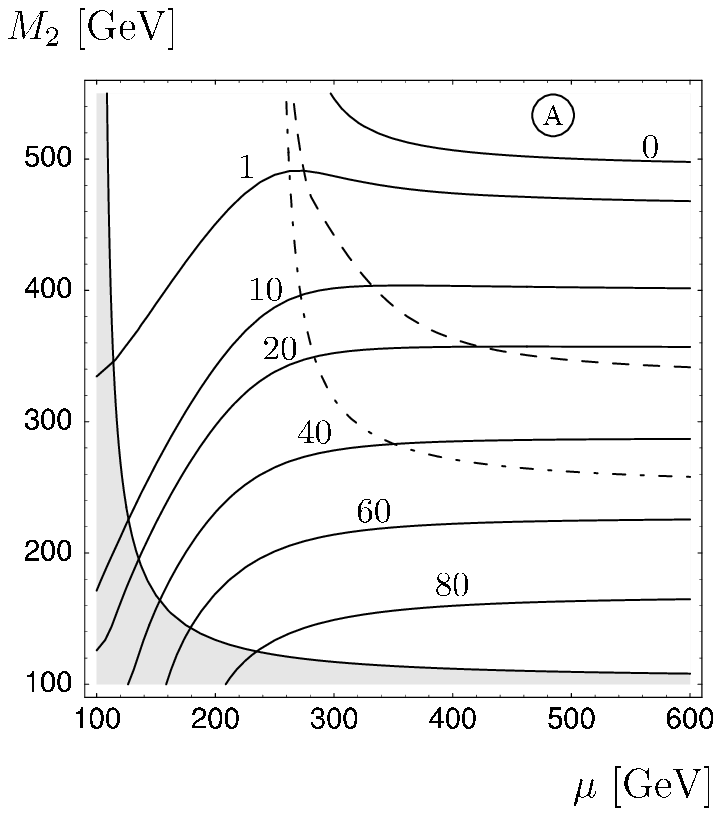}}
\put(1.,8.8){(a)}
   \put(3.5,-4.5){\includegraphics{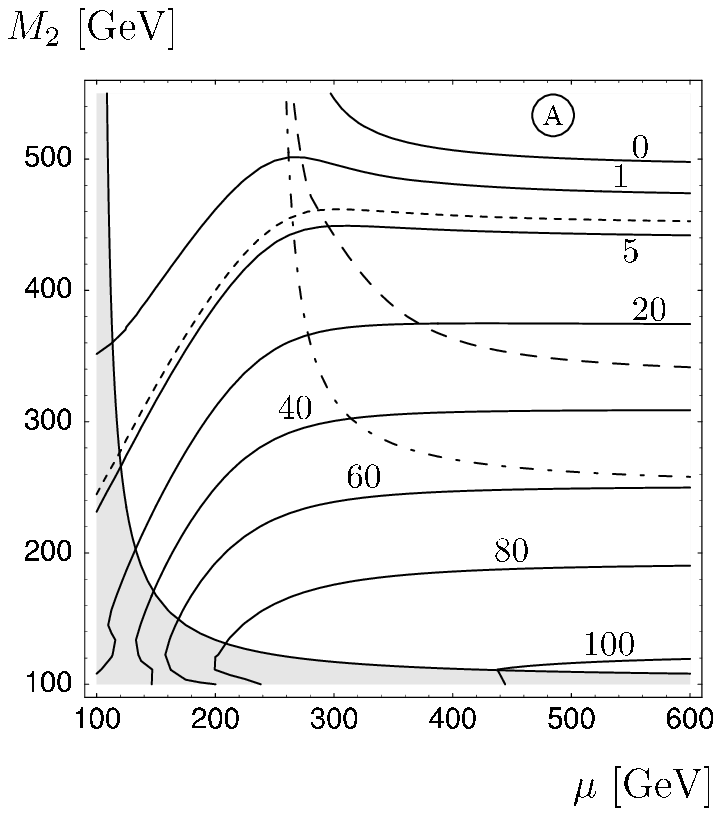}}
        \put(11.,16.5){\fbox{$S=\frac{\sigma}{\sqrt{\sigma+
\sigma_{\rm B}}}\sqrt{\mathcal{L}} $} }
\put(9.,8.8){(b)}
        \put(-4.5,-13.5){\includegraphics{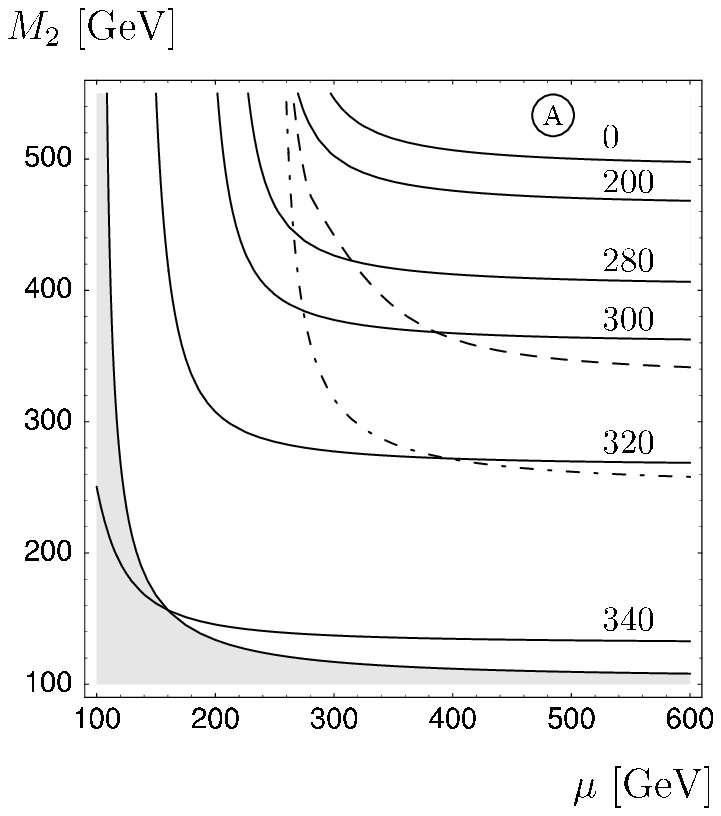}}
        \put(2.2,7.5){\fbox{$\sigma_{\rm B}(e^+e^-\to\nu\bar\nu
                      \gamma)$ in fb }}
\put(1.0,-0.3){(c)}
   \put(3.5,-13.5){\includegraphics{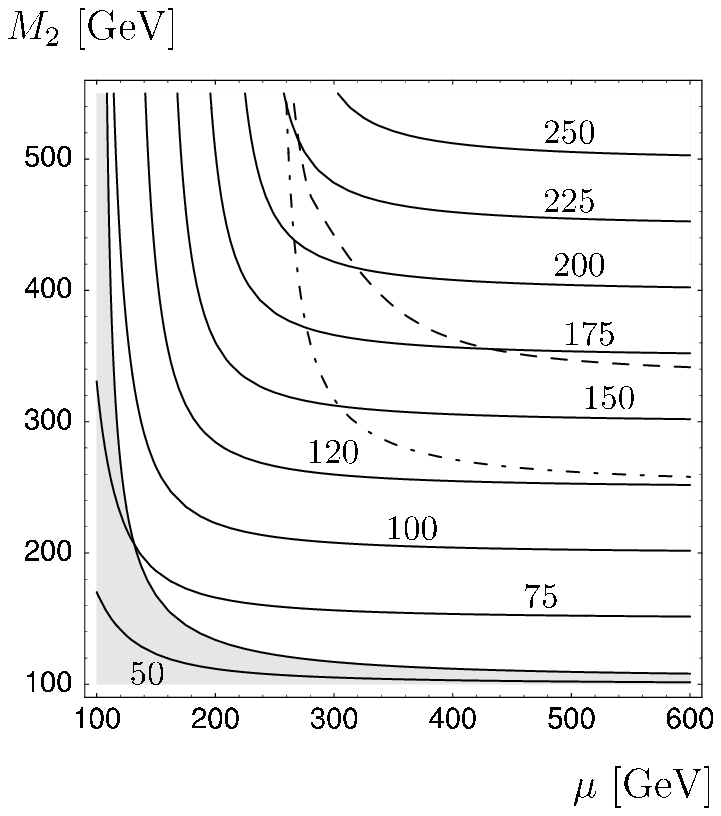}}
        \put(11.2,7.5){\fbox{$m_{\chi_1^0}$ in GeV}}
\put(9.0,-0.3){(d)}
\end{picture}
\vspace*{-1.5cm}
\caption{%
        Contour lines (solid) of (a) the cross section 
        $\sigma(e^+e^- \to \tilde\chi^0_1\tilde\chi^0_1\gamma)$, 
        (b) the significance $S$, (c) the neutrino background
        $\sigma_{\rm B}(e^+e^-\to\nu\bar\nu\gamma)$, and (d) the 
        neutralino mass $m_{\chi_1^0}$ in the $\mu$-$M_2$ plane
        for $\sqrt{s}=500$~GeV, $(P_{e^-},P_{e^+})=(0.8,-0.6)$,
        ${\mathcal{L}}=500~{\rm fb}^{-1}$, with $\tan\beta = 10$, 
        $m_0=100$~GeV, and RGEs for the selectron masses, see 
        Eqs.~(\ref{sleptonR}), (\ref{sleptonL}). The grey area is 
        excluded by $m_{\chi_1^\pm}<104$~GeV. The dashed line
        indicates the kinematical limit $m_{\chi_1^0}+m_{\chi_2^0} 
        =\sqrt{s}$, and the dot-dashed line the kinematical limit 
        $2 m_{\chi_1^\pm} =\sqrt{s}$. Along the dotted line in (b) 
        the signal to background ratio is $\sigma/\sigma_{\rm B}= 
        0.01$. The area A is kinematically forbidden by the cut on 
        the photon energy $E_\gamma$, see Eq.~(\ref{cuts}). 
        \label{CrossSectionMuM2}}
\end{figure}

In Fig.~\ref{CrossSectionMuM2}(a), I show contour lines of the cross
section $\sigma(e^+e^- \to \tilde\chi^0_1\tilde\chi^0_1\gamma)$ in fb
in the $(\mu,\,M_2)$-plane. For $\mu \gsim 300\GeV$ the signal and the
background cross sections are nearly independent of $\mu$, and
consequently also the significance, which is shown in
Fig.~\ref{CrossSectionMuM2}(b). In addition, the dependence of the
neutralino mass on $\mu$ is fairly weak for $\mu \gsim 300\GeV$, as
can be seen in Fig.~\ref{CrossSectionMuM2}(d). Also the couplings have a
rather mild $\mu$-dependence in this parameter region.
 
The cross section $\sigma_{\rm B}(e^+e^- \to \nu\bar\nu\gamma)$ of the
SM background process due to radiative neutrino production, shown in
Fig.~\ref{CrossSectionMuM2}(c), can reach more than $340\fb$ and is
considerably reduced due to the upper cut on the photon energy $x^{\rm
  max}$, see Eq.~(\ref{cuts}).  Without this cut I would have
$\sigma_{\rm B}= 825\fb$.  Thus the signal can be observed with high
statistical significance $S$, see Fig.~\ref{CrossSectionMuM2}(b).  Due
to the large integrated luminosity ${\mathcal{L}}=500~{\rm fb}^{-1}$
of the ILC, I have $S\gsim 25$ with $N_{\rm S}/N_{\rm B}\gsim 1/4$
for $M_2\lsim 350\GeV$.  For $\mu<0$, I get similar results for the
cross sections in shape and size, since the dependence of $N_{11}$
on the sign of $\mu$, see Eq.~(\ref{eq:coefficients}), is weak due to
the large value of $\tan\beta=10$.

In Fig.~\ref{CrossSectionMuM2}, I also indicate the kinematical
limits of the lightest observable associated neutralino production
process, $e^+e^- \to \tilde\chi^0_1\tilde\chi^0_2$ (dashed), and those
of the lightest chargino production process,
$e^+e^-\to\tilde\chi^+_1\tilde \chi^-_1$ (dot-dashed). In the region
above these lines $\mu,M_2\gsim 300\GeV$, heavier neutralinos and charginos
are too heavy to be pair-produced at the first stage of the ILC with
$\sqrt{s}=500\GeV$. In this case radiative neutralino production
$e^+e^-\to\tilde\chi^0_1\tilde\chi^0 _1\gamma$ will be the only
channel to study the gaugino sector. Here significances of $5<S\lsim
25$ can be obtained for $350~{\rm GeV}\lsim M_2 \lsim 450\GeV$, see
Fig.~\ref{CrossSectionMuM2}(b). Note that the production of right
sleptons $e^+e^- \to\tilde\ell^+_R\tilde\ell^-_R$, $\tilde\ell=\tilde
e,\tilde\mu$, and in particular the production of the lighter staus
$e^+e^ - \to \tilde\tau^+_1\tilde\tau^-_1$, due to mixing in the stau
sector~\cite{Bartl:2002bh}, are still open channels to study the
direct production of SUSY particles for $M_2\lsim 500\GeV$ in our GUT
scenario with $m_0=100\GeV$.

\subsection{Dependence on the Selectron Masses
       \label{selectronmassdependence}}

%
\begin{figure}
\setlength{\unitlength}{1cm}
 \begin{picture}(20,20)(0,-2)
        \put(2.2,16.5){\fbox{$\sigma(e^+e^- \to \tilde\chi^0_1\tilde
\chi^0_1\gamma)$ in fb}}
\put(-4.5,-4.5){\includegraphics{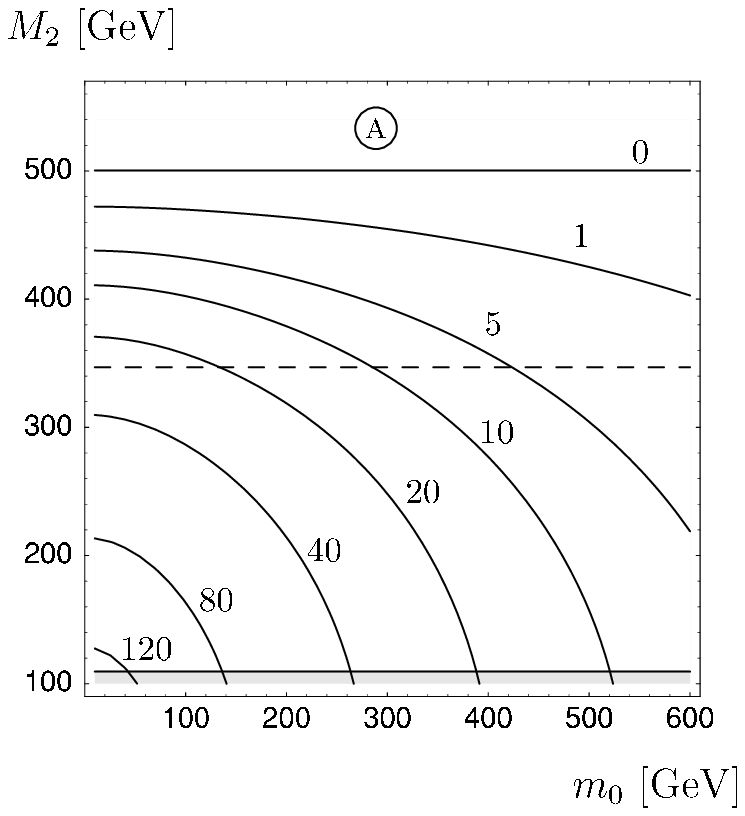}}
\put(1.,8.8){(a)}
   \put(3.5,-4.5){\includegraphics{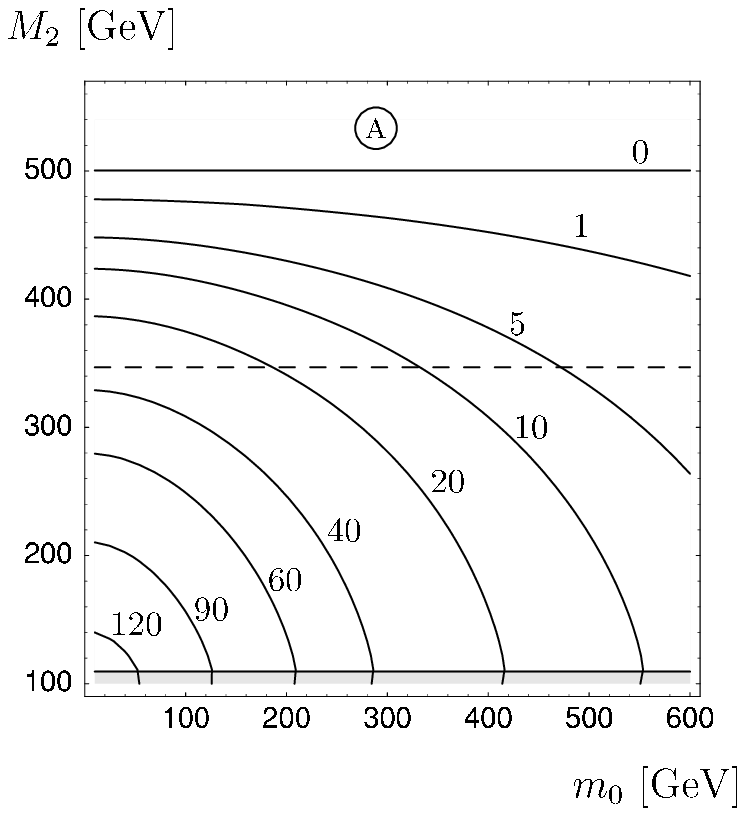}}
        \put(11.,16.5){\fbox{$S=\frac{\sigma}{\sqrt{\sigma+\sigma_{\rm B}}}\sqrt{\mathcal{L}} $} }
\put(9.,8.8){(b)}
        \put(-4.5,-13.5){\includegraphics{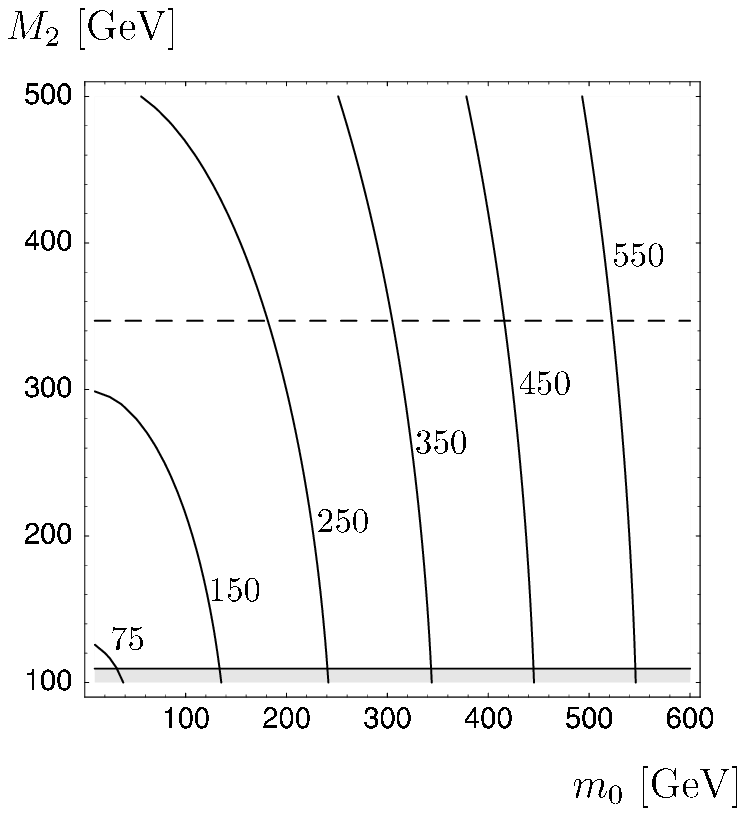}}
        \put(3.2,7.5){\fbox{$ m_{\tilde e_R}$ in GeV }}
\put(1.0,-0.3){(c)}
   \put(3.5,-13.5){\includegraphics{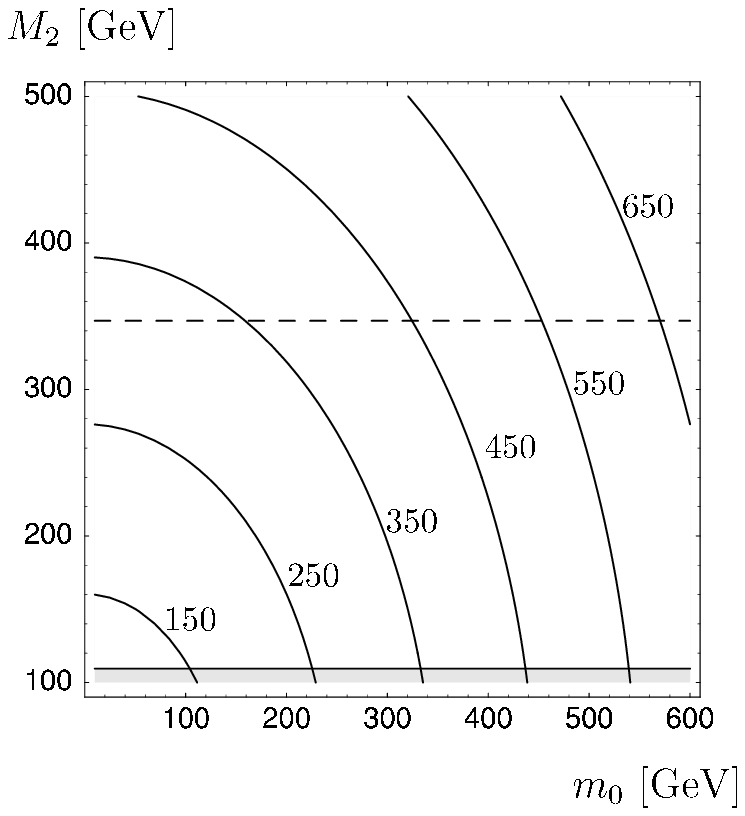}}
        \put(11.2,7.5){\fbox{$m_{\tilde e_L}$ in GeV}}
 \put(9.0,-0.3){(d)}
\end{picture}
\vspace*{-1.5cm}
\caption{%
        (a) Contour lines of the cross section $\sigma(e^+e^- \to 
        \tilde\chi^0_1\tilde\chi^0_1\gamma)$, (b) the significance $S$,
        and (c), (d) the selectron masses $ m_{\tilde e_R}$, $ m_{\tilde e_L}$,
        respectively, in the $m_0$-$M_2$ plane for $\sqrt{s}=500$~GeV, 
        $(P_{e^-},P_{e^+})=(0.8,-0.6)$, ${\mathcal{L}}=500~{\rm fb}^{-1}$, 
        with $\mu=500$~GeV, $\tan\beta = 10$, and RGEs for the selectron 
        masses, see Eqs.~(\ref{sleptonR}), (\ref{sleptonL}). The dashed line 
        indicates the kinematical limit $m_{\chi_1^0}+m_{\chi_2^0} =\sqrt{s}$.
        The grey area is excluded by $m_{\chi_1^\pm}<104$~GeV, the area A is 
        kinematically forbidden. 
\label{CrossSectionM0M2}}
\end{figure}

The cross section for radiative neutralino production $\sigma(e^+e^-
\to\tilde\chi^0_1\tilde\chi^0_1\gamma)$ proceeds mainly via selectron
$\tilde e_{R,L}$ exchange in the $t$ and $u$-channels. Besides the
beam polarisations, which enhance $\tilde e_{R}$ or $\tilde e_{L}$
exchange, the cross section is also very sensitive to the selectron
masses. In the mSUGRA universal supersymmetry breaking 
scenario~\cite{weldon}, the masses are parametrised by $m_0$ and $M_2$, 
besides $\tan\beta$, which enter the RGEs, see Eqs.~(\ref{sleptonR}) and
(\ref{sleptonL}).  I show the contour lines of the selectron masses
$\tilde e_{R,L}$ in the $m_0$-$ M_2$ plane in
Fig.~\ref{CrossSectionM0M2}(c) and \ref{CrossSectionM0M2}(d),
respectively. 
The selectron masses increase with increasing $m_0$ and $M_2$.

For the polarisations $(P_{e^-},P_{e^+})=(0.8,-0.6)$, 
the cross section is dominated by
$\tilde{e}_R$ exchange, as discussed in Sec.~\ref{beampoldep}. 
In Fig.~\ref{CrossSectionM0M2}(a) and~\ref{CrossSectionM0M2}(b), I
show the $m_0$ and $M_2$ dependence of the cross section and the
significance $S$, Eq.~(\ref{significance}). With increasing $m_0$ and
$M_2$ the cross section and the significance decrease, due to  
the increasing mass of $\tilde{e}_R$.
In Fig.~\ref{CrossSectionMuM2}(d), I see that for $\mu\gsim 7/10\, M_2$, 
the neutralino mass $m_{\chi^0_1}$ is practically independent 
of $\mu$ and rises with $M_2$.
Thus for increasing $M_2$, and thereby increasing neutralino mass,
the cross section $\sigma(e^+e^- \to\tilde\chi^0_1\tilde\chi^0_1 \gamma)$
reaches the kinematical limit at $M_2\approx500$~GeV for $\sqrt s =500$~GeV.
A potential background from radiative sneutrino production is only
relevant for $M_2\lsim 200\GeV$, $m_0\lsim 200\GeV$.  For larger values
the production is kinematically forbidden.

\begin{figure}[t]
\setlength{\unitlength}{1cm}
\begin{center}
\begin{picture}(10,8)(0,0)
   \put(-6.,-12.5){\includegraphics{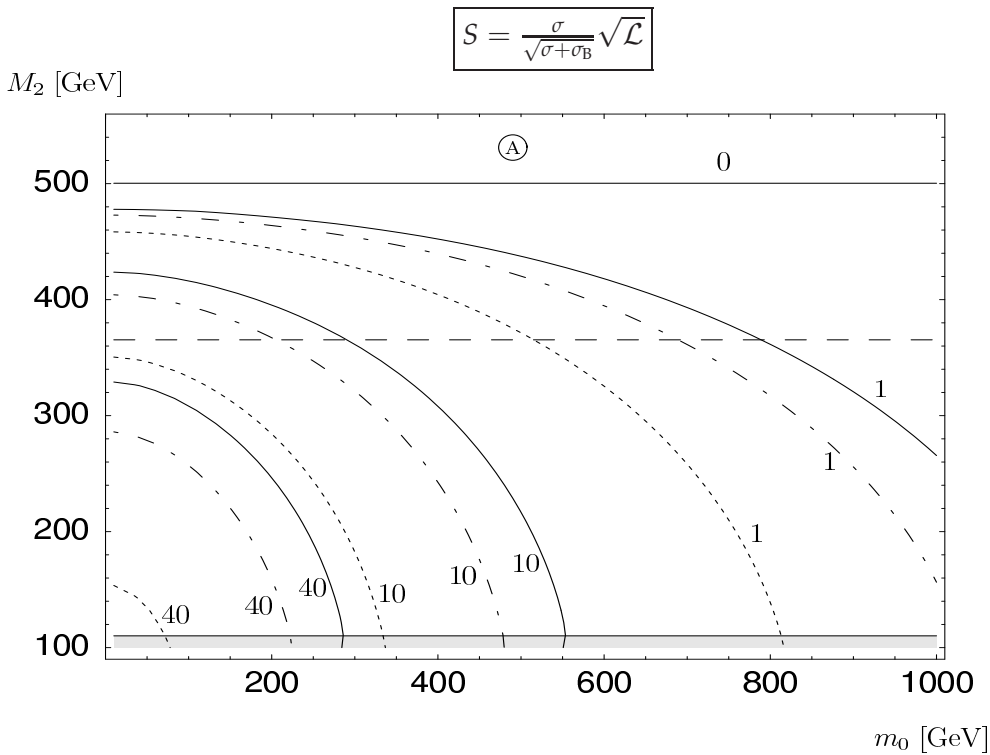}}
        \put(4.,7.4){\fbox{$S=\frac{\sigma}{\sqrt{\sigma+
        \sigma_{\rm B}}}\sqrt{\mathcal{L}} $} }
\end{picture}
\end{center}
\vspace*{1.5cm}
\caption{Contour lines of the significance $S$ 
        in the $m_0$-$M_2$ plane
        for different beam polarisations
        $(P_{e^-},P_{e^+})=(0.8,-0.6)$ (solid),
        $(P_{e^-},P_{e^+})=(0.8,0)$ (dot-dashed),
        and $(P_{e^-},P_{e^+})=(0,0)$ (dotted),
        for $\sqrt{s}=500$~GeV, ${\mathcal{L}}=500~{\rm fb}^{-1}$,
        $\mu=500$~GeV, $\tan\beta = 10$,
        and RGEs for the selectron masses, see Eqs.~(\ref{sleptonR}), (\ref{sleptonL}).
        The dashed line indicates the kinematical limit 
        $m_{\chi_1^0}+m_{\chi_2^0} =\sqrt{s}$.
        The grey area is excluded by $m_{\chi_1^\pm}<104$~GeV,
        the area A is kinematically forbidden.
        \label{plotDiffSig}}
\end{figure}

In Fig.~\ref{CrossSectionM0M2}, I also indicate the kinematical limit
of associated neutralino pair production $m_{\chi_1^0}+m_{\chi_2^0}=
\sqrt{s}=500 $~GeV, reached for $M_2\approx350$~GeV. If in addition $
m_0>200$~GeV, also selectron and smuon pairs cannot be produced at $
\sqrt{s}=500 \GeV$ due to $m_{\tilde \ell_R}>250\GeV$. Thus, in this
parameter range where $M_2>350$~GeV and $m_0> 200\GeV$, radiative
production of neutralinos will be the \textit{only} possible production
process of SUSY particles, if I neglect stau mixing.  
A statistical significance of $S>1$ can
be obtained for selectron masses not larger than $m_{\tilde e_R}
\approx 50 0$~GeV, corresponding to $m_0\lsim500$~GeV and $M_2\lsim450
$~GeV. Thus radiative neutralino production extends the discovery
potential of the ILC in the parameter range $m_0\in[200,500]$~GeV and
$M_2\in[350,450]$~GeV. Here, the beam polarisations will be essential,
see Fig.~\ref{plotDiffSig}. I show contour lines of the statistical
significance $S$ for three different sets of $(P_{e^-},P_{e^+})$.  The
first set has both beams polarised, $(P_{e^-},P_{e^+})=(0.8,-0.6)$,
the second one has only electron beam polarisation, $(P_{e^-},P_{e^+})
=(0.8,0)$, and the third has zero beam polarisations $(P_{e^-},P_{e^+}
)=(0,0)$.  The beam polarisations significantly enhance
the discovery potential of the ILC. At least electron polarisation
$P_{e^-}=0.8$ is needed to extend an exploration of the $m_0$-$M_2$
parameter space.

%

%

\subsection{Note on LEP2 \label{sec:collider}}

I have also calculated the unpolarised cross sections and the
significances for radiative neutralino production at LEP2 energies $
\sqrt{s} = 200\enspace \mathrm{GeV}$, for a luminosity of $\mathcal{L
}=100\pb^{-1}$. I have used the cuts $|\cos\theta_\gamma|\le0.95$ and
$0.2 \le x\le1-m_{\chi_1^0}^2/E_{\rm beam}^2$, cf.  Eq.~(\ref{cuts}).
Even for rather small selectron masses $m_{\tilde e_{R,L}}=80\GeV$,
the cross sections are not larger than $100\fb$. Even if I alter the GUT
relation, Eq.~(\ref{eq:gutrelation}), to $M_1 = r_{12}\,M_2$, and vary $r_{12}$
within the range $0.01 < r_{12} < 0.5$,
I only obtain statistical significances of $S<0.2$. These values
have also been reported by other theoretical studies at LEP2 energies,
see for example Ref.~\cite{Ambrosanio:1995it}.

If I drop the GUT relation, $M_1$ is a free parameter. For
\begin{eqnarray}
M_1 = \frac{M_2 m_Z^2 \sin(2\beta)\sw[2]}{\mu M_2 - m_Z^2\sin(2\beta)\cw[2]}
\end{eqnarray}
the neutralino is massless~\cite{Gogoladze:2002xp} at tree-level and
is apparently experimentally allowed~\cite{lightneutralino}. A
massless neutralino should enlarge the cross section for radiative
neutralino production due to the larger phase space, although the
coupling is also modified to almost pure bino. However, I still find
$S=\mathcal {O}( 10^{-1})$ at most. This is in accordance with the
experimental SUSY searches in photon events with missing energy at
LEP~\cite{Abbiendi:2002vz,Heister:2002ut,Abdallah:2003np,Achard:2003tx,Abbiendi:2000hh},
where no evidence of SUSY particles was found.

\section{The Role of Beam polarization for Radiative Neutralino Production at the ILC}
\label{sec:beampol}
\subsection{Introduction}

Detailed measurements of the masses, decay widths, couplings,
and spins of the discovered particles are only possible at the
international linear collider (ILC)~\cite{Weiglein:2004hn,
Aguilar-Saavedra:2001rg, Abe:2001nn,Abe:2001gc}. In the first stage of
the ILC, the center-of-mass energy will be $\sqrt{s} = 500\GeV$ and
the luminosity, $\mathcal{L}$, will be $500\fb^{-1}$ per year.

\medskip

In preparing for the ILC, there is an on-going debate over the extent
of beam polarization to be included in the initial design~\cite{talk1,
Moortgat-Pick:2005cw,talk2,talk3}. It is clear that there will be at
least 80\% polarization of the electron beam, possibly even 
90\%~\cite{TDR}. A polarized positron beam is technically and financially
more involved. However, it is possible to achieve 30\% polarization
already through the undulator based production of the 
positrons~\cite{talk3}. In light of this discussion, it is the purpose of
this section to reconsider the effect of various degrees of electron
and positron polarization on a particular supersymmetric production
process, namely the radiative production of the lightest neutralino
mass eigenstate $\tilde\chi^0_1$
\begin{equation}
e^+ + e^-\rightarrow \tilde\chi^0_1+\tilde\chi_1^0+\gamma.
\label{eeXXg}
\end{equation}
I shall focus on specific regions of the supersymmetric parameter
space. The signal is a single, highly energetic photon and missing
energy, carried by the neutralinos.

\medskip

The process~(\ref{eeXXg}) was previously studied within the MSSM and
with general neutralino mixing in Refs.~\cite{Datta:1994ac,
  Datta:1996ur, Datta:2002jh,Ambrosanio:1995it}. The additional effect
of polarized beams was considered in Refs.~\cite{Choi:1999bs,
  Baer:2001ia,Dreiner:2006sb}. In Ref.~\cite{Dreiner:2006sb}, it was
shown that polarized beams significantly enhance the signal and
simultaneously suppress the main SM photon background from radiative
neutrino production, 
\begin{equation}
e^++e^- \to \nu +\bar\nu +\gamma\,.
\label{eenng}
\end{equation} 
Moreover, it was pointed out that for certain regions of the MSSM
parameter space, the process~(\ref{eeXXg}) is kinematically the
\textit{only} accessible SUSY production mechanism in the first stage
of the ILC at $\sqrt s=500\GeV$~\cite{Dreiner:2006sb}. Here the
heavier electroweak gauginos and the sleptons are too heavy to be pair
produced, \textit{i.e.} their masses are above 250 GeV.

\medskip

Other than the standard center-of-mass energy, $\sqrt{s}=500\GeV$, at
the ILC, also lower energies are of particular interest, namely for
Higgs and top physics. Higgs strahlung,
\begin{equation}
  e^++e^- \rightarrow Z +h,
\end{equation}
can be well studied at the threshold energy $\sqrt{s} = m_h + m_Z$,
which is $\sqrt{s}\approx220\GeV$, for a Higgs boson mass of $m_h
\approx 130\GeV$. The CP-quantum number and the spin of the Higgs
boson can be determined from an energy scan of the production cross
section near the threshold~\cite{Dova:2003py}.

\medskip

From a scan at the threshold energy of top pair production,
$\sqrt{s}=2m_t\approx350\GeV$, the top mass $m_t$ can be determined
with an error $\delta m_t<0.1\GeV$~\cite{ilc:2006va}. Thus the present
error on the top mass, $\delta m_t\approx 3\GeV$~\cite{Yao:2006px},
and the foreseen error from LHC measurements, $\delta m_t\approx1\GeV
$~\cite{Borjanovic:2004ce}, can be reduced by one order of magnitude.
Also the top width, $\Gamma_t$, and the strong coupling constant,
$\alpha_s$, can be precisely determined by a multi parameter fit of
the cross section, top momentum distribution, and forward-backward
charge asymmetry near threshold~\cite{Martinez:2002st}.

\medskip

In this section, I take these physics questions as a motivation to
study the role of polarized beams in radiative neutralino production
at the energies $\sqrt{s} = 220\GeV$, $350\GeV$, and $500\GeV$ at the
ILC. For each beam energy, I shall focus on a specific supersymmetric
parameter set within the context of minimal supergravity grand
unification (mSUGRA)~\cite{msugra}. I thus consider three mSUGRA
scenarios, which I label A, B and C, respectively, and which are
listed below in Table~\ref{tab:scenarios} together with the resulting
spectra in Table~\ref{tab:masses}. I restrict myself to mSUGRA
scenarios, in order to reduce the number of free parameters and since
I find it suffcient to illustrate my point. The specific scenarios
are chosen such that radiative neutralino production is the
\textit{only} supersymmetric production mechanism which is
kinematically accessible at the given center-of-mass energy.  It is
thus of particular interest to learn as much about supersymmetry as is
possible through this mechanism. As I shall see, beam polarization is
very helpful in this respect.

\medskip

In Sect.~\ref{subsec:cuts}, I define the significance, the signal to
background ratio and define a first set of experimental cuts. In 
Sect.~\ref{numerics}, I study numerically the dependence of the signal cross
sec\-tion and the SM background, the significance, and the signal to
background ratio on the beam polarization.  In particular, I compare
the results for different sets of beam polarizations, $(P_{e^+}|P_
{e^-}) = (0|0)$, $(0|0.8)$, $(-0.3|0.8)$, $(-0.6|0.8)$, $
{(0|0.9)}$ and ${(-0.3|0.9)}$.  I summarize and conclude in
Sect.~\ref{Summary}.

\subsection{Numerical results \label{numerics}}

I choose the three scenarios in such a
way, that only the lightest neutralinos can be radiatively produced
for each of the $\sqrt s$ values, respectively. The other SUSY
particles, \textit{i.e.} the heavier neutralinos and charginos, as
well as the sleptons and squarks are too heavy to be pair produced at
the ILC. It is thus of paramount interest to have an optimal
understanding of the signature~(\ref{eeXXg}), in order to learn as
much as possible about SUSY at a given ILC beam energy. Note that in
the three scenarios (A,B,C) the squark and gluino masses are below
$\{600,800,1000\}$ GeV, respectively and should be observable at the
LHC~\cite{Weiglein:2004hn}. 

\medskip

Scenario A is related to the Snowmass point SPS1a~\cite{
Aguilar-Saavedra:2005pw,Allanach:2002nj,Ghodbane:2002kg} by scaling
the common scalar mass $M_0$, the unified gaugino mass $M_ {1/2}$, and
the common trilinear coupling $A_0$ by $0.9$. Thus the slope $M_0 =
-A_0 = 0.4\, M_{1/2}$ remains unchanged. For scenarios B and C, I
also choose $M_0 = - A_0$, however I change the slopes to $M_0 =
0.42\, M_{1/2}$ in scenario B, and $M_0 = 0.48\, M_{1/2}$ in scenario
C. For all scenarios, I fix the ratio $\tan\beta = 10$ of the vacuum
expectation values of the two neutral Higgs fields. In
Table~\ref{tab:scenarios}, I explicitly give the relevant low energy
mSUGRA parameters for all scenarios. These are the $U(1)$ and $SU(2)$
gaugino mass parameters $M_1$ and $M_2$, respectively, and the
Higgsino mass parameter $\mu$. The masses of the light neutralinos,
charginos, and sleptons are given in Table~\ref{tab:masses}.  All
parameters and masses are calculated at one-loop order with the
computer code SPheno~\cite{Porod:2003um}.

\medskip

\begin{center}
\begin{figure}[ht]
\setlength{\unitlength}{1cm}
\subfigure[Signal cross section $\sigma(e^+e^- \to \tilde\chi^0_1\tilde\chi^0_1\gamma)$ in fb.
\label{fig:sigma220}]%
	{\scalebox{0.53}{\includegraphics{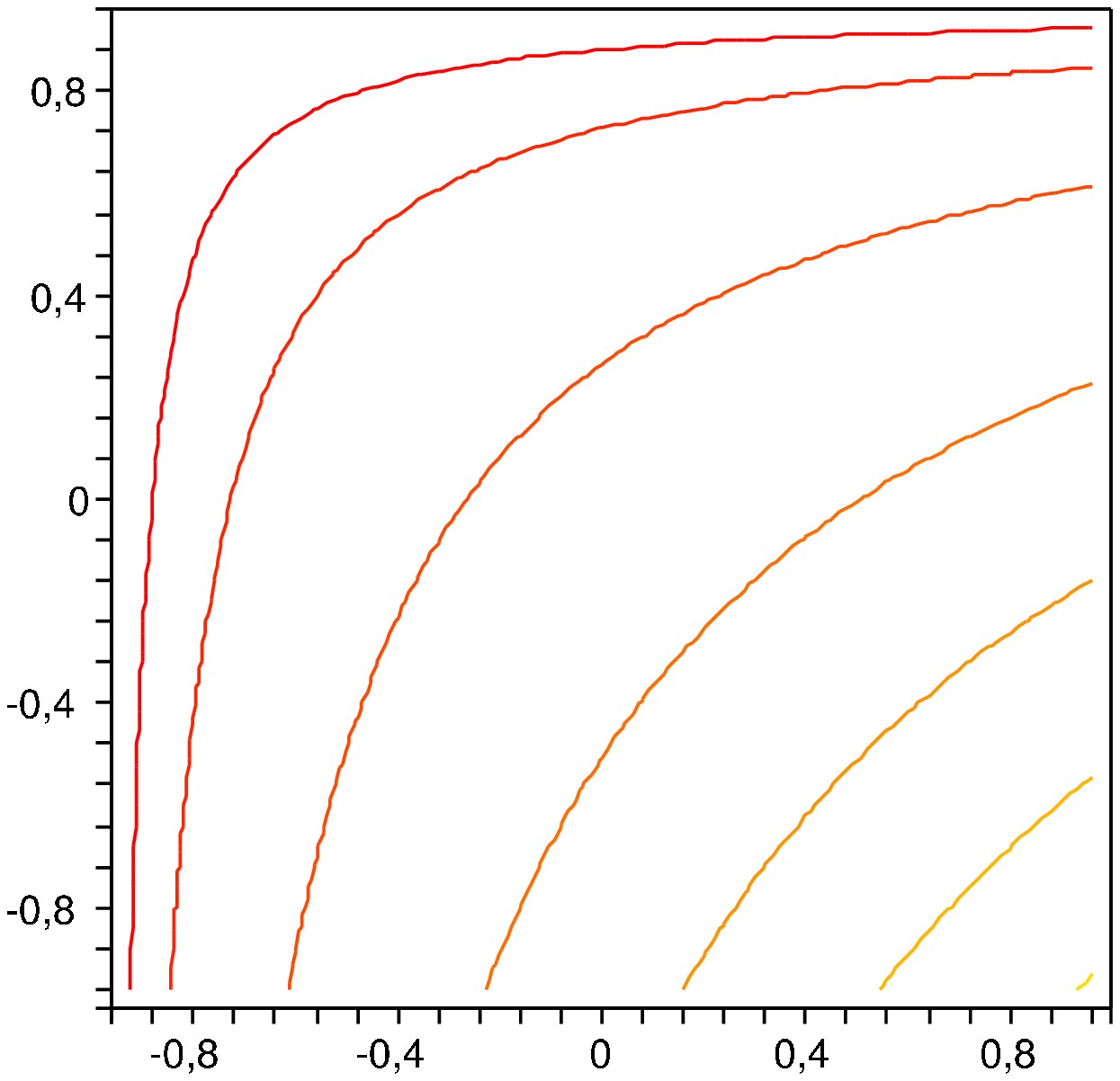}}
\put(-1.2,0.0){$P_{e^-}$}
\put(-8.0,6.5){$P_{e^+}$}
\put(-1.2,1.0){${\scriptstyle{25}}$}
\put(-1.8,1.5){${\scriptstyle{20}}$}
\put(-2.6,2.2){${\scriptstyle{15}}$}
\put(-3.4,3.1){${\scriptstyle{10}}$}
\put(-4.5,4.2){${\scriptstyle{5}}$}
\put(-5.5,5.2){${\scriptstyle{2}}$}
\put(-6.1,5.9){${\scriptstyle{1}}$}
}
\hspace{1mm}
\subfigure[Background cross section $\sigma(e^+e^- \to \nu \bar\nu \gamma)$ in fb.\label{fig:back220}]%
	{\scalebox{0.53}{\includegraphics{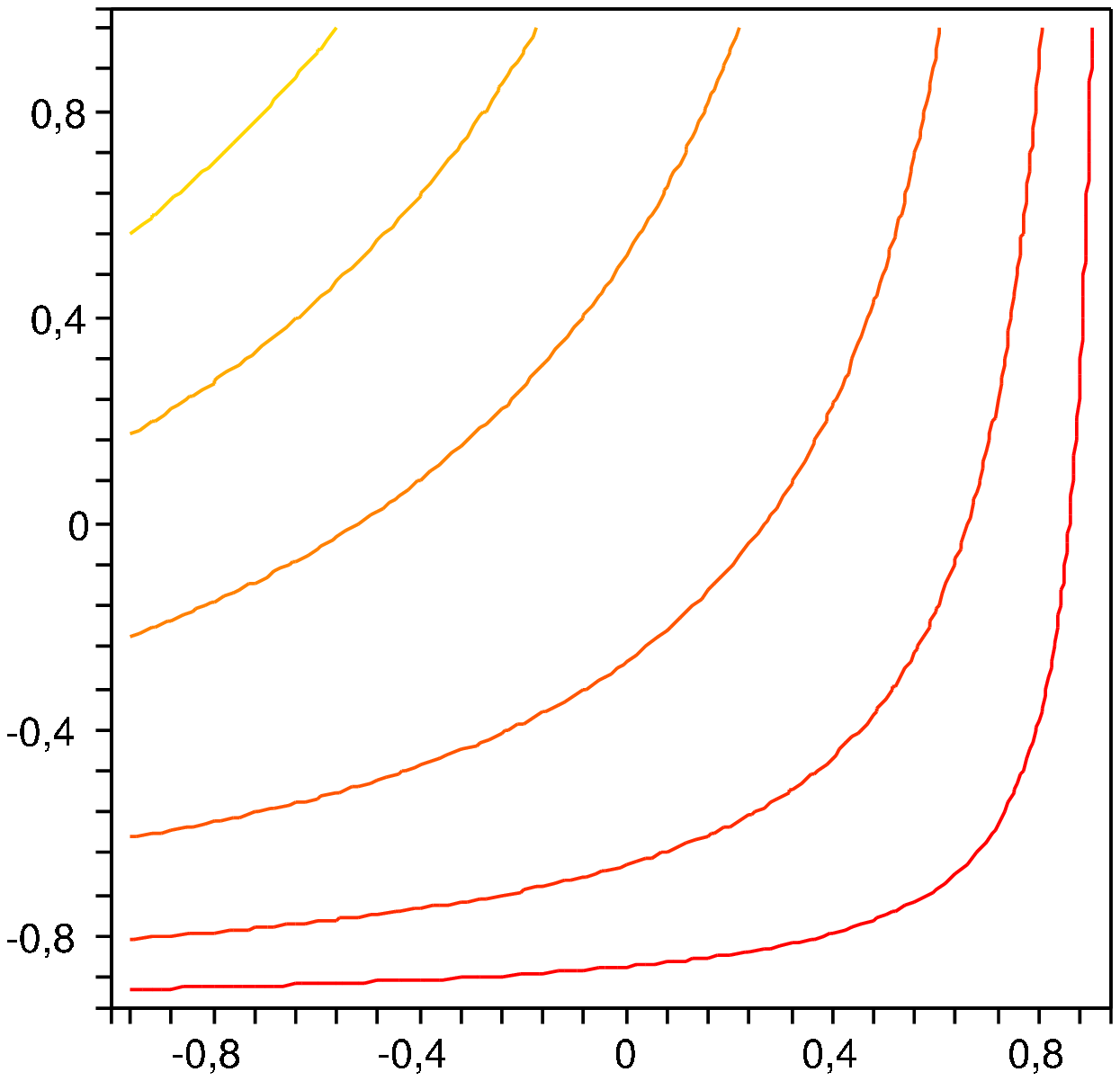}}
\put(-1.2,0){$P_{e^-}$}
\put(-8,6.5){$P_{e^+}$}
\put(-2.0,1.7){${\scriptstyle{500}}$}
\put(-2.9,2.4){${\scriptstyle{1000}}$}
\put(-3.8,3.2){${\scriptstyle{2000}}$}
\put(-4.9,4.4){${\scriptstyle{4000}}$}
\put(-5.8,5.3){${\scriptstyle{6000}}$}
\put(-6.4,6.1){${\scriptstyle{8000}}$}
}

\subfigure[Significance $S$.\label{fig:signi220}]{\scalebox{0.53}{\includegraphics{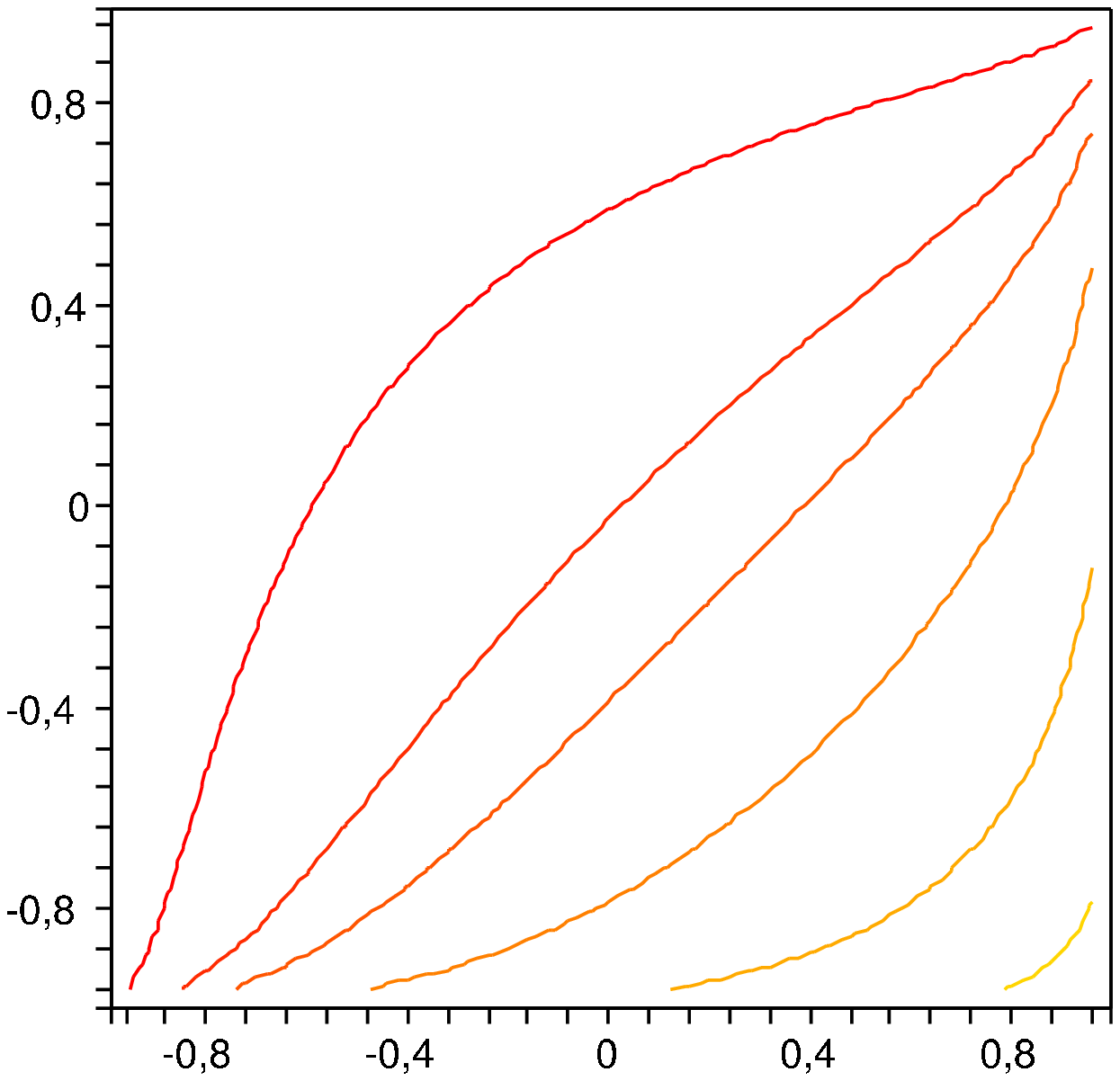}}
\put(-1.2,0){$P_{e^-}$}
\put(-8,6.5){$P_{e^+}$}
\put(-1.3,1.1){${\scriptstyle{30}}$}
\put(-2.0,1.6){${\scriptstyle{20}}$}
\put(-2.9,2.2){${\scriptstyle{10}}$}
\put(-3.7,2.8){${\scriptstyle{5}}$}
\put(-4.2,3.4){${\scriptstyle{3}}$}
\put(-5.3,4.4){${\scriptstyle{1}}$}
}
\hspace{1mm}
\subfigure[Signal to background ratio $r$ in \%. \label{fig:reli220}]{\scalebox{0.53}{\includegraphics{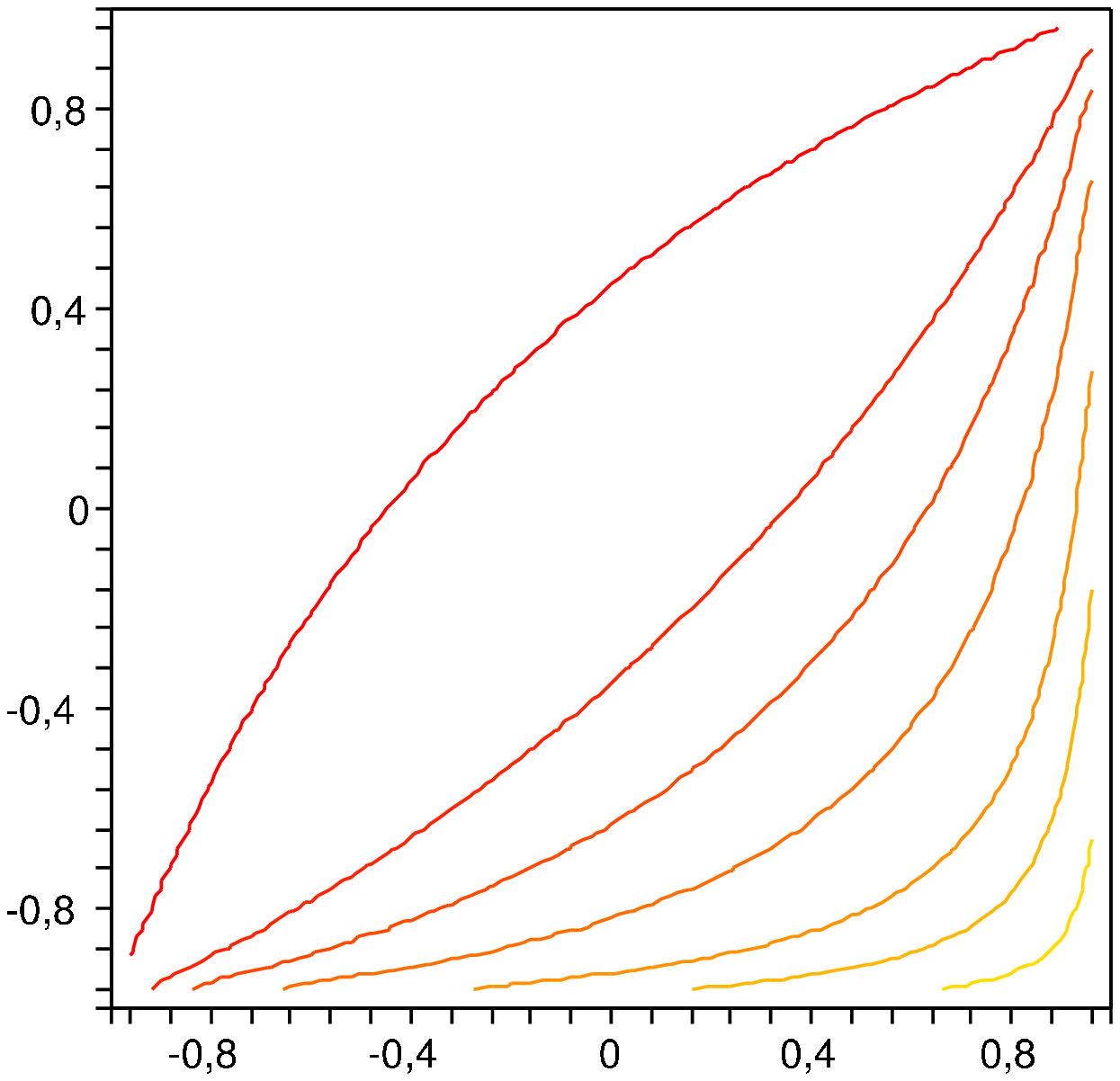}}
\put(-1.2,0){$P_{e^-}$}
\put(-8,6.5){$P_{e^+}$}
\put(-1.3,1.1){${\scriptstyle{8}}$}
\put(-1.7,1.3){${\scriptstyle{6}}$}
\put(-2.1,1.6){${\scriptstyle{4}}$}
\put(-2.6,2.0){${\scriptstyle{2}}$}
\put(-3.1,2.5){${\scriptstyle{1}}$}
\put(-3.7,3.1){${\scriptstyle{0.5}}$}
\put(-5.0,4.4){${\scriptstyle{0.1}}$}
}
\caption{Signal cross section (a), background cross section (b), 
significance (c), and signal to background ratio (d) for $\sqrt{s} =
220\GeV$, and an integrated luminosity $\mathcal{L}=500\fb^{-1}$ for
scenario A: $M_0 = 90\GeV$, $M_{1/2} = 225\GeV$, $A_0 = -90\GeV$, and
$\tan\beta = 10$, see Tables~\ref{tab:scenarios} and~\ref{tab:masses}.}
\label{fig:scenarioA}
\end{figure}
\end{center}

\begin{figure}[ht]
\setlength{\unitlength}{1cm}
\subfigure[Signal cross section $\sigma(e^+e^- \to \tilde\chi^0_1\tilde\chi^0_1\gamma)$ in fb.
\label{fig:sigma350}]{\scalebox{0.53}{\includegraphics{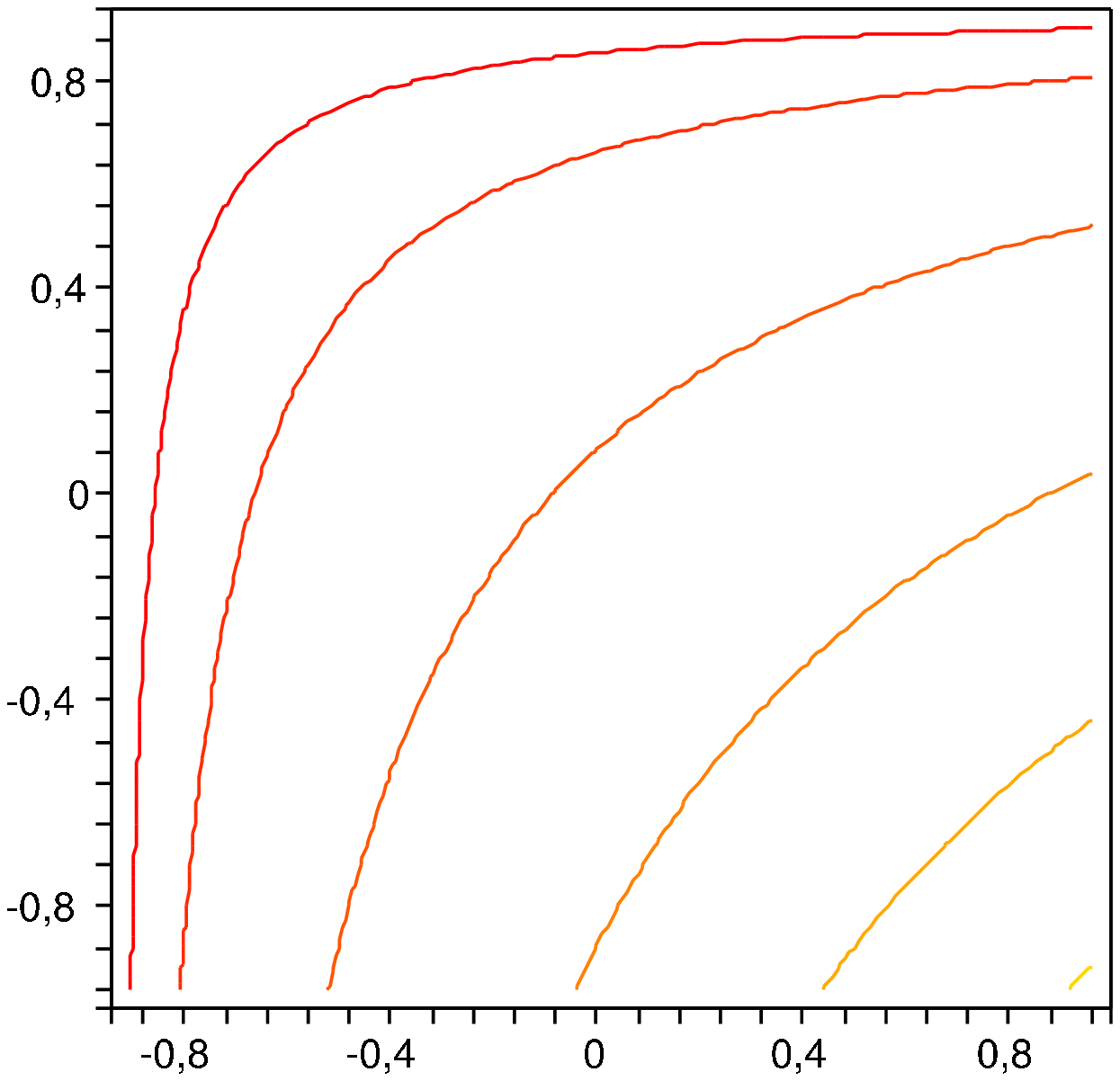}}%
\put(-1.2,0.0){$P_{e^-}$}
\put(-8.0,6.5){$P_{e^+}$}
\put(-1.2,1.0){${\scriptstyle{20}}$}
\put(-2.1,1.7){${\scriptstyle{15}}$}
\put(-3.0,2.7){${\scriptstyle{10}}$}
\put(-4.2,3.9){${\scriptstyle{5}}$}
\put(-5.4,5.1){${\scriptstyle{2}}$}
\put(-6.0,5.8){${\scriptstyle{1}}$}
}
\hspace{1mm}
\subfigure[Background cross section $\sigma(e^+e^- \to \nu \bar\nu \gamma)$ in fb.\label{fig:back350}]
	{\scalebox{0.53}{\includegraphics{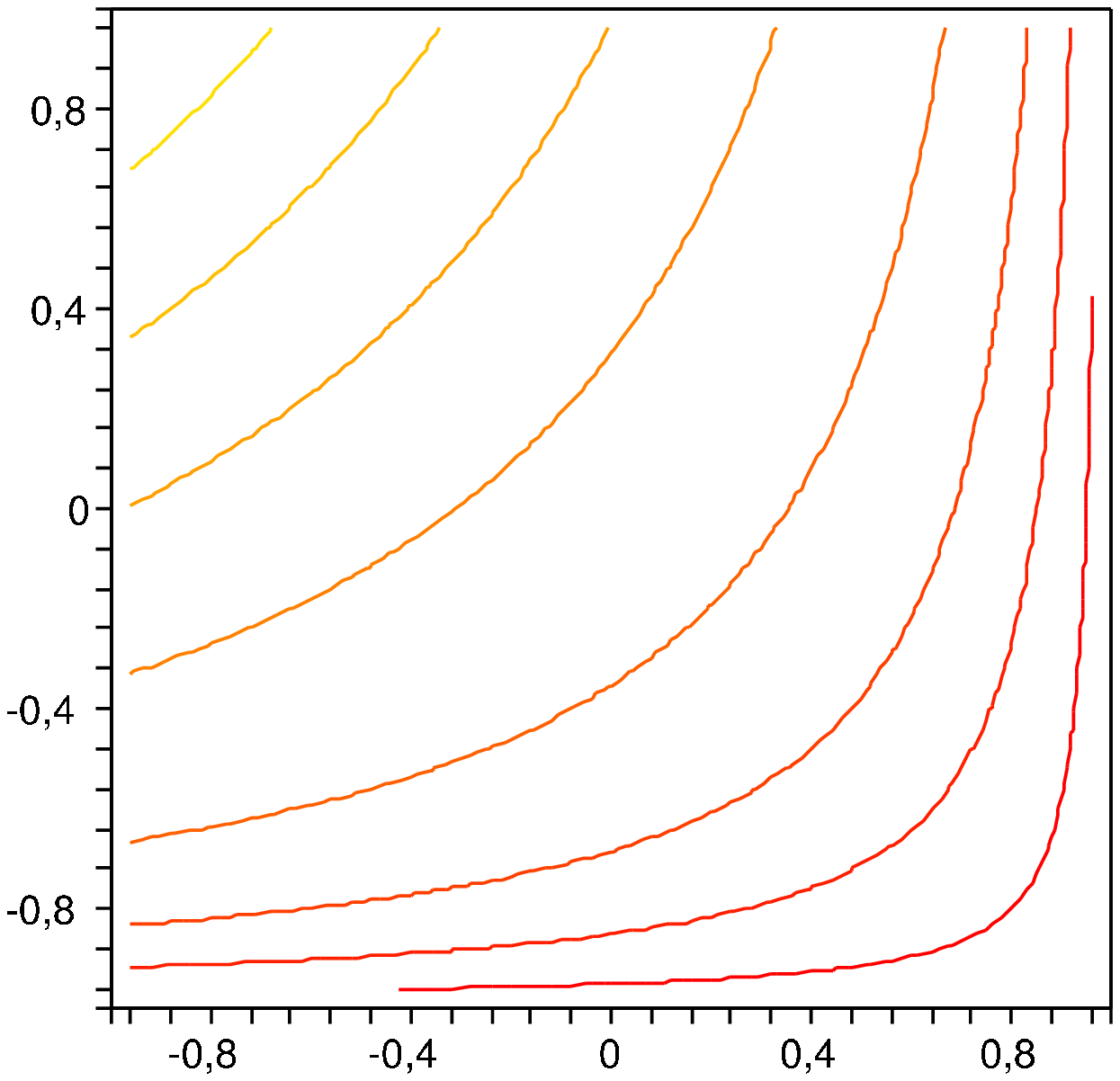}}
\put(-1.2,0){$P_{e^-}$}
\put(-8,6.5){$P_{e^+}$}
\put(-1.7,1.4){${\scriptstyle{200}}$}
\put(-2.3,2.0){${\scriptstyle{500}}$}
\put(-3.0,2.5){${\scriptstyle{1000}}$}
\put(-3.7,3.2){${\scriptstyle{2000}}$}
\put(-4.8,4.2){${\scriptstyle{4000}}$}
\put(-5.4,5.1){${\scriptstyle{6000}}$}
\put(-6.1,5.7){${\scriptstyle{8000}}$}
\put(-6.6,6.3){${\scriptstyle{10000}}$}
}

\subfigure[Significance $S$.\label{fig:signi350}]{\scalebox{0.53}{\includegraphics{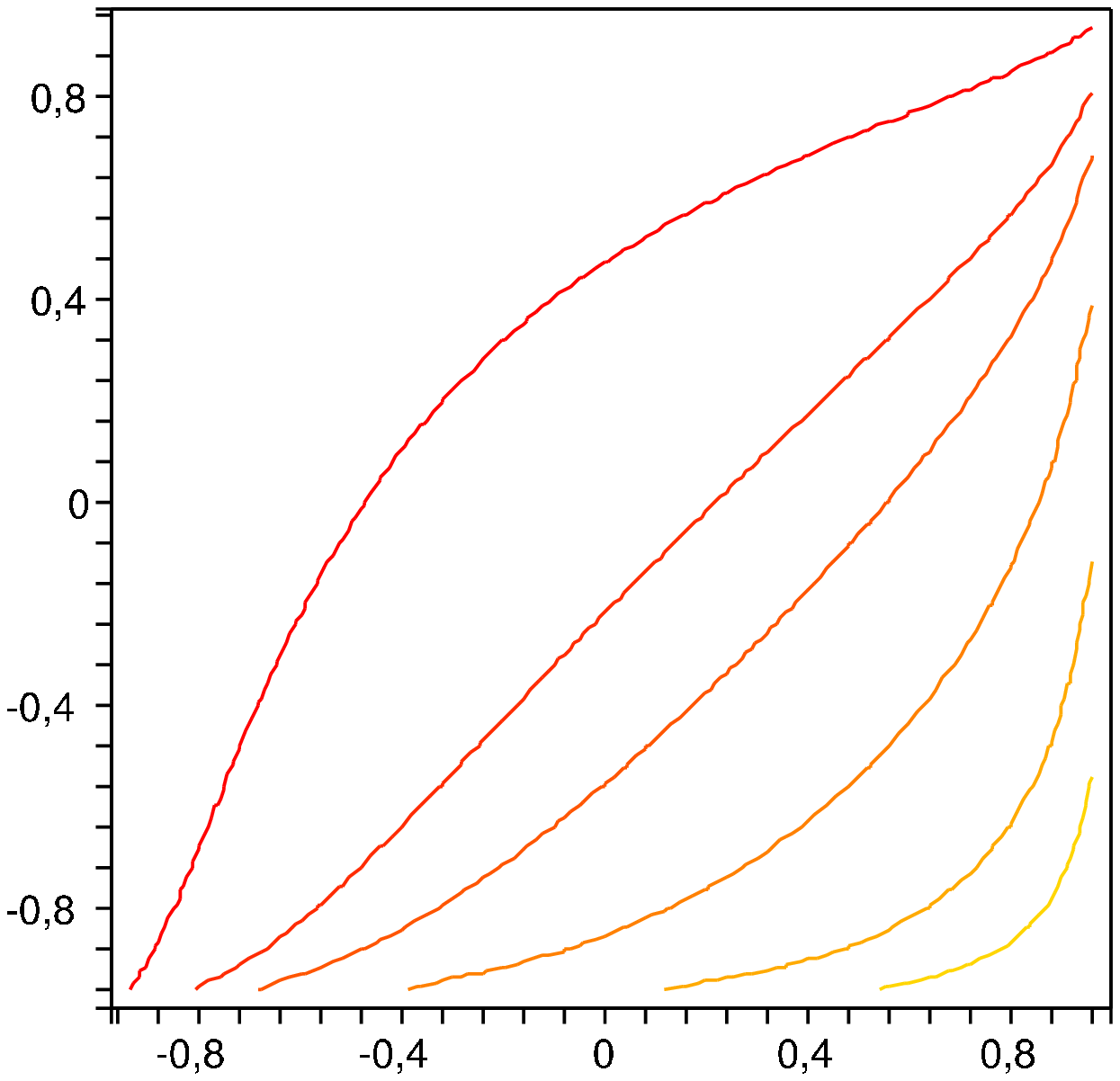}}
\put(-1.2,0){$P_{e^-}$}
\put(-8,6.5){$P_{e^+}$}
\put(-1.3,1.0){${\scriptstyle{30}}$}
\put(-1.9,1.5){${\scriptstyle{20}}$}
\put(-2.7,2.0){${\scriptstyle{10}}$}
\put(-3.4,2.6){${\scriptstyle{5}}$}
\put(-3.9,3.1){${\scriptstyle{3}}$}
\put(-5.1,4.2){${\scriptstyle{1}}$}
}
\hspace{1mm}
\subfigure[Signal to background ratio $r$ in \%.\label{fig:reli350}]{\scalebox{0.53}{\includegraphics{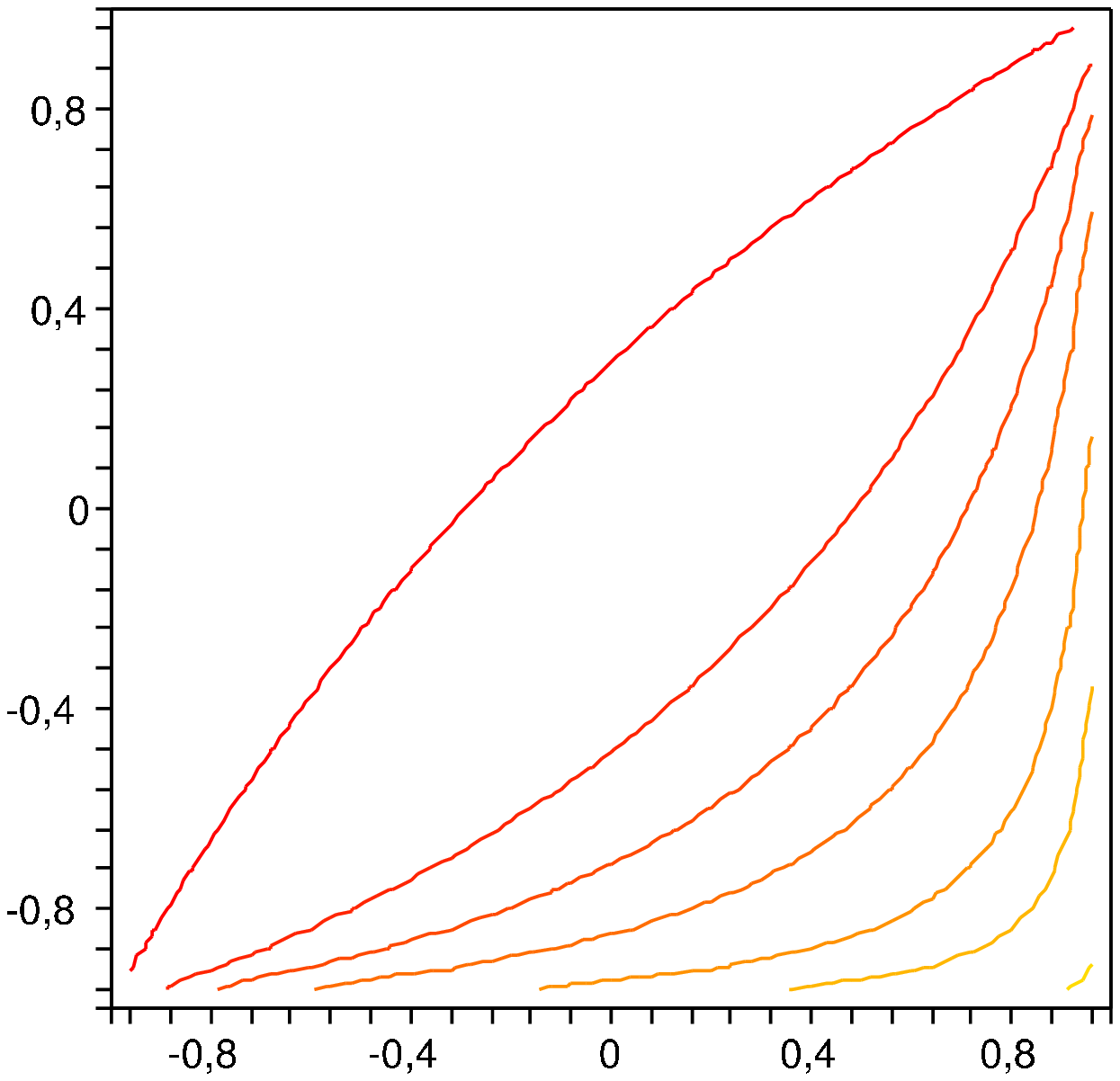}}
\put(-1.2,0){$P_{e^-}$}
\put(-8,6.5){$P_{e^+}$}
\put(-1.2,0.9){${\scriptstyle{20}}$}
\put(-1.7,1.2){${\scriptstyle{10}}$}
\put(-2.0,1.4){${\scriptstyle{5}}$}
\put(-2.5,1.9){${\scriptstyle{2}}$}
\put(-2.9,2.2){${\scriptstyle{1}}$}
\put(-3.6,2.7){${\scriptstyle{0.5}}$}
\put(-4.8,4.0){${\scriptstyle{0.1}}$}
}
\caption{Signal cross section (a), background cross section (b), significance (c), 
and signal to background ratio (d) for $\sqrt{s} = 350\GeV$, and an
integrated luminosity $\mathcal{L}=500\fb^{-1}$ for scenario B: $M_0 =
135\GeV$, $M_{1/2} = 325\GeV$, $A_0 = -135\GeV$, and $\tan\beta = 10$,
see Tables~\ref{tab:scenarios} and~\ref{tab:masses}.  }
\label{fig:scenarioB}
\end{figure}

\begin{figure}[ht]
\setlength{\unitlength}{1cm}
\subfigure[Signal cross section $\sigma(e^+e^- \to \tilde\chi^0_1\tilde\chi^0_1\gamma)$ in fb.\label{fig:sigma500}]
	{\scalebox{0.53}{\includegraphics{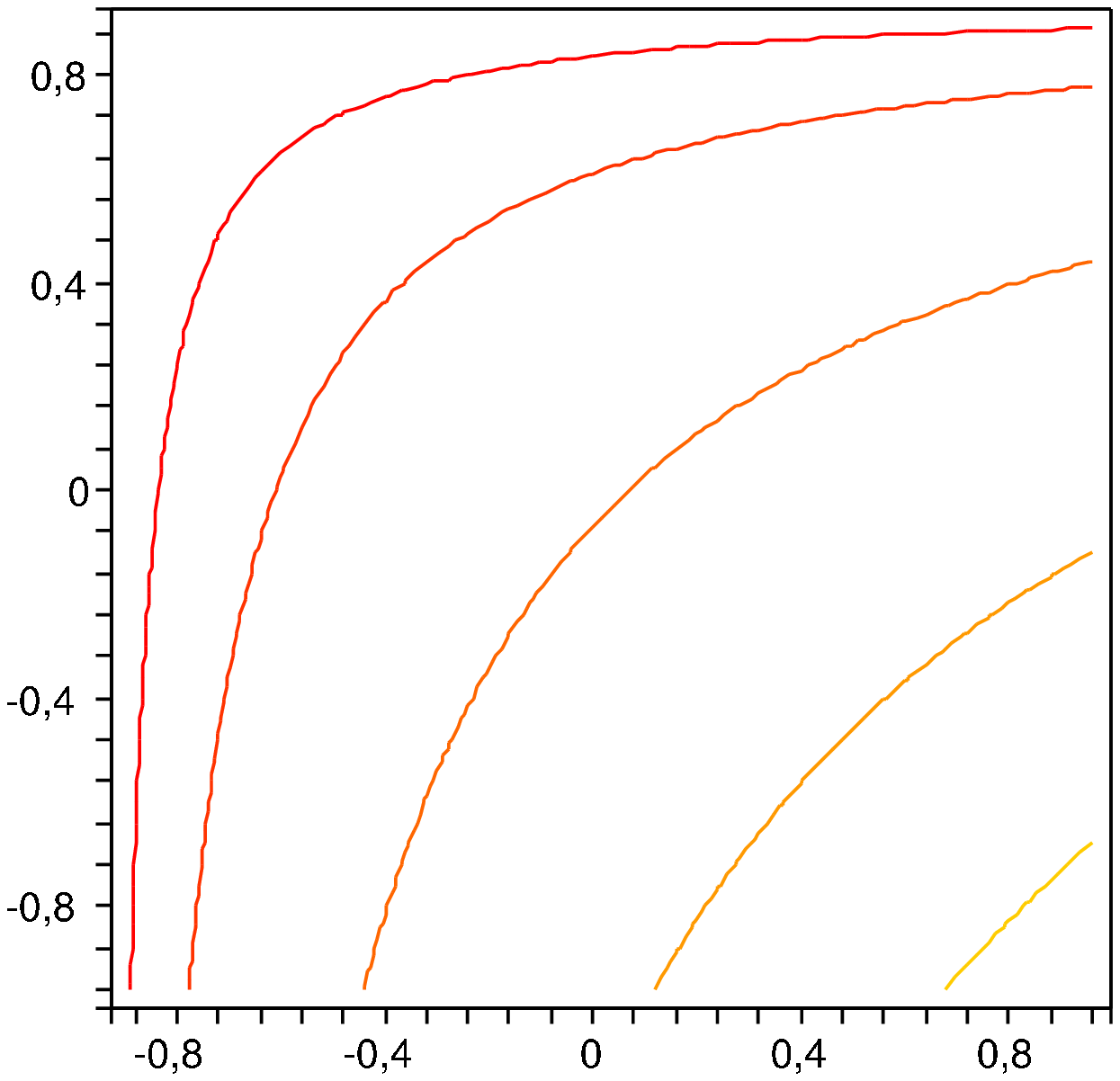}}%
\put(-1.2,0.0){$P_{e^-}$}
\put(-8.0,6.5){$P_{e^+}$}
\put(-1.6,1.3){${\scriptstyle{15}}$}
\put(-2.7,2.4){${\scriptstyle{10}}$}
\put(-4.0,3.7){${\scriptstyle{5}}$}
\put(-5.3,5.1){${\scriptstyle{2}}$}
\put(-6.0,5.8){${\scriptstyle{1}}$}	
}
\hspace{1mm}
\subfigure[Background cross section $\sigma(e^+e^- \to \nu \bar\nu \gamma)$ in fb.\label{fig:back500}]
	{\scalebox{0.53}{\includegraphics{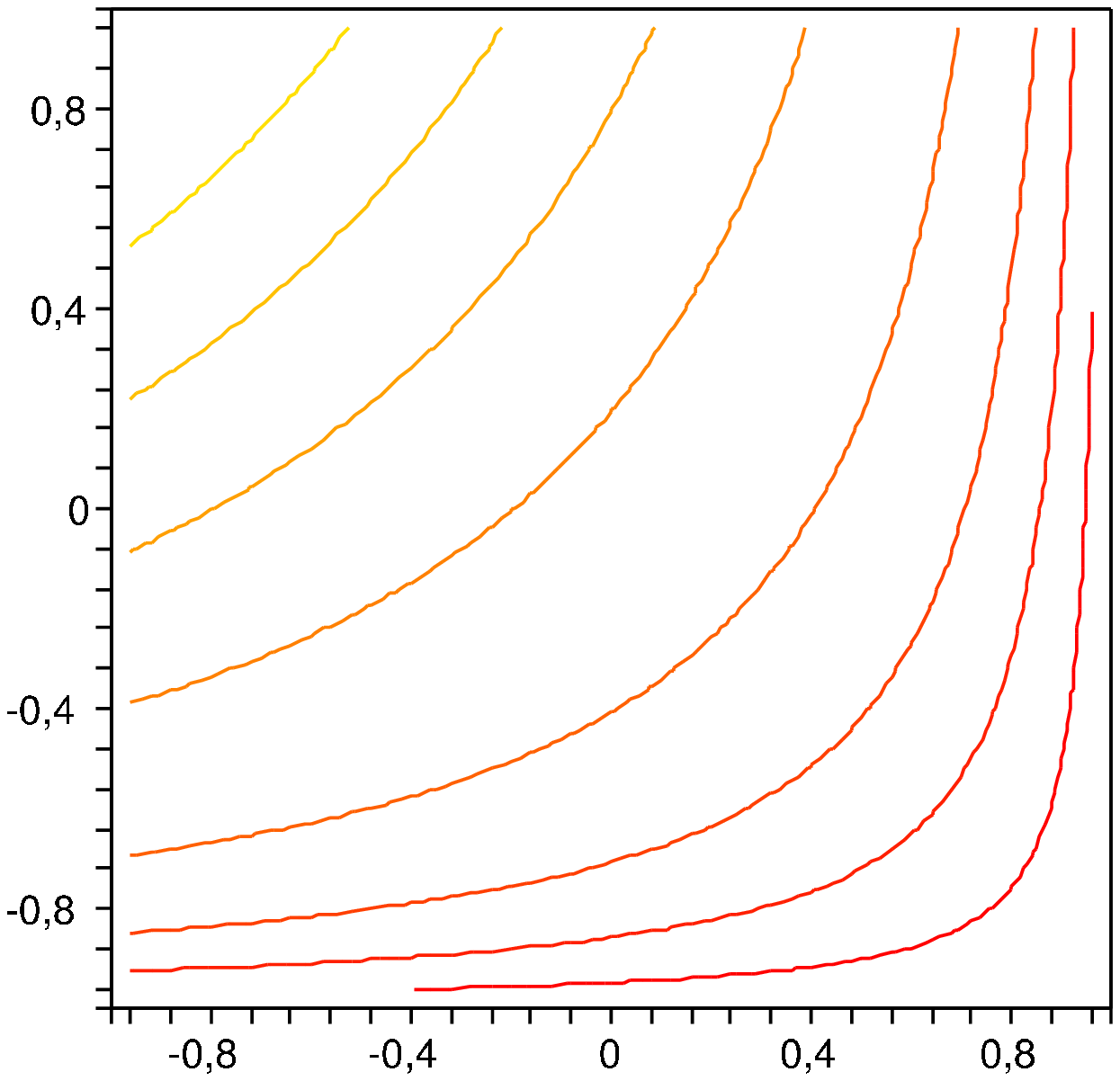}}
\put(-1.2,0){$P_{e^-}$}
\put(-8,6.5){$P_{e^+}$}
\put(-1.8,1.5 ){${\scriptstyle{200}}$}
\put(-2.2,2.0){${\scriptstyle{500}}$}
\put(-2.9,2.5){${\scriptstyle{1000}}$}
\put(-3.6,3.2){${\scriptstyle{2000}}$}
\put(-4.4,4.2){${\scriptstyle{4000}}$}
\put(-5.2,4.9){${\scriptstyle{6000}}$}
\put(-5.8,5.5){${\scriptstyle{8000}}$}
\put(-6.5,6.1){${\scriptstyle{10000}}$}
}

\subfigure[Significance $S$.\label{fig:signi500}]{\scalebox{0.53}{\includegraphics{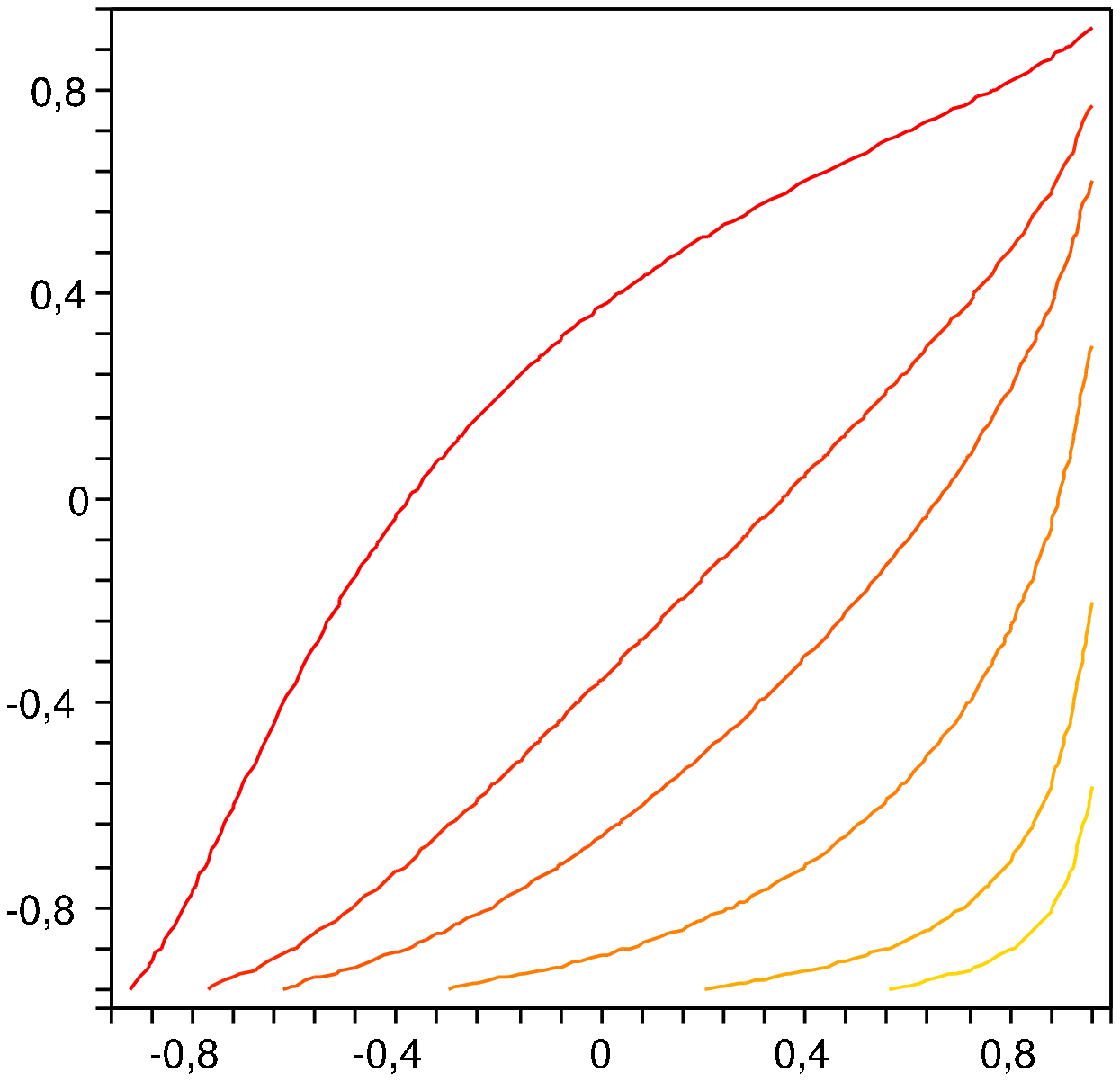}}
\put(-1.2,0){$P_{e^-}$}
\put(-8,6.5){$P_{e^+}$}
\put(-1.3,1.0){${\scriptstyle{30}}$}
\put(-1.7,1.6){${\scriptstyle{20}}$}
\put(-2.4,2.0){${\scriptstyle{10}}$}
\put(-3.1,2.5){${\scriptstyle{5}}$}
\put(-3.8,3.0){${\scriptstyle{3}}$}
\put(-5.0,4.0){${\scriptstyle{1}}$}
}
\hspace{1mm}
\subfigure[Signal to background ratio $r$ in \%.\label{fig:reli500}]{\scalebox{0.53}{\includegraphics{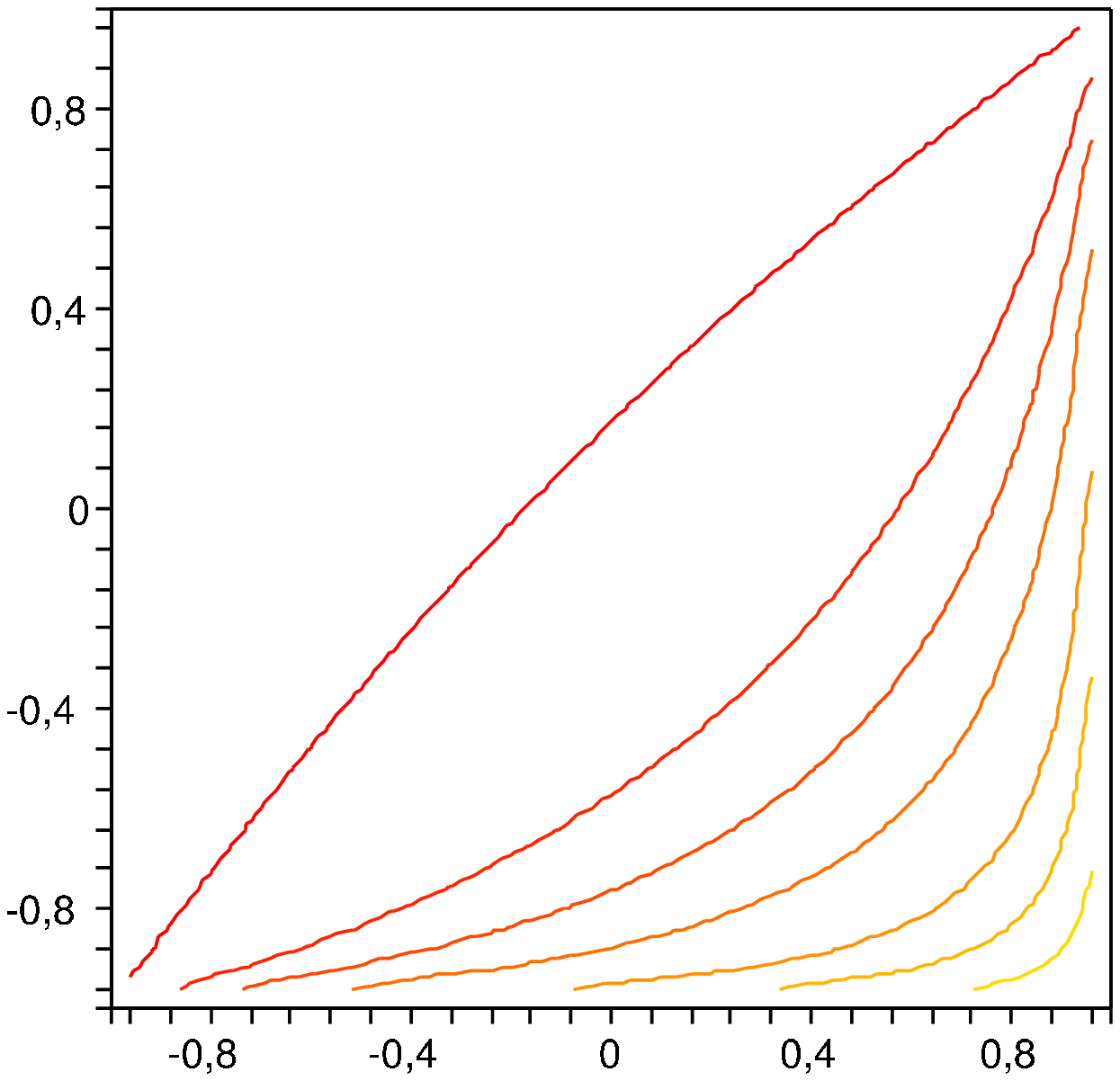}}
\put(-1.2,0){$P_{e^-}$}
\put(-8,6.5){$P_{e^+}$}
\put(-1.2,0.9){${\scriptstyle{20}}$}
\put(-1.7,1.2){${\scriptstyle{10}}$}
\put(-2.0,1.4){${\scriptstyle{5}}$}
\put(-2.4,1.8){${\scriptstyle{2}}$}
\put(-2.8,2.1){${\scriptstyle{1}}$}
\put(-3.4,2.6){${\scriptstyle{0.5}}$}
\put(-4.6,3.8){${\scriptstyle{0.1}}$}
}
\caption{Signal cross section (a), background cross section (b), significance (c), 
and signal to background ratio (d) for $\sqrt{s} = 500\GeV$, and an
integrated luminosity$\mathcal{L}=500\fb^{-1}$ for scenario C: $M_0 =
200\GeV$, $M_{1/2} = 415\GeV$, $A_0 = -200\GeV$, and $\tan\beta = 10$,
see Tables~\ref{tab:scenarios} and~\ref{tab:masses}. }
\label{fig:scenarioC}
\end{figure}

Note that the lightest neutralino, $\tilde\chi_1^0$, is mostly bino in
all three scenarios; $98\%$ in scenario A, $99.1\%$ in scenario B, and
$99.5\%$ in scenario C.  Thus in my scenarios, radiative neutralino
production proceeds mainly via right selectron exchange in the $t$ and
$u$ channel. Left selectron exchange and $Z$ boson exchange are
severely suppressed~\cite{Dreiner:2006sb}.  The background process
$e^+e^- \to \nu \bar\nu \gamma$ mainly proceeds via $W$ boson
exchange.  Thus positive electron beam polarization $P_{e^-}>0$ and
negative positron beam polarization $P_{e^+}<0$ should enhance the
signal rate and reduce the background at the same
time~\cite{Choi:1999bs,Dreiner:2006sb}. This effect is clearly
observed in Figs.~\ref{fig:scenarioA}, \ref{fig:scenarioB}, and
\ref{fig:scenarioC} for all scenarios. The signal cross section and
the background vary by more than one order of magnitude over the full
polarization range. 

\medskip

\begin{table}[htb]
\setlength{\belowcaptionskip}{11pt}
\centering
\caption{\small Definition of the mSUGRA scenarios A, B, and C. All values 
are given in GeV. I have fixed $\tan\beta=10$. For
completeness I have included the corresponding value of $\sqrt{s}$
for each scenario.}
\label{tab:scenarios}
\begin{tabular}{cc|ccc|ccc}
\hline 
scenario&$\sqrt{s}$ & $M_0$ & $M_{1/2}$ & $A_0$ & $M_1$ & $M_2$ &
$\mu$ \\[2mm]
\hline 
A & $220$& $90$  & $225$ & $-90$  & $97.5$ &$188$ & $316$ \\[2mm]
B & $350$& $135$ & $325$ & $-135$ & $143$  &$272$ & $444$ \\[2mm]
C & $500$& $200$ & $415$ & $-200$ & $184$  &$349$ & $560$ \\[2mm]
\hline 
\end{tabular}
\end{table}

\begin{table}[htb]
\setlength{\belowcaptionskip}{11pt}
\centering
\caption{\small Spectrum of the lighter SUSY particles for scenarios 
A, B, and C, calculated with SPheno~\cite{Porod:2003um}. All values
are given in GeV. For completeness I have included the corresponding
value of $\sqrt{s}$ for each scenario.}
\label{tab:masses}
\begin{tabular}{cc|ccccccc}
\hline
scenario&$\sqrt{s}$&$m_{\chi_1^0}$ &$m_{\chi_2^0}$ &$m_{\chi_1^\pm}$ & 
 $m_{\tilde{\tau}_1}$   &$m_{\tilde{e}_R}$ & $m_{\tilde{e}_L}$ &$m_{\tilde{\nu}}$ \\[2mm]
\hline 
A & $220$& $92.4$ & $172$ & $172$ & $124$ & $133$ & $189$ & $171$ \\[2mm]
B & $350$& $138$  & $263$ & $263$ & $183$ & $191$ & $270$ & $258$ \\[2mm]
C & $500$& $180$  & $344$ & $344$ & $253$ & $261$ & $356$ & $347$ \\[2mm]
\hline 
\end{tabular}
\end{table} 
\medskip

\begin{table}[t]
\setlength{\belowcaptionskip}{11pt}
\centering
\caption{Cross sections $\sigma$, significance $S$, and signal to 
background ratio $r$ for different beam polarizations $(P_{e^-}|
P_{e^+})$ for Scenario A at $\sqrt{s} = 220\GeV$, with
$\mathcal{L}=500\fb^{-1}$.}
\label{tab:xa}
\begin{tabular}{c|cccccc}
\hline
 Scenario A &$(0|0)$&$(0|0.8)$&$(-0.3|0.8)$&$(-0.6|0.8)$&$(0|0.9)$&$(-0.3|0.9)$\\[1mm]
\hline
$\sigma(\signal)$&$6.7\fb$&$12\fb$&$16\fb$&$19\fb$&$13\fb$&$16\fb$\\[1mm]
$\sigma(\Backgroundnu)$& $2685\fb$ & $652\fb$ & $534\fb$ & $416\fb$&$398\fb$&$360\fb$\\[1mm] 
$S$ & $2.9$ &$10$ & $15$ & $20$&$14$& $19$\\[1mm]
$r$ & $0.3\%$ & $1.8\%$ & $2.9\%$ & $4.6\%$&$3.2\%$&$4.6\%$\\[1mm]
\hline 
\end{tabular}
\end{table} 

\begin{table}[ht]
\setlength{\belowcaptionskip}{11pt}
\centering
\caption{Cross sections $\sigma$, significance $S$, and signal to 
background ratio $r$ for different beam polarizations $(P_{e^-}|P_{
e^+})$ for Scenario B at $\sqrt{s} = 350\GeV$, with $\mathcal{L}=
500\fb^{-1}$.}
\label{tab:xb}
\begin{tabular}{c|cccccc}
\hline
Scenario B &$(0|0)$&$(0|0.8)$&$(-0.3|0.8)$&$(-0.6|0.8)$&$(0|0.9)$&$(-0.3|0.9)$\\[1mm]
\hline
$\sigma(\signal)$&$5.5\fb$&$9.6\fb$&$13\fb$&$15\fb$&$10.2\fb$&$13.3\fb$\\[1mm]
$\sigma(\Backgroundnu)$& $3064\fb$ & $651\fb$ & $481\fb$ & $312\fb$&$350\fb$&$272\fb$\\[1mm] 
$S$ & $2.2$ &$8.4$ & $13$ & $19$ & $12$ & $18$\\[1mm]
$r$ & $0.2\%$ & $1.5\%$ & $2.6\%$ & $4.9\%$& $2.9\%$ & $4.9\%$\\[1mm]
\hline 
\end{tabular}

\end{table} 
\begin{table}[h!]
\setlength{\belowcaptionskip}{11pt}
\centering
\caption{Cross sections $\sigma$, significance $S$, and signal to background ratio $r$ 
for different beam polarizations $(P_{e^-}|P_{e^+})$ for Scenario C at $\sqrt{s} = 500\GeV$, with
$\mathcal{L}=500\fb^{-1}$.}
\label{tab:xc}
\begin{tabular}{c|cccccc}
\hline
Scenario C &$(0|0)$&$(0|0.8)$&$(-0.3|0.8)$&$(-0.6|0.8)$&$(0|0.9)$&$(-0.3|0.9)$\\[1mm]
\hline
$\sigma(\signal)$&$4.7\fb$&$8.2\fb$&$11\fb$&$13\fb$& $8.6\fb$&$11.2\fb$\\[1mm]
$\sigma(\Backgroundnu)$& $3354\fb$ & $689\fb$ & $495\fb$ & $301\fb$& $356\fb$&$263\fb$\\[1mm] 
$S$ & $1.8$ &$7$ & $11$ & $17$& $10$ & $15$\\[1mm]
$r$ & $0.1\%$ & $1.2\%$ & $2.2\%$ & $4.4\%$ & $2.4\%$ & $4.3\%$\\[1mm]
\hline 
\end{tabular}
\end{table} 

For scenario A, I show the beam polarization dependence of the signal
cross section $\sigma(e^+e^- \to \tilde\chi^0_1\tilde\chi^0_1\gamma)$
in Fig.~\ref{fig:sigma220}, and the dependence of the background cross
section $\sigma(e^+e^- \to \nu \bar\nu \gamma)$ in
Fig.~\ref{fig:back220}. In both cases I have implemented the cuts of
Eq.~(\ref{cuts}). The cont lines in the $P_{e^-}
$-$P_{e^+}$ plane of the significance $S$, Eq.~(\ref{significance}),
and the signal to background ratio $r$, Eq.~(\ref{sigbck}), are shown
in Figs.~\ref{fig:signi220} and \ref{fig:reli220} respectively. The
results for scenario B are shown in Fig.~\ref{fig:scenarioB}, and
those for scenario C are shown in Fig.~\ref{fig:scenarioC}.

\medskip

In order to quantify the behaviour, I give the values for the signal
and background cross sections, the significance $S$ and the signal to
background ratio $r$ for a specific set of beam polarizations $(P_{e^+
}|P_{e^-}) = (0|0)$, $(0|0.8)$, $(-0.3|0.8)$, $(-0.6|0.8)$, $(0|0.9)$,
and $(-0.3|0.9)$ in Tables~\ref{tab:xa}, \ref{tab:xb}, \ref{tab:xc} for
the scenarios A, B, and C, respectively.  I find that an additional
positron polarization $P_{e^+} = -30\%$ enhances the significance $S$
by factors $\{ 1.5,1.5,1.6\}$ in scenarios $\{{\rm A,B,C}\}$,
respectively, compared to beams with only $e^-$ polarization $(P_{e^+}
|P_{e^-}) =(0|0.8)$, and by factors $\{1.4, 1.5, 1.5\}$ in scenarios
$\{{\rm A,B,C}\}$, respectively, for $(P_{e^+}|P_{e^-}) =(-0.3|0.9)$
compared to $(P_{e^+}|P_{e^-}) =(0|0.9)$.  The signal to background
ratio $r$ is enhanced by $\{ 1.7,1.7,1.8\}$ for $(P_{e^+}|P_{e^-})=
(-0.3|0.8)$ compared to $(P_{e^+}|P_{e^-})= (0|0.8)$ and by
$\{1.4,1.7,1.8\}$ for $(P_{e^+}|P_{e^-})= (-0.3|0.9)$ compared to
$(P_{e^+}|P_{e^-})= (0|0.9)$.  If the positron beams would be
polarized by $P_{e^+} = -60\%$, the enhancement factors for $S$ are
$\{ 2,2.3,2.4\}$, and for $r$ they are $\{ 2.5,3.2,3.6\}$. For
$P_{e^-}= 0.8$, it is only with positron polarization that I obtain
values of $r$ clearly above 1\%. If I have $P_{e^-}= 0.9$, then $r$
exceeds $1\%$ without positron beam polarization.

\medskip

Since the neutralinos are mainly bino, the signal cross section also
depends sensitively on the mass $m_{\tilde{e}_R}$ of the right
selectron.  In scenarios $\{{\rm A,B,C}\}$ the masses are
$m_{\tilde{e}_R}=\{133,191,261\}$~GeV, respectively, see
Table~\ref{tab:masses}.  For larger masses, the signal to background
ratio drops below $r<1\%$. With $(P_{e^+}|P_{e^-}) =(-0.3|0.8)$, this
happens for $m_{\tilde{e}_R}=\{214,300,390\}$~GeV, and the
significance would be $S<5$. These selectron masses correspond to the
mSUGRA parameter $M_0=\{190,270,350\}$~GeV.

\subsection{Summary and Conclusions\label{Summary}}

I have studied radiative neutralino production $e^+e^- \to\tilde\chi
^0_1\tilde\chi^0_1\gamma$ at the ILC with longitudinally polarized
beams. For the center-of-mass energies $\sqrt{s} = 220\GeV$, $350\GeV
$, and $500\GeV$, I have considered three specific mSUGRA inspired
scenarios. In my scenarios, only radiative neutralino production is
kinematically accessible, since the other supersymmetric particles are
too heavy to be pair produced. I have investigated the beam
polarization dependence of the cross section from radiative neutralino
production and the background form radiative neutrino production
$e^+e^- \to \nu\bar\nu \gamma$.

\medskip

I have shown that polarized beams enhance the signal and suppress the
background simultaneously and significantly. In my scenarios, the
signal cross section for $(P_{e^+}|P_{e^-}) = (-0.3|0.8)$ is larger
than $10\fb$, the significance $S>10$, and the signal to background
ratio is about $2-3\%$. The background cross section can be reduced to
$500\fb$.  Increasing the positron beam polarization to $P_{e^+} =
-0.6$, both the signal cross section and the significance increase by
about $25\%$, in my scenarios. For $(P_{e^+}|P_{e^-}) = (0.0|0.9)$
the radiative neutralino production signature is observable at the ILC
but both the significance and the signal to background ratio are
considerable improved for $(P_{e^+}|P_{e^-}) = (-0.3|0.9)$, making
more detailed investigations possible. The electron and positron beam
polarization at the ILC are thus essential tools to observe radiative
neutralino production. For unpolarized beams this process cannot be
measured.

\medskip

I conclude that radiative neutralino production can and should 
be studied at $\sqrt{s} =500\GeV$, as well as at the lower energies
$\sqrt{s} = 220\GeV$ and  $\sqrt{s} =350\GeV$,
which are relevant for Higgs and top physics.
I have shown that for these energies there are scenarios, 
where other SUSY particles 
like heavier neutralinos, charginos and sleptons are too heavy to
be pair produced. 
In any case, a pair of radiatively produced neutralinos
is the lightest accessible state of SUSY particles to be produced at 
the linear collider.

%

\section{Summary and Conclusions \label{sec:conclusion}}

I have studied radiative neutralino production $e^+e^- \to
\tilde\chi^0_1 \tilde\chi^0_1\gamma$ at the linear collider with
polarised beams.  I have considered the Standard Model background
process $e^+e^- \to \nu \bar\nu \gamma$ and the SUSY background
$e^+e^- \to \tilde\nu \tilde\nu^\ast \gamma$, which also has the
signature of a high energetic photon and missing energy, if the
sneutrinos decay invisibly. For these processes I have given the
complete tree-level amplitudes and the full squared matrix elements
including longitudinal polarisations from the electron and positron
beam.  In the MSSM, I have studied the dependence of the cross
sections on the beam polarisations, on the gaugino and higgsino mass
parameters $M_2$ and $\mu$, as well as the dependence on the selectron
masses.  Finally, in order to quantify whether an excess of signal
photons, $N_{\mathrm{S}}$, can be measured over the background
photons, $N_{\rm B}$, from radiative neutrino production, I have
analysed the theoretical statistical significance $S=N_{\rm S}/\sqrt{
N_{\rm S} + N_{\rm B}}$ and the signal to background ratio $N_{\rm S}
/ N_{\rm B}$.  Our results can be summarised as follows.
 
\begin{itemize}
\item The cross section for $e^+e^- \to \tilde\chi^0_1 \tilde\chi^0_1
\gamma$ reaches up to $100\fb$ in the $\mu$-$M_2$ and the $m_0$-$M_2$ 
plane at $\sqrt{s} = 500\GeV$. The significance can be as large as
$120$, for a luminosity of $\mathcal{L} = 500\fb^{-1}$, such that
radiative neutralino production should be well accessible at the ILC.

\item At the ILC, electron and positron beam polarisations can be used
to significantly enhance the signal and suppress the background
simultaneously.  I have shown that the significance can then be
increased almost by an order of magnitude, e.g., with $(P_{e^-},P_{e^+}
)=(0.8,-0.6)$ compared to $(P_{e^-},P_{e^+})=(0,0)$.  In the SPS~1a 
scenario the cross section $\sigma(\signal)$ increases from $25\fb$ to
$70\fb$ with polarised beams, whereas the background $\sigma(e^+e^-\to
\nu \bar\nu\gamma)$ is reduced from $3600\fb$ to $330\fb$.  Although a
polarised positron beam is not essential to study radiative neutralino 
production at the ILC, it will help to increase statistics.
\item I note that charginos and heavier neutralinos could be too heavy to be
pair-produced at the ILC in the first stage at $\sqrt{s} = 500$~GeV.
If only slepton pairs are accessible, the radiative production of the lightest
neutralino might be the only SUSY process to study the neutralino
sector.  Even in the regions of the parameter space near the
kinematical limits of $\tilde{\chi}^0_1$ - $\tilde{\chi}^0_2$ pair production 
I find a cross section of about $20\fb$ and
corresponding significances up to $20$.
\item Finally I want to remark that my given values for the 
statistical significance $S$ can only be seen as rough estimates,
since I do not include a detector simulation. However, since I have
obtained large values up to $S\approx 120$, I hope that my results
encourage further experimental studies, including detailed Monte Carlo
simulations.

\end{itemize}
\newpage  


%
\chapter{Magic Neutralino Squares}
\label{ch:magic}

\section{Introduction}
If supersymmetric particles are discovered,
the underlying SUSY parameters can  be determined from measurements of cross
sections, particle masses, decay widths, and branching ratios.
Many authors developed methods and programs to extract the parameters
from these measurements. In the following, I shall give an overview
over the methods, concentrating on the gaugino sector.  
\medskip

Choi et al. analyse in~\cite{Choi:2000ta} the chargino system. They present
an analytical method to extract the parameters $M_2$, $\mu$, and $\tan\beta$ 
of the chargino mixing matrix
from chargino pair production in $e^+e^-$ annihilation with polarized beams.
The absolute errors on $M_2$ and $\mu$ are of the order of $\GeV$, and the error on
$\tan\beta$ is $\mathcal{O}(1)$ if $\tan\beta$ is not too large.
\medskip

In~\cite{Choi:2001ww} the analysis has been extended to the neutralino system
to obtain the bino mass parameter $M_1$. As I shall demonstrate later, this method
demands chargino parameters and neutralino masses measured with an accuracy 
$\mathcal{O}(0.1\GeV)$, which is not feasible in the first run of the ILC.
\medskip 

Desch et al. present in~\cite{Desch:2003vw,Bechtle:2005vt,Bechtle:2005qj} 
a study to determine the parameters
$M_1$, $M_2$, $\mu$, and $\tan\beta$ from a fit of the light and heavy
neutralino and chargino masses to LHC and 
LC\footnote{In 2002, the terminology was "LC". Today, we are talking about the "ILC".} data. 
They present formulae to determine $M_2$, $\mu$, and $\tan\beta$ from the chargino 
masses and from cross sections with left $(P_+|P_-)=(0.6|-0.8)$ and right  
$(P_+|P_-)=(-0.6|0.8)$ longitudinally polarised beams. $M_1$ is obtained from 
the polarised cross sections $\sigma(e^+e^- \rightarrow \tilde{\chi}_1^0\tilde{\chi}_2^0)$
and $\sigma(e^+e^- \rightarrow \tilde{\chi}_2^0\tilde{\chi}_2^0)$. 
They simulated the parameter determination with an LC measurement at the SPS1a 
point~\cite{Ghodbane:2002kg,Allanach:2002nj}.
They could recover the input data with absolute errors $\mathcal{O}(0.1\GeV)$ for $M_1$ 
and $M_2$, $\mathcal{O}(1-10\GeV)$ for $\mu$, and $\mathcal{O}(1)$ for $\tan\beta$.
Combining the analysis with LHC data reduces the errors on these parameters 
by a factor of about $2$.
\medskip

Bechtle et al. present in~\cite{Bechtle:2004pc} the program {\texttt{Fittino}}.
It performs non-linear fits to observables such as masses, cross sections, branching fractions,
widths and edges in mass spectra to determine the SUSY parameters.
More about non-linear fits can be found in Ref.~\cite{nonlinear}.
The authors implemented an iterative fitting technique and the simulated annealing
algorithm to obtain the fit parameters. In~\cite{Bechtle:2005qj} they present
an example calculation using the SPS1a point. They recovered the input data with errors
of about $\mathcal{O}(0.01\GeV)$ for $M_1$ and $M_2$, $\mathcal{O}(1\GeV)$ for $\mu$,
and $\mathcal{O}(0.1)$ for $\tan\beta$.
\medskip

In~\cite{Rolbiecki:2006tx} the authors discuss the parameter determination in a focus
point inspired scenario. The slepton and squark  masses are about $2000\GeV$,
which is even heavier than the particle spectrum
of the SPS2 point. The determination of $M_1$ and $M_2$ succeeds with an error
of $\mathcal{O}(0.1-1\GeV)$, but the errors on $\mu$ and $\tan\beta$ are $\mathcal{O}(10-100\GeV)$
and $\mathcal{O}(10)$, respectively.  
\medskip

\texttt{Sfitter}~\cite{Lafaye:2004cn} is another program to extract SUSY parameters
from particle masses using a fit or multi-dimensional grid or both.
The fit is performed assuming the mSUGRA parameters. 
In a simulation with the SPS1a
point they get errors of about $\mathcal{O}(1 \GeV)$ for $M_1$, $M_2$, and $\mu$;
the error on $\tan\beta$ is $\approx 3$.
\medskip

In~\cite{Lester:2005je} the authors use Markov chain techniques to determine
SUSY mass measurements from simulated ATLAS data.
\medskip

The Supersymmetry Parameter Anaylsis project (SPA)~\cite{Aguilar-Saavedra:2005pw}
provides a common framework for parameter determination. The authors  present a list
of computational tools to perform the required calculation: These are tools to
translate between calculational schemes, spectrum calculators, calculators
for cross sections, decay widths etc., and event generators, parameter analysis programs,
RGE programs, and auxiliary programs.  
The authors define the tasks of the SPA project as follows: 
promoting higher order SUSY calculation, 
improving the understanding of the $\overline{\text{DR}}$ scheme, 
improving experimental and theoretical precision,
improving coherent analyses from LHC and future ILC data
and determining SUSY parameters,
determining and clarifying the nature of dark matter,
and the study of extended SUSY scenarios. 
\medskip

Allanach et al. \cite{Allanach:2004my} use genetic algorithms to distinguish
between different SUSY models.
\medskip

All the methods above determine mass parameters and fundamental SUSY parameters such
as $M_1$, $M_2$, $\mu$, and $\tan\beta$. The errors on the gaugino parameters are 
small, the errors on the higgsino mass parameter and on $\tan\beta$ are somewhat 
larger.  
\medskip

I present a method that determines the {\emph{couplings}} of the lightest neutralino. 
From theses couplings, I calculate the corresponding elements of the neutralino 
diagonalisation matrix, assuming unitarity. 
The absolute errors on the elements of the diagonalisation matrix are of 
order $0.001 - 0.01$.
With the knowledge of the neutralino masses, I then obtain the values
of $M_1$, $M_2$, $\mu$, and $\tan\beta$. The errors are $0.4\GeV$, $4\GeV$,
$2.5\GeV$, and $7$, respectively.  
This method is complementary to the methods described above. This allows for cross checks. 
\medskip

\section{The circle method}
The authors of~\cite{Choi:2001ww} present a method to calculate $M_1$ and $\phi_{M_{1}}$
for the $CP$ violating extension of the MSSM 
from the characteristic polynomial of the matrix~(\ref{eq:neutralinomatrix}). 
$M_1$ and $\mu$ are here complex parameters, cf Eq.~(\ref{eq:complex}):
\begin{equation}
\label{eq:circle}
0 = \det(M^+ M - m_{\x{i}}^2)  = m_{\x{i}}^8 -a m_{\x{i}}^6 + b m_{\x{i}}^4 - c m_{\x{i}}^2 + d, 
\end{equation}
with the polynomial coefficients given by
\begin{eqnarray}
\label{eq:aa}
a &=& |M_1|^2 + M_2^2 + 2|\mu|^2 + 2 m_z^2,\\[2mm]
\label{eq:bb}
b &=& |M_1|^2 M_2^2 + 2|\mu|^2(|M_1|^2 + M_2^2) +(|\mu|^2 + m_Z^2)^2 \notag\\[1mm]
 	&& +2 m_Z^2 \left\{
	|M_1|^2\cw[2] + M_2^2\sw[2] - \right. \notag\\[1mm]
    && \left.|\mu|\sin 2\beta
 	\left[ |M_1|\sw[2]\cos(\phi_1 + \phi_\mu) + M_2 \cw[2]\cos\phi_\mu \right]	 
 	\right\},\\[2mm]
\label{eq:cc}	
c &=& |\mu|^2\left\{|\mu|^2(|M_1|^2 + M_2^2) + 2 |M_1|^2 M_2^2 + m_Z \sin^2 2\beta
	+ 2 m_Z (|M_1|^2\cw[2] + M_2^2\sw[2])\right\}\notag\\[1mm]
   && - 2 m_Z^2|\mu|\sin2\beta\left\{|M_1|(M_2^2 + |\mu|^2)\sw[2] \cos(\phi_1 + \phi_\mu)
  	+ M_2 (|M_1|^2 + |\mu|^2 \cw[2]) \cos\phi_\mu \right\}\notag\\[1mm]
   &&  + m_Z^4 \left\{|M_1|^2\cw[4] + 2 |M_1|M_2 \sw[2]\cw[2]\cos\phi_1 + M_2^2\sw[4]
	\right\},\\[2mm]
\label{eq:dd}	
d &=& |M_1|^2 M_2^2 |\mu|^4 - 2 m_Z^2|\mu|^3|M_1|M_2 \sin 2\beta\left\{
	|M_1|^2\cw[2] \cos\phi_\mu + M_2\sw[2]\cos(\phi_1 + \phi_\mu) \right\}\notag\\[1mm]
   &&	+m_Z^4|\mu|^2\sin^2\beta\left\{|M_1|^2\cw[4] + 2 |M_1|M_2 \sw[2]\cw[2]\cos\phi_1 + 
	M_2^2\sw[4]\right\}.
\end{eqnarray}
Note that the matrix $M$ is symmetric but not hermitian.
So one has to diagonalise $M^+ M$ to get the singular values of $M$ which are
the physical masses.
Eq.~(\ref{eq:circle}) is quadratic in $\real M_1$ and $\imag M_1$ 
for fixed $m_{\x{i}}$, $i = 1\ldots 4$, and for fixed parameters
$M_2$, $\mu$, and $\tan\beta$. 
So it describes four circles in the  $\real M_1$ - $\imag M_1$ -plane, which should intersect 
in one point $(\real M_1, \imag M_1)$. 
In general, four circles do not intersect in only one point. There may be no solution
or even two solutions, in which case all midpoints have to be located on a straight line.
From the coordinates of this  point one can calculate $|M_1|$ and $\phi_{M_1} = \arg(M_1)$:
\begin{eqnarray}
\label{eq:circleequation}
0 &=& \det(M^+ M - m_{\x{i}}^2) \notag \\[2mm] 
	&=& A(m_{\x{i}}, M_2, \mu, \tan\beta) X^2 + A(m_{\x{i}}, M_2, \mu, \tan\beta) Y^2 \notag \\[1mm] 
	&&+ B_1(m_{\x{i}}, M_2, \mu, \tan\beta)  X + B_2(m_{\x{i}}, M_2, \mu, \tan\beta) Y 
	- C(m_{\x{i}}, M_2, \mu, \tan\beta)  
\end{eqnarray}
with $X = \real M_1$, $Y = \imag M_1$.
The radii $r_i$ and the midpoints $m_i$  of the circles described by Eq.~(\ref{eq:circleequation})
are given by
\begin{eqnarray}
\label{eq:radius}
r_i &=& \frac{C}{A} + \left(\frac{B_1}{2 A}\right)^2 + \left(\frac{B_2}{2 A}\right)^2,\\[2mm]
\label{eq:midpoint}
m_i(m_x|m_y) &=& M_i\left(-\frac{B_1}{2 A}\right|\left.-\frac{B_2}{2 A}\right)\enspace .
\end{eqnarray}

\begin{figure}[ht]
\setlength{\unitlength}{1cm}
\psfrag{X}{{}}
\subfigure[Exact circles\label{fig:exact}]{\scalebox{0.53}{\includegraphics{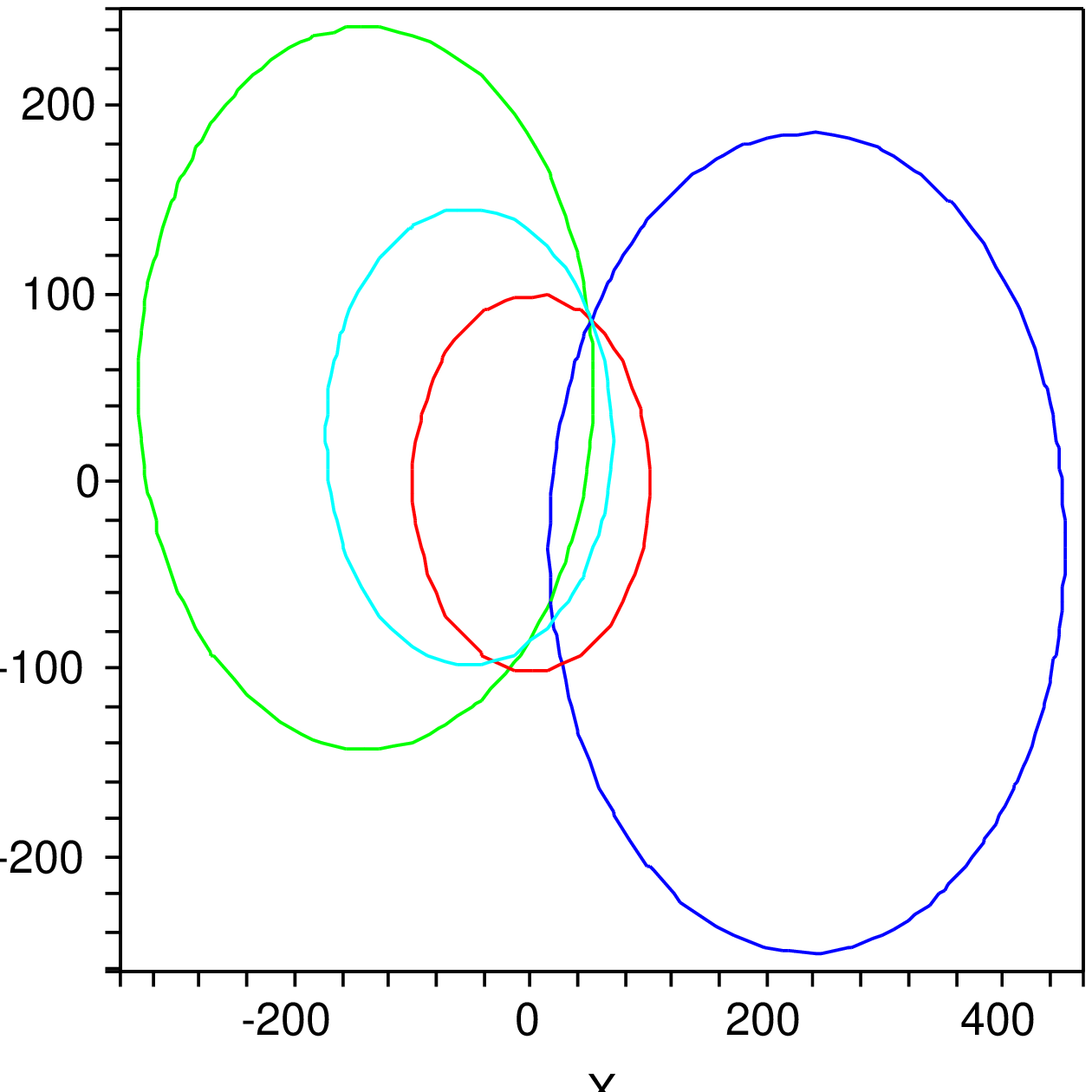}}
\put(-4,0){$\real M_1$}
\put(-7.6,3.4){\rotatebox{90}{$\imag M_1$}}
\put(-6,6){$\scriptstyle{\x{4}}$}
\put(-5.8,4){$\scriptstyle{\x{2}}$}
\put(-3.4,4){$\scriptstyle{\x{1}}$}
\put(-2.4,5.6){$\scriptstyle{\x{3}}$}
}
\hspace{2mm}
\psfrag{X}{{}}
\subfigure[Circles with $\Delta \mu = +0.5\GeV$ 
     \label{fig:deltamu}]{\scalebox{0.53}{\includegraphics{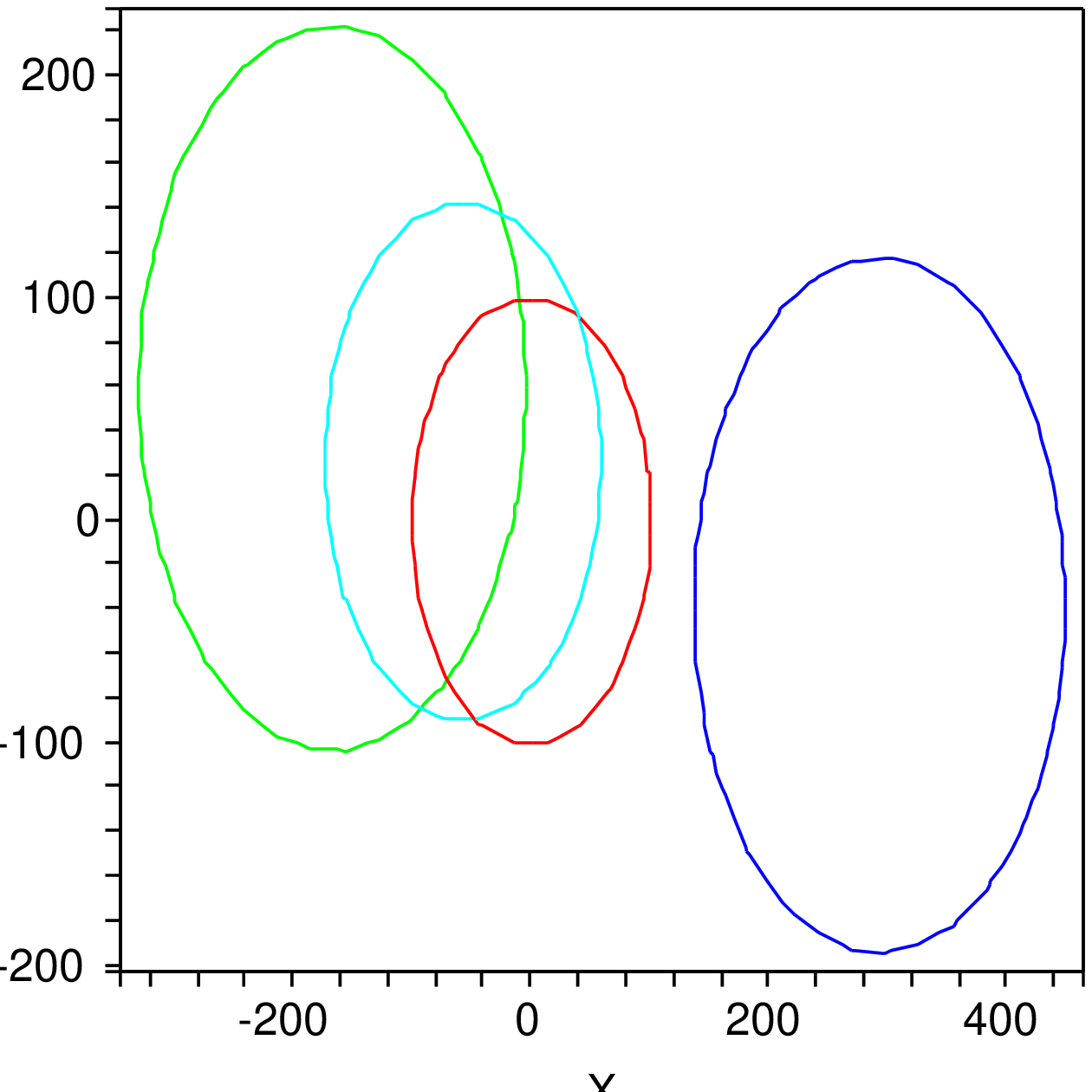}}
\put(-4,0){$\real M_1$}
\put(-7.6,3.1){\rotatebox{90}{$\imag M_1$}}
\put(-6,6.1){$\scriptstyle{\x{4}}$}
\put(-5.8,4){$\scriptstyle{\x{2}}$}
\put(-4.8,4){$\scriptstyle{\x{1}}$}
\put(-2.4,4.8){$\scriptstyle{\x{3}}$}
}
\caption{Influence of a small error in $\mu$ on the intersection point of the four circles in the
$\real M_1$ - $\imag M_1$ - plane. The colors mean: red: $\tilde{\chi}_1^0$, cyan: $\tilde{\chi}_2^0$,
	  blue: $\tilde{\chi}_3^0$, green: $\tilde{\chi}_4^0$.}
\label{fig:circle}
\end{figure}

In Fig.~\ref{fig:exact}, I show the four circles for all neutralino mass $m_{\x{i}}$. 
As input data I have taken the RP'' model of Ref.~\cite{Choi:2001ww}:
\begin{eqnarray}
\label{eq:inputrp}
\begin{pmatrix}
|M_1|,\phi_1,M_2,\mu,\phi_\mu,\tan\beta
\end{pmatrix}
=
\begin{pmatrix}
100.5\GeV, \frac{\pi}{3}, 190.8\GeV, 365.1\GeV, \frac{\pi}{4},10
\end{pmatrix},
\end{eqnarray}
leading to the following neutralino masses
\begin{eqnarray}
\label{eq:rpmasses}
\begin{pmatrix}
m_{\x{1}},m_{\x{2}},m_{\x{3}},m_{\x{4}}
\end{pmatrix}
=
\begin{pmatrix}
99.15\GeV,177.07\GeV,372.0\GeV,387.41\GeV
\end{pmatrix}\enspace .
\end{eqnarray}
It is clear that the four circles intersect in one point: $X = 52.8\GeV$, $Y = 85.5\GeV$.
This yields $|M_1| = 100.5\GeV$ and $\phi_1 = 1.02 \approx \pi/3$. 
These two values agree with the input data~(\ref{eq:inputrp}).  
The algebraic form of the coefficients $A$, $B_1$, $B_2$, and $C$ follow from
the Eqs~(\ref{eq:circle})-(\ref{eq:dd}).
For their numerical values one needs the values of
the parameters $M_2$, $\tan\beta$, $|\mu|$, and $\phi_\mu$, 
which can be determined from the chargino system, see Ref~\cite{Choi:2000ta,Choi:2001ww}, 
and at least three neutralino masses form LHC/ILC measurements.

\begin{figure}[p]
\setlength{\unitlength}{1cm}
\centering
\subfigure[Radius $r_4$ as a function of $\mu$ for the corresponding neutralino mass circle. The exact value for $\mu$
	  is $365.1\GeV$, and the mass of \x{4} is $m_{\x{4}}=387.4\GeV$, see Eqs~(\ref{eq:inputrp}) and (\ref{eq:rpmasses}).
	  The other parameters are kept on their exact values.\label{fig:singumu}]
	  {\scalebox{0.55}{\includegraphics{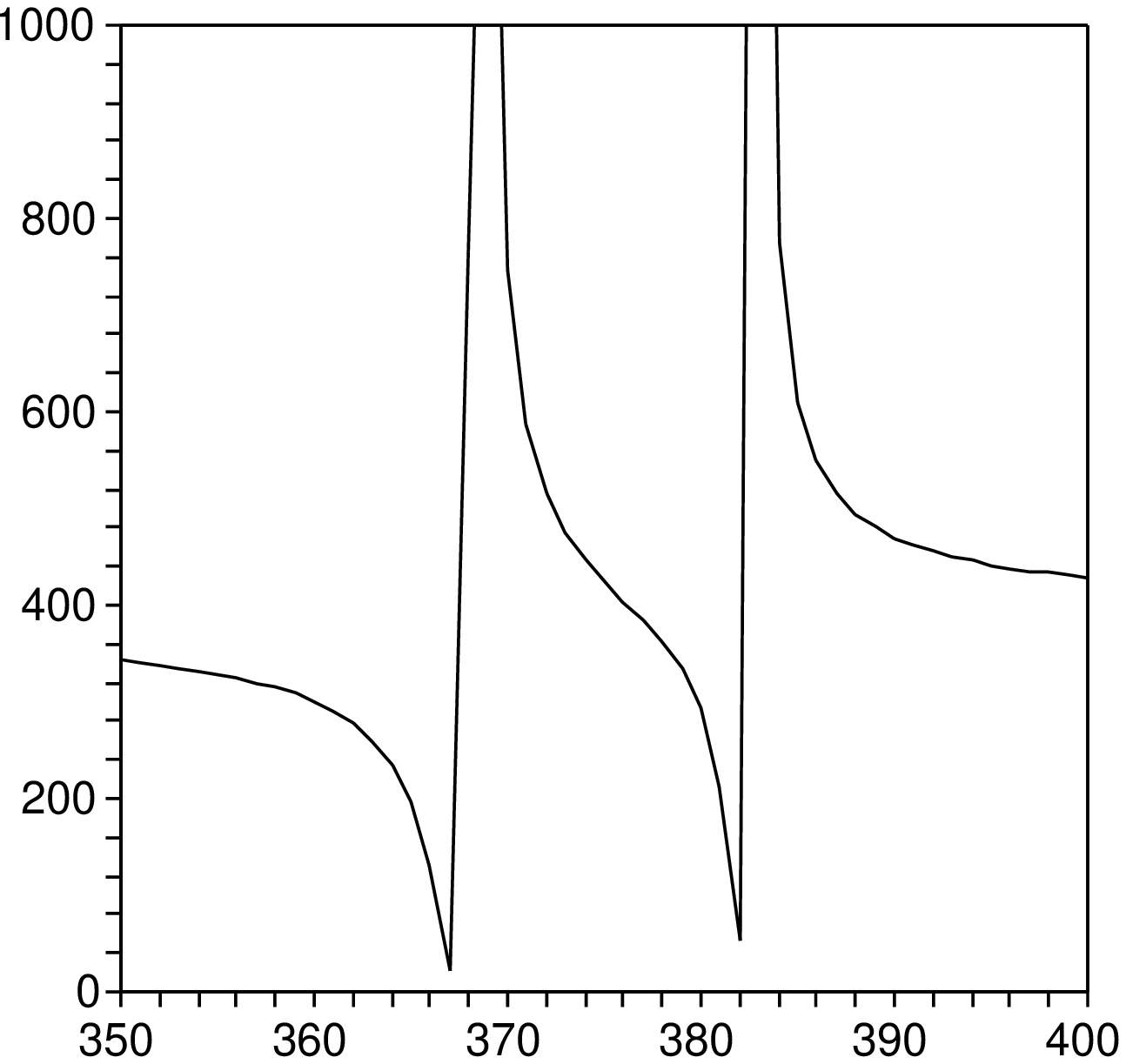}}}
\put(-4.1,0){$\mu\, [\!\!\GeV]$}
\put(-8.3,4.5){\rotatebox{90}{$r_4\,[\!\!\GeV]$}}
\hspace*{5mm}
\subfigure[Radius $r_4$ as a function of $M_2$ for the corresponding neutralino mass circle. The exact value for $M_2$
	  is $190.8\GeV$, and the mass of \x{4} is $m_{\x{4}}=387.4\GeV$, see Eqs~(\ref{eq:inputrp}) and (\ref{eq:rpmasses}).
	  The other parameters are kept on their exact values.\label{fig:singuM2}]
	  {\scalebox{0.55}{\includegraphics{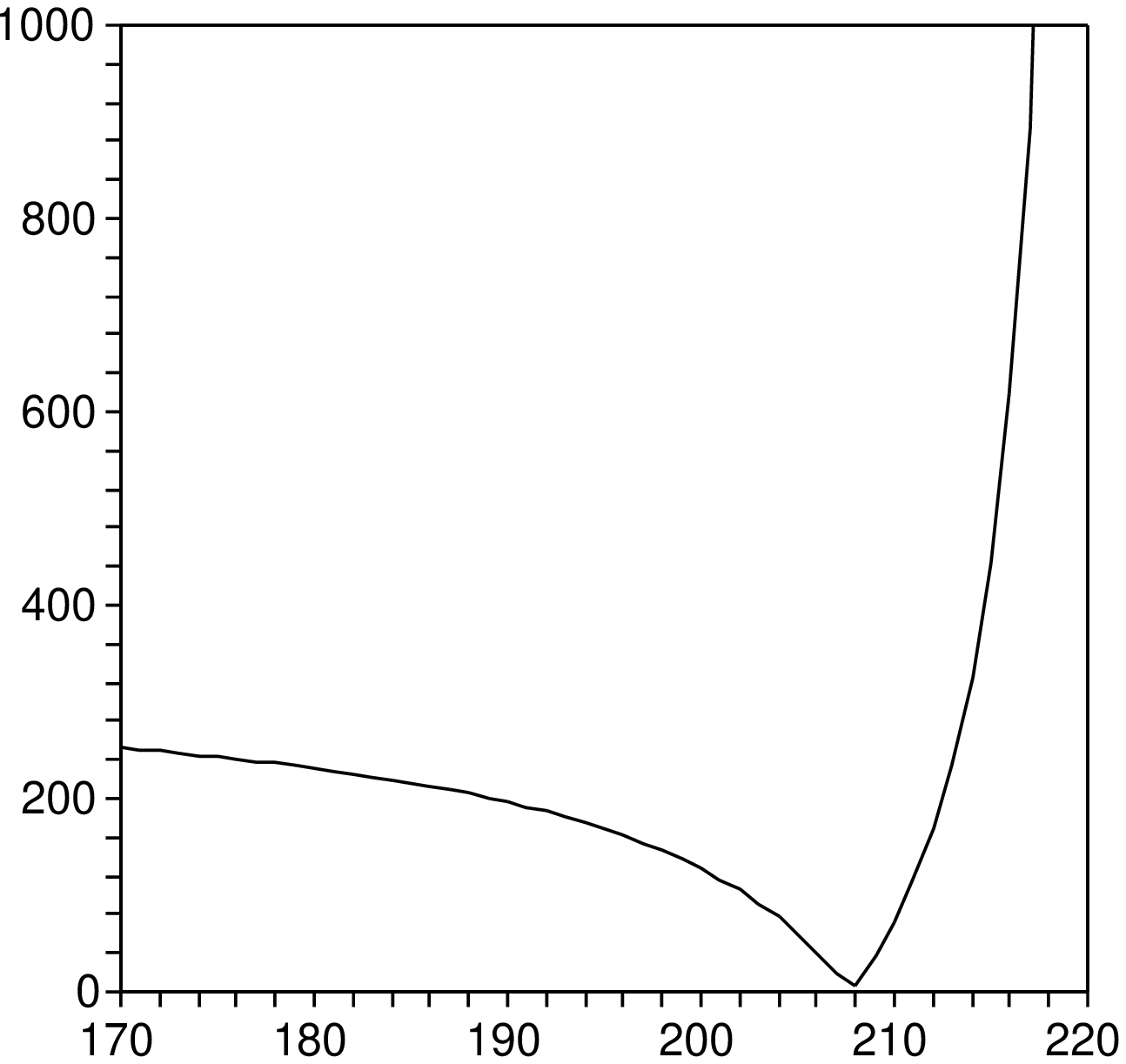}}}
\put(-4.1,0){$M_2\, [\!\!\GeV]$}
\put(-8.3,4.5){\rotatebox{90}{$r_4\,[\!\!\GeV]$}}
	  
\subfigure[Radius $r_4$ as a function of $m_{\x{4}}$ for the corresponding neutralino mass circle. 
	  The mass of \x{4} is $m_{\x{4}}=387.4\GeV$, see Eqs~(\ref{eq:inputrp}) and (\ref{eq:rpmasses}).
	  The other parameters are kept on their exact values.\label{fig:singum4}]
	  {\scalebox{0.55}{\includegraphics{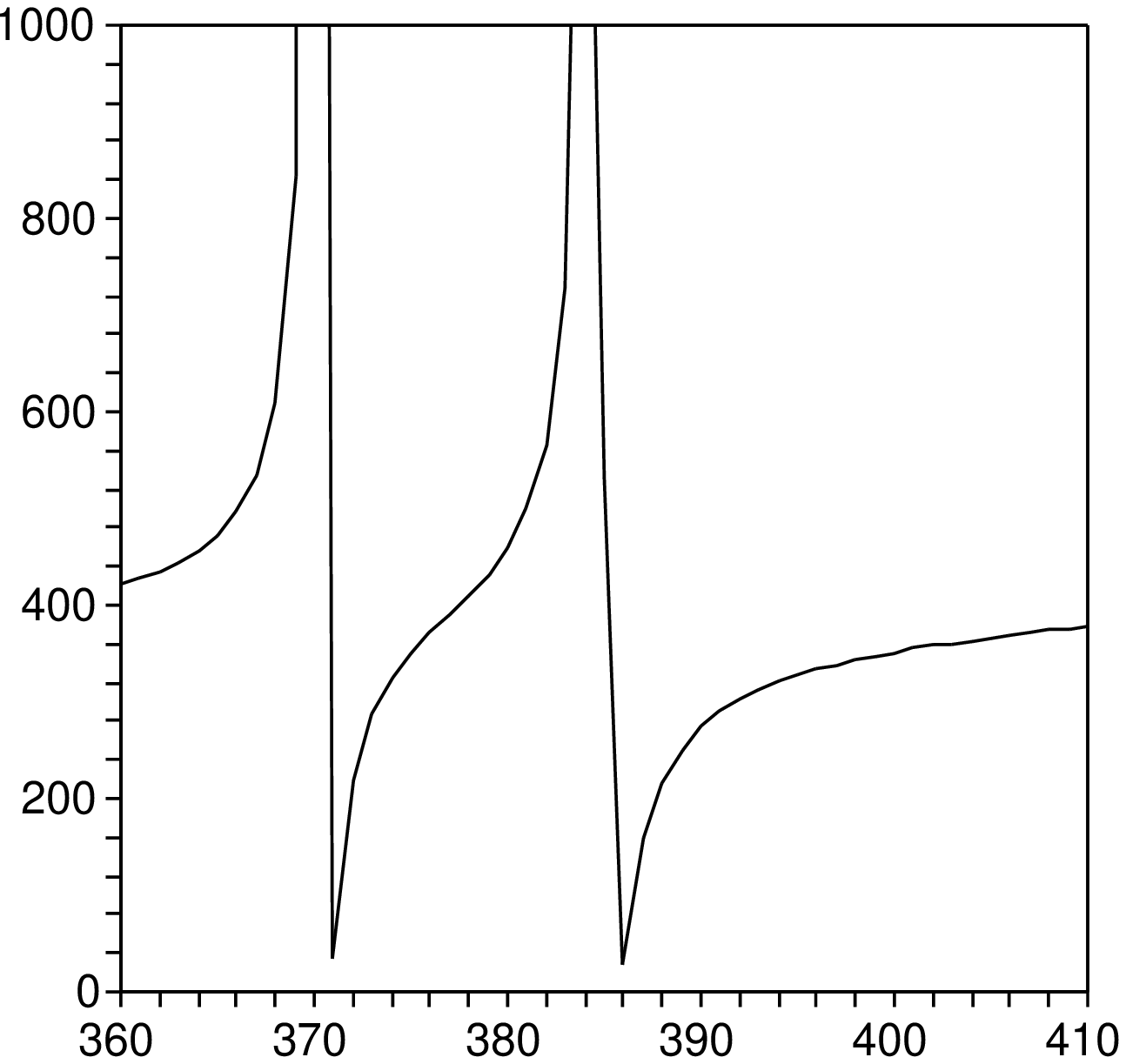}}}
\put(-4.1,0){$m_{\x{4}}\, [\!\!\GeV]$}
\put(-8.3,4.5){\rotatebox{90}{$r_4\,[\!\!\GeV]$}}
\hspace*{5mm}
\subfigure[Radius $r_4$ as a function of $\phi_\mu$ for the corresponding neutralino mass circle. The exact value for 
	  $\phi$ is $\pi/4$, see Eqs~(\ref{eq:inputrp}).
	  The other parameters are kept on their exact values.\label{fig:singuphi3}]
	  {\scalebox{0.55}{\includegraphics{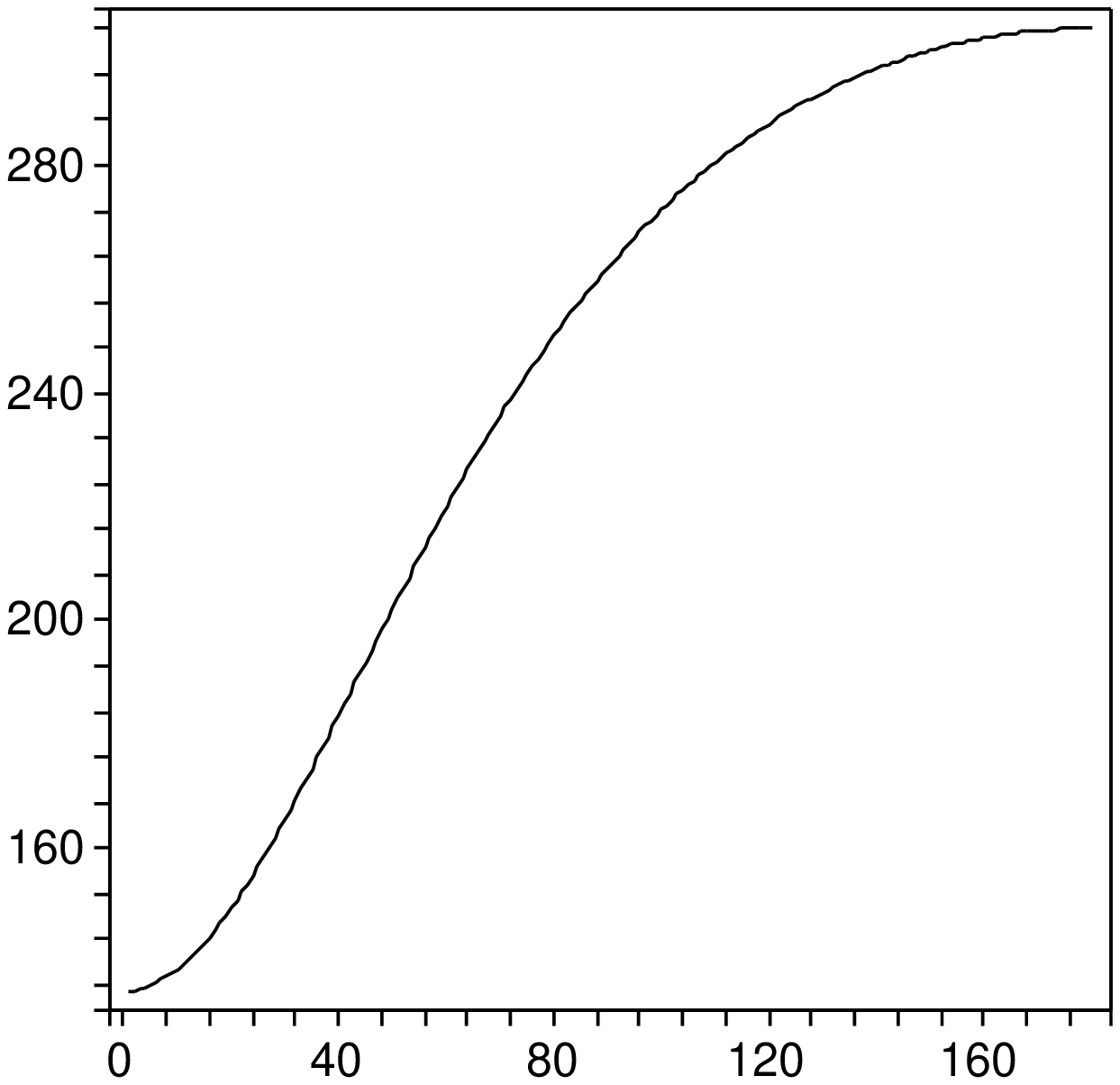}}}
\put(-4.1,0){$\phi_\mu$}
\put(-8.3,4.5){\rotatebox{90}{$r_4\,[\!\!\GeV]$}}
	  
\caption{Dependence of the radius $r_4$ on $\mu$, $M_2$, $m_{\x{4}}$, and $\phi_\mu$.}
\label{fig:singularities}
\end{figure}

Measurements of masses and cross sections have unavoidable errors. 
These errors influence the radii and the midpoints of the circles Eq.~(\ref{eq:circleequation}). 
I have analysed how errors in the parameters $M_2$, $\tan\beta$, $|\mu|$, and $\phi_\mu$
and the neutralino masses $m_{\x{i}}$ influence the circles of the example given 
in~\cite{Choi:2001ww} and found that the circles belonging 
to the neutralinos $\x{2}$ - $\x{4}$ drift away considerably from their exact position even for small errors. 
This behaviour is due to zeros and poles of the coefficients $a = a(M_2, \mu, \tan\beta)$, 
$b_1 = b_1(M_2, \mu, \tan\beta)$, $b_2 = b_2(M_2, \mu, \tan\beta)$, and $c = c(M_2, \mu, \tan\beta)$
which determine the radius and the midpoint of the neutralino circles.
In Fig.~\ref{fig:singumu}-\ref{fig:singuphi3}, I illustrate this behaviour of 
the radius $r_4$ for the neutralino mass circles of \x{4}. 
The poles and the zeros are located very close to each other and close to the neutralino masses. 
This leads to the messy situation that small errors in the input data blow up to large 
errors in the radii and coordinates of the midpoints.
The neutralino mass matrix (\ref{eq:neutralinomatrix}) can be decomposed in main diagonal 
and off-diagonal blocks. The off-diagonal blocks are proportional to $m_Z$.
Then, at zeroth order in $m_Z$, the eigenvalues of $M$ are given by   
$m_{\x{1}} \approx |M_1|$, $m_{\x{2}} \approx M_2$, and $m_{\x{3/4}} \approx \pm|\mu|$, 
these eigenvalues are not necessarily mass ordered. This relation helps to understand
why the mass circle of the fourth neutralino reacts so strongly on the errors on $m_{\x{4}}$ 
and $\mu$. 

In Fig.~\ref{fig:deltamu} I show for the RP'' model ~\cite{Choi:2001ww}
how the circles drift away if the measured value of $\mu$ is $0.5\GeV$ larger than the ``true" value.
This corresponds to a $0.1\%$ error!
There is neither an intersection point nor a small region where all possible pairs of neutralino
circles intersect. 

\begin{figure}
\vspace*{-7mm}
\setlength{\unitlength}{1cm}
\centering
\subfigure[$\Delta m_{\x{4}} = 0.1\GeV$\label{fig:delta01}]{\scalebox{0.33}{\includegraphics{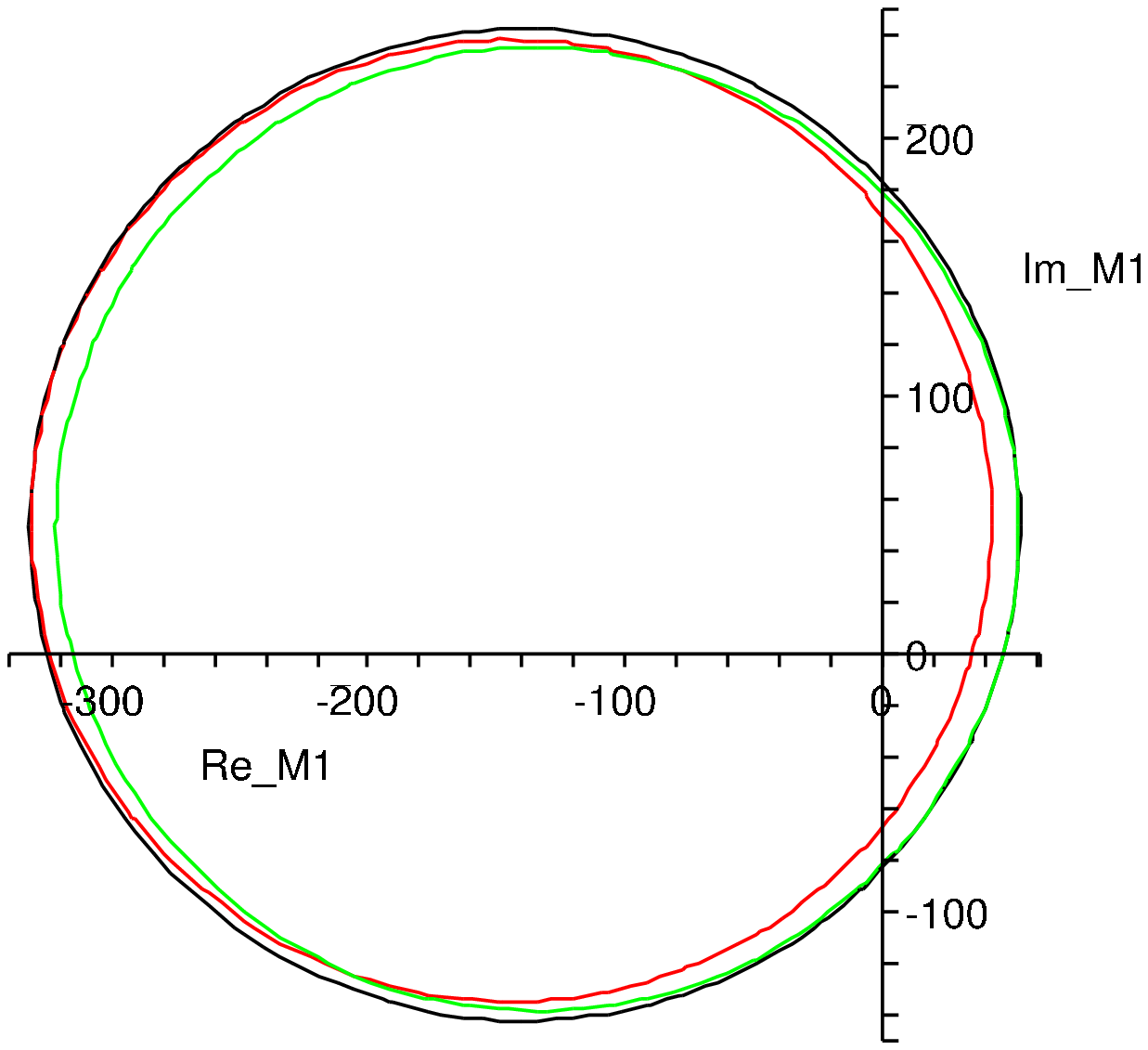}}
\psfrag{X}{$\real M_1$}
}
\hspace{2mm}
\subfigure[$\Delta M_2 = 0.1\GeV$\label{fig:deltaM2_0p1}]
{\scalebox{0.33}{\includegraphics{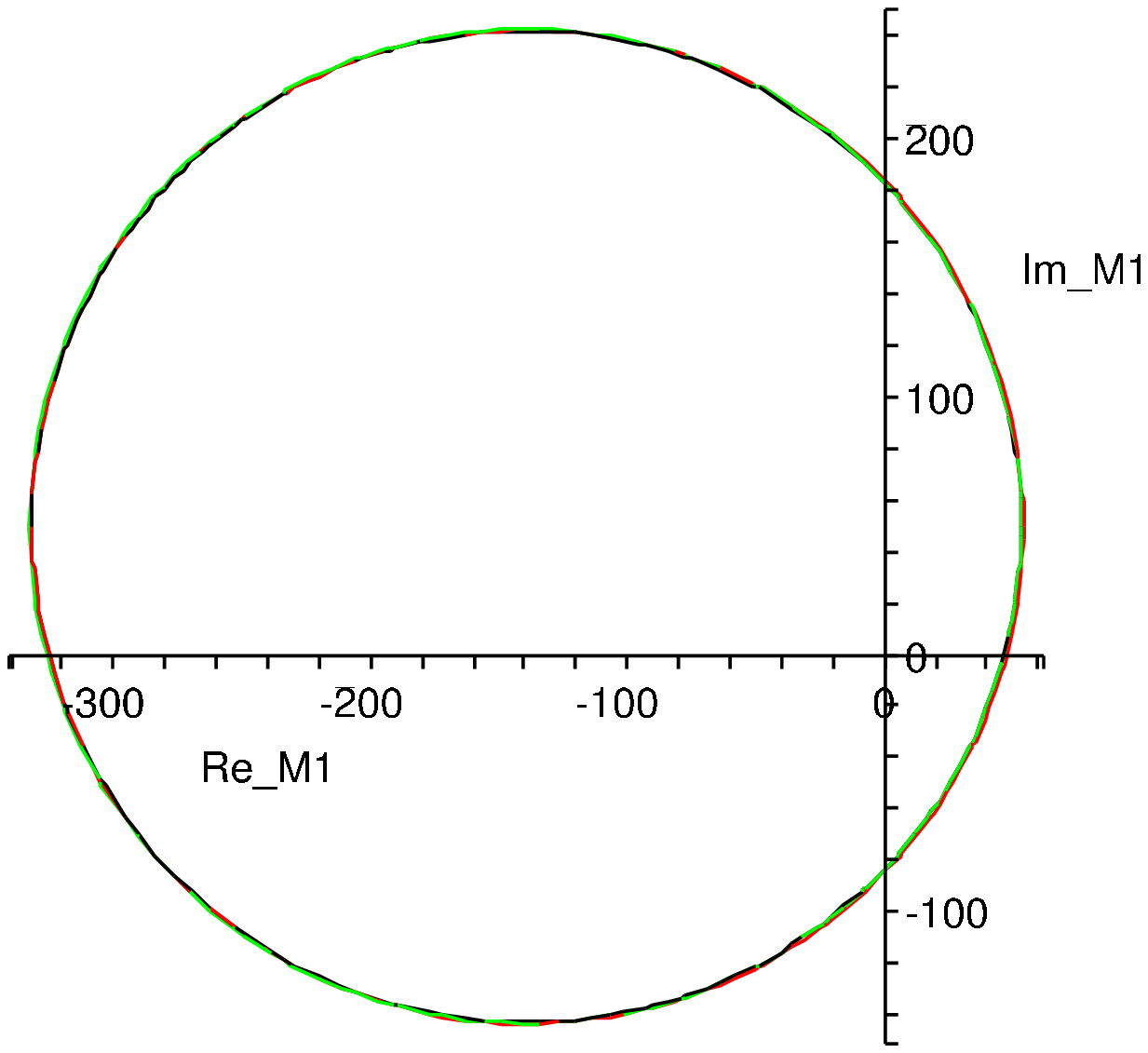}}
\psfrag{X}{$\real M_1$}
}
\hspace{2mm}
\subfigure[$\Delta \mu = 0.1\GeV$\label{fig:deltamu_0p1}]
{\scalebox{0.33}{\includegraphics{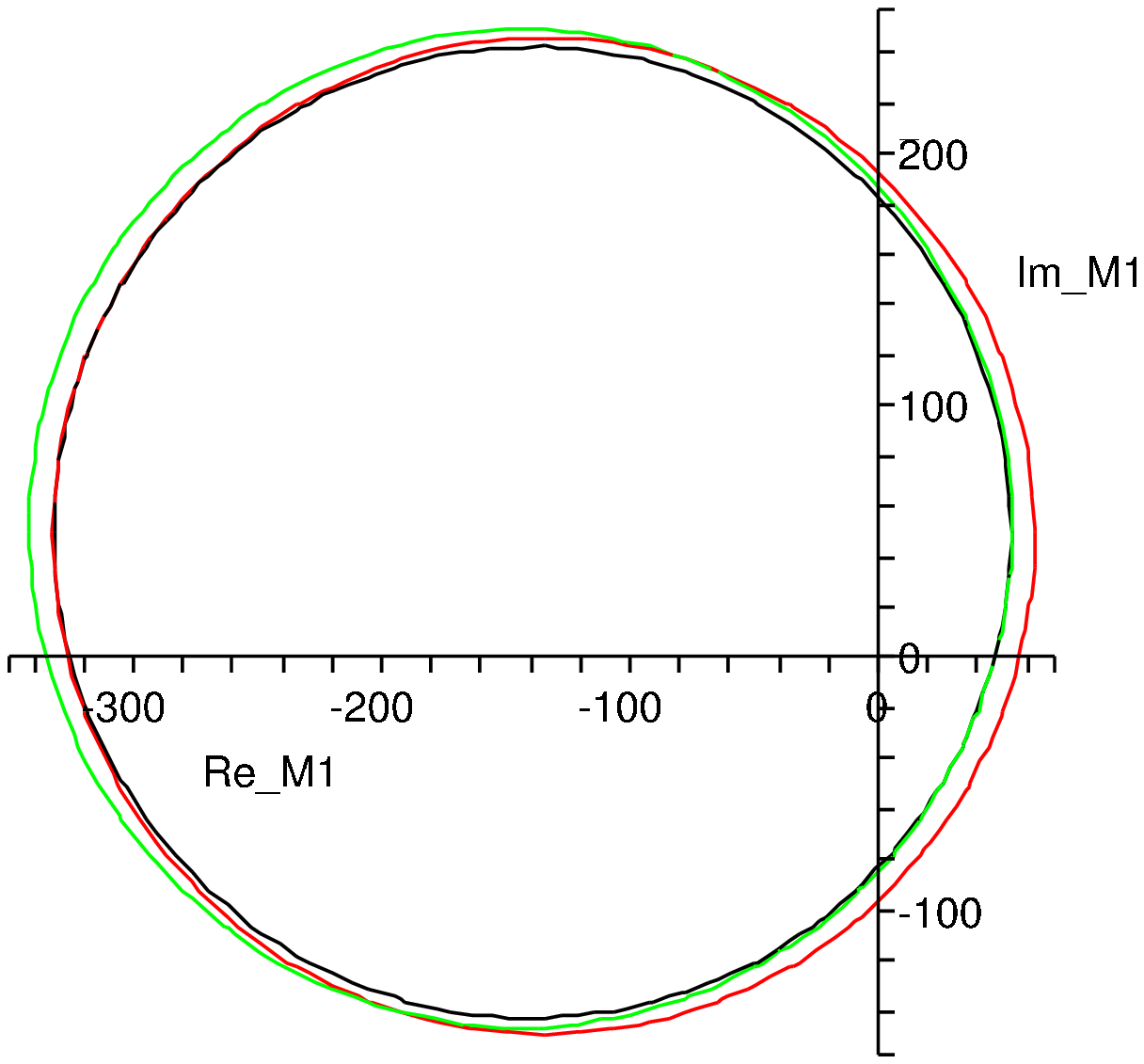}}
	\psfrag{X}{$\real M_1$}
}
\vspace*{-7mm}

\subfigure[$\Delta m_{\x{4}} = 0.5\GeV$\label{fig:delta05}]{\scalebox{0.33}{\includegraphics{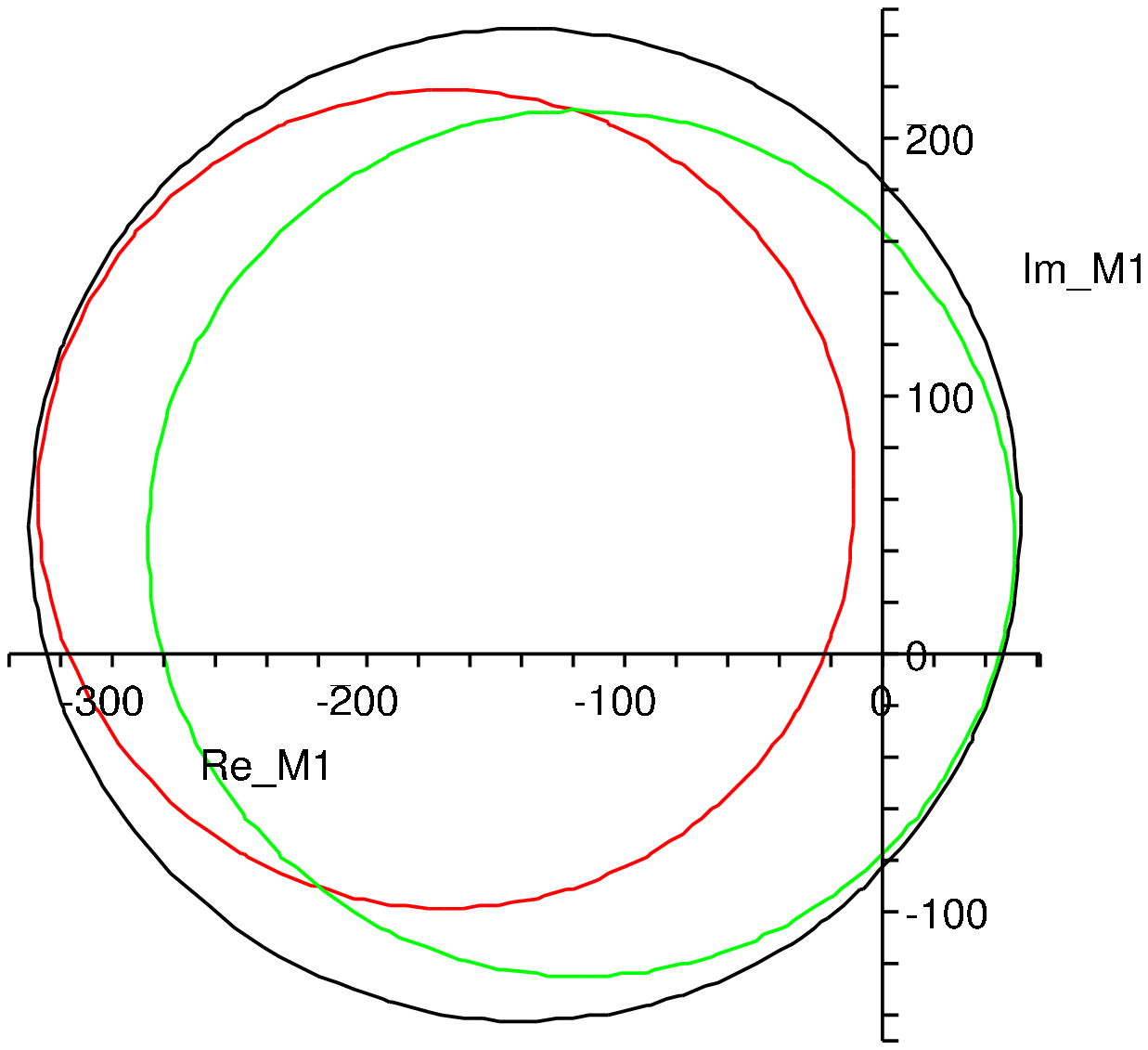}}
}
\hspace{2mm}
\subfigure[$\Delta M_2 = 0.5\GeV$\label{fig:deltaM2_0p5}]
{\scalebox{0.33}{\includegraphics{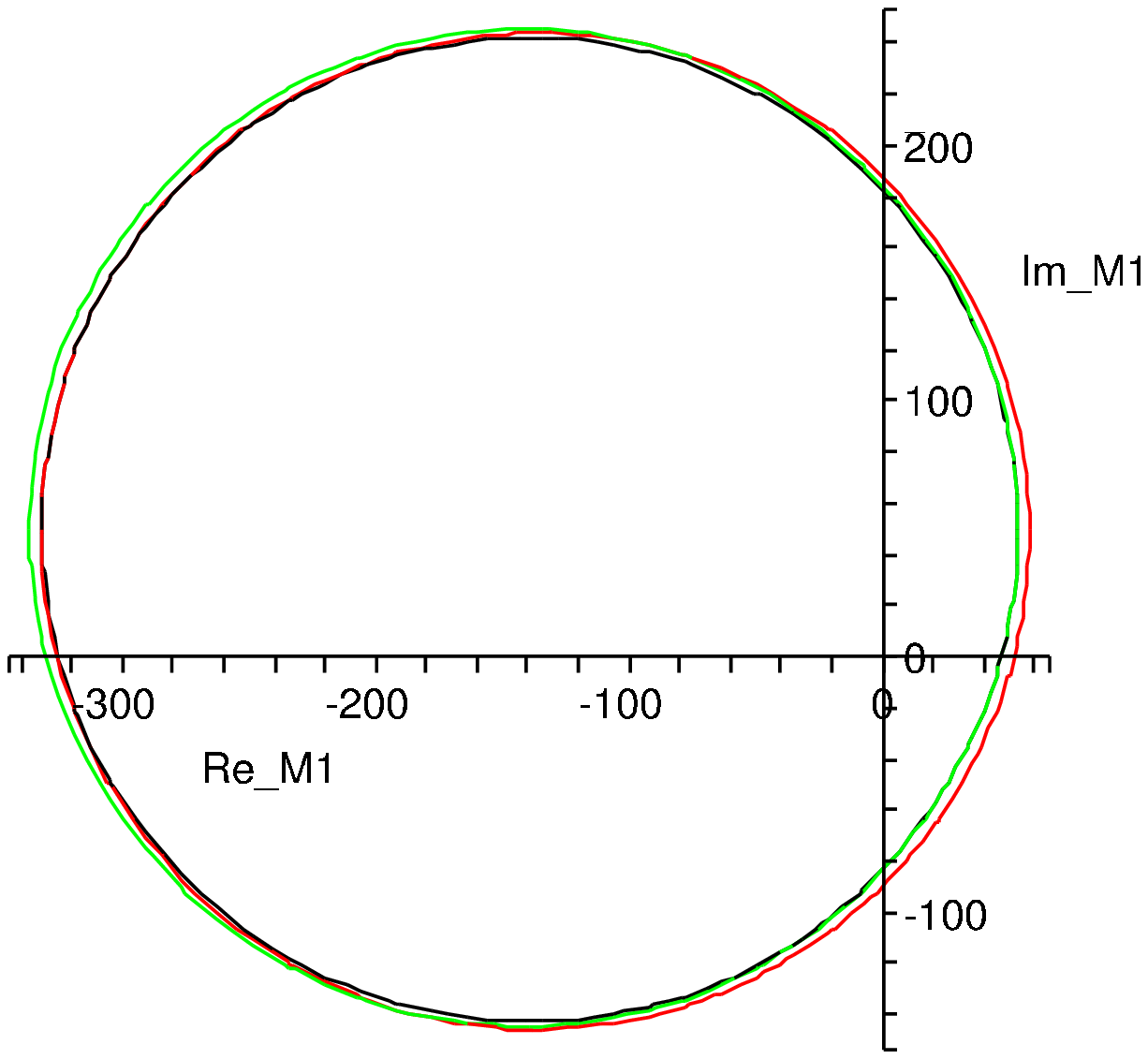}}
}
\hspace{2mm}
\subfigure[$\Delta \mu = 0.5\GeV$ \label{fig:deltamu_0p5}]
{\scalebox{0.33}{\includegraphics{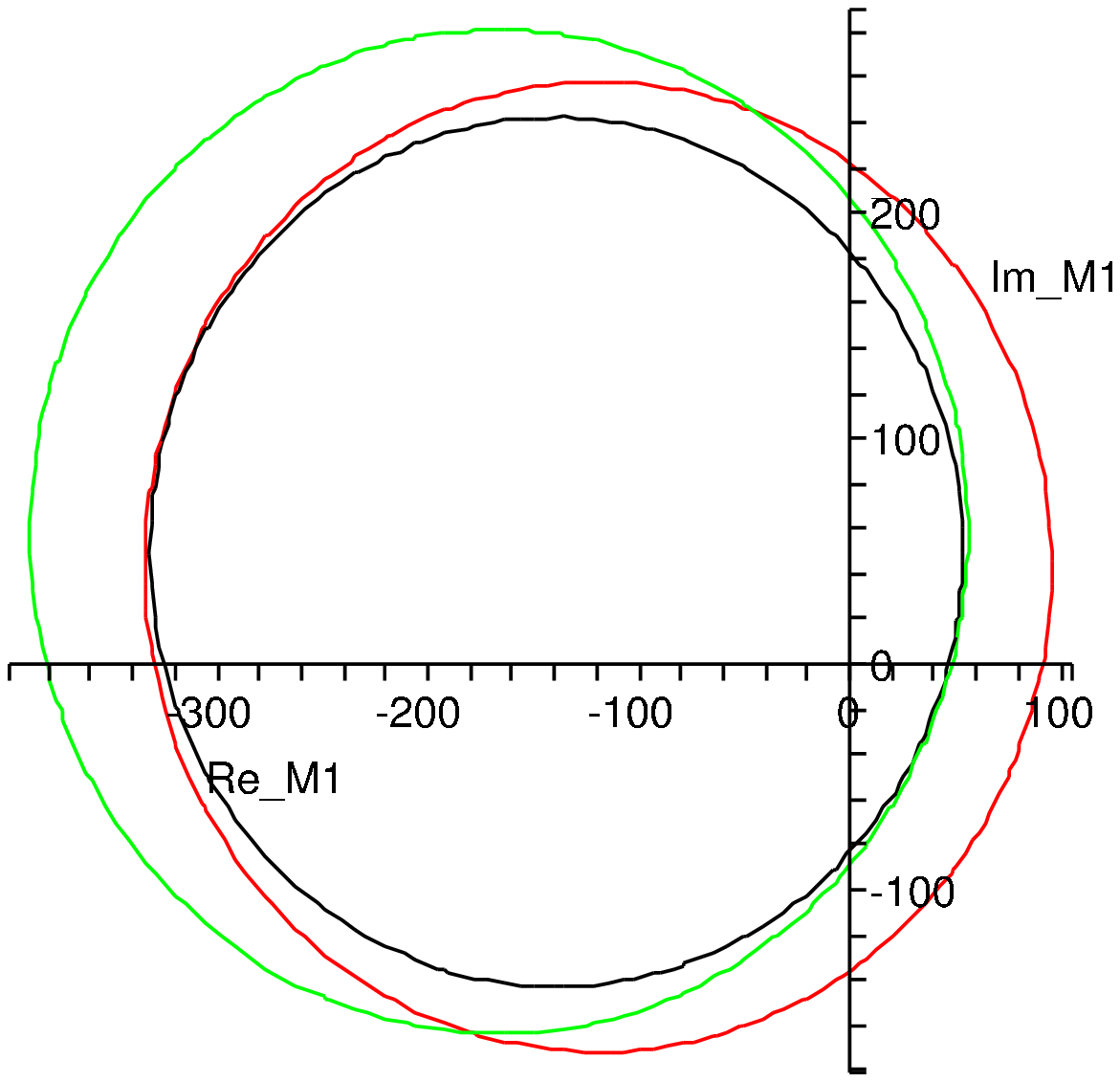}}
}
\vspace*{-7mm}

\subfigure[$\Delta m_{\x{4}} = 1.0\GeV$ \label{fig:delta10}]{\scalebox{0.33}{\includegraphics{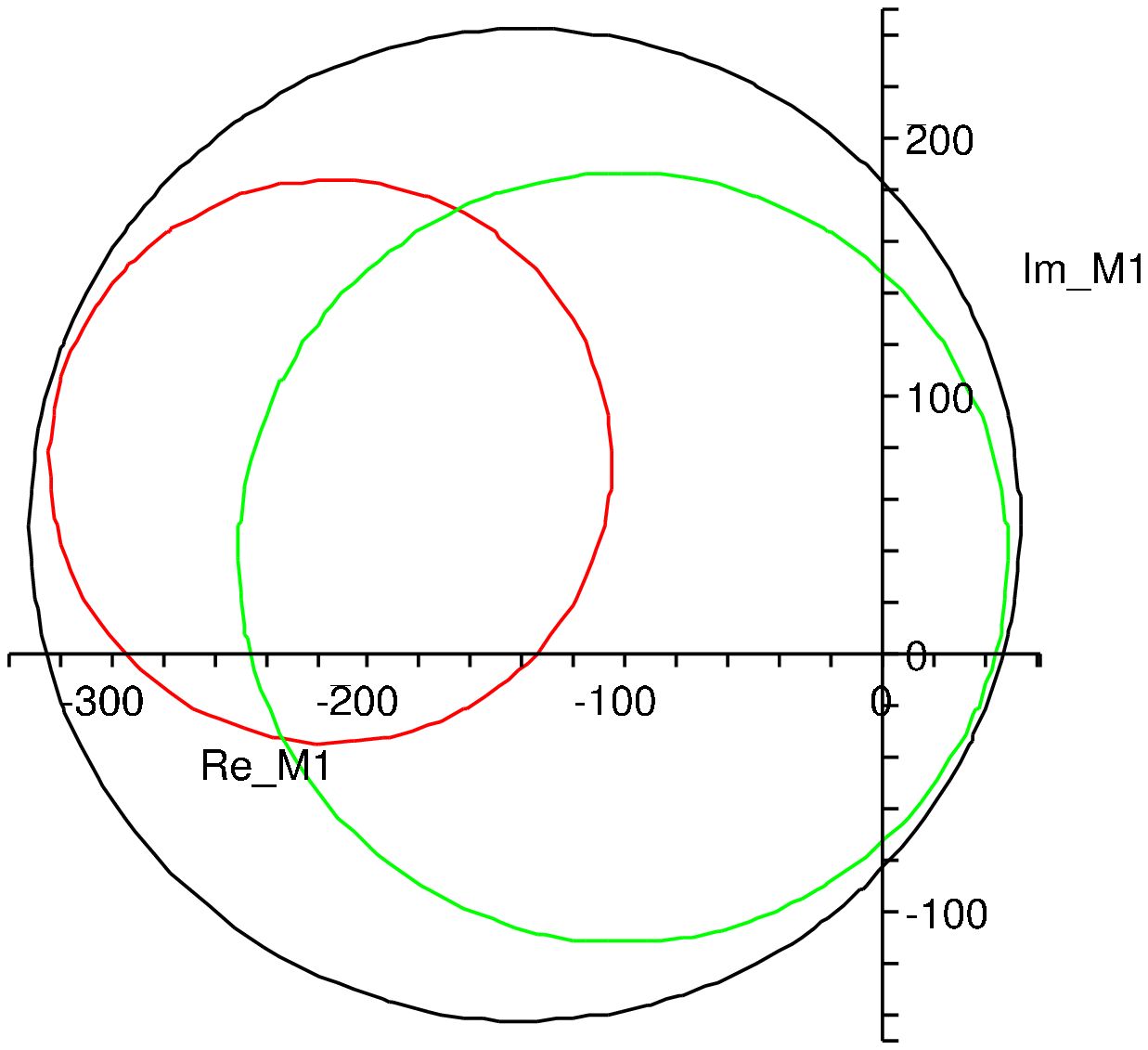}}
}
\hspace{2mm}
\subfigure[$\Delta M_2 = 1.0\GeV$ \label{fig:deltaM2_1p0}]
	{\scalebox{0.33}{\includegraphics{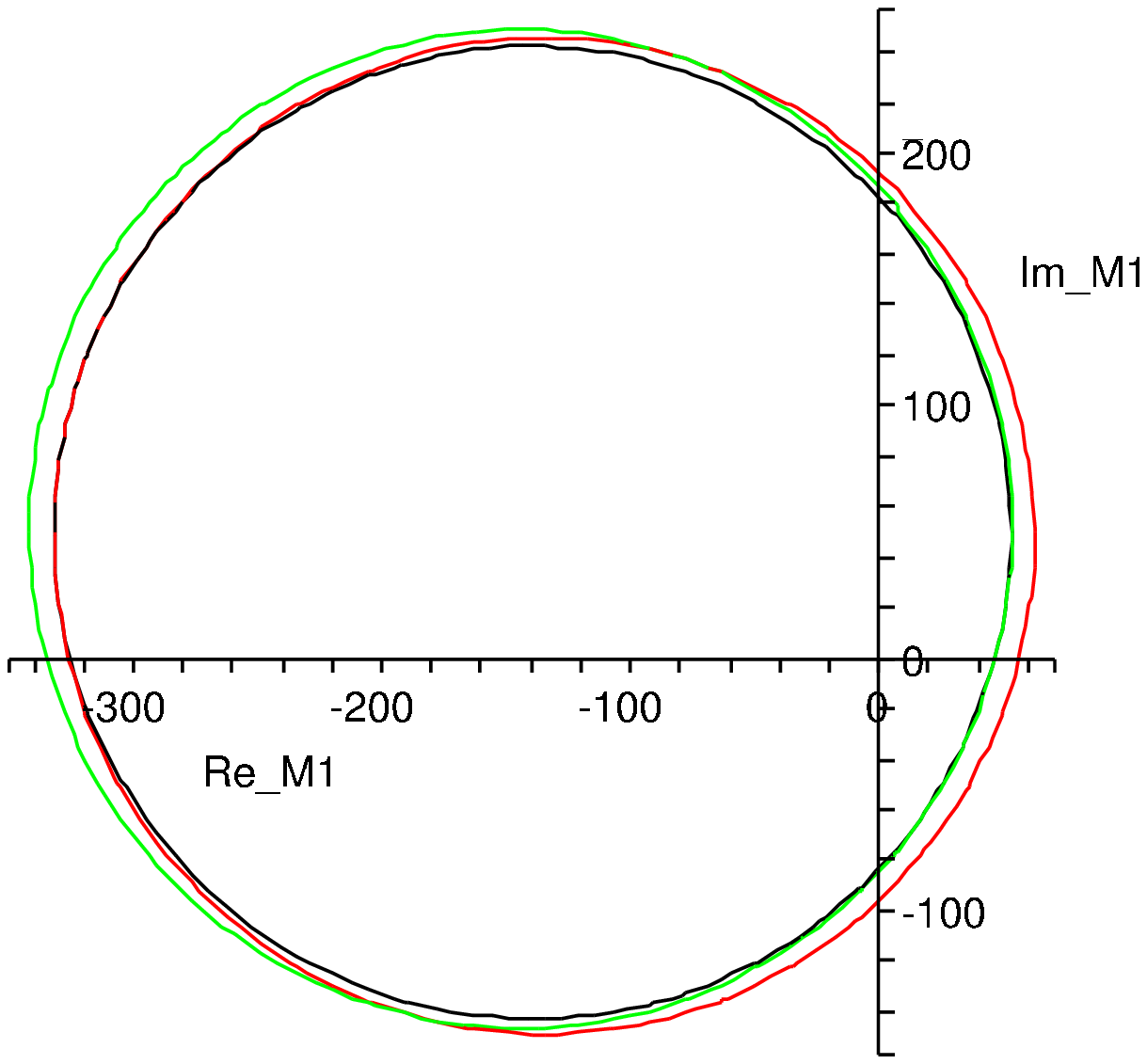}}
}
\hspace{2mm}
\subfigure[$\Delta \mu = 1.0\GeV$ \label{fig:deltamu_1p0}]
	{\scalebox{0.33}{\includegraphics{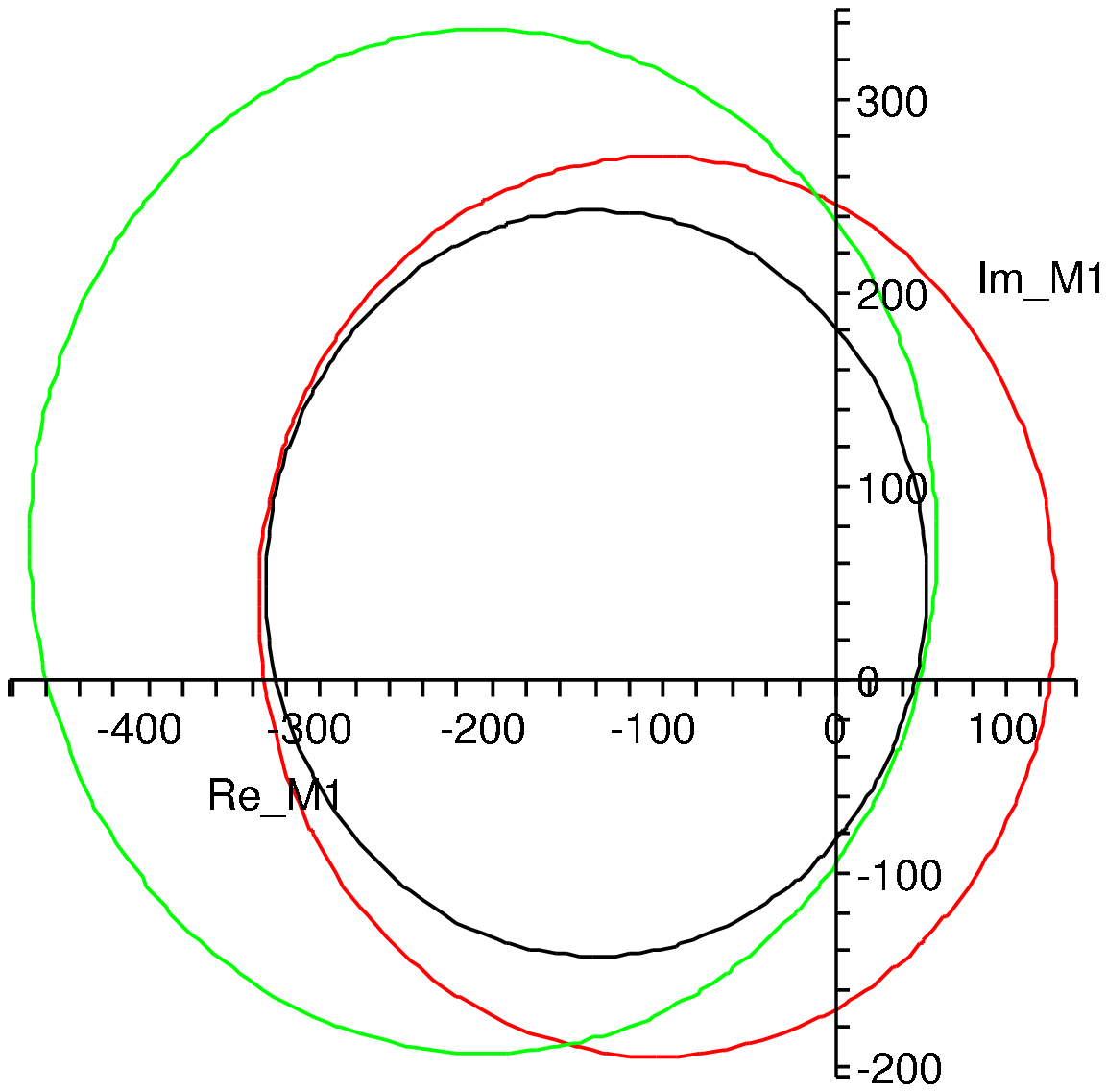}}
}
\vspace*{-29mm}

\subfigure[$\Delta m_{\x{4}} = 2.0\GeV$\label{fig:right}]{\scalebox{0.33}{\includegraphics{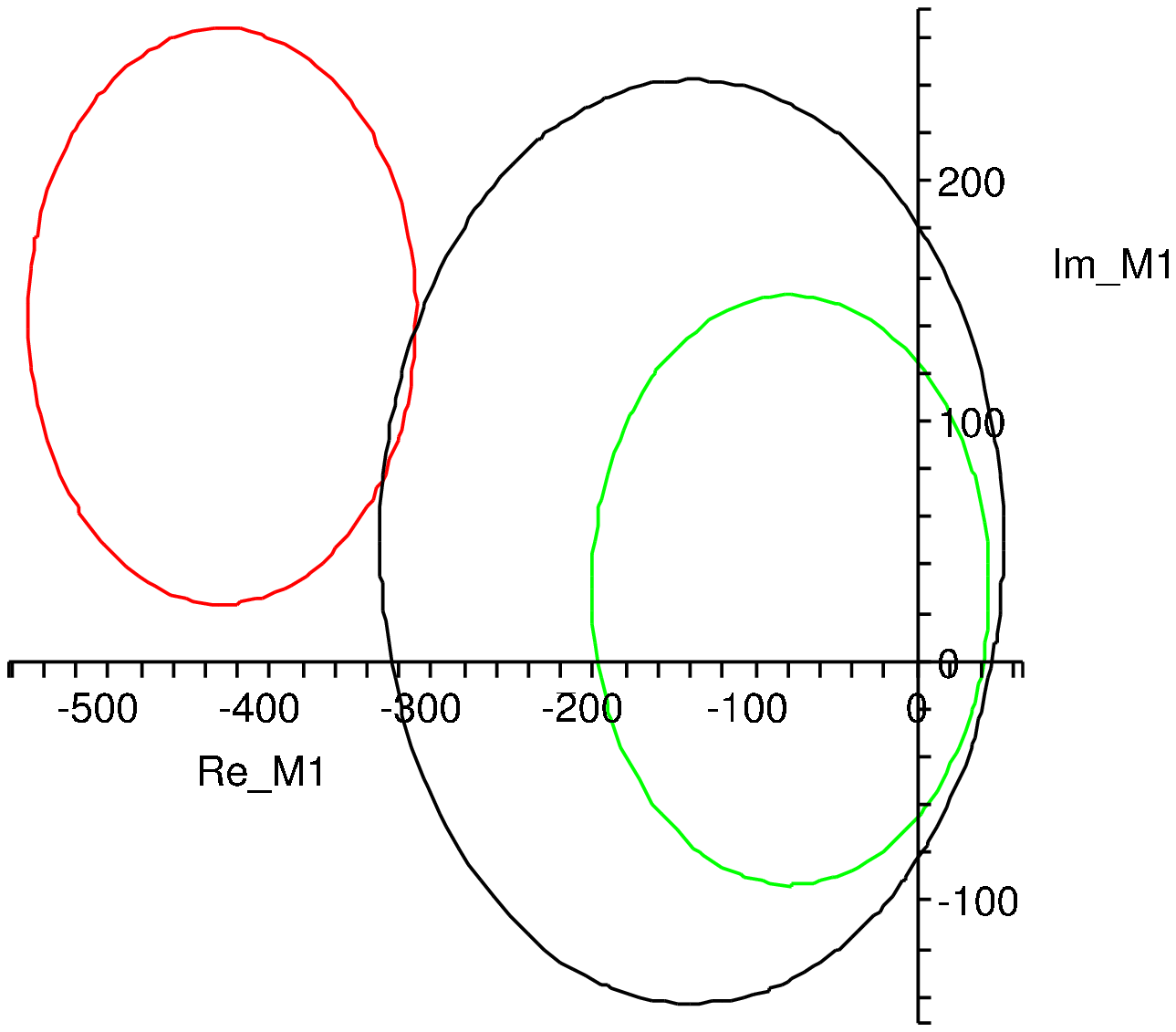}}
\put(-3.5,3.9){{\small r}}
\put(-1.8,3.6){{\small b}}
\put(-1.4,2.9){{\small g}}
\put(-8.9,5,5){\rotatebox{90}{$M_2[\mathrm{GeV}]$}}
}
\hspace{2mm}
\subfigure[$\Delta M_2 = 2.0\GeV$\label{fig:deltaM2_2p0}]
	{\scalebox{0.33}{\includegraphics{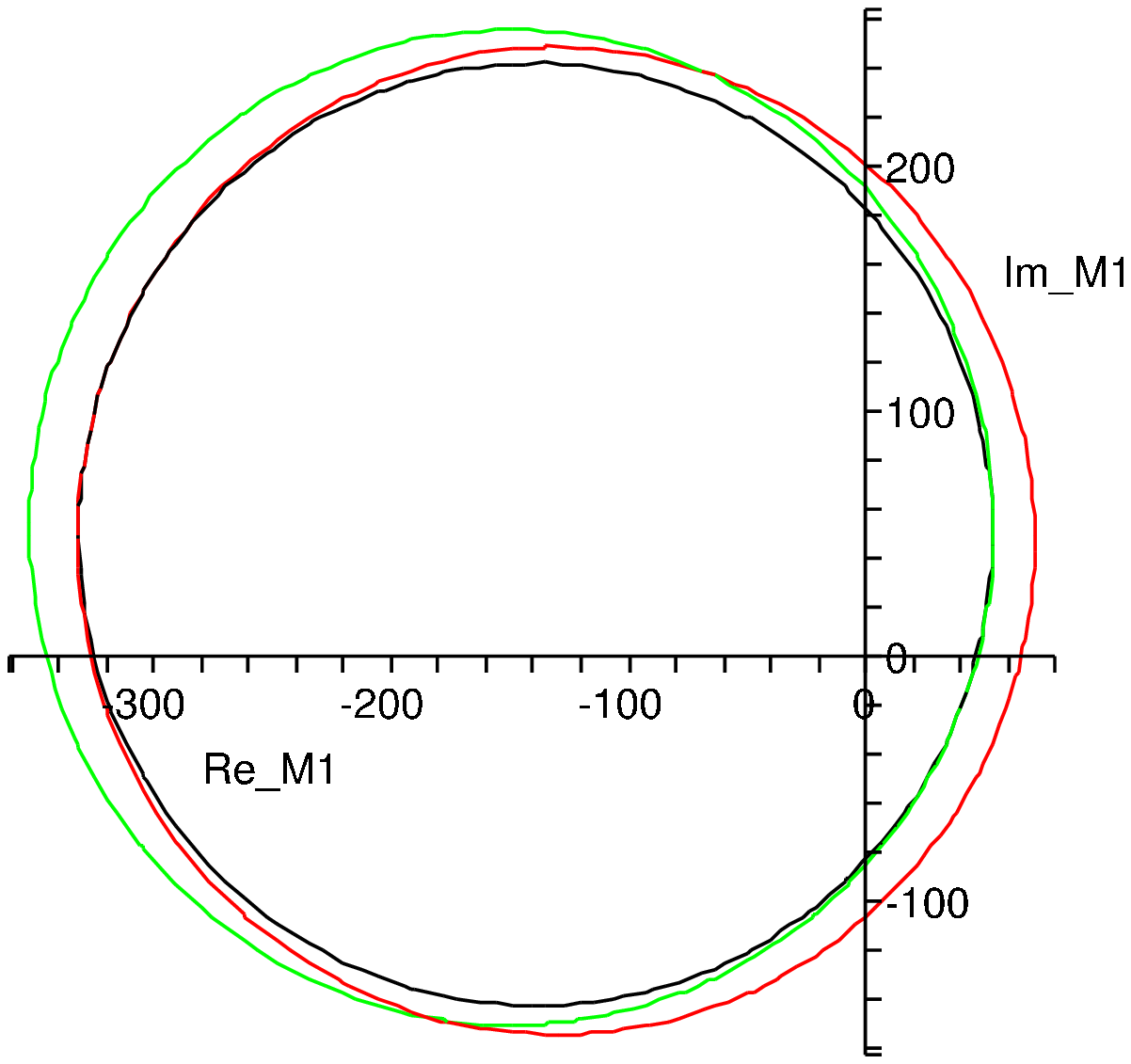}}
}
\hspace{2mm}
\subfigure[$\Delta \mu = 2.0\GeV$\label{fig:deltamu_2p0}]
{\scalebox{0.33}{\includegraphics{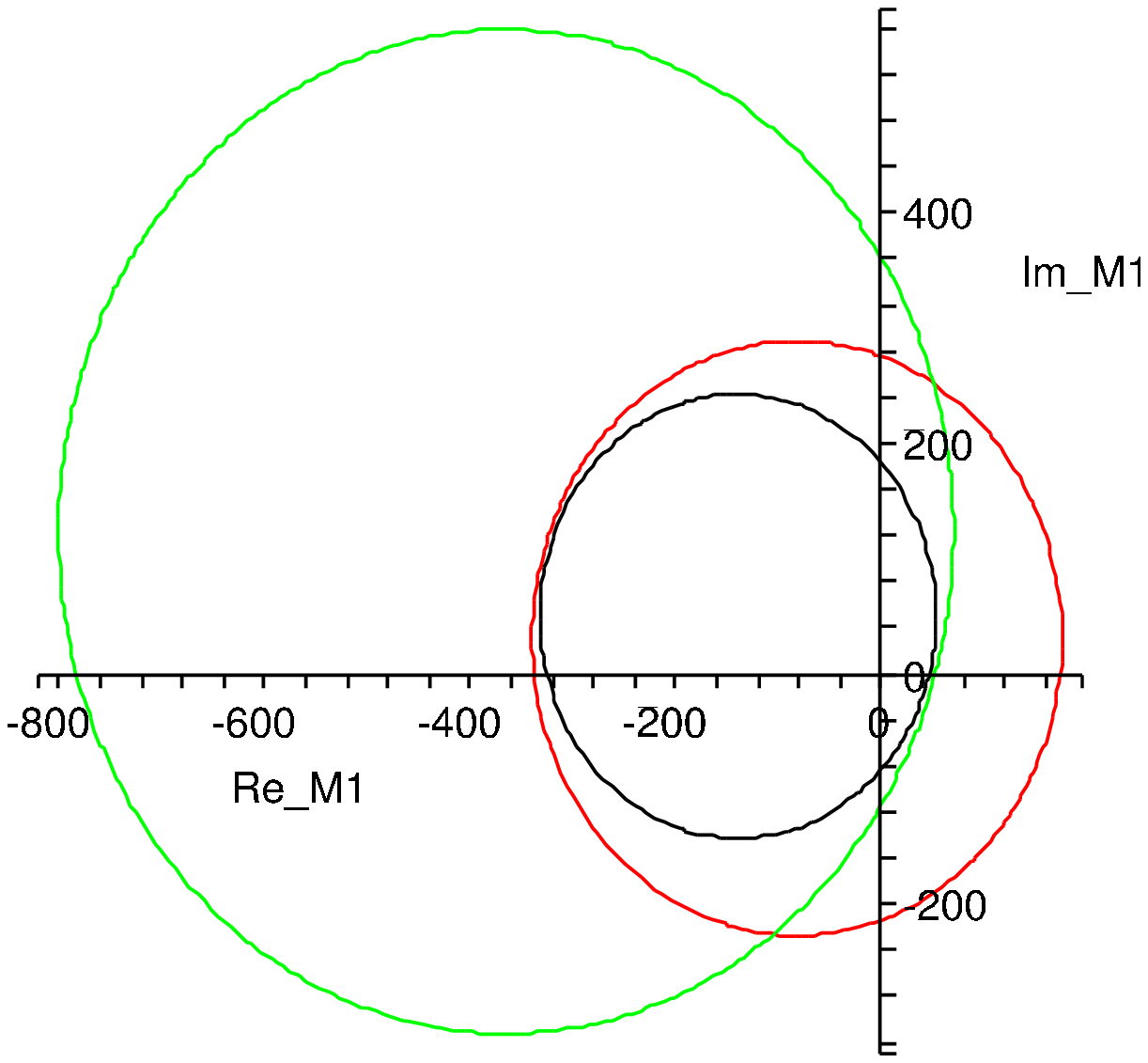}}
}
\vspace*{-6mm}
\caption{\small Influence of small perturbations of $m_{\x{4}}$ (left column), $M_2$ (middle column),
	and on $\mu$ (right column) on the circle of neutralino \x{4}, black: unperturbed circle, 
	green: $+ \Delta (m_{\x{4}},M_2,\mu)$, red: $-\Delta (m_{\x{4}},M_2,\mu)$. In black-white-printing,
	black = black, red = dark-grey, green = light grey. The color code is for all figures the same as 
	in Fig. \ref{fig:right}.}
\label{fig:circles1}
\end{figure}
\begin{figure}
\setlength{\unitlength}{1cm}
\subfigure[$\Delta \phi_\mu = 0.01\pi$\label{fig:deltamuphi_0p01}]
	{\scalebox{0.33}{\includegraphics{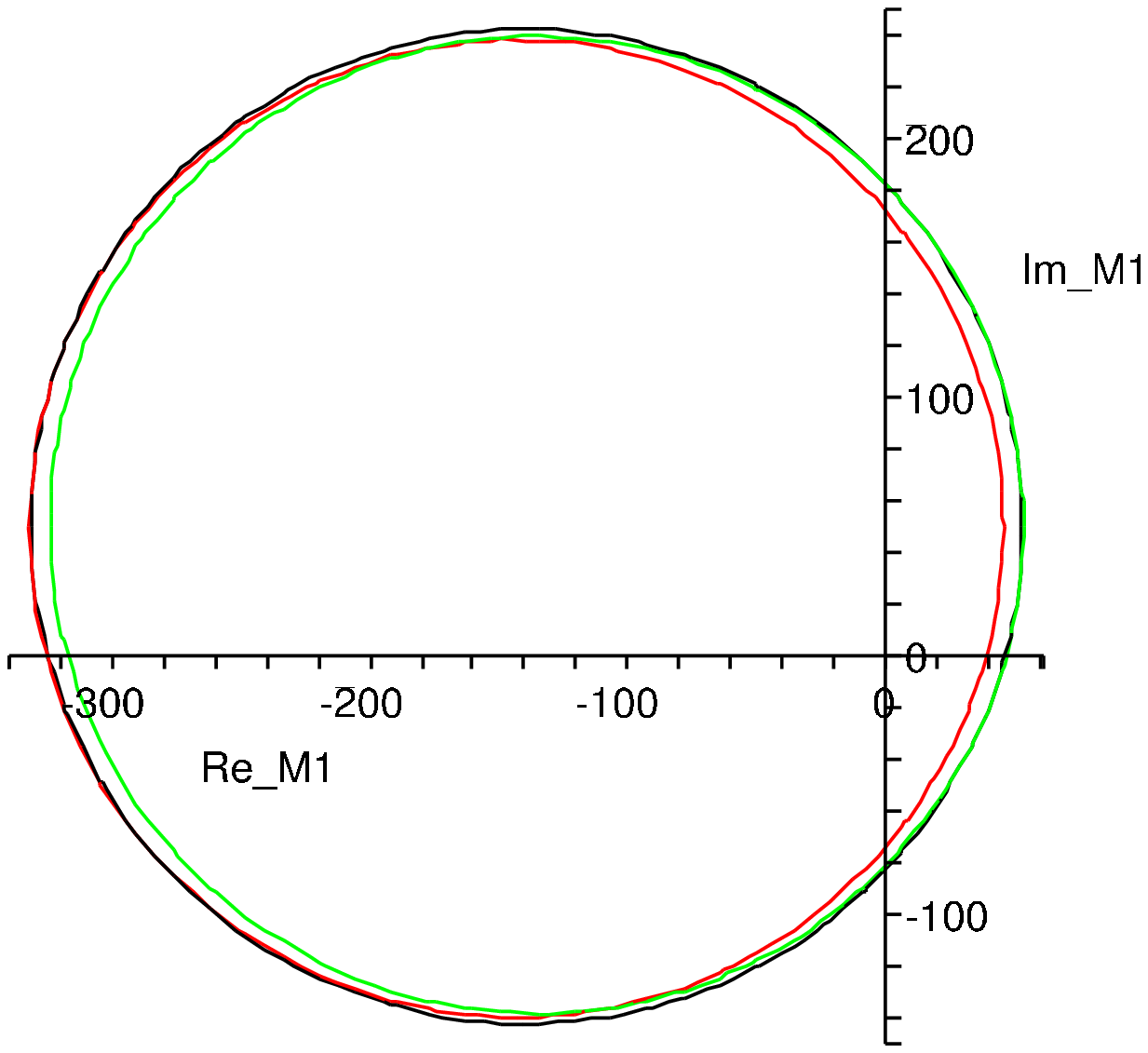}}
}
\hspace{2mm}
\subfigure[$\Delta \tan\beta \approx 0.32$
	\label{fig:deltabeta_0p001}]{\scalebox{0.33}{\includegraphics{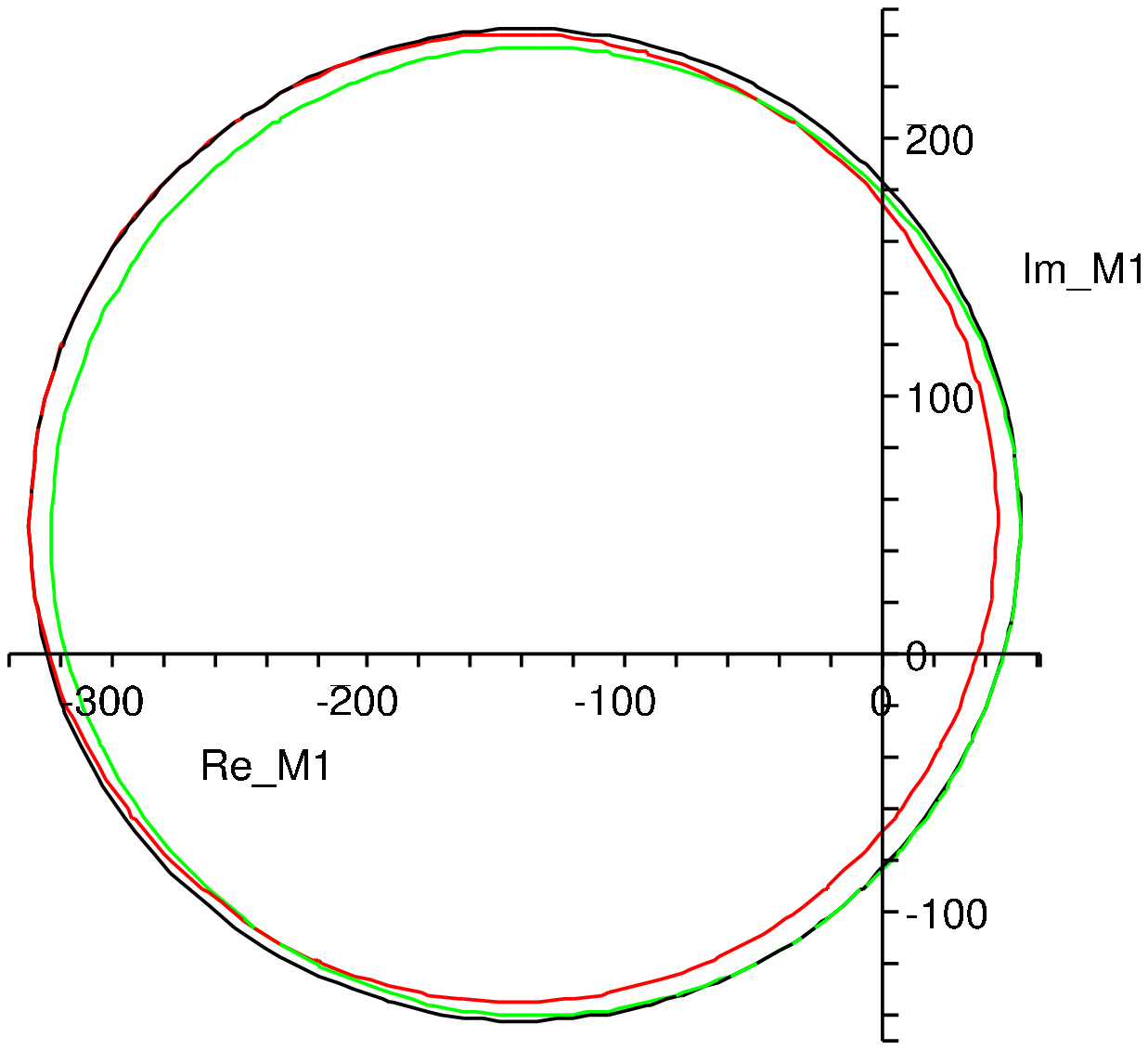}}
}
\vspace{-7mm}

\subfigure[$\Delta \phi_\mu = 0.02\pi$\label{fig:deltaphimu_0p02}]
	{\scalebox{0.33}{\includegraphics{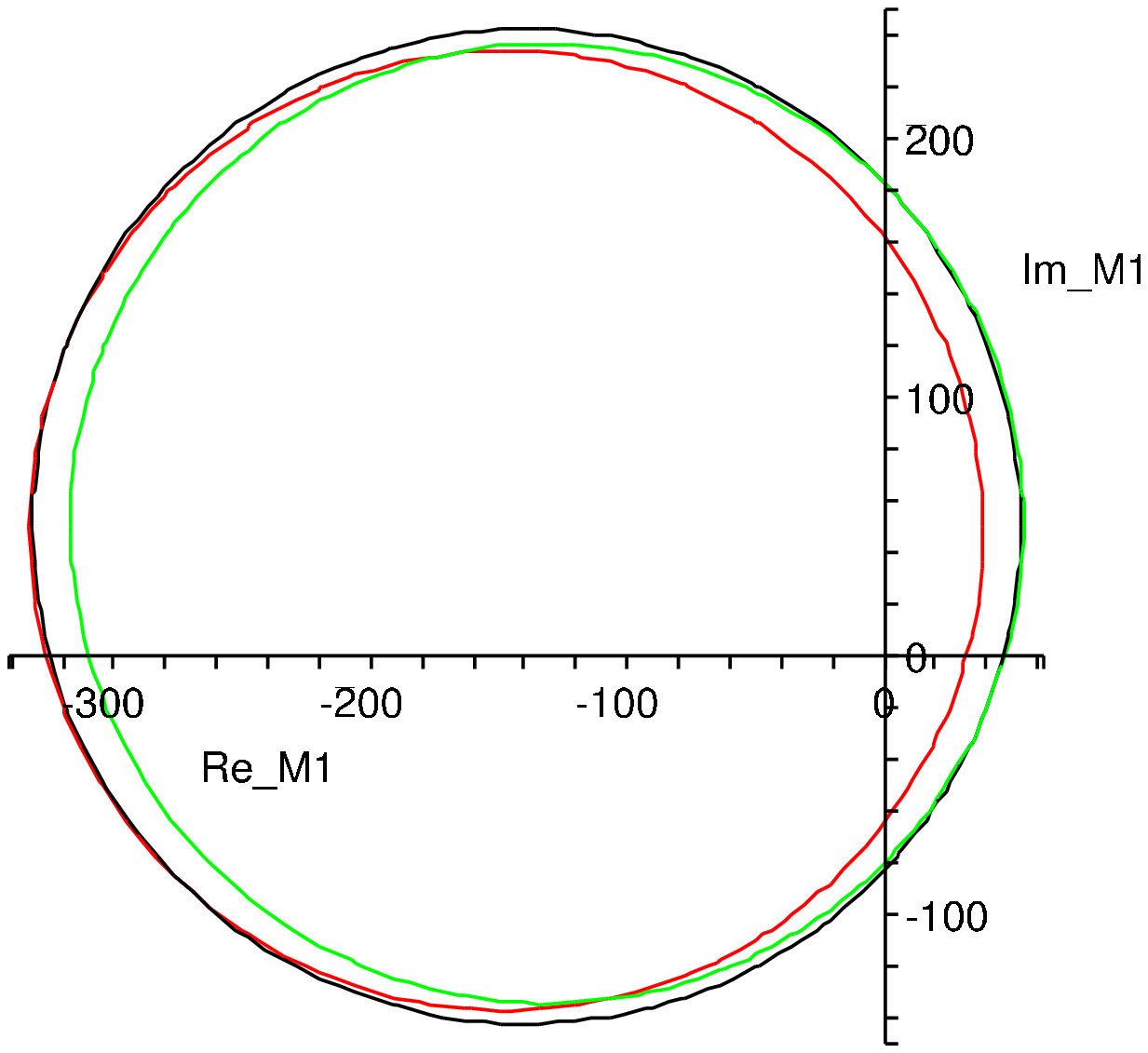}}
}
\hspace{2mm}
\subfigure[$\Delta \tan\beta \approx 0.63$
	\label{fig:deltabeta_0p002}]{\scalebox{0.33}{\includegraphics{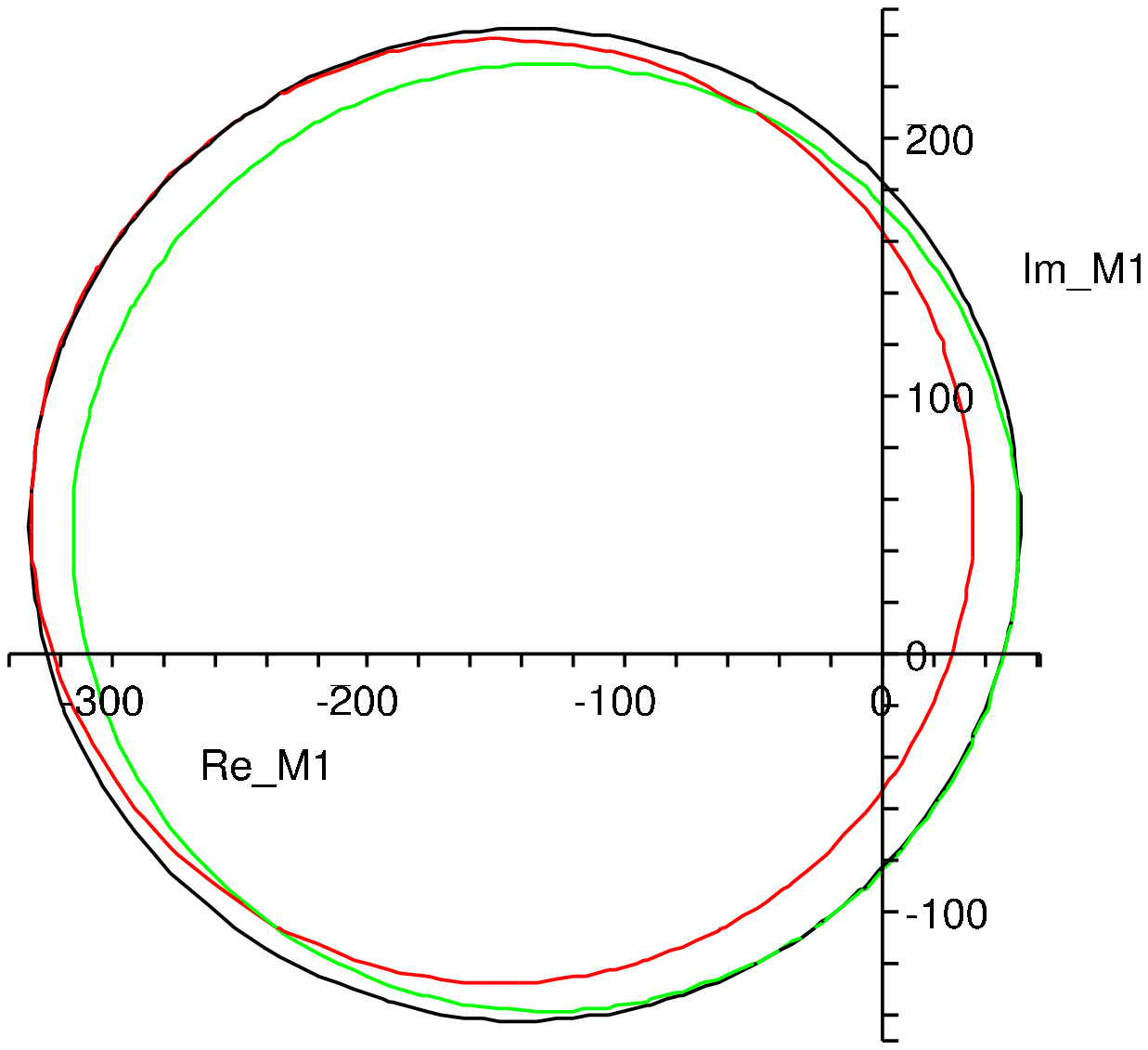}}
}
\vspace{-7mm}

\subfigure[$\Delta \phi_\mu = 0.05\pi$ \label{fig:deltaphimu_0p05}]
	{\scalebox{0.33}{\includegraphics{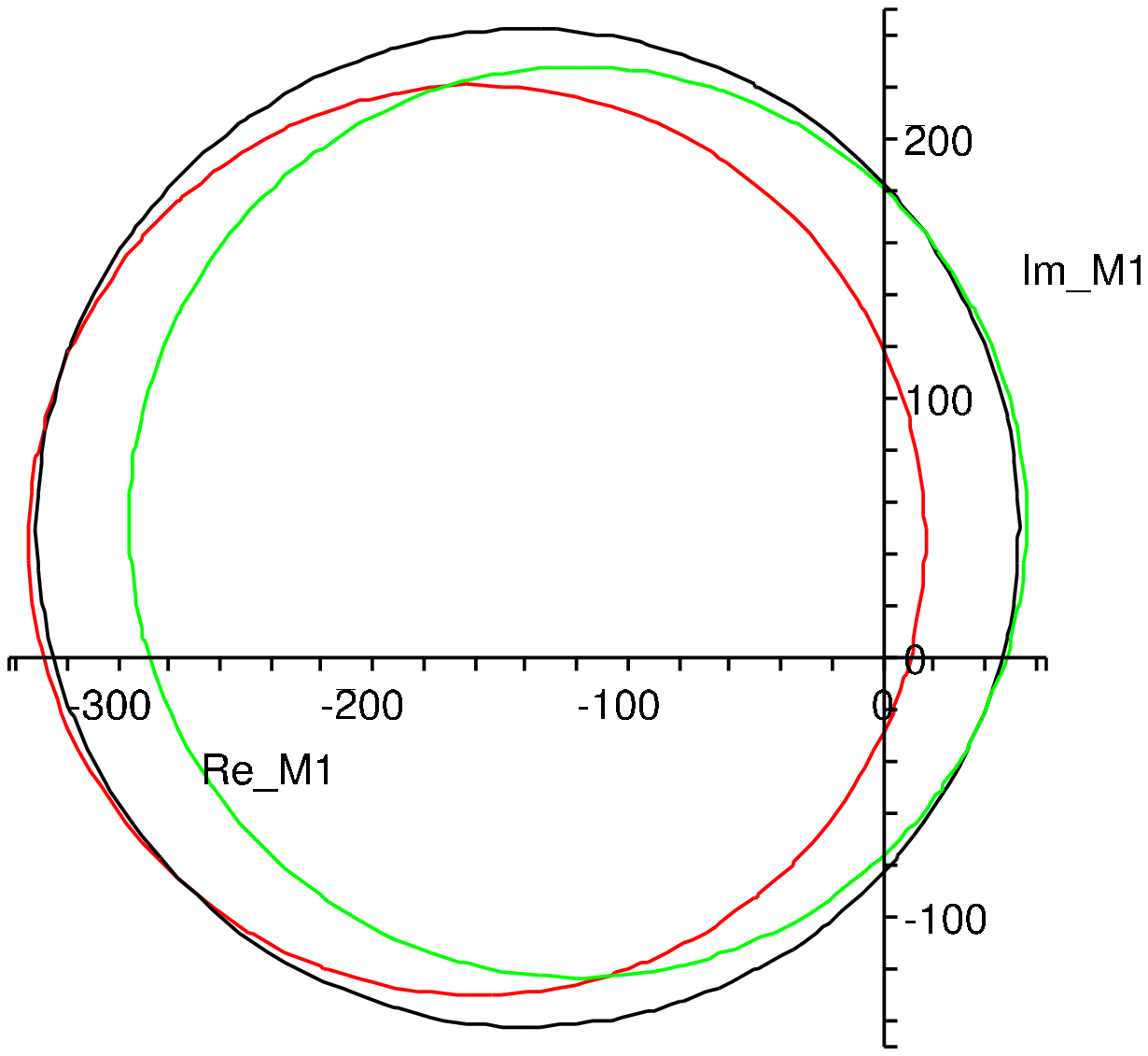}}
}
\hspace{2mm}
\subfigure[$\Delta \tan\beta \approx 1.58$ 
	\label{fig:deltabeta_0p005}]{\scalebox{0.33}{\includegraphics{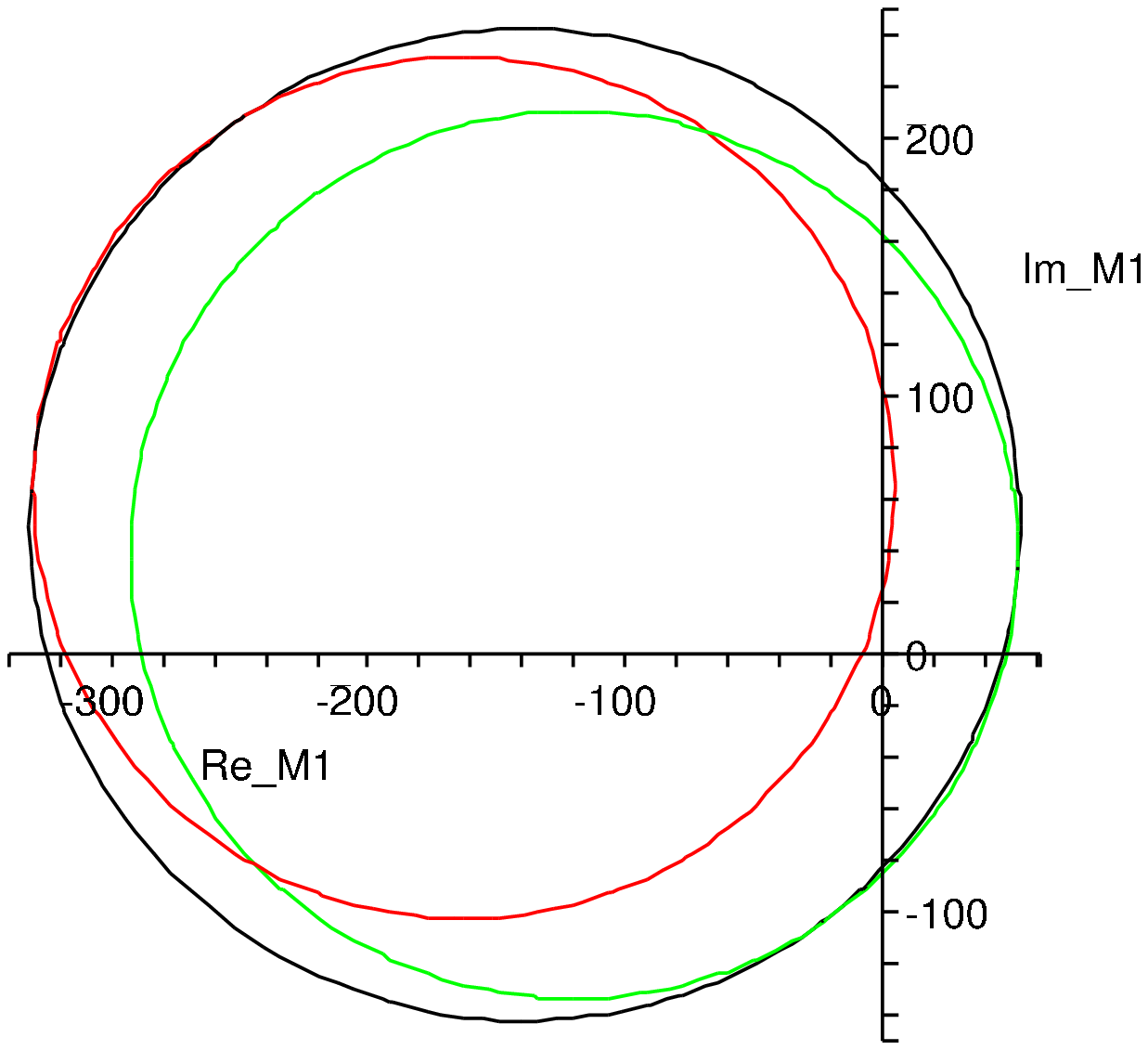}}
}
\vspace{-7mm}

\subfigure[$\Delta \phi_\mu = 0.1\pi$\label{fig:deltaphimu_0p1}]
	{\scalebox{0.33}{\includegraphics{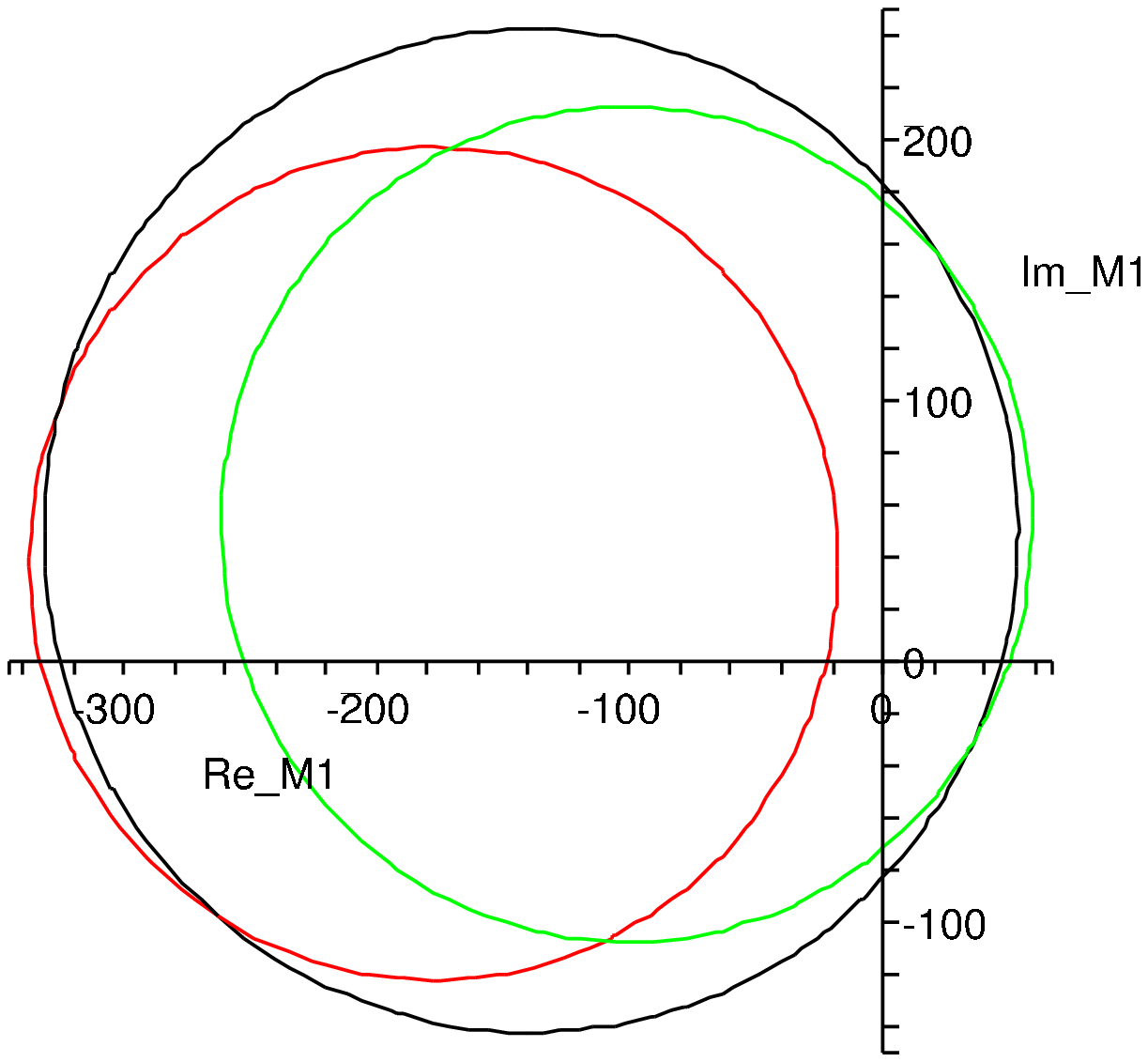}}
}
\subfigure[$\Delta \tan\beta \approx 3.17$
	\label{fig:deltabeta_p01}]{\scalebox{0.33}{\includegraphics{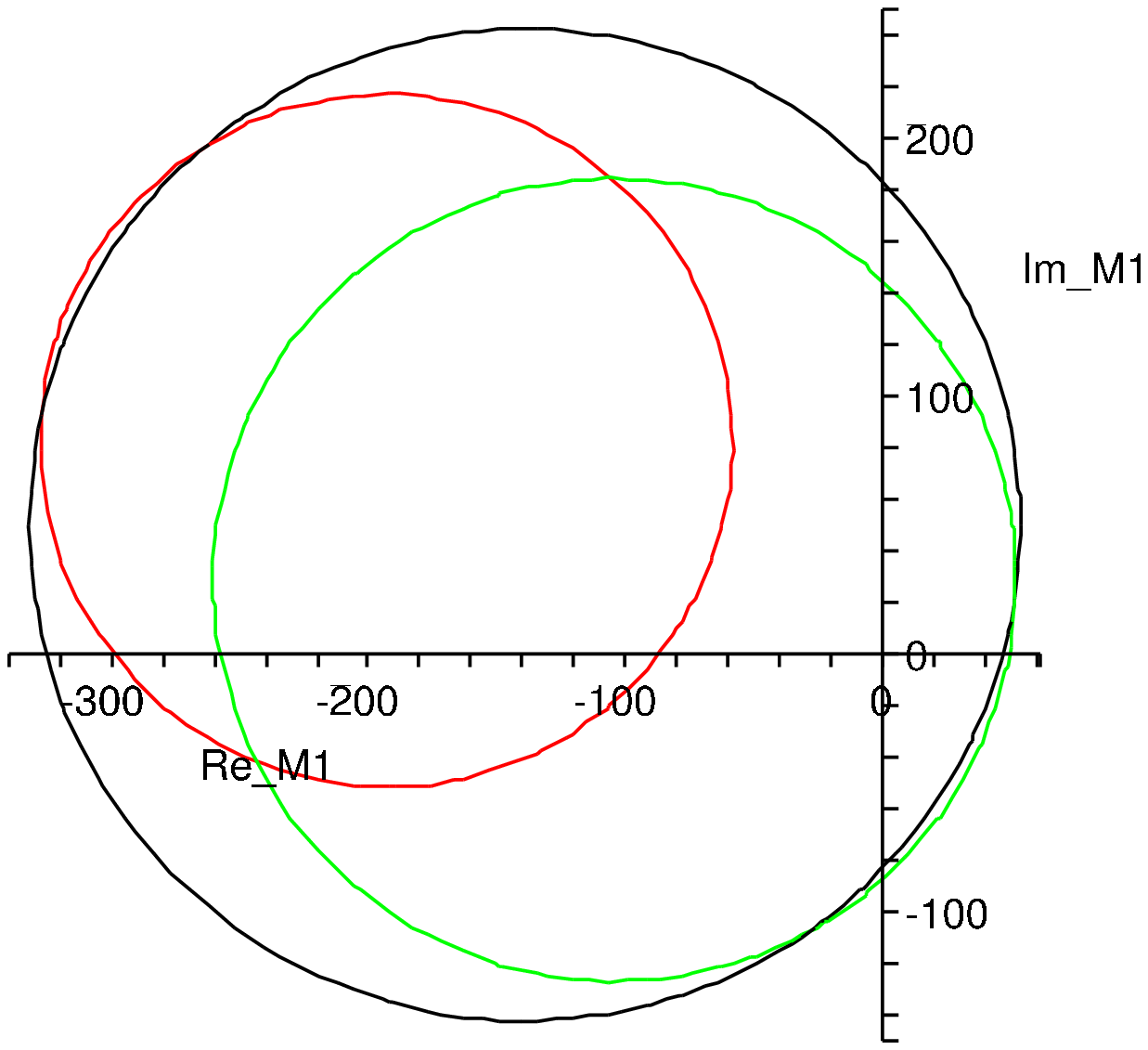}}
}
\caption{These figures show how errors on $\phi_\mu$ and $\tan\beta$ influence the position 
	of the circle of neutralino \x{4}. 
	black: unperturbed circle, green: $+ \Delta (\phi_\mu, \tan\beta)$, 
	red: $-\Delta (\phi_\mu, \tan\beta)$.}
\label{fig:circles2}	
\end{figure}

In Fig.~\ref{fig:circles1} and Fig.~\ref{fig:circles2},
I show how the circle of the {\it fourth} neutralino is disturbed
by errors in the values of $m_{\chi_4^0}$, $M_2$, $\mu$, $\tan\beta$, and $\phi_\mu$. One
of these five parameters is varied, the others are kept at their exact values.

In each picture of Fig.~\ref{fig:circles1} and Fig.~\ref{fig:circles2},
I show three circles: two perturbed circles and one 
unperturbed (black) circle. In one case the parameter is a little bit too large (green circle) , 
in the other case a little bit too low (red circle).  
The other parameters are not varied.
The unperturbed circle (black) is thus the same in all Figs (note however the scale change).

From these figures I conclude for the circle of the fourth neutralino:
\begin{itemize}
\item $m_{\x{4}}$ must be measured very precisely, small errors lead to large deviations from 
	the unperturbed circle.
	The upper bound on this error is $\Delta m_{\x{4}} \approx 0.1\GeV$.	 
\item The dependence on $M_2$ is weaker. A maximal error on about $\Delta M_2 \approx 1 \GeV$ is allowed.
	The reason for the weak dependence is that $m_{\x{4}}$ does not depend on $M_2$ to
	zeroth order.
\item The  experimental error on $\mu$ should be small: $\Delta \mu \lsim 0.1\GeV$.
	This is due to the strong dependence of $m_{\x{4}}$ on $\mu$. For the error on its phase, 
	I find: $\Delta \phi_\mu \leq 0.01\pi$.
\item For the error on $\tan\beta$, I find $\Delta \tan\beta \leq 0.3$.
\item The expected experimental errors of $M_2$ and $\mu$ at the ILC are $\mathcal{O}(1\GeV)$, 
      see Ref~\cite{Choi:2000ta}.
      This error is too large for the method described in Ref~\cite{Choi:2001ww}.    
\end{itemize}
 
The situation is similar or worse for the circles of the neutralinos \x{2} and \x{3}. 
For the circles of the second 
neutralino the influence of the error on $M_2$ is disastrous. 
I conclude that it is not possible to determine $|M_1|$ and $\phi_1$ with the circle method 
from Ref~\cite{Choi:2001ww}. In the following section I propose an alternative method.

\section{Determining Neutralino Couplings} 

I will show in this sections how radiative neutralino production together with 
neutralino pair production 
\begin{eqnarray}
e^+e^- \rightarrow \x{i}\x{j}
\end{eqnarray} 
can be used to determine 
the couplings of the neutralinos to the $Z$ boson and to $\tilde{e}_{R/L}$. 
My method is not only designed for the MSSM, it is applicable to every model with 
measurable cross sections. 
The idea is as follows: Write the cross section $\sigma$ of an arbitrary process as 
\begin{eqnarray}
\label{eq:idee}
\sigma = \sum_i c_i(a_i,b_i,f_{i}) X_i\enspace,
\end{eqnarray}  
where the $c_i$ are functions of the unknown couplings $a_i$, $b_i$, and $f_{ij}$,
the functions $c_i$ are not necessarily linear. 
The $X_i$ are calculable factors, depending only on the neutralino and selectron masses.
If there are $n$ cross section measurements, $n \ge $ number of couplings, 
then one can perform a least square fit to determine the couplings $a_i$, $b_i$, and $f_{ij}$.
If the Eq.~(\ref{eq:idee}) is non-linear in the couplings one can either linearize Eq.~(\ref{eq:idee})
and use the custom linear least square functions provided by \texttt{Maple} or \texttt{Mathematica}
or one can use techniques for non-linear fits like \texttt{Minuit}\cite{minuit}. 
I have chosen the first approach because the linearized equations are not too complicated.
The method is best illustrated by an example.

\subsection{Mathematical Structure of the cross section and the couplings}
To determine the couplings of the neutralinos to the selectrons and the $Z^0$-boson, I use 
the cross sections of the following reactions:
\begin{subequations}
\label{eq:system}
\begin{align}
\label{x:11gamma}
e^+e^- &\rightarrow \tilde{\chi}_1^0 \tilde{\chi}_1^0 \gamma \\
\label{x:1i}
e^+e^- &\rightarrow \tilde{\chi}_1^0 \tilde{\chi}_i^0,\enspace i = 2,3,4\enspace ,\\
\label{x:22}
e^+e^- &\rightarrow \tilde{\chi}_2^0 \tilde{\chi}_2^0 \enspace .
\end{align}
\end{subequations}
It is straightforward to include further reactions $e^+e^- \rightarrow \tilde{\chi}_i^0 \tilde{\chi}_j^0$,
${ij} = {23}, {24}, {33}, {34}, {44}$, if they are measurable.  
Their cross sections can be decomposed as follows:
\begin{subequations}
\begin{align}
\label{eq:x1x1g}
\sigma(e^+e^- \rightarrow \tilde{\chi}_1^0 \tilde{\chi}_1^0 \gamma) &\equiv \sigma_{11\gamma} = 
	a_1^4 X + b_1^4 Y + F_{1} Z,\hspace*{0mm}\\[2mm]
\label{eq:x1xj}
\sigma(e^+e^- \rightarrow \tilde{\chi}_1^0 \tilde{\chi}_i^0) &\equiv \sigma_{1i}= 
 a_1^2 a_i^2 X_{1i} + b_1^2 b_i^2 Y_{1i} + a_1 a_i f_{1i} X_{2i} - b_1 b_i f_{1i}Y_{2i} + f_{1i}^2 Z_i,\\[2mm]
\label{eq:x2x2}
\sigma(e^+e^- \rightarrow \tilde{\chi}_2^0 \tilde{\chi}_2^0) &\equiv \sigma_{22}= 
	a_2^4 X_{1,22} + b_2^4 Y_{1,22} + a_2^2  f_{22} X_{2,22} -b_2^2 f_{22} Y_{2,22} + f_{22}^2 Z_{22}\enspace .
\end{align}
\end{subequations}
The factors $a_i$, $b_i$, $f_{ij}$ are associated with the couplings to the right-selectron, left-selectron,
and $Z^0$-boson, respectively, and are given by
\begin{subequations}
\begin{align}
\label{eq:a}
a_i &= -\frac{N_{i1}}{\cw}, \\[2mm]
\label{eq:b}
b_i & =  \half\left(\frac{N_{i2}}{\sw} + \frac{N_{i1}}{\cw}\right),\\[2mm]
\label{eq:f}
f_{ij} &= \half\left(N_{i3}N_{j3} - N_{i4}N_{j4} \right),\\[2mm]
\label{eq:F}
F_1 &=f_{11}^2 \enspace .
\end{align}
\end{subequations}
$f_{11}$ appears only quadratically in Eq.~(\ref{eq:x1x1g}) and may be small, its error 
from a least square fit however may be large. 
The optimal value for $f_{11}^2$ can become negative, which is unphysical.
Therefore, the use of $F_1$ instead of $f_{11}$ as a fit parameter secures the convergence of the 
iteration. 
Due to this problem, the value of $f_{11}$ is not used further. 
The $X_i$, $Y_i$, $Z_i$ are functions of the right selectron mass, left selectron mass, and $Z^0$-mass, 
respectively. They all depend on the neutralino masses, the center of mass energy and the longitudinal 
polarisation of the electron-positron-beam. Their explicit form can be found 
in~\cite{Moortgat-Pick:1999di}.
They need to be calculated only once. So one does not have the problem that the iterations in the program that 
tries to find the minimum of $\chi^2$ does not converge due to errors occurring when integrating out 
the $X_i$, $Y_i$, and $Z_i$ by Monte Carlo integration.     

This set of equations is nonlinear in the couplings parameters. The equations can be expanded in
a Taylor series up to first order:
\begin{eqnarray}
\sigma(p) \approx \sigma(p_0) + \sigma'(p_0)(p - p_0),
\end{eqnarray} 
where $p$ is a vector, collecting the parameters $a_i$, $b_i$, $f_{ij}$.
The linearized equations can be solved by a least square fit recursively.
$p_0$ is a first guess of the solution.

\medskip
From the parameters $a_i$, $b_i$, and $f_{ij}$ one can determine the entries of the neutralino
diagonalisation matrix $N$.
\begin{subequations}
\begin{align}
\label{eq:N11}
N_{i1} & =  -\cw a_i, \enspace i = 1\ldots 4, \\[3mm]
\label{eq:N12}
N_{i2} & =  \sw (2 b_i + a_i),\enspace  i = 1\ldots 4, \\[3mm]
\label{eq:N23}
N_{23} & =  \pm \sqrt{\half(1 - N_{21}^2 - N_ {22}^2 + 2 f_{22})},\\[2mm]
\label{eq:N24}
N_{24} & = \pm \sqrt{\half(1 - N_{21}^2 - N_ {22}^2 - 2 f_{22})},
=  \pm \sqrt{\half(1 - N_{21}^2 - N_ {22}^2 - N_{23}^2)},\\[2mm]
\label{eq:N13}
N_{13} & =  \frac{2 f_{12} - N_{11} N_{21} - N_{12}N_{22}}{2N_{23}},\\[2mm]
\label{eq:N14}
N_{14} & =  -\frac{2 f_{12} + N_{11} N_{21} + N_{12}N_{22}}{2N_{24}},\\[2mm]
\label{eq:N33}
N_{33} & =  \frac{2 f_{13} - N_{11}N_{31} - N_{12}N_{32}}{2 N_{13}},\\[2mm]
\label{eq:N34}
N_{34} & = \pm\sqrt{1 - N_{31}^2 - N_{32}^2 - N_{33}^2}\, ,\\[2mm]
\label{eq:N43}
N_{43} & =  \frac{2 f_{14} - N_{11}N_{41} - N_{12}N_{42}}{2 N_{13}},\\[2mm]
\label{eq:N44}
N_{44} & = \pm\sqrt{1 - N_{41}^2 - N_{42}^2 - N_{43}^2}\, ,
\end{align}
\end{subequations}
Eqs.~(\ref{eq:N11})-~(\ref{eq:N12}) are derived from Eqs.~(\ref{eq:a})-(\ref{eq:b}),
$N_{23}$ and $N_{24}$ are obtained from the unitarity relation of the matrix $N$; 
$f_{22}$, $N_{13}$ and $N_{14}$ are
constructed in such a way, that $\widetilde{\chi}^0_1 = (N_{11},N_{12},N_{13}, N_{14})$ is orthogonal to  
$\widetilde{\chi}^0_2 =(N_{21},N_{22},N_{23}, N_{24})$; the elements $N_{33}$, $N_{43}$  
are calculated from  $f_{13}$ and $f_{14}$ as well as the orthogonality relation 
$\widetilde{\chi}^0_{3/4} \cdot \widetilde{\chi}^0_1 = 0$.
We use unitarity to calculate $N_{43}$ and $N_{44}$, because this leads to a smaller error 
for these elements.  
The sign is choosen in such a way, that $\tilde{\chi}^0_{3,4}$ are orthogonal
to $\tilde{\chi}^0_{1}$.
\subsection{The cross sections for Neutralino pair production}
\begin{figure}[p]
\centering
\setlength{\unitlength}{1cm}
\subfigure[Unpolarised cross sections\label{fig:xsections}]{\scalebox{0.33}
{\includegraphics{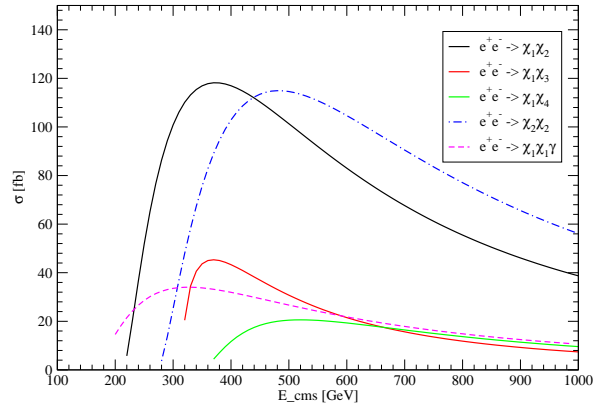}}}
\vspace*{10mm}

\subfigure[Beam polarisation $(P_+|P_-) = (-0.6|0.8)$ \label{fig:xsectionspp}]{\scalebox{0.33}
{\includegraphics{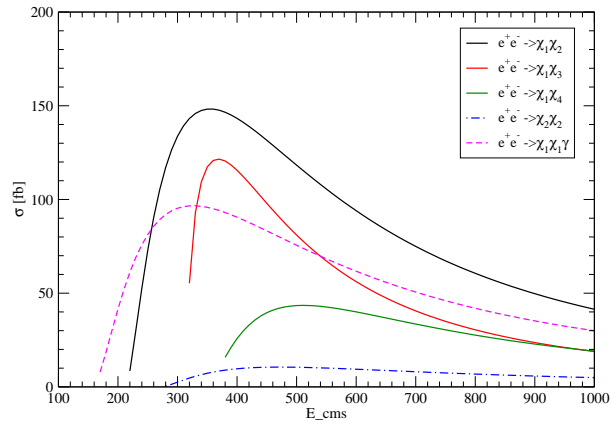}}}
\vspace*{10mm}

\subfigure[Beam polarisation $(P_+|P_-) = (0.6|-0.8)$ \label{fig:xsectionsnp}]{\scalebox{0.33}
{\includegraphics{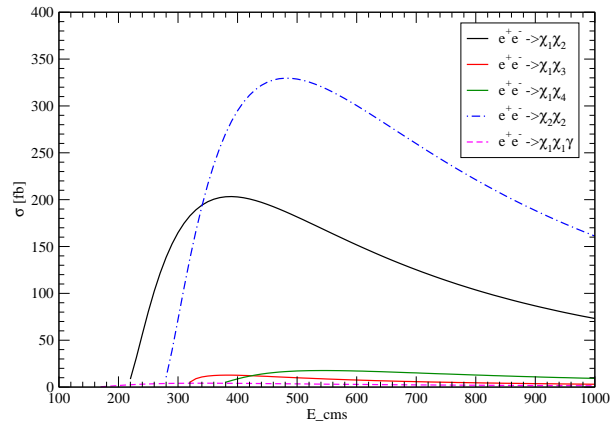}}}
\caption{Comparison of cross sections for different beam polarisations.}
\label{fig:xsec}
\end{figure}

In Fig.~\ref{fig:xsec}, I show the cross sections for 
$e^+ e^- \rightarrow \tilde{\chi}_1^0 \tilde{\chi}_i^0$, $i = 2 \ldots 4$ and \signal[{}] 
for cms - energies from 
$200\GeV$ - $1000\GeV$ for three different polarisations: $(P_+|P_-) = (0|0)$ in 
Fig.~\ref{fig:xsections}, $(P_+|P_-) = (-0.6|0.8)$ in Fig.~\ref{fig:xsectionspp}, 
and $(P_+|P_-) = (0.6|-0.8)$ in Fig.~\ref{fig:xsectionsnp}.
The figures show, how suitable beam polarisation enhances cross sections.
$(P_+|P_-) = (0.6|-0.8)$ enhances $\tilde{\chi}_1^0\tilde{\chi}_2^0$, and
$\tilde{\chi}_2^0\tilde{\chi}_2^0$, 
pair production compared to unpolarised beams, the opposite
beam polarisation enhances radiative neutralino production, 
$\tilde{\chi}_1^0\tilde{\chi}_2^0$ and $\tilde{\chi}_1^0\tilde{\chi}_3^0$
and $\tilde{\chi}_1^0\tilde{\chi}_4^0$.
In my example model which I will present below in detail,
the $\chi_1^0$ is mainly bino ($\approx 95\%$) which couples mostly to right handed sleptons, 
the $\chi_2^0$ mainly wino ($\approx 85\%$) which couples preferably to left handed slepton; so 
cross sections with $\chi_1^0$ involved are enhanced by right handed beam polarisation,
and cross sections with $\chi_2^0$ are enhanced by left handed beam polarisation.
   
Polarised beams are essential for the described method to determine parameters, since
they enhance couplings either between right handed particles or left handed particles.  

\section{An example}
\subsection{The model}
In order to demonstrate my method,
I choose an example  model with light neutralinos, so that at least the processes
\signal, $e^+e^-\rightarrow \tilde{\chi}_1^0\tilde{\chi}_i^0$, $i = 2,3,4$,  
and  $e^+e^-\rightarrow \tilde{\chi}_2^0\tilde{\chi}_2^0$ are kinematically accessible at the ILC
and the cross sections exceed $\mathcal{O}(10\fb)$ for both polarisations. 
The higgsino components of $\tilde{\chi}^0_1$
should not be too small, so that there might be a chance to determine the  
$\tilde{\chi}^0_1$-$\tilde{\chi}^0_1$-$Z$ coupling.

The input data and the derived neutralino and selectron masses are listed in the first row of
Tab.~\ref{tab:exampledata}. For comparison I also list the values for the SPS1a 
scenario.
\begin{table}[h!]
\begin{center}
\begin{tabular}{lccccc|cccccc}
\toprule
&$M_2$ & $M_1$ & $\mu$ & $M_0$ & $\tan\beta$ & $m_{\x{1}}$ & $m_{\x{2}}$ & $m_{\x{3}}$ & $m_{\x{4}}$&
						$m_{\tilde{e}_R}$ & $m_{\tilde{e}_L}$\\[1mm]
\hline
my model&165 & 82.5 & 230 & 75 & 11 & 76.3 & 136.8 & -239.2 & 273.2& 117.3 & 171.3\\[1mm]
SPS1a& 192 & 102 & 352 & 100 & 10 & 99 & 175 & 348 & 369 & 145 & 204 \\[1mm]  
\bottomrule
\end{tabular}
\caption{Input and mass parameters of the example model. The minus-sign appearing in the 
	third neutralino mass denotes the $CP$-eigenvalue of this particle. All masses are 
	given in GeV.}
\label{tab:exampledata}	
\end{center}
\end{table}
In my model the cross section for the process $e^+e^-\rightarrow \tilde{\chi}_2^0\tilde{\chi}_3^0$
is larger than $10\fb$ for both beam polarisations and  
the cross section of the process $e^+e^-\rightarrow \tilde{\chi}_2^0\tilde{\chi}_4^0$ is larger than $10\fb$
for left beam polarisation. Therefore, I shall later present a study
including these processes. The fairly light particle spectrum and, additionally, 
the large wino component in the latter process are the reasons for the large cross sections.    
I do not include them from the beginning.   

With these values the neutralino diagonalisation matrix follows as 
\begin{eqnarray}
N &=& 
\begin{pmatrix}
\phantom{-}0.953 & -0.117 & \phantom{-}0.257 & -0.106 \\[1mm]
-0.241 & -0.844 & \phantom{-}0.402 & -0.262 \\[1mm]
-0.089 & \phantom{-}0.129 &\phantom{-} 0.682 &\phantom{-} 0.715 \\[1mm]
-0.158 & \phantom{-}0.508 &\phantom{-}0.554 &-0.640
\end{pmatrix}.
\end{eqnarray}

The neutralino couplings to $\tilde{e}_R$,  $\tilde{e}_L$, and $Z^0$ are listed in 
Tab.~\ref{tab:thcouplings}.
\begin{table}[h!]
\begin{center}
\begin{tabular}{lll}
\toprule
couplings to $\tilde{e}_R$ & couplings to $\tilde{e}_L$ & couplings to $Z^0$\\
\hline
$a_1 = -0.740$ & $b_1 = \phantom{-}0.287$ & $f_{11} = \phantom{-} 0.065 $\\[1mm]
$a_2 = \phantom{-} 0.187$ & $b_2 = -0.690$ & $f_{12}= \phantom{-} 0.038$\\[1mm]
$a_3 = \phantom{-}0.069$ & $b_3 = \phantom{-}0.057$ & $f_{13} = \phantom{-}0.126$\\[1mm]
$a_4 = \phantom{-}0.123$ & $b_4 = \phantom{-}0.398$& $f_{14} = \phantom{-}0.037$\\[1mm]
&&$f_{22} = \phantom{-}0.047$\\
\bottomrule
\end{tabular}
\caption{Theoretical couplings.}
\label{tab:thcouplings}
\end{center}
\end{table}

I assume that eight cross section measurements of each process are available:
Four different cms energies of the beam ($500\GeV$, $550\GeV$, $600\GeV$, $650\GeV$)
are combined  with two different longitudinal beam polarisations 
$(P_+|P_-) = (-0.6|+ 0.8)$ and $(P_+|P_-) = (+0.6|-0.8)$. 

Each measurement has an error. The error on the cross sections consists of the statistical
Poisson error and the systematic error. I consider only the statistical error. 
To simulate the statistical error on a measurement, 
I calculate the exact cross section and add a Gaussian distributed random number 
with zero mean and variance $V = (\delta\sigma)^2$.  
The statistical error $\delta \sigma$ follows from
\begin{eqnarray}
\label{eq:delta11g}
N_{\signal} &=& N_{e^+e^-\to \ssl{E}\gamma} - N_{\Backgroundnu}, \qquad 
      N_{\mathrm{process}} = \sigma(\mathrm{process}) \mathcal{L},\notag\\[1mm]
(\delta N_{\signal})^2 &=&  (\delta N_{e^+e^-\to \ssl{E}\gamma})^2 + (\delta N_{\Backgroundnu})^2\notag\\[1mm]
&=&  N_{e^+e^-\to \ssl{E}\gamma} +  N_{\Backgroundnu}\notag\\[2mm]
&=&  N_{\signal} +  2 N_{\Backgroundnu},\notag\\[2mm]
\delta \sigma(\signal) &=& \frac{\delta N_{\signal}}{\mathcal{L}}
         = \sqrt{\frac{\sigma(\signal) + 2\sigma(\Backgroundnu)}{\mathcal{L}}},\\[2mm]
\label{eq:deltaij}
\delta \sigma(e^+e^-\rightarrow \tilde{\chi}_i^0  \tilde{\chi}_j^0) &=& 
	  \frac{\delta N_{e^+e^-\to \x{i}\x{j}}}{\mathcal{L}} =
	\sqrt{\frac{\sigma(e^+e^-\rightarrow \tilde{\chi}_i^0  \tilde{\chi}_j^0)}{\mathcal{L}}}
\end{eqnarray}
with $\mathcal{L}$ and $N_{\mathrm{process}}$ denoting the integrated luminosity in $\fb^{-1}$ and the 
number of events of the process, respectively. 
Eq.~(\ref{eq:delta11g}) accounts for the fact that the 
number of events for radiative neutralino production is calculated as the difference from all events
"photon plus missing energy" and the radiative neutrino background. Eq.~(\ref{eq:deltaij}) is the Poisson
error for neutralino pair production.


\medskip\noindent
The values for the cross sections and their errors are:
{\normalsize
\begin{equation}
\label{eq:inputcross}
\begin{array}{lcrrrrr}
\hline
\vspace{0mm}
\sqrt{s}& \text{Polarisation}& \sigma_{11\gamma} &\sigma_{12}&\sigma_{13} & \sigma_{14} 
&\sigma_{22}\\[1mm]
[\text{GeV}]&\text{in}\enspace \% &\multicolumn{5}{c}{\text{all cross sections in} \fb}\\[2mm]
\hline
 500 &(-0.6|+0.8) & 75.6 \pm 1.0  & 118.3 \pm 0.5 & 81.0 \pm 0.4 & 43.4 \pm 0.3 & 10.5\pm 0.1\\[1mm]
 500 &(+0.6|-0.8) & (3.4 \pm 4.5) & 181.5 \pm 0.6 & 10.0 \pm 0.1 & 17.2 \pm 0.2 &328.7\pm 0.8\\[1mm]
 550 &(-0.6|+0.8) & 68.3 \pm 1.0  & 105.6 \pm 0.5 & 67.2 \pm 0.4 & 42.6 \pm 0.3 & 10.1\pm 0.1\\[1mm]
 550 &(+0.6|-0.8) &(3.11 \pm 4.6) & 166.5 \pm 0.6 &  8.6 \pm 0.1 & 17.7 \pm 0.2 &317.9\pm 0.8\\[1mm]
 600 &(-0.6|+0.8) & 61.7 \pm 1.0  &  94.0 \pm 0.4 & 56.2 \pm 0.3 & 40.0 \pm 0.3 &  9.4\pm 0.1\\[1mm]
 600 &(+0.6|-0.8) & (2.8 \pm 4.6) & 151.7 \pm 0.5 &  7.5 \pm 0.1 & 17.2 \pm 0.2 &300.4\pm 0.8\\[1mm]
 650 &(-0.6|+0.8) & 55.8 \pm 0.9  &  83.9 \pm 0.4 & 47.5 \pm 0.3 & 36.8 \pm 0.3 &  8.8\pm 0.1\\[1mm]
 650 &(+0.6|-0.8) & (2.6 \pm 4.6) & 137.8 \pm 0.5 &  6.6 \pm 0.1 & 16.2 \pm 0.2 &280.3\pm 0.8\\[1mm]     
\hline
\end{array}
\end{equation}
}
The values in brackets are not used for further calculations as they are too small. 
These input data from Tab.~\ref{eq:inputcross} lead to $36$ equations for $13$ fit parameters.
The diagonalisation matrix has 16 entries, but only six of them are independent because of unitarity. 
There are ten relations 
between the matrix elements together with 13 equations from the fit parameters.
The system is overdetermined. I choose the equations to determine the elements of the 
diagonalisation matrix such, that the error on the elements is as small as possible.
In principle, unitarity can 
be tested by the first and the second column. 
So it can be tested if there is an additional singlino field.

\medskip
For the couplings I get as a result of the least square fit:
\begin{table}[h!]
\begin{center}
\begin{tabular}{lll}
\toprule
couplings to $\tilde{e}_R$ & couplings to $\tilde{e}_L$ & couplings to $Z^0$\\
\hline
$a_1 = -0.7397 \pm 0.007$ & $b_1 = \phantom{-}0.2882 \pm 0.003$ & $(f_{11}^2 = -0.035) $\\[1mm]
$a_2 = \phantom{-} 0.1879\pm 0.003$ & $b_2 = -0.6918\pm 0.002$ & $f_{12}= \phantom{-} 0.036\pm 0.006$\\[1mm]
$a_3 = \phantom{-}0.0667\pm 0.001$ & $b_3 = \phantom{-}0.064\pm 0.002$ & $f_{13} = \phantom{-}0.131\pm 0.002$\\[1mm]
$a_4 = \phantom{-}0.122\pm 0.002$ & $b_4 = \phantom{-}0.303\pm 0.006$& $f_{14} = \phantom{-}0.043 \pm 0.002$\\[1mm]
&&$f_{22} = \phantom{-}0.051 \pm 0.009$\\
\bottomrule
\end{tabular}
\caption{Result of the fit on the couplings}
\label{tab:fittedcouplings}
\end{center}
\end{table}

The obtained $\chi^2$ value is
\begin{eqnarray}
\label{eq:chi2}
\chi^2 & = & 20.9,\enspace
\frac{\chi^2}{36-13} = 0.91\enspace .
\end{eqnarray}

With these data I get as a neutralino mixing matrix:
\begin{eqnarray}
N & =& 
\left(\begin {array}{cccc}  
\phantom{-}0.953& -0.116& \phantom{-}0.252&- 0.124\\\noalign{\medskip}
- 0.242& -0.845 & \phantom{-}0.405& -0.249\\\noalign{\medskip}
 -0.086& \phantom{-}0.138& \phantom{-}0.714& \phantom{-}0.681\\\noalign{\medskip}
- 0.157& \phantom{-}0.514& \phantom{-}0.585&- 0.607
\end {array} \right) 
\pm 
\left(\begin {array}{cccc}  
0.009 & 0.005 & 0.01 & 0.05\\ \noalign{\medskip} 
0.004 & 0.002 & 0.015 & 0.04\\ \noalign{\medskip}
0.001 & 0.003 & 0.030 & 0.031\\ \noalign{\medskip} 
0.002 & 0.008 & 0.035 & 0.030
\end {array} \right) .
\end{eqnarray}
The error matrix shows that the bino and the wino components can be determined with high accuracy. 
The higgsino components have large(r) errors, especially the ones of the heavy neutralinos.

\medskip         
The parameters $M_1$, $M_2$, and $\mu$ follow from
\begin{subequations}
\begin{align}
M &= N^T \diag(m_1, m_2, m_3, m_{\x{4}}) N =
\left( \begin {array}{cccc}  
 82.8& 0.38&- 5.35& 39.2\\\noalign{\medskip} 
  0.38& 166.5& 9.4&- 77.8\\\noalign{\medskip}
- 5.35& 9.43&- 1.06& - 229.6\\\noalign{\medskip}
 39.2&- 77.8&- 229.6& -0.43
\end {array}\right)\enspace , \\[2mm]
M_1 & = \phantom{-}M_{11}= \phantom{0}82.8\pm 1.2\GeV,\\[2mm]
M_2 & = \phantom{-}M_{22}= 166.5\pm 2.3\GeV,\\[2mm]
\mu & = -M_{34} = 229.6 \pm 2.5\GeV\enspace .
\end{align} 
\end{subequations}  
I do not derive any value for $\tan\beta$ because the result is not very reliable.
The relative error on $N_{13}$, $N_{14}$, $N_{23}$, and $N_{24}$ are about $10\%$
(error propagation in this $2\times 2$ sub-block), and together with the bad behaviour
of the $\tan$ function near its poles this leads to a result with a large error. 

\subsection{How much does radiative neutralino production improve the measurements?}
If I omit the data from the process \signal, and repeat the calculation, then I get for
the couplings the following result:
\begin{table}[h!]
\begin{center}
\begin{tabular}{lll}
\toprule
couplings to $\tilde{e}_R$ & couplings to $\tilde{e}_L$ & couplings to $Z^0$\\
\hline
$a_1 = -0.736 \pm 0.039$ & $b_1 = \phantom{-}0.278 \pm 0.005$ & \\[1mm]
$a_2 = \phantom{-} 0.180\pm 0.009$ & $b_2 = -0.694\pm 0.005$ & $f_{12}= \phantom{-} 0.056\pm 0.007$\\[1mm]
$a_3 = \phantom{-}0.058\pm 0.003$ & $b_3 = \phantom{-}0.099\pm 0.003$ & $f_{13} = \phantom{-}0.154\pm 0.002$\\[1mm]
$a_4 = \phantom{-}0.128\pm 0.007$ & $b_4 = \phantom{-}0.296\pm 0.008$& $f_{14} = \phantom{-}0.026 \pm 0.005$\\[1mm]
&&$f_{22} = \phantom{-}0.025 \pm 0.017$\\
\bottomrule
\end{tabular}
\caption{Result of the fit on the couplings without the data from radiative neutralino production.}
\label{tab:fittedcouplingsohne}
\end{center}
\end{table} 

The errors on the elements of the neutralino mixing matrix are increased by about a factor of $2-5$.
This leads to the following values of the gaugino parameters
\begin{subequations}
\begin{align}
M_1 & = \phantom{0}82.5\pm 6.3\GeV,\\[2mm]
M_2 & = 160.5\pm 3.1\GeV,\\[2mm]
\mu & = 220.6 \pm 6.2\GeV\enspace .
\end{align} 
\end{subequations}  
The error on $M_1$ is five times larger than the error of the case that includes
radiative neutralino production. The error on $M_2$ is enlarged only by a small amount and the error
of $\mu$ is more than doubled. 

Including the cross section of radiative neutralino production leads to smaller errors on $M_1$, 
$M_2$, and $\mu$. So it is worthwhile to examine radiative neutralino production
at a future linear collider.    
  

\subsection{The effect of including the production of further neutralino pairs}
\begin{figure}[p]
\centering
\setlength{\unitlength}{1cm}
\subfigure[Unpolarised cross section\label{fig:heavy}]{\scalebox{0.33}
{\includegraphics{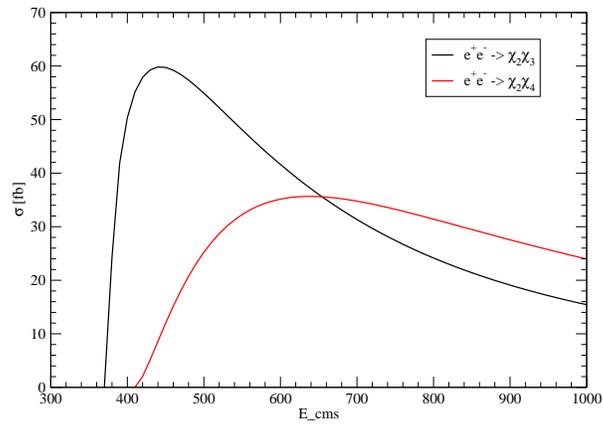}}}
\vspace*{10mm}

\subfigure[Beam polarisation $(P_+|P_-) = (-0.6|0.8)$ \label{fig:heavymp}]{\scalebox{0.33}
{\includegraphics{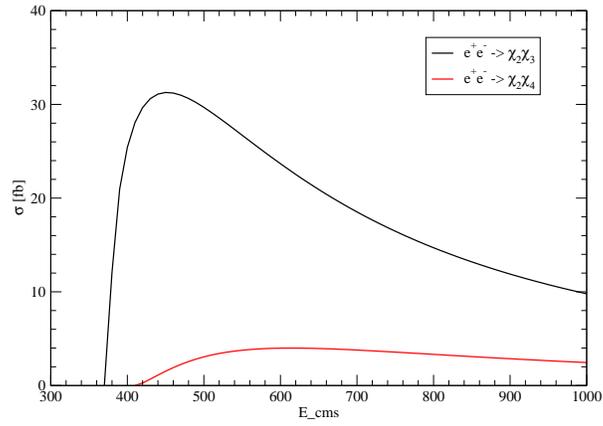}}}
\vspace*{10mm}

\subfigure[Beam polarisation $(P_+|P_-) = (0.6|-0.8)$ \label{fig:heavypm}]{\scalebox{0.33}
{\includegraphics{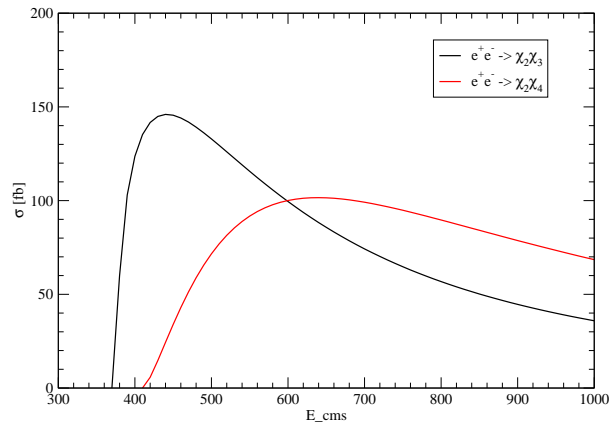}}}
\caption{Comparison of cross sections for different beam polarisations.}
\label{fig:heavypair}
\end{figure}

As I mentioned at the beginning of this section, the cross sections for 
\begin{eqnarray}
e^+e^-&\rightarrow &\tilde{\chi}_2^0\tilde{\chi}_3^0\\[2mm] 
\text{and}\quad e^+e^-&\rightarrow &\tilde{\chi}_2^0\tilde{\chi}_4^0
\end{eqnarray}
can exceed $10\fb$, see Fig.~\ref{fig:heavypair}. 
These two processes introduce two further couplings, $f_{23}$ and $f_{24}$, for their definition
see Eq.~(\ref{eq:f}). They are not necessary for calculating the elements of the neutralino 
diagonalisation matrix and I do not use them for further calculations. Nevertheless,
they provide useful information about signs, see the next subsection.
$N_{33}$, $N_{34}$, $N_{43}$ and $N_{44}$ can be expressed as
\begin{subequations}
\begin{align}
N_{33} =  \phantom{-} \frac{2 f_{23} - N_{21}N_{31} - N_{22}N_{32}}{N_{23}}&, \\[2mm]
N_{34} = - \frac{2 f_{23} + N_{21}N_{31} + N_{22}N_{32}}{N_{24}}& ,\\[2mm]
N_{43} =  \phantom{-}\frac{2 f_{24} - N_{21}N_{41} - N_{22}N_{42}}{N_{23}}&,  \\[2mm]
N_{44} = - \frac{2 f_{24} + N_{21}N_{41} + N_{22}N_{42}}{N_{24}}& \enspace .
\end{align}
\end{subequations}     

The additional processes reduce the errors on $a_i$ and $b_i$, $i = 2,3,4$. As a consequence
the errors on $M_1$, $M_1$, and $\mu$ are reduced by $20\%$, respectively. 

\subsection{Resolving Ambiguities}
The system of Eq.~(\ref{eq:system}) has eight fix points. The starting point determines
that fix point to which the iteration will converge. Some of these fix points do not fulfill
the unitarity condition, so these points are to be discarded.
The signs of $N_{11}$ is choosen as $+1$. 
With this choice the signs of $a_i$, $i > 1$, are fixed, if the solution is physical.
The other signs of $N_{ij}$ are choosen such that the eigenvectors are orthogonal
to each other. 
 

\subsection{Unitarity}
Throughout this method, I assumed unitarity, which means that the neutralino system is complete.
This assumption can be dropped to test if there is an additional singlino field~\cite{Ellwanger:1996gw}.
The eigenvector of the first and the second neutralino are suitable candidates to test unitarity.
But a detailed analysis is beyond the scope of this study.  

\subsection{Further Studies}
The presented method can be extended to
\begin{itemize}
\item the MSSM with a $CP$ violating gaugino sector, 
\item NMSSM to test unitarity,
\item the chargino sector to determine the matrices.
\item The circle method could be used for precision measurements of the chargino 
	parameters.
\end{itemize}

\section{Conclusion and Summary}
In this chapter
I presented a method to determine the $\tilde{\chi}_1^0$-$\tilde{\chi}_i^0$-$\tilde{e}_{R/L}$ and
$\tilde{\chi}_1^0$-$\tilde{\chi}_i^0$-$Z^0$ couplings. 
\begin{itemize}
\item The method from~\cite{Choi:2001ww} to determine $M_1$ does not work because the circles
	are too sensitive to errors of the input data.
\item It is possible to determine the discussed couplings from the polarised cross sections of
	radiative neutralino production and neutralino pair production with errors
	$\mathcal{O}(0.001 - 0.01)$. The masses of the neutralinos and the selectrons
	must be known from LHC/ILC measurements.
\item From the couplings one can determine the neutralino diagonalisation matrix.
	The errors on the elements are about  $\mathcal{O}(0.001 - 0.01)$.
\item From the neutralino diagonalisation matrix and the neutralino masses one can determine
	the neutralino mass matrix. The errors of $M_1$, $M_2$, and $\mu$ are about
	$\mathcal{O}(1\GeV$). It is difficult to determine $\tan\beta$ with my method.
\item Omitting the cross sections of radiative neutralino productions enlarges the errors
	on $M_1$, $M_2$, and $\mu$. Additional processes such as 
	$e^+e^-\rightarrow\tilde{\chi}^0_2\tilde{\chi}^0_{3/4}$
	reduce the error on the corresponding couplings and on $M_1$, $M_2$, and $\mu$. 
\item The differences to a global fit approach are as follows:
	In {\tt{Fittino}}~\cite{Bechtle:2004pc}, the parameters $M_1$, $M_2$, $\mu$, and
	$\tan\beta$ are fitted directly to the data, and the couplings are obtained as a by-product.
	In {\tt{Fittino}}, up to $24$ SUSY parameters can be fitted to cross sections,
	edge positions, branching fractions, cross sections times branching fractions
	and Standard Model parameters from LHC/ILC measurements.	
	The authors recover
	$M_1$ and $M_2$ with absolute errors of the order $\mathcal{O}(0.01 - 0.1\GeV)$,
	and $\mu$ with an absolute error of the order $\mathcal{O}(1\GeV)$ ($129$ degrees of freedom).
			
	In my method, the couplings are the fit parameters to cross sections from
	ILC measurements. 
	From the couplings, I obtain $M_1$, $M_2$, and $\mu$.
	The couplings enter the tree level cross sections of the considered
	particles, which are in my case neutralinos. 
	I recover the input parameters of my model with 
	absolute errors of the order $\mathcal{O}(1\GeV)$ ($23$ degrees of freedom).
	The coupling independent terms of the cross sections need to be computed only once.
	The cross sections need not to be approximated as in {\tt{Fittino}}.
	I do not fit all MSSM parameters to the cross sections. 
	I assume that all masses are known from LHC/ILC measurements. 
\end{itemize}
 

\appendix
\chapter{Radiative Neutralino Production}
\label{sec:app:chifore}
\section{Lagrangian and Couplings}
\begin{table}[t!]
\begin{center}
\caption{Vertex factors with parameters $a$, $b$, $c$, $d$, $f$, and
$g$ defined in Eqs.~(\ref{eq:coefficients}), (\ref{eq:Zcoefficients}), with $e>0.$}
\vspace*{5mm}
\begin{tabular}{cl}
\toprule
\vspace{2mm}
Vertex & Factor
\vspace*{1mm}\\
\midrule
\hspace{-10mm}
{%
\unitlength=1.0 pt
\SetScale{1.0}
\SetWidth{0.7}      
\scriptsize    
{} \qquad\allowbreak
\begin{picture}(95,79)(0,0)
\Text(15.0,60.0)[r]{$\tilde e_R$}
\DashArrowLine(16.0,60.0)(58.0,60.0){1.0} 
\Text(80.0,70.0)[l]{$\tilde\chi^0_1$}
\Line(58.0,60.0)(79.0,70.0) 
\Text(80.0,50.0)[l]{$e$}
\ArrowLine(58.0,60.0)(79.0,50.0) 
\end{picture} \ 
}
& \raisebox{2cm}{$\ie e \sqrt{2} a P_L$}\\[-10mm]
\hspace{-10mm}
{%
\unitlength=1.0 pt
\SetScale{1.0}
\SetWidth{0.7}      
\scriptsize    
{} \qquad\allowbreak
\begin{picture}(95,79)(0,0)
\Text(15.0,60.0)[r]{$\tilde{e}_L$}
\DashArrowLine(16.0,60.0)(58.0,60.0){1.0} 
\Text(80.0,70.0)[l]{$\tilde{\chi}^0_1$}
\Line(58.0,60.0)(79.0,70.0) 
\Text(80.0,50.0)[l]{$e$}
\ArrowLine(58.0,60.0)(79.0,50.0) 
\end{picture} \ 
}
&\raisebox{2cm}{$\ie e \sqrt{2} b P_R$}\\[-10mm]
\hspace{-10mm}
{%
\unitlength=1.0 pt
\SetScale{1.0}
\SetWidth{0.7}      
\scriptsize    
{} \qquad\allowbreak
\begin{picture}(95,79)(0,0)
\Text(15.0,60.0)[r]{$\gamma$}
\Photon(16.0,60.0)(58.0,60.0){1.0}{5} 
\Text(80.0,70.0)[l]{$e$}
\ArrowLine(58.0,60.0)(79.0,70.0) 
\Text(80.0,50.0)[l]{$e$}
\ArrowLine(79.0,50.0)(58.0,60.0) 
\end{picture} \ 
}
&\raisebox{2cm}{$ \ie e \gamma^\mu$}\\[-10mm]
\hspace{-10mm}
{%
\unitlength=1.0 pt
\SetScale{1.0}
\SetWidth{0.7}      
\scriptsize    
{} \qquad\allowbreak
\begin{picture}(95,79)(0,0)
\Text(15.0,60.0)[r]{$\gamma$}
\Photon(16.0,60.0)(58.0,60.0){1.0}{5} 
\Text(80.0,70.0)[l]{$\tilde{e}_{L,R}$}
\DashArrowLine(58.0,60.0)(79.0,70.0){1.0} 
\Text(80.0,50.0)[l]{$\tilde{e}_{L,R}^\ast$}
\DashArrowLine(79.0,50.0)(58.0,60.0){1.0} 
\Text(68,71)[c]{\rotatebox{30}{$\rightarrow p_1$}}
\Text(68,49)[c]{\rotatebox{-30}{$\leftarrow p_2$}}
\end{picture} \ 
}
&\raisebox{2cm}{$ \ie e (p_1 + p_2)^\mu$}\\[-10mm]
\hspace{-10mm}
{%
\unitlength=1.0 pt
\SetScale{1.0}
\SetWidth{0.7}      
\scriptsize    
{} \qquad\allowbreak
\begin{picture}(95,79)(0,0)
\Text(15.0,60.0)[r]{$Z$}
\DashLine(16.0,60.0)(58.0,60.0){3.0} 
\Text(80.0,70.0)[l]{$e$}
\ArrowLine(58.0,60.0)(79.0,70.0) 
\Text(80.0,50.0)[l]{$e$}
\ArrowLine(79.0,50.0)(58.0,60.0) 
\end{picture} \ 
}
&\raisebox{2cm}{$ \ie {e}\gamma^\mu\left(c P_L + d P_R\right)$}\\[-10mm]
\hspace{-10mm}
{%
\unitlength=1.0 pt
\SetScale{1.0}
\SetWidth{0.7}      
\scriptsize    
{} \qquad\allowbreak
\begin{picture}(95,79)(0,0)
\Text(15.0,60.0)[r]{$Z$}
\DashLine(16.0,60.0)(58.0,60.0){3.0} 
\Text(80.0,70.0)[l]{$\tilde{\chi}^0_1$}
\Line(58.0,60.0)(79.0,70.0) 
\Text(80.0,50.0)[l]{$\tilde{\chi}^0_1$}
\Line(58.0,60.0)(79.0,50.0) 
\end{picture} \ 
}
&\raisebox{2cm}{$\displaystyle{\frac{\ie {e}}{2}} \gamma^\mu \left(g P_L + fP_R
\right)$}\\[-15mm]
\bottomrule
\end{tabular}
\label{tab:vertex}
\end{center}
\end{table}

For radiative neutralino production  
\begin{eqnarray}
e^-(p_1) + e^+(p_2) \rightarrow \tilde{\chi}_1^0(k_1) + \tilde{\chi}_1^0(k_2) 
+ \gamma(q), 
\end{eqnarray} 
the SUSY Lagrangian is given by~\cite{Haber:1984rc}
\begin{eqnarray}
\label{eq:lagrangian}
{\mathcal L}    &=& \sqrt{2}e a \bar{f}_eP_L\tilde{\chi}^0_1\tilde{e}_R
                +{\sqrt{2}}e b \bar{f}_e P_R\tilde{\chi}^0_1\tilde{e}_L
                +\half e Z_\mu \bar{\tilde{\chi}}_1^0\gamma^\mu\big[g P_L + f P_R\big]\tilde{\chi}_1^0
                + \mathrm{h.~c.},
\end{eqnarray} 
with the electron, selectron, neutralino and $Z$ boson fields $f_e$, $\tilde{e}_{L,R}$, 
$\tilde{\chi}_1^0$, and $Z_\mu$, respectively, 
and $P_{L,R} = \left(1 \mp \gamma^5\right)/2$.
The couplings are
\begin{equation}
\label{eq:coefficients}
\begin{array}{llllll} 
a &= & -\frac{1}{\cw}N_{11}^\ast,\qquad&
\;\; b&=& \phantom{-}\frac{1}{2\sw} (N_{12} + \tw N_{11}),\\[3mm] 
g &=&  -\frac{1}{2\sw\cw}\left(|N_{13}|^2 - |N_{14}|^2\right),\;\;\qquad&
\;\;f &=& -g,
\end{array}
\end{equation}
see the Feynman rules in Tab.~\ref{tab:vertex}. 
The $Z$-$e$-$e$ couplings are
\begin{equation}
\label{eq:Zcoefficients}
\begin{array}{llllll}
c &=& \phantom{-}\frac{1}{\sw\cw}\left(\frac{1}{2} - \sw[2]\right),\qquad&
\;\; d &=& - \tw.
\end{array}
\end{equation}
\section{Amplitudes for Radiative Neutralino Production}
I define the selectron and $Z$ boson propagators as
\begin{eqnarray}
\Delta_{\tilde{e}_{L,R}}(p_i,k_j) & \equiv & 
\frac{1}{m_{\tilde{e}_{\scriptscriptstyle{L,R}}}^2 - m_{\chi_1^0}^2 + 
2\mink{p_i}{k_j}},\\ 
\Delta_Z(k_1,k_2)& \equiv & \frac{1}{m_Z^2 -  2 m_{\chi_1^0}^2 -
  2\mink{k_1}{k_2} - \ie \Gamma_Z m_Z}\,.
\end{eqnarray}
The tree-level amplitudes for right selectron exchange in the $t$-channel, see the
diagrams 1-3 in Fig.~\ref{fig:diagrams}, are
\begin{eqnarray}
\M_1 &\!=\!& 2 \ie e^3 |a|^2 \, \Big[\uu(k_1) P_R \,\frac{(\ssl{p}_1 -
\ssl{q})}{2 \mink{p_1}{q}}\,\ssl{\epsilon}^\ast u(p_1)\Big]\, 
         \Big[\vv(p_2) P_L v(k_2)\Big]\Delta_{\tilde{e}_{R}}(p_2,k_2)\,,
\label{eq:m1}\\[2mm]
\M_2 &\!=\!& 2 \ie e^3 |a|^2 \, \Big[\uu(k_1) P_R  u(p_1) \Big]
          \Big[\vv(p_2)\ssl{\epsilon}^\ast\,\frac{(\ssl{q} -\ssl{p}_2)}{2 
\mink{p_2}{q}}\, P_L v(k_2)\Big]\,
          \Delta_{\tilde{e}_{R}}(p_1,k_1)\label{eq:m2}\,,\\[2mm] 
\M_3 &\!=\!& 2 \ie e^3 |a|^2 \, \Big[\uu(k_1) P_R u(p_1)\Big]\,\Big[\vv(p_2) 
P_L v(k_2)\Big]
{\mink{(2 p_1 - 2 k_1 -q)}{\epsilon^\ast}}
              \Delta_{\tilde{e}_{R}}(p_1,k_1)\Delta_{\tilde{e}_{R}}(p_2,k_2)
\notag\,.\\[-1mm]
&&\label{eq:m3}
\end{eqnarray}
The amplitudes for $u$-channel $\tilde e_R$ exchange, 
see the diagrams 4-6 in Fig.~\ref{fig:diagrams}, are
\begin{eqnarray}
\M_4 &\!=\!& -2 \ie e^3 |a|^2 \, \Big[\uu(k_2) P_R \,\frac{(\ssl{p}_1 -
\ssl{q})}{2 \mink{p_1}{q}}\,\ssl{\epsilon}^\ast u(p_1)\Big]\,
         \Big[\vv(p_2) P_L v(k_1) \Big]\Delta_{\tilde{e}_{R}}(p_2,k_1)
\label{eq:m4}\,,\\[2mm] 
\M_5 &\!=\!& -2 \ie e^3 |a|^2 \, \Big[\uu(k_2) P_R  u(p_1)\Big]
         \Big[\vv(p_2)\ssl{\epsilon}^\ast\,\frac{(\ssl{q} -\ssl{p}_2)}{2 
\mink{p_2}{q}}\, P_L v(k_1)\Big]\, 
           \Delta_{\tilde{e}_{R}}(p_1,k_2)\label{eq:m5}\,,\\[2mm]
\M_6 &\!=\!& -2 \ie e^3 |a|^2 \, \Big[\uu(k_2) P_R u(p_1)\Big]\,
\Big[\vv(p_2) P_L v(k_1)\Big] 
{\mink{(2 p_1 - 2 k_2 -q)}{\epsilon^\ast}}\Delta_{\tilde{e}_{R}}
(p_1,k_2)\Delta_{\tilde{e}_{R}}(p_2,k_1)\notag\,. \\[-1mm]
&&\label{eq:m6}
\end{eqnarray}
In the photino limit, my amplitudes $\M_1$-$\M_6$, Eqs.~\eqref{eq:m1}-\eqref{eq:m6}, 
agree with those given in Ref.~\cite{Grassie:1983kq}.

The amplitudes for $Z$ boson exchange, see the diagrams 7 and 8 in
Fig.~\ref{fig:diagrams}, are
\begin{eqnarray}
\M_7 &\!=\!& {\ie e^3}\Big[\vv(p_2) \gamma^\mu \left(c P_L + d P_R \right)  
             \frac{(\ssl{p}_1 -\ssl{q})}{2 \mink{p_1}{q}}\,\ssl{\epsilon}^
\ast u(p_1)\Big]\,
           \Big[\uu(k_1) \gamma_\mu  \left(g P_L + f P_R \right) \,v(k_2)
\Big] \Delta_Z(k_1,k_2)\,,\notag\\[-1mm]
&&\label{eq:m7}\\
\M_8 &\!=\!& {\ie e^3}\Big[ \vv(p_2)\ssl{\epsilon}^\ast\frac{(\ssl{q} -
\ssl{p}_2)}{2 \mink{p_2}{q}}
               \gamma^\mu \left(c P_L + d P_R \right)  u(p_1)\Big]\, 
             \Big[\uu(k_1) \gamma_\mu  \left(g P_L + f P_R \right) \,
v(k_2)\Big] \Delta_Z(k_1,k_2)\,. \notag\\[-1mm]
&&\label{eq:m8}
\end{eqnarray}
Note that additional sign factors appear in the amplitudes
$\M_4$-$\M_6$ and $\M_7$-$\M_8$, compared to $\M_1$-$\M_3$.
They stem from the reordering of fermionic operators in the Wick 
expansion of the $S$ matrix. For radiative neutralino production
$e^+e^- \to \tilde\chi_1^0\tilde\chi_1^0\gamma$, such sign factors
appear since the two external neutralinos are 
fermions.\footnote{ 
Note that in Ref.~\cite{Bayer} the relative
sign in the amplitudes for $Z$ boson and $t$-channel $\tilde e_R$ exchange is missing.}
For details 
see Refs.~\cite{Bartl:1986hp,Fraas:1991ky}. 
In addition an extra factor 2 is obtained in the Wick 
expansion of the $S$ matrix, since the Majorana field $\tilde\chi_1^0$ contains both
creation and annihilation operators. 
In my conventions I follow here closely  Ref.~\cite{Bartl:1986hp}.
Other authors take care of this
factor by multiplying the $Z\tilde\chi_1^0\tilde\chi_1^0$ vertex 
by a factor 2 already~\cite{Haber:1984rc}.
For more details of this subtlety, see Ref.~\cite{Haber:1984rc}.

The amplitudes $\M_9-\M_{14}$ for left selectron exchange, see the diagrams 
9-14 in Fig.~\ref{fig:diagrams},  are obtained from the 
$\tilde e_R $ amplitudes by substituting
\begin{eqnarray}
a \rightarrow b, \qquad P_L \rightarrow P_R,\qquad  P_R \rightarrow P_L, \qquad
\Delta_{\tilde{e}_{R}} \rightarrow \Delta_{\tilde{e}_{L}}.
\end{eqnarray}
For $\tilde e_L$ exchange in the $t$-channel they read
\begin{eqnarray}
\M_9 &\!=\!& 2 \ie e^3 |b|^2 \, \Big[\uu(k_1) P_L \,\frac{(\ssl{p}_1 -
\ssl{q})}{2 \mink{p_1}{q}}\,\ssl{\epsilon}^\ast u(p_1)\Big]\, 
         \Big[\vv(p_2) P_R v(k_2)\Big]\Delta_{\tilde{e}_{L}}(p_2,k_2)
\,,\label{eq:m9}\\[2mm]
\M_{10} &\!=\!& 2 \ie e^3 |b|^2 \,\Big[\uu(k_1) P_L  u(p_1)\Big] 
          \Big[\vv(p_2)\ssl{\epsilon}^\ast\,\frac{(\ssl{q} -\ssl{p}_2)}
{2 \mink{p_2}{q}}\, P_R v(k_2)\Big]\,
          \Delta_{\tilde{e}_{L}}(p_1,k_1)\,,\label{eq:m10}\\[2mm] 
\M_{11} &\!=\!& 2 \ie e^3 |b|^2 \, \Big[\uu(k_1) P_L u(p_1)\Big]\,
\Big[\vv(p_2) P_R v(k_2)\Big]
{\mink{(2 p_1 - 2 k_1 -q)}{\epsilon^\ast}}\Delta_{\tilde{e}_{L}}
(p_1,k_1)\Delta_{\tilde{e}_{L}}(p_2,k_2)\,.\notag\\[-1mm]
&&\label{eq:m11} 
\end{eqnarray}
The $u$-channel $\tilde e_L$ exchange amplitudes are
\begin{eqnarray}
\M_{12} &\!=\!& -2 \ie e^3 |b|^2 \,  \Big[\uu(k_2) P_L \,\frac{(\ssl{p}_1 -
\ssl{q})}{2 \mink{p_1}{q}}\,\ssl{\epsilon}^\ast u(p_1)\Big]\,
         \Big[\vv(p_2) P_R v(k_1)\Big]\Delta_{\tilde{e}_L}(p_2,k_1)\,, 
\label{eq:m12}\\[2mm] 
\M_{13} &\!=\!& -2 \ie e^3 |b|^2 \, \big[\uu(k_2) P_L  u(p_1)\Big]
         \Big[\vv(p_2)\ssl{\epsilon}^\ast\,\frac{(\ssl{q} -\ssl{p}_2)}
{2 \mink{p_2}{q}}\, P_R v(k_1)\Big]\, 
           \Delta_{\tilde{e}_{L}}(p_1,k_2)\,,\label{eq:m13}\\[2mm]
\M_{14} &\!=\!& -2 \ie e^3 |b|^2 \,\Big[ \uu(k_2) P_L u(p_1)\Big]\,
\Big[\vv(p_2) P_R v(k_1)\Big] 
            \mink{(2 p_1 - 2 k_2 -q)}{\epsilon^\ast}\Delta_{\tilde{e}_{L}}
(p_1,k_1)\Delta_{\tilde{e}_{L}}(p_2,k_2)\,.\notag\\[-1mm]
&&\label{eq:m14}
\end{eqnarray} 
Our amplitudes $\M_1$-$\M_{14}$ agree with those given in  
Ref.~\cite{Weidner,Fraas:1991ky}, however there is an obvious 
misprint in amplitude $T_5$ of Ref.~\cite{Weidner}.
In addition I have checked that the amplitudes $\M_i=
\epsilon_\mu\M^\mu_i$ for $i=1,\dots,14$ fulfill the Ward identity $q_\mu
(\sum_i\M^\mu_i)=0$, as done in Refs.~\cite{Bayer, Weidner}. 
I find  $q_\mu(\M^\mu_1+\M^\mu_2+\M^\mu_3)=0$ for $t$-channel $\tilde e_R$ exchange,
$q_\mu(\M^\mu_4+\M^\mu_5+\M^\mu_6)=0$ for $u$-channel $\tilde e_R$
exchange, $q_\mu(\M^\mu_7+\M^\mu_8)=0$ for $Z$ boson exchange,
and analog relations for the $\tilde e_L$ exchange amplitudes.

\section{Spin Formalism and Squared Matrix Elements}
I include the longitudinal beam polarisations of electron, $P_{e^-}$, and positron, 
$P_{e^+}$, with $ -1 \le P_{e^\pm}\le +1$ in their density matrices
\begin{eqnarray}
\label{eq:density1}
\rho^{-}_{\lambda_{-} \lambda_{-}^\prime}  &=& 
     \half\left(\delta_{\lambda_{-} \lambda_{-}^\prime} + 
      P_{e^-}\sigma^3_{\lambda_{-} \lambda_{-}^\prime}\right),\\[2mm]
\label{eq:density2}
\rho^{+}_{\lambda_{+} \lambda_{+}^\prime}  &=& 
     \half\left(\delta_{\lambda_{+} \lambda_{+}^\prime} + 
   P_{e^+}\sigma^3_{\lambda_{+} \lambda_{+}^\prime}\right),
\end{eqnarray}
where  $\sigma^3$ is the third Pauli matrix and  $\lambda_{-},\lambda^\prime_{-}$ 
and $\lambda_{+},\lambda^\prime_{+}$
are the helicity indices of electron and positron, respectively.
The squared matrix elements are then obtained by
\begin{eqnarray}
T_{ii} &=& \sum_{\lambda_{-},\lambda_{-}^\prime,\lambda_{+},\lambda_{+}^\prime}
             \rho^{-}_{\lambda_{-} \lambda_{-}^\prime}  \rho^{+}_{\lambda_{+} \lambda_{+}^\prime} 
               \M_i^{\lambda_{-} \lambda_{+}}{\M_i^{\ast}}^{\lambda_{-}^\prime \lambda_{+}^\prime},
\label{Tii}\\[2mm]
T_{ij} &=& 2\, \real\left( 
                \sum_{\lambda_{-},\lambda_{-}^\prime,\lambda_{+},\lambda_{+}^\prime}
               \rho^{-}_{\lambda_{-} \lambda_{-}^\prime}  \rho^{+}_{\lambda_{+} \lambda_{+}^\prime} 
               \M_i^{\lambda_{-} \lambda_{+}} {\M_j^{\ast}}^{\lambda_{-}^\prime \lambda_{+}^\prime} \right),
	      	\enspace i \neq j , 
\label{Tij}\\[2mm]
|\M|^2 & = & \sum_{i \leq j} T_{ij} ,
\end{eqnarray}
where an internal sum over the helicities of the outgoing neutralinos, as well as a sum over 
the polarisations of the photon is included.
Note that I suppress the electron and positron helicity indices of the amplitudes 
$\M_i^{\lambda_{-} \lambda_{+}}$ in the formulas~(\ref{eq:m1})-(\ref{eq:m14}).
The product of the amplitudes in Eqs.~(\ref{Tii}) and (\ref{Tij}) contains
the projectors
\begin{eqnarray}
\label{eq:beam1}
u(p,\lambda_{-})\uu(p,\lambda_{-}^\prime) &=& 
                   \half\left(\delta_{\lambda_{-} \lambda_{-}^\prime} + 
                   \gamma^5\sigma^3_{\lambda_{-}\lambda_{-}^\prime} \right)\ssl{p},\\[2mm]
\label{eq:beam2}
v(p,\lambda_{+}^\prime)\vv(p,\lambda_{+}) &=& 
                   \half\left(\delta_{\lambda_{+} \lambda_{+}^\prime} + 
                   \gamma^5\sigma^3_{\lambda_{+}\lambda_{+}^\prime} \right)\ssl{p}.
\end{eqnarray} 
The contraction with the density matrices of the electron and positron beams leads finally to
\begin{eqnarray}
\sum_{\lambda_{-}, \lambda_{-}^\prime}
\rho^{-}_{\lambda_{-} \lambda_{-}^\prime} u(p,\lambda_{-})\uu(p,\lambda_{-}^\prime) &=& 
\left(\frac{1-P_{e^-}}{2}P_L + \frac{1+P_{e^-}}{2} P_R\right)\ssl{p},\\[2mm]
\sum_{\lambda_{+}, \lambda_{+}^\prime}
\rho^{+}_{\lambda_{+} \lambda_{+}^\prime} v(p,\lambda_{+}^\prime)\vv(p,\lambda_{+}) &=& 
\left(\frac{1+P_{e^+}}{2}P_L + \frac{1-P_{e^+}}{2} P_R\right)\ssl{p}.
\end{eqnarray}
In the squared amplitudes, I include the electron and positron 
beam polarisations in the coefficients
\begin{eqnarray}
C_R = (1 + P_{e^-})(1-P_{e^+}),\qquad C_L = (1 - P_{e^-})(1 + P_{e^+}).
\end{eqnarray}
In the following I give the squared amplitudes $T_{ij}$, 
as defined in Eqs.~(\ref{Tii}) and (\ref{Tij}),
for $\tilde e_R$ and $Z$ exchange.
To obtain the corresponding squared amplitudes for $\tilde e_L$ exchange,
one has to substitute
\begin{eqnarray}
a \rightarrow b, \qquad  d \leftrightarrow c, \qquad  f\leftrightarrow g,
\qquad C_R \rightarrow C_L,\qquad \Delta_{\tilde{e}_R}\rightarrow \Delta_{\tilde{e}_L}\,. 
\end{eqnarray}
There is no interference between diagrams with $\tilde e_R$ and
$\tilde e_L$ exchange, since I assume the high energy limit where
ingoing particles are considered massless.  

\begin{eqnarray}
T_{11} &=& 4 e^6 C_R |a|^4 \,\Delta^2_{\tilde{e}_{R}}(p_2,k_2)\,
\frac{\mink{p_2}{k_2} \,\,q \cdot k_1}{\mink{q}{p_1}}\\[4mm] 
T_{22} &=& 4 e^6 C_R |a|^4 \,\Delta^2_{\tilde{e}_{R}}(p_1,k_1)\,\frac{\mink{p_1}{k_1}
\,\,\mink{q}{k_2}}{\mink{q}{p_2}}\\[4mm]
T_{33} &=& 4 e^6 C_R |a|^4 \,\Delta^2_{\tilde{e}_{R}}(p_1,k_1)\Delta^2_{\tilde{e}_{R}}
(p_2,k_2)\,{\mink{p_1}{k_1}\,\,\mink{p_2}{k_2}} 
       \Bigl[-4 m_{\chi_1^0}^2 + 8 \mink{p_1}{k_1} + 4 \mink{q}{p_1} - 4\mink{q}{k_1}
\Bigr]\notag\\[1mm]
&&\\[0mm]
T_{44} &=& 4 e^6 C_R |a|^4 \,\Delta^2_{\tilde{e}_{R}}(p_2,k_1)\,\frac{\mink{p_2}{k_1} 
\,\,\mink{q}{k_2}}{\mink{q}{p_1}}\\[4mm]
T_{55} &=& 4 e^6 C_R |a|^4 \,\Delta^2_{\tilde{e}_{R}}(p_1,k_2)\,\frac{\mink{p_1}{k_2}
\,\,\mink{q}{k_1}}{\mink{q}{p_2}}\\[4mm]
T_{66} &=& 4 e^6 C_R |a|^4 \,\Delta^2_{\tilde{e}_{R}}(p_1,k_2)\Delta^2_{\tilde{e}_{R}}
(p_2,k_1){\mink{p_1}{k_2}\,\,\mink{p_2}{k_1}}
       \Bigl[-4 m_{\chi_1^0}^2 + 8 \mink{p_1}{k_2} + 4 \mink{q}{p_1} - 4
       \mink{q}{k_2}\Bigr]\notag\\[1mm]
&&\\
T_{77} &=& \frac{4e^6}{\mink{q}{p_1}} |\Delta_Z(k_1,k_2)|^2 
               \Bigl[(C_R d^2 f^2 + C_L c^2 g^2)\mink{p_2}{k_1}\mink{q}{k_2}+
                (C_R d^2 g^2 + C_Lc^2 f^2)\mink{p_2}{k_2}\mink{q}{k_1} \Bigr.
\notag\\[2mm]
               &&  \hspace*{30mm}+ \Bigl.f g (C_L c^2 + C_R d^2) m_{\chi_1^0}^2 \,\mink{q}{p_2}\Bigr] \\[4mm] 
T_{88} &=& \frac{4e^6}{\mink{q}{p_2}} |\Delta_Z(k_1,k_2)|^2 
               \Bigl[(C_R d^2 f^2 + C_L c^2 g^2)\mink{p_1}{k_2}\mink{q}{k_1} + 
                 (C_R d^2 g^2 + C_L c^2 f^2)\mink{p_1}{k_1}\mink{q}{k_2} \Bigr.
\notag\\[2mm]
               &&  \hspace*{30mm}\Bigl.+ f g (C_L c^2 + C_R d^2) m_{\chi^0_1}^2 
\,\mink{q}{p_1}\Bigr] \\[4mm]
T_{12} &=& - 4 e^6 C_R |a|^4 \Delta_{\tilde{e}_{R}}(p_1,k_1)\Delta_{\tilde{e}_{R}}(p_2,k_2)
             \frac{1}{\mink{q}{p_1}\mink{q}{p_2}}\notag \\[2mm]
        &&\Bigl[\mink{q}{k_2}\mink{p_1}{k_1}\mink{p_1}{p_2}-\mink{p_1}{k_1}
\mink{q}{p_2}\mink{p_1}{k_2}
         +\mink{p_1}{k_1}\mink{p_2}{k_2}\mink{q}{p_1}\Bigr.
        + \mink{p_1}{p_2}\mink{q}{k_1}\mink{p_2}{k_2}\notag\\[2mm]
        &&\hspace*{5mm}-\mink{q}{p_1}\mink{p_2}{k_2}\mink{p_2}{k_1} 
         + \mink{p_1}{k_1}\mink{p_2}{k_2}\mink{q}{p_2} 
        \Bigl. -2\mink{p_1}{p_2}\mink{p_1}{k_1}\mink{p_2}{k_2}  \Bigr]\\[4mm]
T_{13} &=& 8 e^6  C_R |a|^4  \Delta_{\tilde{e}_{R}}(p_1,k_1)\Delta^2_{\tilde{e}_{R}}(p_2,k_2)
                             \frac{\mink{p_2}{k_2}}{\mink{q}{p_1}}\notag\\[2mm]
        &&\Bigl[m_{\chi^0_1}^2 \mink{q}{p_1} + 2 (\mink{p_1}{k_1})^2 + 
		\mink{p_1}{k_1}\mink{q}{p_1}  
             - 2\mink{p_1}{k_1} \mink{q}{k_1}\Bigr]\\[4mm]
T_{14} &=& - 4 e^6 C_R |a|^4  \Delta_{\tilde{e}_{R}}(p_2,k_1)\Delta_{\tilde{e}_{R}}(p_2,k_2)
              \frac{m_{\chi^0_1}^2 \mink{q}{p_2}}{\mink{q}{p_1}} \\[4mm]
T_{15} &=&   4 e^6 C_R |a|^4  \Delta_{\tilde{e}_{R}}(p_1,k_2)\Delta_{\tilde{e}_{R}}(p_2,k_2) 
             \frac{m_{\chi^0_1}^2\mink{p_1}{p_2}}{\mink{q}{p_1} \mink{q}{p_2}}
    \Bigl[ \mink{q}{p_1} -\mink{p_1}{p_2}+ \mink{q}{p_2} \Bigr]\\[4mm]
T_{16} &=&  4 e^6 C_R |a|^4  \Delta_{\tilde{e}_{R}}(p_1,k_2)\Delta_{\tilde{e}_{R}}(p_2,k_1) 
	\Delta_{\tilde{e}_{R}}(p_2,k_2)
               \frac{m_{\chi^0_1}^2}{\mink{q}{p_1}} \notag\\[2mm]
              &&\Bigl[-2 \mink{p_1}{k_2}\mink{p_1}{p_2}- \mink{q}{p_1}\mink{p_1}{p_2} +
              \mink{q}{k_2}\mink{p_1}{p_2} \Bigr.
              -\mink{q}{p_1}\mink{p_2}{k_2}+\mink{q}{p_2}\mink{p_1}{k_2}\Bigr]\\[4mm]
T_{17} &=& 4 e^6 |a|^2 C_R d \frac{1}{\mink{q}{p_1}}\Delta_{\tilde{e}_{R}}(p_2,k_2)
            \real\{\Delta_Z(k_1,k_2)\} \bigl[2 g  \mink{p_2}{k_2}\mink{q}{k_1}
             + fm_{\chi^0_1}^2 \mink{q}{p_2}\bigr]\\[4mm]
T_{18} &=& - 4 e^6 C_R |a|^2 d\frac{1}{\mink{q}{p_1}\mink{q}{p_2}}
            \Delta_{\tilde{e}_{R}}(p_2,k_2)\real\{\Delta_Z(k_1,k_2)\}\notag\\[2mm]
          &&\Bigl[g \Bigl(-2 \mink{p_1}{p_2}\mink{p_2}{k_2}\mink{p_1}{k_1} 
	+\mink{p_2}{k_2}(\mink{q}{k_1}\mink{p_1}{p_2}
          - \mink{p_2}{k_1}\mink{q}{p_1}
          + \mink{p_1}{k_1}\mink{q}{p_2}) \bigr.\notag\\[2mm]
          &&\bigl.+\mink{p_1}{k_1}(\mink{q}{p_1}\mink{p_2}{k_2} 
           +\bigl.\mink{q}{k_2}\mink{p_1}{p_2} -\mink{q}{p_2}\mink{p_1}{k_2})\Bigr) \notag\\[2mm] 
          && - f m_{\chi^0_1}^2 \mink{p_1}{p_2} 
          \bigl(\mink{p_1}{p_2}-\mink{q}{p_2}-\mink{q}{p_1}
          \bigr)\Bigl]
\\[4mm]
T_{23} &=& 8 e^6  C_R |a|^4 \frac{\mink{p_1}{k_1}}{\mink{q}{p_2}}
         \Delta^2_{\tilde{e}_{R}}(p_1,k_1)\Delta_{\tilde{e}_{R}}(p_2,k_2)\notag \\[2mm]
      && \hspace*{0mm}\Bigl[m_{\chi^0_1}^2 \mink{q}{p_2} + 2 (\mink{p_2}{k_2})^2 + 
	\mink{p_2}{k_2}\mink{q}{p_2}  
         - 2\mink{p_2}{k_2} \mink{q}{k_2}\Bigr]\\[4mm] 
T_{24} &=& 4 e^6  C_R |a|^4 \Delta_{\tilde{e}_{R}}(p_1,k_1)\Delta_{\tilde{e}_{R}}(p_2,k_1) 
         \frac{m_{\chi^0_1}^2\mink{p_1}{p_2}}{\mink{q}{p_1} \mink{q}{p_2}}
    \Bigl( \mink{q}{p_1} -\mink{p_1}{p_2}+ \mink{q}{p_2} \Bigr)\\[4mm]
T_{25} &=& 4 e^6  C_R |a|^4 \Delta_{\tilde{e}_{R}}(p_1,k_1)\Delta_{\tilde{e}_{R}}
(p_1,k_2)
             \frac{m_{\chi^0_1}^2  \mink{q}{p_1}}{\mink{q}{p_2}}\\[4mm]
T_{26} &=&  4 e^6 C_R |a|^4 \Delta_{\tilde{e}_{R}}(p_1,k_2)\Delta_{\tilde{e}_{R}}(p_2,k_1)
	\Delta_{\tilde{e}_{R}}(p_1,k_1)
              \frac{m_{\chi^0_1}^2}{\mink{q}{p_2}}\notag\\[2mm]
              &&\Bigl[-2 \mink{p_2}{k_1}\mink{p_1}{p_2}- \mink{q}{p_2}\mink{p_1}{p_2} +
              \mink{q}{k_1}\mink{p_1}{p_2} 
              - \mink{q}{p_2}\mink{p_1}{k_1}+\mink{q}{p_1}\mink{p_2}{k_1}\Bigr]\\[4mm] 
T_{27} &=&  \frac{ 4 e^6 C_R |a|^2 d}{\mink{q}{p_1}\mink{q}{p_2}}
               \Delta_{\tilde{e}_{R}}(p_1,k_1)\real\{\Delta_Z(k_1,k_2)\}\notag\\[2mm] 
          &&\Bigl[g \Bigl(2 \mink{p_1}{p_2}\mink{p_2}{k_2}\mink{p_1}{k_1} +\mink{p_2}
{k_2}(-\mink{q}{k_1}\mink{p_1}{p_2}
          + \mink{p_2}{k_1}\mink{q}{p_1}- \mink{p_1}{k_1}\mink{q}{p_2}) \bigr.\notag\\[2mm] 
          &&+ \mink{p_1}{k_1}(-\mink{q}{p_1}\mink{p_2}{k_2}  
          \bigl. - \mink{q}{k_2}\mink{p_1}{p_2} + \mink{q}{p_2}\mink{p_1}{k_2})  
          \Bigr) \Bigr.\notag\\[2mm] 
           &&+ f m_{\chi^0_1}^2 \mink{p_1}{p_2} 
          \bigl(\mink{p_1}{p_2}-\mink{q}{p_2}-\mink{q}{p_1}
          \bigr)\Bigl]
\\[4mm]
T_{28} &=&  \frac{4 e^6 C_R |a|^2 d}{\mink{q}{p_2}}\Delta_{\tilde{e}_{R}}(p_1,k_1)\real\{\Delta_Z(k_1,k_2)\}
                \Bigl[2 g  \mink{p_1}{k_1}\mink{q}{k_2}
                           + f m_{\chi^0_1}^2 \mink{q}{p_1}\Bigr]\\[4mm]
T_{34} &=& - 4 e^6  C_R |a|^4 \frac{m_{\chi^0_1}^2}{\mink{q}{p_1}}
               \Delta_{\tilde{e}_{R}}(p_1,k_1)\Delta_{\tilde{e}_{R}}(p_2,k_1)\Delta_{\tilde{e}_{R}}(p_2,k_2)\notag\\[2mm]
           &&\Bigl[2\mink{p_1}{p_2}\mink{p_1}{k_1} + \mink{p_1}{p_2}\mink{q}{p_1}-\mink{p_1}{k_1}\mink{q}{p_2}
            +\mink{p_2}{k_1}\mink{q}{p_1} -\mink{p_1}{p_2}\mink{q}{k_1}\Bigr]\\[4mm]
T_{35} &=&  -4 e^6  C_R |a|^4 \frac{m_{\chi^0_1}^2}{\mink{q}{p_2}}
             \Delta_{\tilde{e}_{R}}(p_1,k_1)\Delta_{\tilde{e}_{R}}(p_1,k_2)\Delta_{\tilde{e}_{R}}(p_2,k_2)\notag\\[2mm]
          &&\Bigl[2\mink{p_1}{p_2}\mink{p_2}{k_2}- \mink{p_1}{p_2}\mink{q}{k_2}+
          \mink{p_1}{p_2}\mink{q}{p_2}- \mink{p_2}{k_2} \mink{q}{p_1} +\mink{p_1}{k_2}\mink{q}{p_2}\Bigr]\\[4mm]
T_{36} &=& 8 e^6  C_R |a|^4 \Delta_{\tilde{e}_{R}}(p_1,k_1)\Delta_{\tilde{e}_{R}}(p_1,k_2)
                          \Delta_{\tilde{e}_{R}}(p_2,k_1)\Delta_{\tilde{e}_{R}}(p_2,k_2) \notag\\[2mm]
                &&        m_{\chi^0_1}^2 \mink{p_1}{p_2}
         \Bigl[- 2\mink{p_1}{k_1} - 2\mink{q}{p_1} -2\mink{p_1}{k_2} + 2 \mink{k_1}{k_2} +  \mink{q}{k_2}
         + \mink{q}{k_1}\Bigr]\\[4mm] 
T_{37} &=& \frac{ 4e^6 C_R |a|^2 d}{\mink{q}{p_1}}
                   \Delta_{\tilde{e}_{R}}(p_1,k_1)\Delta_{\tilde{e}_{R}}(p_2,k_2) \real\{\Delta_Z(k_1,k_2)\}\notag\\[2mm]
              && \Bigl[ 2 g\mink{p_2}{k_2} \bigl( 
                  \mink{q}{p_1}\mink{p_1}{k_1}-2\mink{p_1}{k_1}\mink{q}{k_1} + 2(\!\mink{p_1}{k_1}\!)^2
                  + m_{\chi^0_1}^2\mink{q}{p_1}\bigr) \Bigr.\notag\\[2mm] 
          &&\Bigl.+ f m_{\chi^0_1}^2\bigl(
                  2 \mink{p_1}{p_2}\mink{p_1}{k_1}+\mink{p_1}{p_2}\mink{q}{p_1}-\mink{p_1}{p_2}\mink{q}{k_1}  
                  -\mink{p_1}{k_1}\mink{q}{p_2}+\mink{q}{p_1}\mink{p_2}{k_1}
                  \bigr) \Bigr]
                   \\[4mm]
T_{38} &=& \frac{4 e^6 C_R |a|^2 d}{\mink{q}{p_2}}
           \Delta_{\tilde{e}_{R}}(p_1,k_1)\Delta_{\tilde{e}_{R}}(p_2,k_2)\real\{\Delta_Z(k_1,k_2)\}\notag\\[2mm]
           &&\Bigl[ 2 g\mink{p_1}{k_1}
             \bigl(2 (\!\mink{p_2}{k_2}\!)^2 + \mink{p_2}{k_2} \mink{q}{p_2} - 2\mink{p_2}{k_2} \mink{q}{k_2}
             + m_{\chi^0_1}^2  \mink{q}{p_2}  \bigr)\Bigr.\notag\\[2mm] 
             &&\bigl.+ f m_{\chi^0_1}^2  \bigl(2 \mink{p_1}{p_2}\mink{p_2}{k_2} 
             +\mink{p_1}{p_2}\mink{q}{p_2} - \mink{p_1}{p_2}\mink{q}{k_2} + \mink{q}{p_2}\mink{p_1}{k_2} - 
             \mink{q}{p_1}\mink{p_2}{k_2} \bigr) \Bigr]
           \\[4mm]
T_{45} &=& - \frac{4 e^6 C_R |a|^4} {\mink{q}{p_1} \mink{q}{p_2}} 
          \Delta_{\tilde{e}_{R}}(p_1,k_2) \Delta_{\tilde{e}_{R}}(p_2,k_1)\notag \\[2mm] 
        &&\Bigl[\mink{q}{k_1}\mink{p_1}{k_2}\mink{p_1}{p_2}-\mink{p_1}{k_2}\mink{q}{p_2}\mink{p_1}{k_1}
        +\mink{p_1}{k_2}\mink{p_2}{k_1}\mink{q}{p_1}
        + \mink{p_1}{p_2}\mink{q}{k_2}\mink{p_2}{k_1}\Bigr.\notag \\[2mm]        
       && -\mink{q}{p_1}\mink{p_2}{k_1}\mink{p_2}{k_2}         
        + \mink{p_1}{k_2}\mink{p_2}{k_1}\mink{q}{p_2} 
        -2\mink{p_1}{p_2}\mink{p_1}{k_2}\mink{p_2}{k_1}  \Bigr]\\[4mm] 
T_{46} &=&  8 e^6  C_R |a|^4 \frac{\mink{p_2}{k_1}}{\mink{q}{p_1}}
                    \Delta_{\tilde{e}_{R}}(p_1,k_2) \Delta^2_{\tilde{e}_{R}}(p_2,k_1)\notag \\[2mm]
           &&\hspace*{0mm} \left[m_{\chi^0_1}^2 \mink{q}{p_1} + 2 (\mink{p_1}{k_2})^2 + \mink{p_1}{k_2}\mink{q}{p_1}  
            - 2\mink{p_1}{k_2} \mink{q}{k_2}\right]\\[4mm] 
T_{47} &=&  - \frac{4 e^6 C_R |a|^2 d}{\mink{q}{p_1}}
              \Delta_{\tilde{e}_{R}}(p_2,k_1)\real\{\Delta_Z(k_1,k_2)\}
        \Bigl[2 g  \mink{p_2}{k_1}\mink{q}{k_2} + f m_{\chi^0_1}^2 \mink{q}{p_2}\Bigr]\\[4mm]
T_{48} &=&  \frac{ -4 e^6 C_R |a|^2 d}{\mink{q}{p_1}\mink{q}{p_2}}
             \Delta_{\tilde{e}_{R}}(p_2,k_1)\real\{\Delta_Z(k_1,k_2)\}\notag\\[2mm]
          &&\Bigl[g \Bigl(2 \mink{p_1}{p_2}\mink{p_1}{k_2}\mink{p_2}{k_1} 
            +\mink{p_2}{k_1} \big( -\mink{q}{k_2}\mink{p_1}{p_2} - \mink{p_1}{k_2}\mink{q}{p_2}
            + \mink{p_2}{k_2}\mink{q}{p_1}\big)  \notag\\[2mm]
           && +\mink{p_1}{k_2}(-\mink{q}{p_1}\mink{p_2}{k_1} 
           +\mink{q}{p_2}\mink{p_1}{k_1} - \mink{q}{k_1}\mink{p_1}{p_2}) \Bigr)  \notag\\[2mm]
          && + f m_{\chi^0_1}^2 \mink{p_1}{p_2}
          \bigl(\mink{p_1}{p_2}-\mink{q}{p_2} - \mink{q}{p_1} 
          \bigr)\Bigl]
             \\[4mm]
T_{56} &=&  8 e^6 C_R |a|^4 \frac{\mink{p_1}{k_2}}{\mink{q}{p_2}} 
                \Delta^2_{\tilde{e}_{R}}(p_1,k_2) \Delta_{\tilde{e}_{R}}(p_2,k_1) \notag\\[2mm] 
          &&\hspace*{0mm} \Bigl[\mink{p_2}{k_1}\mink{q}{p_2} - 2 \mink{p_2}{k_1}\mink{q}{k_1}
           + 2 (\!\mink{p_2}{k_1} \!)^2 + m_{\chi^0_1}^2 \mink{q}{p_2}\Bigr]\\[4mm] 
T_{57} &=& - \frac{4 e^6 C_R |a|^2 d}{\mink{q}{p_2}\mink{q}{p_1}}
                \Delta_{\tilde{e}_{R}}(p_1,k_2)\real\{\Delta_Z(k_1,k_2)\}\notag\\[2mm]
                     &&\Bigl[ g \Bigl(
                     2\mink{p_1}{p_2}\mink{p_1}{k_2}\mink{p_2}{k_1} 
                     + \mink{p_1}{k_2} \bigl(-\mink{p_1}{p_2}\mink{q}{k_1}+\mink{p_1}{k_1}\mink{q}{p_2} 
                     -\mink{q}{p_1}\mink{p_2}{k_1}\!\bigr) \notag\\[2mm] 
                     &&+ \mink{p_2}{k_1} \bigl(-\mink{p_1}{p_2}\mink{q}{k_2} 
                     -\mink{p_1}{k_2}\mink{q}{p_2} + \mink{q}{p_1}\mink{p_2}{k_2} \bigr)\Bigr) \Bigl. \notag\\[2mm]
                     &&+ f m_{\chi^0_1}^2 \mink{p_1}{p_2} 
                     \bigl(\mink{p_1}{p_2}-\mink{q}{p_2} - \mink{q}{p_1}\bigr) \Bigr]
                \\[4mm]             
T_{58} &=& - \frac{ 4 e^6 C_R |a|^2 d}{\mink{q}{p_2}}\Delta_{\tilde{e}_{R}}(p_1,k_2)\real\{\Delta_Z(k_1,k_2)\}
                    \Bigl[2 g \mink{p_1}{k_2}\mink{q}{k_1} + f m_{\chi^0_1}^2 \mink{q}{p_1}\Bigr] \\[4mm]
T_{67} &=&   - \frac{ 4 e^6 C_R |a|^2 d }{\mink{q}{p_1}}
               \Delta_{\tilde{e}_{R}}(p_1,k_2)\Delta_{\tilde{e}_{R}}(p_2,k_1)
               \real\{\Delta_Z(k_1,k_2)\}\notag\\[2mm]
               &&\Bigl[ 2 g \mink{p_2}{k_1}\bigl(\mink{p_1}{k_2}\mink{q}{p_1} -
                 2\mink{q}{k_2}\mink{p_1}{k_2} 
                 +2(\!\mink{p_1}{k_2}\!)^2 + m_{\chi^0_1}^2 
                 \mink{q}{p_1}\bigr)\Bigr.\notag \\[2mm] 
                 && \Bigl.\bigl. + f m_{\chi^0_1}^2 \bigl(2\mink{p_1}{k_2}\mink{p_1}{p_2} +
                 \mink{q}{p_1}\mink{p_1}{p_2} -\mink{q}{k_2}\mink{p_1}{p_2} - 
                 \mink{q}{p_2}\mink{p_1}{k_2}  +\mink{q}{p_1}\mink{p_2}{k_2} \bigr)\Bigr]
               \\[4mm]
T_{68} &=&  - \frac{4 e^6 C_R |a|^2 d}{\mink{q}{p_2}}
                  \Delta_{\tilde{e}_{R}}(p_1,k_2)\Delta_{\tilde{e}_{R}}(p_2,k_1)\real\{\Delta_Z(k_1,k_2)\}\notag\\[2mm]
             &&\Bigl[2g\mink{p_1}{k_2}\bigl(2 (\!\mink{p_2}{k_1}\!)^2 + \mink{q}{p_2}\mink{p_2}{k_1}-
                    2 \mink{p_2}{k_1} \mink{q}{k_1} + m_{\chi^0_1}^2\mink{q}{p_2}\bigr) \bigr.\Bigr.\notag \\[2mm] 
       && + f m_{\chi^0_1}^2\bigl(
                      2\mink{p_1}{p_2} \mink{p_2}{k_1} + \mink{p_1}{p_2} \mink{q}{p_2}- \mink{p_2}{k_1} \mink{q}{p_1} 
                      +\mink{p_1}{k_1} \mink{q}{p_2} -\mink{p_1}{p_2}
                       \mink{q}{k_1}\bigr) \Bigr] 
                  \\[4mm]
T_{78} &=&  \frac{ 4e^6}{\mink{q}{p_2}\mink{q}{p_1}}|\Delta_Z(k_1,k_2)|^2 
            \Bigl[(C_R g^2 d^2 + C_L f^2 c^2)\Bigl(2\mink{p_1}{p_2}\mink{p_1}{k_1} \mink{p_2}{k_2}\notag \\[2mm] 
              &&\hspace*{40mm} +\mink{p_1}{k_1}\bigl( \mink{p_1}{k_2}\mink{q}{p_2} - \mink{p_1}{p_2}\mink{q}{k_2} - 
                               \mink{p_2}{k_2} \mink{q}{p_2} \bigr)\Bigr) \notag \\[2mm]     
&&\hspace*{40mm}+\mink{p_2}{k_2}\bigl( \mink{p_2}{k_1}\mink{q}{p_1} - \mink{p_1}{p_2}\mink{q}{k_1} - 
                               \mink{p_1}{k_1} \mink{q}{p_1} \bigr)\Bigr) \notag \\[2mm]
     &&  + (C_L g^2 c^2 + C_R f^2 d^2)\Bigl(2\mink{p_1}{p_2}\mink{p_1}{k_2} \mink{p_2}{k_1} \notag\\[2mm] 
               &&\hspace*{40mm} +\mink{p_1}{k_2}\bigl( \mink{p_1}{k_1}\mink{q}{p_2} - \mink{p_1}{p_2}\mink{q}{k_1} - 
                               \mink{p_2}{k_1} \mink{q}{p_2} \bigr)\Bigr) \notag \\[2mm]
     &&   \hspace*{40mm}+\mink{p_2}{k_1}\bigl( \mink{p_2}{k_2}\mink{q}{p_1} - \mink{p_1}{p_2}\mink{q}{k_2} - 
                               \mink{p_1}{k_2} \mink{q}{p_1} \bigr)\Bigr) \notag \\[2mm]                               
     &&  + \Bigl. 2 g f (C_L c^2 + C_R d^2) m_{\chi_1^0}^2\mink{p_1}{p_2}  
                               \bigl(\mink{p_1}{p_2}-\mink{q}{p_2} - \mink{q}{p_1}\bigr) \Bigr]
\end{eqnarray}
I have calculated the squared amplitudes with \texttt{FeynCalc}~\cite{Kublbeck:1992mt}.
When integrating the squared amplitude over the phase space, see
Appendix~\ref{sec:phasespace}, the $s$-$t$-interference terms cancel
the $s$-$u$-interference terms due to a symmetry in these channels,
caused by the Majorana properties of the neutralinos~\cite{
  Choi:1999bs}.  
Note that in principle also terms proportional to 
$\eps \imag \{\Delta_Z\}$
would appear in the squared amplitudes $T_{ij}$,
due to the inclusion of the $Z$ width 
to regularise the pole of the propagator $\Delta_Z$. 
However, since this is a higher order effect
which is small far off the $Z$ resonance,
I neglect such terms.
In addition they would
vanish after performing a complete phase
space integration.

\chapter{Amplitudes for Radiative Neutrino Production \label{sec:app:nuback}}

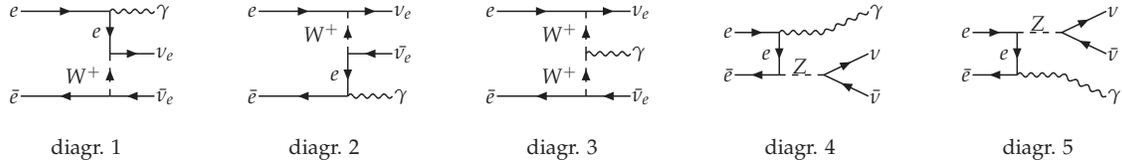
\begin{figure}[t]
{%
\unitlength=0.8 pt
\SetScale{0.8}
\SetWidth{0.7}      
\scriptsize    
\allowbreak
\begin{picture}(95,79)(0,0)
\Text(15.0,70.0)[r]{$e$}
\ArrowLine(16.0,70.0)(58.0,70.0) 
\Text(80.0,70.0)[l]{$\gamma$}
\Photon(58.0,70.0)(79.0,70.0){1.0}{4} 
\Text(54.0,60.0)[r]{$e$}
\ArrowLine(58.0,70.0)(58.0,50.0) 
\Text(80.0,50.0)[l]{$\nu_e$}
\ArrowLine(58.0,50.0)(79.0,50.0) 
\Text(54.0,40.0)[r]{$W^+$}
\DashArrowLine(58.0,30.0)(58.0,50.0){3.0} 
\Text(15.0,30.0)[r]{$\bar{e}$}
\ArrowLine(58.0,30.0)(16.0,30.0) 
\Text(80.0,30.0)[l]{$\bar{\nu}_e$}
\ArrowLine(79.0,30.0)(58.0,30.0) 
\Text(47,0)[b] {diagr. 1}
\end{picture} \ 
{} \quad
\begin{picture}(95,79)(0,0)
\Text(15.0,70.0)[r]{$e$}
\ArrowLine(16.0,70.0)(58.0,70.0) 
\Text(80.0,70.0)[l]{$\nu_e$}
\ArrowLine(58.0,70.0)(79.0,70.0) 
\Text(54.0,60.0)[r]{$W^+$}
\DashArrowLine(58.0,50.0)(58.0,70.0){3.0} 
\Text(80.0,50.0)[l]{$\bar{\nu_e}$}
\ArrowLine(79.0,50.0)(58.0,50.0) 
\Text(54.0,40.0)[r]{$e$}
\ArrowLine(58.0,50.0)(58.0,30.0) 
\Text(15.0,30.0)[r]{$\bar{e}$}
\ArrowLine(58.0,30.0)(16.0,30.0) 
\Text(80.0,30.0)[l]{$\gamma$}
\Photon(58.0,30.0)(79.0,30.0){1.0}{4} 
\Text(47,0)[b] {diagr. 2}
\end{picture} \ 
{} \quad
\begin{picture}(95,79)(0,0)
\Text(15.0,70.0)[r]{$e$}
\ArrowLine(16.0,70.0)(58.0,70.0) 
\Text(80.0,70.0)[l]{$\nu_e$}
\ArrowLine(58.0,70.0)(79.0,70.0) 
\Text(54.0,60.0)[r]{$W^+$}
\DashArrowLine(58.0,50.0)(58.0,70.0){3.0} 
\Text(80.0,50.0)[l]{$\gamma$}
\Photon(58.0,50.0)(79.0,50.0){1.0}{4} 
\Text(54.0,40.0)[r]{$W^+$}
\DashArrowLine(58.0,30.0)(58.0,50.0){3.0} 
\Text(15.0,30.0)[r]{$\bar{e}$}
\ArrowLine(58.0,30.0)(16.0,30.0) 
\Text(80.0,30.0)[l]{$\bar{\nu}_e$}
\ArrowLine(79.0,30.0)(58.0,30.0) 
\Text(47,0)[b] {diagr. 3}
\end{picture} \ 
{} \quad
\begin{picture}(95,79)(0,0)
\Text(15.0,60.0)[r]{$e$}
\ArrowLine(16.0,60.0)(37.0,60.0) 
\Photon(37.0,60.0)(58.0,60.0){1.0}{4} 
\Text(80.0,70.0)[l]{$\gamma$}
\Photon(58.0,60.0)(79.0,70.0){1.0}{4} 
\Text(33.0,50.0)[r]{$e$}
\ArrowLine(37.0,60.0)(37.0,40.0) 
\Text(15.0,40.0)[r]{$\bar{e}$}
\ArrowLine(37.0,40.0)(16.0,40.0) 
\Text(47.0,41.0)[b]{$Z$}
\DashLine(37.0,40.0)(58.0,40.0){3.0} 
\Text(80.0,50.0)[l]{$\nu$}
\ArrowLine(58.0,40.0)(79.0,50.0) 
\Text(80.0,30.0)[l]{$\bar{\nu}$}
\ArrowLine(79.0,30.0)(58.0,40.0) 
\Text(47,0)[b] {diagr. 4}
\end{picture} \ 
{} \quad
\begin{picture}(95,79)(0,0)
\Text(15.0,60.0)[r]{$e$}
\ArrowLine(16.0,60.0)(37.0,60.0) 
\Text(47.0,61.0)[b]{$Z$}
\DashLine(37.0,60.0)(58.0,60.0){3.0} 
\Text(80.0,70.0)[l]{$\nu$}
\ArrowLine(58.0,60.0)(79.0,70.0) 
\Text(80.0,50.0)[l]{$\bar{\nu}$}
\ArrowLine(79.0,50.0)(58.0,60.0) 
\Text(33.0,50.0)[r]{$e$}
\ArrowLine(37.0,60.0)(37.0,40.0) 
\Text(15.0,40.0)[r]{$\bar{e}$}
\ArrowLine(37.0,40.0)(16.0,40.0) 
\Photon(37.0,40.0)(58.0,40.0){1.0}{4} 
\Text(80.0,30.0)[l]{$\gamma$}
\Photon(58.0,40.0)(79.0,30.0){1.0}{4} 
\Text(47,0)[b] {diagr. 5}
\end{picture} \ 
}
\caption{Contributing diagrams to 
$e^+e^- \rightarrow {\nu}{\bar{\nu}}\gamma$~\cite{Boos:2004kh}.}
\label{fig:neutrino}
\end{figure}
\noindent

For radiative neutrino production 
\begin{eqnarray}
e^-(p_1) + e^+(p_2) \rightarrow \nu(k_1) + \bar{\nu}(k_2) + \gamma(q),
\end{eqnarray}
I define the $W$ and $Z$ boson propagators as
\begin{eqnarray}
\Delta_W(p_i,k_j) & \equiv & 
\frac{1}{m_W^2 + 2\mink{p_i}{k_j}}\,,\\ 
\Delta_Z(k_1,k_2)& \equiv & \frac{1}{m_Z^2 -
  2\mink{k_1}{k_2} - \ie \Gamma_Z m_Z}\,.
\end{eqnarray}
\begin{table}[t]
\begin{center}
\caption{Vertex factors with the parameters $a$, $c$, $d$, and $f$ 
defined in Eq.~(\ref{eq:ncouplinga}) 
  and (\ref{eq:ncoupling}).}
\vspace*{5mm}
\begin{tabular}{cl}
\toprule
\vspace{2mm}
Vertex & Factor
\vspace*{1mm}\\
\midrule
{
\unitlength=1.25 pt
\SetScale{1.25}
\SetWidth{0.7}      
\scriptsize    
{} \qquad\allowbreak
\begin{picture}(95,79)(0,0)
\Text(15.0,60.0)[r]{$Z$}
\DashLine(16.0,60.0)(58.0,60.0){3.0} 
\Text(80.0,70.0)[l]{$\nu_\ell$}
\ArrowLine(58.0,60.0)(79.0,70.0) 
\Text(80.0,50.0)[l]{$\nu_\ell$}
\ArrowLine(79.0,50.0)(58.0,60.0) 
\end{picture} \ 
}
&\raisebox{2.5cm}{$ - \displaystyle{\frac{\ie {e}}{2}}f \gamma^\mu  
P_L, \quad \ell=e,\mu,\tau$}\\[-15mm]
{
\unitlength=1.25 pt
\SetScale{1.25}
\SetWidth{0.7}      
\scriptsize    
{} \qquad\allowbreak
\begin{picture}(95,79)(0,0)
\Text(15.0,63.0)[r]{$\beta \atop{\displaystyle \gamma}$}
\Photon(16.0,60.0)(58.0,60.0){1.0}{4}
\Text(37,65)[c]{$k_1\to$}
\Text(68,71)[c]{\rotatebox{30}{$\leftarrow k_3$}}
\Text(68,49)[c]{\rotatebox{-30}{$\leftarrow k_2$}}
\Text(80.0,70.0)[l]{$\alpha \atop {W^+}$}
\DashArrowLine(79.0,70.0)(58.0,60.0){3.0} 
\Text(80.0,50.0)[l]{$\mu \atop {W^-}$}
\DashArrowLine(79.0,50.0)(58.0,60.0){3.0} 
\end{picture} \ 
}
&\raisebox{2.5cm}{$- \ie e[(k_1 - k_2)_\alpha g_{\beta\mu} + (k_2 - k_3)_\beta g_{\mu\alpha}+ 
    (k_3 - k_1)_\mu g_{\alpha\beta}]$}\\[-15mm]
{
\unitlength=1.25 pt
\SetScale{1.25}
\SetWidth{0.7}      
\scriptsize    
{} \qquad\allowbreak
\begin{picture}(95,79)(0,0)
\Text(15.0,60.0)[r]{$W^+$}
\DashArrowLine(16.0,60.0)(58.0,60.0){3.0} 
\Text(80.0,70.0)[l]{$\nu_e$}
\ArrowLine(58.0,60.0)(79.0,70.0) 
\Text(80.0,50.0)[l]{$e$}
\ArrowLine(79.0,50.0)(58.0,60.0) 
\end{picture} \ 
}
&\raisebox{2.5cm}{$\displaystyle - \frac{1}{\sqrt{2}}\ie e a \gamma_\mu P_L$}\\[-15mm]
\bottomrule
\end{tabular}
\label{tab:vertexnu}
\end{center} 
\end{table}
The tree-level amplitudes for $W$ boson exchange, see the diagrams 1-3
in Fig.~\ref{fig:neutrino}, are then
\begin{eqnarray}
\M_1 &\!=\!& \frac{\ie e^3 a^2}{4\mink{q}{p_1}}\Delta_W(p_2,k_2)\Big[\vv(p_2)
\gamma^\mu P_L v(k_2)\Big]\,
         \Big[\uu(k_1)\gamma_\mu P_L (\ssl{q} - \ssl{p}_1)\ssl{\epsilon}^\ast 
u(p_1)\Big],\\[2mm]
\M_2 &\!=\!& \frac{\ie e^3 a^2}{4\mink{q}{p_2}}\Delta_W(p_1,k_1)\Big[\uu(k_1)
\gamma^\mu P_L u(p_1)\Big]
         \Big[\vv(p_2)\ssl{\epsilon}^\ast (\ssl{p}_2 - \ssl{q})\gamma_\mu P_L 
v(k_2)\Big],\\[2mm]         
\M_3 &\!=\!& \frac{1}{2}\ie e^3 a^2\Delta_W(p_1,k_1) \Delta_W(p_2,k_2) 
\Big[\uu(k_1)\gamma^\beta P_L u(p_1)\Big]\,
        \Big[\vv(p_2)\gamma^\alpha   P_L v(k_2)\Big] \notag\\[1mm]
    && \big((2 k_1 - 2 p_1 + q)_\mu g_{\alpha \beta}
        +(p_1 - k_1 -2 q)_\beta g_{\mu\alpha}+ (p_1-k_1+q)_\alpha g_{\beta\mu}\big) (\epsilon^\mu)^\ast,
\end{eqnarray}
with the parameter
\begin{eqnarray}
\label{eq:ncouplinga}
a  = \frac{1}{\sw}.
\end{eqnarray}
The amplitudes for $Z$ boson exchange, see diagrams 4 and 5 in
Fig.~\ref{fig:neutrino}, are
\begin{eqnarray}
\M_4 &\!=\!& \frac{\ie e^3 f}{4 \mink{q}{p_1}}\Delta_Z(k_1,k_2)\Big[\uu(k_1)
\gamma^\nu P_L v(k_2)\Big]\,
          \Big[\vv(p_2)\gamma_\nu(c P_L + d P_R)(\ssl{q} - \ssl{p}_1)\ssl{\epsilon}^
\ast u(p_1)\Big],\hspace*{10mm}\\[2mm]
\M_5 &\!=\!&  \frac{\ie e^3 f}{4 \mink{q}{p_2}}\Delta_Z(k_1,k_2)\Big[\uu(k_1)\gamma^
\nu P_L v(k_2)\Big]\,
          \Big[\vv(p_2)\ssl{\epsilon}^\ast(\ssl{p}_2 - \ssl{q})\gamma_\nu
(c P_L + d P_R)u(p_1)\Big], 
\end{eqnarray}
with the parameters 
\begin{eqnarray}
\label{eq:ncoupling}
c = \frac{1}{\sw\cw}\left(\frac{1}{2}-\sw[2]\right),\qquad
d = -\tw,\qquad        f =\frac{1}{\sw\cw}.
\end{eqnarray}
I have checked that the amplitudes $\M_i=
\epsilon_\mu\M^\mu_i$ for $i=1,\dots,5$ fulfill the Ward identity 
$q_\mu(\sum_i\M^\mu_i)=0$.
I find $q_\mu(\M^\mu_1+\M^\mu_2+\M^\mu_3)=0$ for  $W$ exchange and
$q_\mu(\M^\mu_4+\M^\mu_5)=0$ for $Z$ exchange.

I obtain the squared amplitudes $T_{ii}$ and $T_{ij}$ 
as defined in Eqs.~(\ref{Tii}) and (\ref{Tij}):
\begin{eqnarray}    
T_{11}& =& \frac{e^6 C_L a^4}{\mink{q}{p_1}}\Delta_W^2(p_2,k_2) 
        \mink{p_2}{k_1}  \mink{q}{k_2}\\[2mm]
T_{22} &=& \frac{e^6 C_L a^4}{\mink{q}{p_2}} \Delta_W^2(p_1,k_1) 
        \mink{p_1}{k_2}  \mink{q}{k_1}\\[2mm]   
T_{33} &=& e^6 C_L a^4  \Delta_W^2(p_2,k_2) \Delta_W^2(p_1,k_1) 
        \Big[\mink{p_2}{k_2}\mink{p_1}{k_1}\mink{p_1}{k_1} + 
        (\mink{p_2}{k_1}(7\mink{p_1}{k_2} -6\mink{q}{k_2}) + \notag\\[1mm] 
         &&\mink{p_2}{k_2}(\mink{q}{k_1} - \mink{q}{p_1}) - 
        \mink{q}{k_2}(\mink{p_1}{p_2} + 2\mink{q}{p_2})+ 
        \mink{p_1}{k_2}(\mink{p_1}{p_2} + 
        6\mink{q}{p_2}))\mink{p_1}{k_1} + \notag\\[1mm] 
         &&\mink{p_2}{k_1}\mink{q}{k_1}\mink{p_1}{k_2} - 
        3\mink{q}{k_1}\mink{p_1}{k_2}\mink{p_1}{p_2} +  
        \mink{q}{k_1}\mink{q}{k_2}\mink{p_1}{p_2} -
        \mink{p_2}{k_1}\mink{p_1}{k_2}\mink{q}{p_1} + \notag\\[1mm] 
        &&\mink{q}{k_1}\mink{p_2}{k_2}\mink{q}{p_1}+ 
        2\mink{p_2}{k_1}\mink{q}{k_2}\mink{q}{p_1} +   
        2\mink{q}{k_1}\mink{p_1}{k_2}\mink{q}{p_2} + 
        \mink{k_1}{k_2}\big(-2\mink{q}{k_1}\mink{p_1}{p_2} + \notag\\[1mm]  
        &&\mink{p_1}{k_1}(\mink{p_2}{k_1} - \mink{p_1}{p_2}+
        \mink{q}{p_2}) + \mink{q}{p_1}(3\mink{p_2}{k_1} +  
        2 \mink{p_1}{p_2} + \mink{q}{p_2})\big)\Big]\\[2mm]    
T_{44} &=& 3\frac{e^6 f^2}{\mink{q}{p_1}} |\Delta_Z(k_1,k_2)|^2  
        (C_L c^2 \mink{p_2}{k_1}\mink{q}{k_2} + 
        C_R d^2 \mink{p_2}{k_2}  \mink{q}{k_1})\\[2mm] 
T_{55} &=& 3\frac{e^6 f^2}{\mink{q}{p_2}}|\Delta_Z(k_1,k_2)|^2 
        (C_L c^2 \mink{p_1}{k_2}\mink{q}{k_1} + 
        C_R d^2 \mink{p_1}{k_1}\mink{q}{k_2})\\[2mm] 
T_{12} &=& \frac{e^6 C_L a^4}{\mink{q}{p_1}\mink{q}{p_2}}\Delta_W(p_1,k_1)\Delta_W(p_2,k_2) \notag\\[1mm]  
        &&\Big[2\mink{p_2}{k_1}\mink{p_1}{k_2}\mink{p_1}{p_2} - 
        \mink{q}{k_1}\mink{p_1}{k_2}\mink{p_1}{p_2} - 
        \mink{p_2}{k_1}\mink{q}{k_2}\mink{p_1}{p_2} -
        \mink{p_2}{k_1}\mink{p_1}{k_2}\mink{q}{p_1}  + \notag\\[1mm]   
        &&\mink{p_2}{k_1}\mink{p_2}{k_2}\mink{q}{p_1} + 
        \mink{p_1}{k_1}\mink{p_1}{k_2}\mink{q}{p_2}- 
       \mink{p_2}{k_1}\mink{p_1}{k_2}\mink{q}{p_2}\Big] \\[2mm] 
T_{13} &=& \frac{e^6 C_L a^4}{\mink{q}{p_1}} \Delta_W^2(p_2,k_2) \Delta_W(p_1,k_1) \notag\\[1mm]  
     &&\Big[4\mink{p_1}{k_1}\mink{p_2}{k_1}\mink{p_1}{k_2} -
        \mink{p_2}{k_1}\mink{q}{k_1}\mink{p_1}{k_2}-
        3\mink{q}{k_1}\mink{p_1}{p_2}\mink{p_1}{k_2}+
        3\mink{p_1}{k_1}\mink{q}{p_2}\mink{p_1}{k_2}-\notag\\[1mm]
        &&3\mink{p_1}{k_1}\mink{p_2}{k_1}\mink{q}{k_2}+
        \mink{q}{k_1}\mink{q}{k_2}\mink{p_1}{p_2}+
        \mink{k_1}{k_2}\mink{p_2}{k_1}\mink{q}{p_1}-
        \mink{p_1}{k_1}\mink{p_2}{k_2}\mink{q}{p_1}+\notag\\[1mm] 
        &&3 \mink{p_2}{k_1}\mink{q}{k_2}\mink{q}{p_1}+ 
        \mink{k_1}{k_2}\mink{p_1}{p_2}\mink{q}{p_1}-
        \mink{p_1}{k_1}\mink{q}{k_2}\mink{q}{p_2}\Big]\\[2mm]
T_{14} &=& - \frac{2e^6 C_L c f a^2}{\mink{q}{p_1}} \Delta_W(p_2,k_2)\real \{\Delta_Z(k_1,k_2)\}  
        \mink{p_2}{k_1} \mink{q}{k_2}\\[2mm]    
T_{15} &=& -\frac{e^6 C_L c f a^2}{\mink{q}{p_1}\mink{q}{p_2}}\Delta_W(p_2,k_2)\real \{\Delta_Z(k_1,k_2)\} \notag\\[1mm]  
        &&\Big[2\mink{p_2}{k_1}\mink{p_1}{k_2}\mink{p_1}{p_2} - 
        \mink{q}{k_1}\mink{p_1}{k_2}\mink{p_1}{p_2} -  
        \mink{p_2}{k_1}\mink{q}{k_2}\mink{p_1}{p_2} - \notag\\[1mm] 
       && \mink{p_2}{k_1}\mink{p_1}{k_2}\mink{q}{p_1} + 
        \mink{p_2}{k_1}\mink{p_2}{k_2}\mink{q}{p_1} + 
        \mink{p_1}{k_1}\mink{p_1}{k_2}\mink{q}{p_2} -  
        \mink{p_2}{k_1}\mink{p_1}{k_2}\mink{q}{p_2}\Big] 
       \\[2mm]
T_{23} &=& \frac{e^6 C_L a^4}{\mink{q}{p_2}}\Delta_W^2(p_1,k_1) \Delta_W(p_2,k_2)\notag\\[1mm]  
         && \Big[-3\mink{p_1}{k_2}\mink{p_2}{k_1}\mink{p_2}{k_1} +
             3\mink{q}{k_1}\mink{p_1}{k_2}\mink{p_2}{k_1} -
              \mink{p_1}{k_1}\mink{p_2}{k_1}\mink{p_2}{k_2}  + 
          \mink{k_1}{k_2}\mink{p_1}{p_2}\mink{p_2}{k_1} +\notag\\[1mm]
            &&  2\mink{p_1}{k_2}\mink{p_1}{p_2}\mink{p_2}{k_1} -
             \mink{q}{k_2}\mink{p_1}{p_2}\mink{p_2}{k_1}  - 
           2\mink{p_1}{k_2}\mink{q}{p_1}\mink{p_2}{k_1}+
             \mink{p_2}{k_2}\mink{q}{p_1}\mink{p_2}{k_1} -\notag\\[1mm]
           &&  3\mink{p_1}{k_2}\mink{q}{p_2}\mink{p_2}{k_1}+ 
          \mink{p_1}{k_1}\mink{q}{k_1}\mink{p_2}{k_2}-
            \mink{k_1}{k_2}\mink{q}{k_1}\mink{p_1}{p_2} -
            \mink{q}{k_1}\mink{q}{k_2}\mink{p_1}{p_2}+ \notag\\[1mm]
          && \mink{q}{k_1}\mink{p_2}{k_2}\mink{q}{p_1}+
             2\mink{p_1}{k_1}\mink{p_1}{k_2}\mink{q}{p_2}+
             3 \mink{q}{k_1}\mink{p_1}{k_2}\mink{q}{p_2}\Big] \\[2mm]
T_{24} &=& -\frac{e^6 C_L c f a^2}{\mink{q}{p_1}\mink{q}{p_2}}\Delta_W(p_1,k_1)
	\real \{\Delta_Z(k_1,k_2)\}\notag\\[1mm]  
        &&\Big[2\mink{p_2}{k_1}\mink{p_1}{k_2}\mink{p_1}{p_2} - 
        \mink{q}{k_1}\mink{p_1}{k_2}\mink{p_1}{p_2} -  
        \mink{p_2}{k_1}\mink{q}{k_2}\mink{p_1}{p_2} - 
        \mink{p_2}{k_1}\mink{p_1}{k_2}\mink{q}{p_1}  + \notag\\[1mm]  
        &&\mink{p_2}{k_1}\mink{p_2}{k_2}\mink{q}{p_1} + 
        \mink{p_1}{k_1}\mink{p_1}{k_2}\mink{q}{p_2} -  
        \mink{p_2}{k_1}\mink{p_1}{k_2}\mink{q}{p_2}\Big] 
          \\[2mm]
T_{25} &=& - \frac{2e^6 C_L c f a^2}{\mink{q}{p_2}}\Delta_W(p_1,k_1)\real \{\Delta_Z(k_1,k_2)\} 
        \mink{p_1}{k_2} \mink{q}{k_1} \\[2mm]
T_{34} &=& - \frac{e^6 C_L c f a^2}{\mink{q}{p_1}} \Delta_W(p_1,k_1)\Delta_W(p_2,k_2)\real \{\Delta_Z(k_1,k_2)\} \notag\\[1mm]  
         &&\Big[4\mink{p_1}{k_1}\mink{p_2}{k_1}\mink{p_1}{k_2} -  
             \mink{p_2}{k_1}\mink{q}{k_1}\mink{p_1}{k_2}-
             3\mink{q}{k_1}\mink{p_1}{p_2}\mink{p_1}{k_2}+ 
          3\mink{p_1}{k_1}\mink{q}{p_2}\mink{p_1}{k_2} - \notag\\[1mm]
            && 3\mink{p_1}{k_1}\mink{p_2}{k_1}\mink{q}{k_2}+
             \mink{q}{k_1} \mink{q}{k_2}\mink{p_1}{p_2}+
          \mink{k_1}{k_2}\mink{p_2}{k_1}\mink{q}{p_1}-
             \mink{p_1}{k_1}\mink{p_2}{k_2}\mink{q}{p_1}+ \notag\\[1mm]
          &&    3\mink{p_2}{k_1}\mink{q}{k_2}\mink{q}{p_1}+
          \mink{k_1}{k_2}\mink{p_1}{p_2}\mink{q}{p_1}-
             \mink{p_1}{k_1}\mink{q}{k_2}\mink{q}{p_2}\Big]
               \\[2mm] 
T_{35} &=& -\frac{e^6 C_L c f a^2}{\mink{q}{p_2}}\Delta_W(p_1,k_1)\Delta_W(p_2,k_2) 
		\real \{\Delta_Z(k_1,k_2)\}\notag\\[1mm]  
    && \Big[-3\mink{p_1}{k_2}\mink{p_2}{k_1}\mink{p_2}{k_1}+
        3\mink{q}{k_1}\mink{p_1}{k_2}\mink{p_2}{k_1}-
        \mink{p_1}{k_1}\mink{p_2}{k_2}\mink{p_2}{k_1}+
     \mink{k_1}{k_2}\mink{p_1}{p_2}\mink{p_2}{k_1}+ \notag\\[1mm]   
     && 2\mink{p_1}{k_2}\mink{p_1}{p_2}\mink{p_2}{k_1} -
        \mink{q}{k_2}\mink{p_1}{p_2}\mink{p_2}{k_1}-
      2\mink{p_1}{k_2}\mink{q}{p_1}\mink{p_2}{k_1}+
        \mink{p_2}{k_2}\mink{q}{p_1}\mink{p_2}{k_1}-\notag\\[1mm]   
        && 3\mink{p_1}{k_2}\mink{q}{p_2}\mink{p_2}{k_1}+
       \mink{p_1}{k_1}\mink{q}{k_1}\mink{p_2}{k_2}-
        \mink{k_1}{k_2}\mink{p_1}{p_2}\mink{q}{k_1}-
        \mink{q}{k_1}\mink{q}{k_2}\mink{p_1}{p_2}+\notag\\[1mm]   
      &&\mink{q}{k_1}\mink{p_2}{k_2}\mink{q}{p_1}+
        2\mink{p_1}{k_1}\mink{p_1}{k_2}\mink{q}{p_2}+
        3\mink{q}{k_1}\mink{p_1}{k_2}\mink{q}{p_2} \Big]
     \\[2mm]
T_{45} &=& \frac{3 e^6f^2}{\mink{q}{p_1}\mink{q}{p_2}} |\Delta_Z(k_1,k_2)|^2  \notag\\[1mm]  
        &&\Big[C_L c^2 \big(2\mink{p_2}{k_1}\mink{p_1}{k_2}\mink{p_1}{p_2} - 
        \mink{q}{k_1}\mink{p_1}{k_2}\mink{p_1}{p_2} -  
        \mink{p_2}{k_1}\mink{q}{k_2}\mink{p_1}{p_2} -\notag\\[1mm] 
        &&\mink{p_2}{k_1}\mink{p_1}{k_2}\mink{q}{p_1}  + 
        \mink{p_2}{k_1}\mink{p_2}{k_2}\mink{q}{p_1} + 
        \mink{p_1}{k_1}\mink{p_1}{k_2}\mink{q}{p_2} - 
        \mink{p_2}{k_1}\mink{p_1}{k_2}\mink{q}{p_2}\big) + \notag\\[1mm] 
        &&C_R d^2  \big(2\mink{p_1}{k_1}\mink{p_2}{k_2}\mink{p_1}{p_2} -  
        \mink{q}{k_1}\mink{p_2}{k_2}\mink{p_1}{p_2} - 
        \mink{p_1}{k_1}\mink{q}{k_2}\mink{p_1}{p_2} -  \notag\\[1mm] 
        &&\mink{p_1}{k_1}\mink{p_2}{k_2}\mink{q}{p_1}  + 
        \mink{p_2}{k_1}\mink{p_2}{k_2}\mink{q}{p_1} +   
        \mink{p_1}{k_1}\mink{p_1}{k_2}\mink{q}{p_2} -
        \mink{p_1}{k_1}\mink{p_2}{k_2}\mink{q}{p_2}\big)\Big]\notag\\
&&    
\end{eqnarray}   
I have calculated the squared amplitudes with \texttt{FeynCalc}~\cite{Kublbeck:1992mt}.
I neglect terms proportional to 
$\eps \imag \{\Delta_Z\}$,
see the discussion at the end of Appendix~\ref{sec:app:chifore}.

\chapter{Amplitudes for Radiative Sneutrino Production}
\label{sec:app:snuback}

\begin{figure}[t]
{%
\unitlength=1.0pt
\SetScale{1.0}
\SetWidth{0.7}      
\scriptsize    
\allowbreak
\begin{picture}(95,79)(0,0)
\Text(15.0,70.0)[r]{$e$}
\ArrowLine(16.0,70.0)(58.0,70.0) 
\Text(80.0,70.0)[l]{$\gamma$}
\Photon(58.0,70.0)(79.0,70.0){1.0}{4} 
\Text(54.0,60.0)[r]{$e$}
\ArrowLine(58.0,70.0)(58.0,50.0) 
\Text(80.0,50.0)[l]{$\widetilde{\nu}_e$}
\DashArrowLine(58.0,50.0)(79.0,50.0){1.0} 
\Text(54.0,40.0)[r]{$\widetilde{\chi}^+_1$}
\ArrowLine(58.0,30.0)(58.0,50.0) 
\Text(15.0,30.0)[r]{$\bar{e}$}
\ArrowLine(58.0,30.0)(16.0,30.0) 
\Text(80.0,30.0)[l]{${\widetilde{\nu}^\ast_e}$}
\DashArrowLine(79.0,30.0)(58.0,30.0){1.0} 
\Text(47,0)[b] {diagr. 1}
\end{picture} \ 
{} \qquad\allowbreak
\begin{picture}(95,79)(0,0)
\Text(15.0,70.0)[r]{$e$}
\ArrowLine(16.0,70.0)(58.0,70.0) 
\Text(80.0,70.0)[l]{$\widetilde{\nu}_e$}
\DashArrowLine(58.0,70.0)(79.0,70.0){1.0} 
\Text(54.0,60.0)[r]{$\widetilde{\chi}^+_1$}
\ArrowLine(58.0,50.0)(58.0,70.0) 
\Text(80.0,50.0)[l]{${\widetilde{\nu}^\ast_e}$}
\DashArrowLine(79.0,50.0)(58.0,50.0){1.0} 
\Text(54.0,40.0)[r]{$e$}
\ArrowLine(58.0,50.0)(58.0,30.0) 
\Text(15.0,30.0)[r]{$\bar{e}$}
\ArrowLine(58.0,30.0)(16.0,30.0) 
\Text(80.0,30.0)[l]{$\gamma$}
\Photon(58.0,30.0)(79.0,30.0){1.0}{4} 
\Text(47,0)[b] {diagr. 2}
\end{picture} \ 
{} \qquad\allowbreak
\begin{picture}(95,79)(0,0)
\Text(15.0,70.0)[r]{$e$}
\ArrowLine(16.0,70.0)(58.0,70.0) 
\Text(80.0,70.0)[l]{$\widetilde{\nu}_e$}
\DashArrowLine(58.0,70.0)(79.0,70.0){1.0} 
\Text(54.0,60.0)[r]{$\widetilde{\chi}^+_1$}
\ArrowLine(58.0,50.0)(58.0,70.0) 
\Text(80.0,50.0)[l]{$\gamma$}
\Photon(58.0,50.0)(79.0,50.0){1.0}{4} 
\Text(54.0,40.0)[r]{$\widetilde{\chi}^+_1$}
\ArrowLine(58.0,30.0)(58.0,50.0) 
\Text(15.0,30.0)[r]{$\bar{e}$}
\ArrowLine(58.0,30.0)(16.0,30.0) 
\Text(80.0,30.0)[l]{$\widetilde{\nu}^\ast_e$}
\DashArrowLine(79.0,30.0)(58.0,30.0){1.0} 
\Text(47,0)[b] {diagr. 3}
\end{picture} \ 
{} \qquad\allowbreak
\begin{picture}(95,79)(0,0)
\Text(15.0,70.0)[r]{$e$}
\ArrowLine(16.0,70.0)(58.0,70.0) 
\Text(80.0,70.0)[l]{$\gamma$}
\Photon(58.0,70.0)(79.0,70.0){1.0}{4} 
\Text(54.0,60.0)[r]{$e$}
\ArrowLine(58.0,70.0)(58.0,50.0) 
\Text(80.0,50.0)[l]{$\widetilde{\nu}_e$}
\DashArrowLine(58.0,50.0)(79.0,50.0){1.0} 
\Text(54.0,40.0)[r]{$\widetilde{\chi}^+_2$}
\ArrowLine(58.0,30.0)(58.0,50.0) 
\Text(15.0,30.0)[r]{$\bar{e}$}
\ArrowLine(58.0,30.0)(16.0,30.0) 
\Text(80.0,30.0)[l]{${\widetilde{\nu}^\ast_e}$}
\DashArrowLine(79.0,30.0)(58.0,30.0){1.0} 
\Text(47,0)[b] {diagr. 4}
\end{picture} \ 
{} \qquad\allowbreak
\begin{picture}(95,79)(0,0)
\Text(15.0,70.0)[r]{$e$}
\ArrowLine(16.0,70.0)(58.0,70.0) 
\Text(80.0,70.0)[l]{$\widetilde{\nu}_e$}
\DashArrowLine(58.0,70.0)(79.0,70.0){1.0} 
\Text(54.0,60.0)[r]{$\widetilde{\chi}^+_2$}
\ArrowLine(58.0,50.0)(58.0,70.0) 
\Text(80.0,50.0)[l]{${\widetilde{\nu}^\ast_e}$}
\DashArrowLine(79.0,50.0)(58.0,50.0){1.0} 
\Text(54.0,40.0)[r]{$e$}
\ArrowLine(58.0,50.0)(58.0,30.0) 
\Text(15.0,30.0)[r]{$\bar{e}$}
\ArrowLine(58.0,30.0)(16.0,30.0) 
\Text(80.0,30.0)[l]{$\gamma$}
\Photon(58.0,30.0)(79.0,30.0){1.0}{5} 
\Text(47,0)[b] {diagr. 5}
\end{picture} \ 
{} \qquad\allowbreak
\begin{picture}(95,79)(0,0)
\Text(15.0,70.0)[r]{$e$}
\ArrowLine(16.0,70.0)(58.0,70.0) 
\Text(80.0,70.0)[l]{$\widetilde{\nu}_e$}
\DashArrowLine(58.0,70.0)(79.0,70.0){1.0} 
\Text(54.0,60.0)[r]{$\widetilde{\chi}^+_2$}
\ArrowLine(58.0,50.0)(58.0,70.0) 
\Text(80.0,50.0)[l]{$\gamma$}
\Photon(58.0,50.0)(79.0,50.0){1.0}{4} 
\Text(54.0,40.0)[r]{$\widetilde{\chi}^+_2$}
\ArrowLine(58.0,30.0)(58.0,50.0) 
\Text(15.0,30.0)[r]{$\bar{e}$}
\ArrowLine(58.0,30.0)(16.0,30.0) 
\Text(80.0,30.0)[l]{${\widetilde{\nu}^\ast}_e$}
\DashArrowLine(79.0,30.0)(58.0,30.0){1.0} 
\Text(47,0)[b] {diagr. 6}
\end{picture} \ 
{} \qquad\allowbreak
\begin{picture}(95,79)(0,0)
\Text(15.0,60.0)[r]{$e$}
\ArrowLine(16.0,60.0)(37.0,60.0) 
\Photon(37.0,60.0)(58.0,60.0){1.0}{4} 
\Text(80.0,70.0)[l]{$\gamma$}
\Photon(58.0,60.0)(79.0,70.0){1.0}{4} 
\Text(33.0,50.0)[r]{$e$}
\ArrowLine(37.0,60.0)(37.0,40.0) 
\Text(15.0,40.0)[r]{$\bar{e}$}
\ArrowLine(37.0,40.0)(16.0,40.0) 
\Text(47.0,41.0)[b]{$Z$}
\DashLine(37.0,40.0)(58.0,40.0){3.0} 
\Text(80.0,50.0)[l]{$\widetilde{\nu}$}
\DashArrowLine(58.0,40.0)(79.0,50.0){1.0} 
\Text(80.0,30.0)[l]{${\widetilde{\nu}^\ast}$}
\DashArrowLine(79.0,30.0)(58.0,40.0){1.0} 
\Text(47,0)[b] {diagr. 7}
\end{picture} \ 
{} \qquad\allowbreak
\begin{picture}(95,79)(0,0)
\Text(15.0,60.0)[r]{$e$}
\ArrowLine(16.0,60.0)(37.0,60.0) 
\Text(47.0,61.0)[b]{$Z$}
\DashLine(37.0,60.0)(58.0,60.0){3.0} 
\Text(80.0,70.0)[l]{$\widetilde{\nu}$}
\DashArrowLine(58.0,60.0)(79.0,70.0){1.0} 
\Text(80.0,50.0)[l]{${\widetilde{\nu}^\ast}$}
\DashArrowLine(79.0,50.0)(58.0,60.0){1.0} 
\Text(33.0,50.0)[r]{$e$}
\ArrowLine(37.0,60.0)(37.0,40.0) 
\Text(15.0,40.0)[r]{$\bar{e}$}
\ArrowLine(37.0,40.0)(16.0,40.0) 
\Photon(37.0,40.0)(58.0,40.0){1.0}{4} 
\Text(80.0,30.0)[l]{$\gamma$}
\Photon(58.0,40.0)(79.0,30.0){1.0}{4} 
\Text(47,0)[b] {diagr. 8}
\end{picture} \
}
\caption{Contributing diagrams to 
$e^+e^- \rightarrow \tilde{\nu}\tilde{\nu}^\ast\gamma$~\cite{Boos:2004kh}.}
\label{fig:sneutrino}
\end{figure}
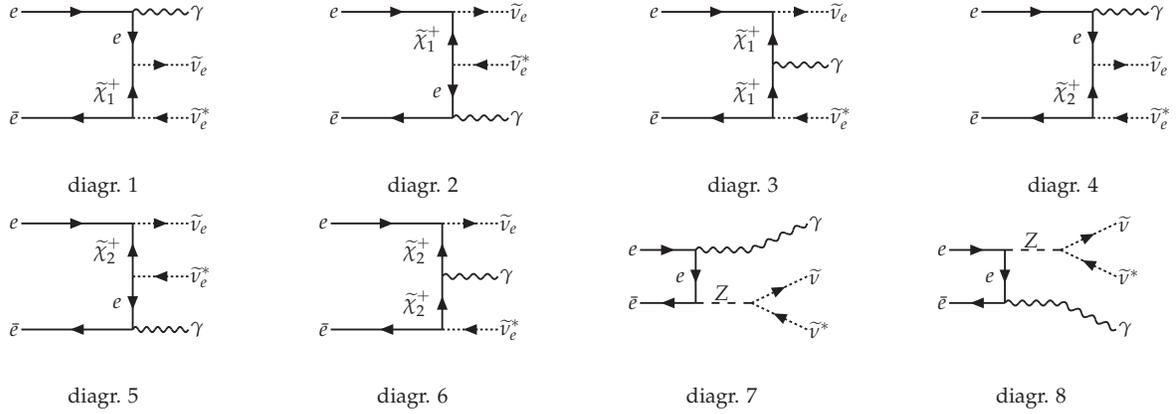
\noindent
For radiative sneutrino production 
\begin{eqnarray}
e^-(p_1) + e^+(p_2) \rightarrow \tilde{\nu}(k_1) + \tilde{\nu}^\ast(k_2) + \gamma(q)
\end{eqnarray}
I define the chargino and $Z$ boson propagators as
\begin{eqnarray}
\Delta_{{\chi}_{1,2}^+}(p_i,k_j) & \equiv & 
\frac{1}{m_{{\chi}_{1,2}^+}^2 - m_{\tilde\nu}^2 + 2\mink{p_i}{k_j}},\\ 
\Delta_Z(k_1,k_2)& \equiv & \frac{1}{m_Z^2 -  2 m_{\tilde\nu}^2 - 2\mink{k_1}{k_2} - \ie \Gamma_Z m_Z}.
\end{eqnarray}

\begin{table}[t]
\begin{center}
\caption{Vertex factors with parameters $a$, $f$ defined in Eqs.~(\ref{eq:ncouplinga}) 
and (\ref{eq:ncoupling}), and $C$ the charge conjugation operator.}
\vspace*{5mm}
\begin{tabular}{cl}
\toprule
\vspace{2mm}
Vertex & Factor
\vspace*{1mm}\\
\midrule
{
\unitlength=1.25 pt
\SetScale{1.25}
\SetWidth{0.7}      
\scriptsize    
{} \qquad\allowbreak
\begin{picture}(95,79)(0,0)
\Text(15.0,60.0)[r]{$Z$}
\DashLine(16.0,60.0)(58.0,60.0){3.0} 
\Text(80.0,70.0)[l]{$\widetilde\nu_\ell$}
\DashArrowLine(58.0,60.0)(79.0,70.0){1.0} 
\Text(80.0,50.0)[l]{$\widetilde\nu^\ast_\ell$}
\DashArrowLine(79.0,50.0)(58.0,60.0){1.0} 
\Text(68,71)[c]{\rotatebox{30}{$\rightarrow p_{\tilde\nu}$}}
\Text(68,49)[c]{\rotatebox{-30}{$\leftarrow p_{\tilde\nu^\ast}$}}
\end{picture} \ 
}
&\raisebox{2.5cm}{$-\half \ie e f (p_{\tilde\nu} + p_{\tilde\nu^\ast})_\mu, 
                   \quad \ell = e,\mu,\tau$}\\[-15mm]
{
\unitlength=1.25 pt
\SetScale{1.25}
\SetWidth{0.7}      
\scriptsize    
{} \qquad\allowbreak
\begin{picture}(95,79)(0,0)
\Text(15.0,60.0)[r]{$\widetilde{\chi}^+_j$}
\ArrowLine(16.0,60.0)(58.0,60.0) 
\Text(80.0,70.0)[l]{$\widetilde\nu_e$}
\DashArrowLine(58.0,60.0)(79.0,70.0){1.0} 
\Text(80.0,50.0)[l]{$ e$}
\ArrowLine(79.0,50.0)(58.0,60.0) 
\end{picture} \ 
}
&\raisebox{2.5cm}{$-\ie e a  V_{j1} P_R C, \quad \tilde\chi^+_j\,{\rm transposed} $}\\[-15mm]
{
\unitlength=1.25 pt
\SetScale{1.25}
\SetWidth{0.7}      
\scriptsize    
{} \qquad\allowbreak
\begin{picture}(95,79)(0,0)
\Text(15.0,60.0)[r]{$\gamma$}
\Photon(16.0,60.0)(58.0,60.0){1.0}{4} 
\Text(80.0,70.0)[l]{$\widetilde\chi^-_j$}
\ArrowLine(58.0,60.0)(79.0,70.0) 
\Text(80.0,50.0)[l]{$\widetilde\chi^+_j$}
\ArrowLine(79.0,50.0)(58.0,60.0) 
\end{picture} \ 
}
&\raisebox{2.5cm}{$ -\ie e \gamma_\mu$}\\[-15mm]
\bottomrule
\end{tabular}
\label{tab:vertexsnu}
\end{center}
\end{table}
The tree-level amplitudes for chargino $\tilde{\chi}_1^\pm$ exchange,
see the contributing diagrams 1-3 in Fig.~\ref{fig:sneutrino}, are
\begin{eqnarray}
\M_1 &=& \frac{\ie e^3 a^2|V_{11}|^2}{2 \mink{q}{p_1}}\Delta_{\chi_1^+}(p_2,k_2) 
    \Big[\vv(p_2) P_R (\ssl{p}_2 - \ssl{k}_2 -  m_{{\chi}_1^+})P_L(\ssl{p}_1 - 
\ssl{q})\ssl{\epsilon}^\ast u(p_1)\Big],
   \hspace*{0mm}\\[2mm]
\M_2 &=&-\frac{\ie e^3 a^2|V_{11}|^2}{2 \mink{q}{p_2}}\Delta_{\chi_1^+}(p_1,k_1)
\Big[\vv(p_2)\ssl{\epsilon}^\ast (\ssl{p}_2 - \ssl{q}) 
           P_R(\ssl{k}_1 - \ssl{p}_1 -  m_{{\chi}_1^+})P_L u(p_1)\Big],\\[3mm]
\M_3 &=& -{\ie e^3 a^2|V_{11}|^2}\Delta_{\chi_1^+}(p_1,k_1)\Delta_{\chi_1^+}(p_2,k_2) 
\notag \\[2mm]
       &&   \Big[\vv(p_2) P_R (\ssl{p}_2 - \ssl{k}_2 -  m_{{\chi}_1^+})
\ssl{\epsilon}^\ast
            (\ssl{k}_1 - \ssl{p}_1 -  m_{{\chi}_1^+})P_L u(p_1)\Big],
\end{eqnarray}
with the parameter $a$ defined in Eq.~(\ref{eq:ncouplinga}).  The
$2\times2$ matrices $U$ and $V$ diagonalise the chargino mass matrix
$X$~\cite{Haber:1984rc}
\begin{eqnarray}
U^\ast X V^{-1} = \mathrm{diag}\begin{pmatrix}m_{\chi_1^+},& m_{\chi_2^+}\end{pmatrix}.
\end{eqnarray}
The amplitudes for  chargino $\tilde{\chi}_2^\pm$ exchange, 
see the contributing diagrams 4-6 in Fig.~\ref{fig:sneutrino}, are
\begin{eqnarray}
\M_4 &=& \frac{\ie e^3 a^2|V_{21}|^2}{2 \mink{q}{p_1}}\Delta_{\chi_2^+}(p_2,k_2) 
\Big[\vv(p_2) P_R
       (\ssl{p}_2 - \ssl{k}_2 -  m_{{\chi}_2^+})P_L(\ssl{p}_1 - \ssl{q})
\ssl{\epsilon}^\ast u(p_1)\Big],\\[2mm] 
\M_5 &=&-\frac{\ie e^3 a^2|V_{21}|^2}{2 \mink{q}{p_1}}\Delta_{\chi_2^+}(p_1,k_1)
\Big[\vv(p_2)\ssl{\epsilon}^\ast (\ssl{p}_2 - \ssl{q})
           P_R(\ssl{k}_1 - \ssl{p}_1 -  m_{{\chi}_2^+})P_L u(p_1)\Big],\\[3mm] 
\M_6 &=& -{\ie e^3 a^2|V_{21}|^2}\Delta_{\chi_2^+}(p_1,k_1)\Delta_{\chi_2^+}
(p_2,k_2) \notag\\[2mm]
       &&   \Big[\vv(p_2) P_R (\ssl{p}_2 - \ssl{k}_2 -  m_{{\chi}_2^+})
\ssl{\epsilon}^\ast 
            (\ssl{k}_1 - \ssl{p}_1 -  m_{{\chi}_2^+})P_L u(p_1)\Big].
\end{eqnarray}
The amplitudes for $Z$ boson exchange, see the diagrams 7 and 8 in 
Fig.~\ref{fig:sneutrino}, read
\begin{eqnarray}
\M_7 &=&  \frac{\ie e^3 f}{4 \mink{q}{p_1}}\Delta_Z(k_1,k_2)\Big[\vv(p_2)
(\ssl{k}_1-\ssl{k}_2) 
          (c P_L + d P_R) (\ssl{p}_1 - \ssl{q})\ssl{\epsilon}^\ast u(p_1)\Big],\\[2mm]
\M_8 &=&  \frac{\ie e^3 f}{4 \mink{q}{p_2}}\Delta_Z(k_1,k_2)\Big[\vv(p_2)
\ssl{\epsilon}^\ast 
          (\ssl{q} - \ssl{p}_2)(\ssl{k}_1-\ssl{k}_2)(c P_L + d P_R)u(p_1)\Big],
\end{eqnarray}
with the parameters $c$, $d$, and $f$ defined in
Eq.~(\ref{eq:ncoupling}).  I have checked that the amplitudes $\M_i=
\epsilon_\mu\M^\mu_i$, $i=1,\dots,8$, fulfill the Ward identity $q_\mu
(\sum_i\M^\mu_i)=0$, as done in Ref.~\cite{Franke:thesis}. I find $q_
\mu(\M^\mu_1+\M^\mu_2+\M^\mu_3)=0$ for $\tilde{\chi}_1^\pm$ exchange,
$q_\mu(\M^\mu_4+\M^\mu_5+\M^\mu_6)=0$ for $\tilde{\chi}_2^\pm$
exchange, and $q_\mu(\M^\mu_7+\M^\mu_8)=0$ for $Z$ boson exchange.
Our amplitudes for chargino and $Z$ boson exchange agree with those
given in Refs.~\cite{Franke:thesis,Franke:1994ph}, and in the limit of
vanishing chargino mixing with those of Ref.~\cite{Chen:1987ux}.
However, there are obvious misprints in the amplitudes $M_2$ and $M_4$
of Ref.~\cite{Franke:1994ph}, see their Eqs.~(7) and (9),
respectively, and in the amplitude $T_5$ of Ref.~\cite{Chen:1987ux},
see their Eq.~(F.3).

I then obtain the squared amplitudes 
$T_{ii}$ and $T_{ij}$ 
as defined in Eqs.~(\ref{Tii}) and (\ref{Tij}):
\begin{eqnarray}
  T_{1 1} &=& \frac{e^6 C_L a^4 |V_{11}|^4}{2 \mink{q}{p_1} } \Delta_{\chi_1^+}^2(p_2,k_2)
  (2\mink{p_2}{k_2} \mink{q}{k_2} - m_{\tilde{\nu}}^2 \mink{q}{p_2})  \\[2mm]
  T_{2 2} &=& \frac{e^6 C_L a^4 |V_{11}|^4}{2 \mink{q}{p_2} } \Delta_{\chi_1^+}^2(p_1,k_1)
  (2\mink{p_1}{k_1} \mink{q}{k_1} - m_{\tilde{\nu}}^2 \mink{q}{p_1})   \\[2mm]
  T_{3 3} &=&{e^6 C_L a^4 |V_{11}|^4}\Delta_{\chi_1^+}^2(p_1,k_1) \Delta_{\chi_1^+}^2(p_2,k_2)
  \big[m_{\chi_1^+}^4 \mink{p_1}{p_2} + 
  4 m_{{\chi_1^+}}^2 \mink{p_1}{k_1} \mink{p_2}{k_2} - \notag\\[1mm]
  && 2m_{\tilde{\nu}}^2 \mink{p_1}{k_1} \mink{p_2}{k_1} +
  4\mink{k_1}{k_2} \mink{p_1}{k_1} \mink{p_2}{k_2} -  
  2m_{\tilde{\nu}}^2 \mink{p_1}{k_2} \mink{p_2}{k_2} + 
  m_{\tilde{\nu}}^4 \mink{p_1}{p_2} \big] \\[2mm] 
  T_{4 4} &=& \frac{e^6 C_L a^4 |V_{21}|^4}{2\mink{q}{p_1}} \Delta_{\chi_2^+}^2(p_2,k_2)
  (2\mink{p_2}{k_2} \mink{q}{k_2} - m_{\tilde{\nu}}^2 \mink{q}{p_2})  \\[2mm] 
  T_{5 5} &=& \frac{e^6 C_L a^4 |V_{21}|^4}{2\mink{q}{p_2} } \Delta_{\chi_2^+}^2(p_1,k_1) 
  (2\mink{p_1}{k_1} \mink{q}{k_1} - m_{\tilde{\nu}}^2 \mink{q}{p_1})  \\[2mm] 
  T_{6 6} &=& {e^6 C_L a^4 |V_{21}|^4}\Delta_{\chi_2^+}^2(p_1,k_1) \Delta_{\chi_2^+}^2(p_2,k_2)
  \big[m_{\chi^+_2}^4 \mink{p_1}{p_2} + 
  4  m_{{\chi_2^+}}^2\mink{p_1}{k_1} \mink{p_2}{k_2} -  \notag\\[1mm]
  &&2m_{\tilde{\nu}}^2 \mink{p_1}{k_1} \mink{p_2}{k_1} +  
  4\mink{k_1}{k_2} \mink{p_1}{k_1} \mink{p_2}{k_2} -  
  2m_{\tilde{\nu}}^2 \mink{p_1}{k_2} \mink{p_2}{k_2} + 
  m_{\tilde{\nu}}^4 \mink{p_1}{p_2}\big]  \\[2mm] 
  T_{7 7} &=& 3\frac{e^6 f^2 (C_L c^2 + C_R d^2)}{4 \mink{q}{p_1}}  |\Delta_Z(k_1,k_2)|^2 \notag\\[1mm]
  &&\big[\mink{p_2}{k_1} \mink{q}{k_1} - \mink{p_2}{k_2} \mink{q}{k_1} - 
  \mink{p_2}{k_1} \mink{q}{k_2} + \mink{p_2}{k_2} \mink{q}{k_2} -
  m_{\tilde{\nu}}^2 \mink{q}{p_2} + \mink{k_1}{k_2} \mink{q}{p_2}\big]  \\[2mm]
  T_{8 8} &=& 3\frac{e^6 f^2 (C_L c^2 + C_R d^2)}{4 \mink{q}{p_2}}  |\Delta_Z(k_1,k_2)|^2 \notag\\[1mm]
  &&\big[\mink{p_1}{k_1} \mink{q}{k_1} - \mink{p_1}{k_2} \mink{q}{k_1} - 
  \mink{p_1}{k_1} \mink{q}{k_2} + \mink{p_1}{k_2} \mink{q}{k_2} -
   m_{\tilde{\nu}}^2 \mink{q}{p_1} + \mink{k_1}{k_2} \mink{q}{p_1}\big] \\[2mm]
  T_{1 2} &=& -\frac{e^6 C_L a^4 |V_{11}|^4}{\mink{q}{p_1} \mink{q}{p_2}}  
  \Delta_{\chi_1^+}(p_1,k_1) \Delta_{\chi_1^+}(p_2,k_2)\notag\\[1mm]
  && \big[-\mink{k_1}{k_2} \mink{p_1}{p_2} \mink{p_1}{p_2} + 
  \mink{p_2}{k_1} \mink{p_1}{k_2} \mink{p_1}{p_2} +  
  \mink{p_1}{k_1} \mink{p_2}{k_2} \mink{p_1}{p_2} -\notag\\[1mm]
 && \mink{q}{k_1} \mink{p_2}{k_2} \mink{p_1}{p_2} -  
  \mink{p_1}{k_1} \mink{q}{k_2} \mink{p_1}{p_2} + 
  \mink{k_1}{k_2} \mink{q}{p_1} \mink{p_1}{p_2} + 
  \mink{k_1}{k_2} \mink{q}{p_2} \mink{p_1}{p_2} - \notag\\[1mm]
 && \mink{p_2}{k_1} \mink{p_1}{k_2} \mink{q}{p_1} +  
  \mink{p_2}{k_1} \mink{p_2}{k_2} \mink{q}{p_1} +
  \mink{p_1}{k_1} \mink{p_1}{k_2} \mink{q}{p_2} -  
  \mink{p_2}{k_1} \mink{p_1}{k_2} \mink{q}{p_2}\big]  \\[2mm]
  T_{1 3} &=& -\frac{e^2 C_L a^4 |V_{11}|^4}{\mink{q}{p_1}}   
  \Delta_{\chi_1^+}(p_1,k_1) \Delta_{\chi_1^+}^2(p_2,k_2) \notag\\[1mm]
  &&\big[m_{\chi^+_1}^2 \mink{q}{k_2} \mink{p_1}{p_2} +  
  m_{\chi^+_1}^2 \mink{q}{p_1} \mink{p_2}{k_2} - 
  m_{\chi^+_1}^2 \mink{q}{p_2} \mink{p_1}{k_2} -
   4\mink{p_1}{k_1} \mink{p_1}{k_2} \mink{p_2}{k_2} + \notag\\[1mm] 
   &&4\mink{p_1}{k_1} \mink{p_2}{k_2} \mink{q}{k_2} +  
   2m_{\tilde{\nu}}^2 \mink{p_1}{k_1} \mink{p_1}{p_2} - \ 
   2m_{\tilde{\nu}}^2 \mink{p_1}{k_1} \mink{q}{p_2}\big]  \\[2mm]
  T_{1 4} &=& \frac{e^6 C_L a^4 |V_{11}|^2 |V_{21}|^2}{\mink{q}{p_1}}
  	\Delta_{\chi_1^+}(p_2,k_2) \Delta_{\chi_2^+}(p_2,k_2) 
  \big[2\mink{p_2}{k_2} \mink{q}{k_2} -  
  m_{\tilde{\nu}}^2 \mink{q}{p_2}\big] \\[2mm]
  T_{1 5} &=& -\frac{e^6 C_L a^4 |V_{11}|^2 |V_{21}|^2}{\mink{q}{p_1} \mink{q}{p_2}}
  \Delta_{\chi_2^+}(p_1,k_1) \Delta_{\chi_1^+}(p_2,k_2)\notag\\[1mm]
  &&\big[-\mink{k_1}{k_2} \mink{p_1}{p_2} \mink{p_1}{p_2} + 
  \mink{p_2}{k_1} \mink{p_1}{k_2} \mink{p_1}{p_2} +  
  \mink{p_1}{k_1} \mink{p_2}{k_2} \mink{p_1}{p_2} -
  \mink{q}{k_1} \mink{p_2}{k_2} \mink{p_1}{p_2} - \notag\\[1mm] 
  &&\mink{p_1}{k_1} \mink{q}{k_2} \mink{p_1}{p_2} + 
  \mink{k_1}{k_2} \mink{q}{p_1} \mink{p_1}{p_2} + 
  \mink{k_1}{k_2} \mink{q}{p_2} \mink{p_1}{p_2} - 
  \mink{p_2}{k_1} \mink{p_1}{k_2} \mink{q}{p_1} +  \notag\\[1mm]
  &&\mink{p_2}{k_1} \mink{p_2}{k_2} \mink{q}{p_1} +
  \mink{p_1}{k_1} \mink{p_1}{k_2} \mink{q}{p_2} -  
  \mink{p_2}{k_1} \mink{p_1}{k_2} \mink{q}{p_2}\big] \\[2mm]
  T_{1 6} &=& -\frac{e^6 C_L a^4 |V_{11}|^2 |V_{21}|^2}{\mink{q}{p_1}} 
  \Delta_{\chi_2^+}(p_1,k_1) \Delta_{\chi_1^+}(p_2,k_2) \Delta_{\chi_2^+}(p_2,k_2)\notag\\[1mm]
  &&\big[m_{\chi^+_2}^2 \mink{q}{k_2} \mink{p_1}{p_2} +  
  m_{\chi^+_2}^2 \mink{q}{p_1} \mink{p_2}{k_2} - 
  m_{\chi^+_2}^2 \mink{q}{p_2} \mink{p_1}{k_2} - 
  4\mink{p_1}{k_1} \mink{p_1}{k_2} \mink{p_2}{k_2} + \notag\\[1mm] 
  &&4\mink{p_1}{k_1} \mink{p_2}{k_2} \mink{q}{k_2} + 
  2m_{\tilde{\nu}}^2 \mink{p_1}{k_1} \mink{p_1}{p_2} - \
  2m_{\tilde{\nu}}^2 \mink{p_1}{k_1} \mink{q}{p_2}\big] \\[2mm] 
  T_{1 7} &=& -\frac{e^6 C_L a^2 c f |V_{11}|^2}{2 \mink{q}{p_1}} 
  \Delta_{\chi_1^+}(p_2,k_2) \real \{\Delta_Z(k_1,k_2)\} \notag\\[1mm]
  &&\big[\mink{q}{k_1} \mink{p_2}{k_2} - 
  2\mink{q}{k_2} \mink{p_2}{k_2} + \mink{p_2}{k_1} \mink{q}{k_2} + m_{\tilde{\nu}}^2 \mink{q}{p_2} 
  - \mink{k_1}{k_2} \mink{q}{p_2}\big] 
  \\[2mm] 
  T_{1 8} &=& -\frac{e^6 C_L a^2 c f |V_{11}|^2}{2\mink{q}{p_1} \mink{q}{p_2}} 
  \Delta_{\chi_1^+}(p_2,k_2) \real \{\Delta_Z(k_1,k_2)\}\notag\\[1mm]
  &&\big[-\mink{q}{p_2} \mink{p_1}{k_2} \mink{p_1}{k_2} + 
  \mink{p_2}{k_1} \mink{p_1}{p_2} \mink{p_1}{k_2} -  
  2\mink{p_2}{k_2} \mink{p_1}{p_2} \mink{p_1}{k_2} +
  \mink{q}{k_2} \mink{p_1}{p_2} \mink{p_1}{k_2} - \notag\\[1mm] 
  &&\mink{p_2}{k_1} \mink{q}{p_1} \mink{p_1}{k_2} + 
  \mink{p_2}{k_2} \mink{q}{p_1} \mink{p_1}{k_2} + 
  \mink{p_1}{k_1} \mink{q}{p_2} \mink{p_1}{k_2} - 
  \mink{p_2}{k_1} \mink{q}{p_2} \mink{p_1}{k_2} + \notag\\[1mm] 
  &&\mink{p_2}{k_2} \mink{q}{p_2} \mink{p_1}{k_2} +
  m_{\tilde{\nu}}^2 \mink{p_1}{p_2} \mink{p_1}{p_2} -  
  \mink{k_1}{k_2} \mink{p_1}{p_2} \mink{p_1}{p_2} + 
  \mink{p_1}{k_1} \mink{p_2}{k_2} \mink{p_1}{p_2} -\notag\\[1mm]
  &&\mink{q}{k_1} \mink{p_2}{k_2} \mink{p_1}{p_2} - 
  \mink{p_1}{k_1} \mink{q}{k_2} \mink{p_1}{p_2} +  
  \mink{p_2}{k_2} \mink{q}{k_2} \mink{p_1}{p_2} -
  \mink{p_2}{k_2} \mink{p_2}{k_2} \mink{q}{p_1} + \notag\\[1mm]
  &&\mink{p_2}{k_1} \mink{p_2}{k_2} \mink{q}{p_1} -  
  m_{\tilde{\nu}}^2 \mink{p_1}{p_2} \mink{q}{p_1} +
  \mink{k_1}{k_2} \mink{p_1}{p_2} \mink{q}{p_1} -  
  m_{\tilde{\nu}}^2 \mink{p_1}{p_2} \mink{q}{p_2} + \notag\\[1mm] 
  &&\mink{k_1}{k_2} \mink{p_1}{p_2} \mink{q}{p_2}\big]
  \\[2mm]
  T_{2 3} &=&-\frac{e^6 C_L a^4 |V_{11}|^4}{\mink{q}{p_2}} \Delta_{\chi_1^+}^2(p_1,k_1) 
  \Delta_{\chi_1^+}(p_2,k_2)\notag\\[1mm]
  &&\big[m_{\chi^+_1}^2 \mink{q}{k_1} \mink{p_1}{p_2} - 
  m_{\chi^+_1}^2 \mink{p_2}{k_1} \mink{q}{p_1} +  
  m_{\chi^+_1}^2 \mink{p_1}{k_1} \mink{q}{p_2} -
  4\mink{p_1}{k_1} \mink{p_2}{k_1} \mink{p_2}{k_2} +  \notag\\[1mm]
  &&4\mink{p_1}{k_1} \mink{q}{k_1} \mink{p_2}{k_2} + 
  2m_{\tilde{\nu}}^2 \mink{p_2}{k_2} \mink{p_1}{p_2} - 
  2m_{\tilde{\nu}}^2 \mink{p_2}{k_2} \mink{q}{p_1}\big] \\[2mm]
  T_{2 4} &=& -\frac{e^6 C_L a^4 |V_{11}|^2 |V_{21}|^2}{\mink{q}{p_1} \mink{q}{p_2}}
  \Delta_{\chi_1^+}(p_1,k_1) \Delta_{\chi_2^+}(p_2,k_2)\notag\\[1mm]
  &&\big[-\mink{k_1}{k_2} \mink{p_1}{p_2} \mink{p_1}{p_2} + 
  \mink{p_2}{k_1} \mink{p_1}{k_2} \mink{p_1}{p_2} +  
  \mink{p_1}{k_1} \mink{p_2}{k_2} \mink{p_1}{p_2} -
  \mink{q}{k_1} \mink{p_2}{k_2} \mink{p_1}{p_2} - \notag\\[1mm] 
  &&\mink{p_1}{k_1} \mink{q}{k_2} \mink{p_1}{p_2} + 
  \mink{k_1}{k_2} \mink{q}{p_1} \mink{p_1}{p_2} + 
  \mink{k_1}{k_2} \mink{q}{p_2} \mink{p_1}{p_2} - 
  \mink{p_2}{k_1} \mink{p_1}{k_2} \mink{q}{p_1} +  \notag\\[1mm]
  &&\mink{p_2}{k_1} \mink{p_2}{k_2} \mink{q}{p_1} +
  \mink{p_1}{k_1} \mink{p_1}{k_2} \mink{q}{p_2} -  
  \mink{p_2}{k_1} \mink{p_1}{k_2} \mink{q}{p_2}\big] \\[2mm]
  T_{2 5} &=& \frac{e^6 C_L a^4 |V_{11}|^2 |V_{21}|^2}{ \mink{q}{p_2}} 
  \Delta_{\chi_1^+}(p_1,k_1) \Delta_{\chi_2^+}(p_1,k_1)
  \big[2\mink{p_1}{k_1} \mink{q}{k_1} - 
  m_{\tilde{\nu}}^2 \mink{q}{p_1}\big] \\[2mm] 
  T_{2 6} &=& - \frac{e^6 C_L a^4 |V_{11}|^2 |V_{21}|^2}{\mink{q}{p_2}}\Delta_{\chi_2^+}(p_1,k_1) 
  \Delta_{\chi_1^+}(p_1,k_1) \Delta_{\chi_2^+}(p_2,k_2)  \notag\\[1mm]
  &&\big[m_{\chi^+_2}^2 \mink{q}{k_1} \mink{p_1}{p_2} - 
  m_{\chi^+_2}^2 \mink{q}{p_1} \mink{p_2}{k_1} +  
  m_{\chi^+_2}^2 \mink{q}{p_2} \mink{p_1}{k_1} -
  4\mink{p_1}{k_1} \mink{p_2}{k_1} \mink{p_2}{k_2} + \notag\\[1mm] 
  &&4\mink{p_1}{k_1} \mink{p_2}{k_2} \mink{q}{k_1} + 
  2m_{\tilde{\nu}}^2 \mink{p_2}{k_2} \mink{p_1}{p_2} -  
  2m_{\tilde{\nu}}^2 \mink{p_2}{k_2} \mink{q}{p_1}\big]\\[2mm]
  T_{2 7} &=& \frac{e^6 C_L a^2 c f |V_{11}|^2}{2 \mink{q}{p_1} \mink{q}{p_2}} 
  \Delta_{\chi_1^+}(p_1,k_1) \real \{\Delta_Z(k_1,k_2)\}\notag\\[1mm]
  &&\big[\mink{q}{p_2} \mink{p_1}{k_1} \mink{p_1}{k_1} + 
  2\mink{p_2}{k_1} \mink{p_1}{p_2} \mink{p_1}{k_1} -  
  \mink{q}{k_1} \mink{p_1}{p_2} \mink{p_1}{k_1} -
  \mink{p_2}{k_2} \mink{p_1}{p_2} \mink{p_1}{k_1} + \notag\\[1mm] 
  &&\mink{q}{k_2} \mink{p_1}{p_2} \mink{p_1}{k_1} - 
  \mink{p_2}{k_1} \mink{q}{p_1} \mink{p_1}{k_1} -  
  \mink{p_2}{k_1} \mink{q}{p_2} \mink{p_1}{k_1} - 
  \mink{p_1}{k_2} \mink{q}{p_2} \mink{p_1}{k_1} - \notag\\[1mm] 
  &&m_{\tilde{\nu}}^2 \mink{p_1}{p_2} \mink{p_1}{p_2} + 
  \mink{k_1}{k_2} \mink{p_1}{p_2} \mink{p_1}{p_2} -  
  \mink{p_2}{k_1} \mink{q}{k_1} \mink{p_1}{p_2} - 
  \mink{p_2}{k_1} \mink{p_1}{k_2} \mink{p_1}{p_2} + \notag\\[1mm] 
  &&\mink{q}{k_1} \mink{p_2}{k_2} \mink{p_1}{p_2} + 
  \mink{p_2}{k_1} \mink{p_2}{k_1} \mink{q}{p_1} +  
  \mink{p_2}{k_1} \mink{p_1}{k_2} \mink{q}{p_1} - 
  \mink{p_2}{k_1} \mink{p_2}{k_2} \mink{q}{p_1} + \notag\\[1mm] 
  &&m_{\tilde{\nu}}^2 \mink{p_1}{p_2} \mink{q}{p_1} - 
  \mink{k_1}{k_2} \mink{p_1}{p_2} \mink{q}{p_1} +  
  \mink{p_2}{k_1} \mink{q}{p_2} \mink{p_1}{k_2} +  
  m_{\tilde{\nu}}^2 \mink{p_1}{p_2} \mink{q}{p_2} - \notag\\[1mm]
  &&\mink{k_1}{k_2} \mink{p_1}{p_2} \mink{q}{p_2}\big]
\\[2mm]
  T_{2 8} &=& \frac{e^6 C_L a^2 c f |V_{11}|^2}{2 \mink{q}{p_2}} \Delta_{\chi_1^+}(p_1,k_1) 
  \real \{\Delta_Z(k_1,k_2)\} \notag\\[1mm]
  &&\big[- \mink{q}{k_1} \mink{p_1}{k_2} + 
  2\mink{q}{k_1} \mink{p_1}{k_1}  - \mink{p_1}{k_1} \mink{q}{k_2} - m_{\tilde{\nu}}^2 \mink{q}{p_1} \
  + \mink{k_1}{k_2} \mink{q}{p_1}\big] 
  \\[2mm] 
  T_{3 4} &=& -\frac{e^6 C_L a^4 |V_{11}|^2 |V_{21}|^2}{ \mink{q}{p_1}} \Delta_{\chi_1^+}(p_1,k_1) 
  \Delta_{\chi_1^+}(p_2,k_2) \Delta_{\chi_2^+}(p_2,k_2)\notag\\[1mm]
  &&\big[m_{\chi^+_1}^2 (\mink{q}{k_2} \mink{p_1}{p_2} + 
                        \mink{p_2}{k_2} \mink{q}{p_1} - 
                       \mink{p_1}{k_2} \mink{q}{p_2}) - 
  4\mink{p_1}{k_1} \mink{p_2}{k_2} (\mink{p_1}{k_2} - \mink{q}{k_2}) + \notag\\[1mm] 
  &&2m_{\tilde{\nu}}^2 \mink{p_1}{k_1} (\mink{p_1}{p_2} - \mink{q}{p_2})\big] \\[2mm]
  T_{3 5} &=&  -\frac{e^6 C_L a^4 |V_{11}|^2 |V_{21}|^2}{\mink{q}{p_2}}\Delta_{\chi_1^+}(p_1,k_1) 
  \Delta_{\chi_1^+}(p_2,k_2) \Delta_{\chi_2^+}(p_1,k_1)\notag\\[1mm]
  &&\big[m_{\chi^+_1}^2 (\mink{q} {k_1} \mink{p_1}{p_2} + 
  \mink{p_1}{k_1} \mink{q}{p_2} - \mink{p_2}{k_1} \mink{q}{p_1}) - 
  4\mink{p_1}{k_1} \mink{p_2}{k_2} (\mink{p_2}{k_1} - \mink{q}{k_1}) + \notag\\[1mm] 
  &&2m_{\tilde{\nu}}^2 \mink{p_2}{k_2} (\mink{p_1}{p_2} - \mink{q}{p_1})\big] \\[2mm]
  T_{3 6} &=& 2{e^6 C_L a^4 |V_{11}|^2 |V_{21}|^2} \Delta_{\chi_1^+}(p_1,k_1) \Delta_{\chi_1^+}(p_2,k_2) 
  \Delta_{\chi_2^+}(p_1,k_1) \Delta_{\chi_2^+}(p_2,k_2)\notag\\[1mm]
  &&\big[2\mink{p_1}{k_1} \mink{p_2}{k_2} (m_{\chi^+_1}^2 + m_{\chi^+_2}^2 + 
  2\mink{k_1}{k_2}) + m_{\chi^+_1}^2 m_{\chi^+_2}^2 \mink{p_1}{p_2} - \notag\\[1mm] 
  &&2m_{\tilde{\nu}}^2 (\mink{p_1}{k_1} \mink{p_2}{k_1} + \mink{p_1}{k_2} \mink{p_2}{k_2}) + 
  m_{\tilde{\nu}}^4 \mink{p_1}{p_2}\big] \\[2mm] 
  T_{3 7} &=& \frac{e^6 C_L a^2 c f |V_{11}|^2}{2\mink{q}{p_1}} \Delta_{\chi_1^+}(p_1,k_1) 
  \Delta_{\chi_1^+}(p_2,k_2) \real \{\Delta_Z(k_1,k_2)\}\notag\\[1mm]
  &&\hspace*{-2mm}\big[m_{\chi^+_1}^2 (\mink{q}{k_1} \mink{p_1}{p_2} - 
  \mink{q}{k_2} \mink{p_1}{p_2} +  
  \mink{p_2}{k_1} \mink{q}{p_1} - 
  \mink{p_2}{k_2} \mink{q}{p_1} - 
  \mink{p_1}{k_1} \mink{q}{p_2} + \notag\\[1mm]
  &&\mink{p_1}{k_2} \mink{q}{p_2}) - 
  2\mink{p_1}{k_1} \mink{p_2}{k_1} \mink{p_1}{k_2} -
  2\mink{p_1}{k_1} \mink{p_1}{k_1} \mink{p_2}{k_2} +  
  2\mink{p_1}{k_1} \mink{p_2}{k_2} \mink{q}{k_1} + \notag\\[1mm]
  &&4\mink{p_1}{k_1} \mink{p_2}{k_2} \mink{p_1}{k_2} + 
  2\mink{p_1}{k_1} \mink{p_2}{k_1} \mink{q}{k_2} - 
  4\mink{p_1}{k_1} \mink{p_2}{k_2} \mink{q}{k_2} -  
  2m_{\tilde{\nu}}^2 \mink{p_1}{p_2} \mink{p_1}{k_1} +\notag\\[1mm]
  &&2\mink{k_1}{k_2} \mink{p_1}{k_1} \mink{p_1}{p_2} +
  2m_{\tilde{\nu}}^2 \mink{p_1}{k_1} \mink{q}{p_2} - 
  2\mink{k_1}{k_2} \mink{p_1}{k_1} \mink{q}{p_2}\big]  
  \\[2mm]
  T_{3 8} &=&  -\frac{e^6 C_L a^2 c f |V_{11}|^2}{2\mink{q}{p_2}}\Delta_{\chi_1^+}(p_1,k_1)
  \Delta_{\chi_1^+}(p_2,k_2) \real\{\Delta_Z(k_1,k_2)\} \notag\\[1mm]
  &&\hspace*{-2mm}\big[m_{\chi^+_1}^2 (\mink{q}{k_1} \mink{p_1}{p_2} - 
  \mink{q}{k_2} \mink{p_1}{p_2} -  
  \mink{p_2}{k_1} \mink{q}{p_1} + 
  \mink{p_2}{k_2} \mink{q}{p_1} + 
  \mink{p_1}{k_1} \mink{q}{p_2} - \notag\\[1mm]
  &&\mink{p_1}{k_2} \mink{q}{p_2}) +   
  2\mink{p_1}{k_1} \mink{p_2}{k_2} \mink{p_2}{k_2} -
  4\mink{p_1}{k_1} \mink{p_2}{k_1} \mink{p_2}{k_2} +  
  4\mink{p_1}{k_1} \mink{p_2}{k_2} \mink{q}{k_1} +\notag\\[1mm] 
  &&2\mink{p_2}{k_1} \mink{p_1}{k_2} \mink{p_2}{k_2} -  
  2\mink{p_1}{k_2} \mink{p_2}{k_2} \mink{q}{k_1} - 
  2\mink{p_1}{k_1} \mink{p_2}{k_2} \mink{q}{k_2} +  
  2m_{\tilde{\nu}}^2 \mink{p_1}{p_2} \mink{p_2}{k_2} -\notag\\[1mm] 
  &&2\mink{k_1}{k_2} \mink{p_2}{k_2} \mink{p_1}{p_2} -  
  2m_{\tilde{\nu}}^2 \mink{p_2}{k_2} \mink{q}{p_1} + 
  2\mink{k_1}{k_2} \mink{p_2}{k_2} \mink{q}{p_1}\big] 
  \\[2mm]
  T_{4 5} &=& -\frac{e^6 C_L a^4 |V_{21}|^4}{\mink{q}{p_1} \mink{q}{p_2}}\
  \Delta_{\chi_2^+}(p_1,k_1) \Delta_{\chi_2^+}(p_2,k_2) \notag\\[1mm]
   &&\big[-\mink{k_1}{k_2} \mink{p_1}{p_2} \mink{p_1}{p_2} + 
  \mink{p_1}{k_2} \mink{p_2}{k_1} \mink{p_1}{p_2} +  
  \mink{p_1}{k_1} \mink{p_2}{k_2} \mink{p_1}{p_2} -
  \mink{q}{k_1} \mink{p_2}{k_2} \mink{p_1}{p_2} -  \notag\\[1mm] 
  &&\mink{p_1}{k_1} \mink{q}{k_2} \mink{p_1}{p_2} + 
  \mink{k_1}{k_2} \mink{q}{p_1} \mink{p_1}{p_2} + 
  \mink{k_1}{k_2} \mink{q}{p_2} \mink{p_1}{p_2} - 
  \mink{p_2}{k_1} \mink{p_1}{k_2} \mink{q}{p_1} + \notag\\[1mm]   
  &&\mink{p_2}{k_1} \mink{p_2}{k_2} \mink{q}{p_1} +
  \mink{p_1}{k_1} \mink{p_1}{k_2} \mink{q}{p_2} -  
  \mink{p_2}{k_1} \mink{p_1}{k_2} \mink{q}{p_2}\big] \\[2mm]
  T_{4 6} &=& -\frac{e^6 C_L a^4 |V_{21}|^4}{\mink{q}{p_1}}
  \Delta_{\chi_2^+}(p_1,k_1) \Delta_{\chi_2^+}^2(p_2,k_2) \notag\\[1mm]
  &&\big[m_{\chi^+_2}^2 \mink{q}{k_2} \mink{p_1}{p_2} + 
  m_{\chi^+_2}^2 \mink{q}{p_1} \mink{p_2}{k_2} -  
  m_{\chi^+_2}^2 \mink{q}{p_2} \mink{p_1}{k_2} -
  4\mink{p_1}{k_1} \mink{p_1}{k_2} \mink{p_2}{k_2} + \notag\\[1mm] 
  &&4\mink{p_1}{k_1} \mink{p_2}{k_2} \mink{q}{k_2} + 
  2m_{\tilde{\nu}}^2 \mink{p_1}{k_1} \mink{p_1}{p_2} - 
  2m_{\tilde{\nu}}^2 \mink{p_1}{k_1} \mink{q}{p_2}\big]  \\[2mm]
  T_{4 7} &=&  -\frac{e^6 C_L a^2 c f |V_{21}|^2}{2 \mink{q}{p_1}}\Delta_{\chi_2^+}(p_2,k_2) 
  \real \{\Delta_Z(k_1,k_2)\}\notag\\[1mm]
  &&[\mink{q}{k_1} \mink{p_2}{k_2} - 
  2\mink{q}{k_2} \mink{p_2}{k_2} + \mink{p_2}{k_1} \mink{q}{k_2} + m_{\tilde{\nu}}^2 \mink{q}{p_2} - 
  \mink{k_1}{k_2} \mink{q}{p_2}] 
  \\[2mm] 
  T_{4 8} &=&  -\frac{e^6 C_L a^2 c f |V_{21}|^2}{2\mink{q}{p_1} \mink{q}{p_2}}\Delta_{\chi_2^+}(p_2,k_2) 
  \real \{\Delta_Z(k_1,k_2)\}\notag\\[1mm]
  &&\big[-\mink{q}{p_2} \mink{p_1}{k_2} \mink{p_1}{k_2} + 
  \mink{p_2}{k_1} \mink{p_1}{p_2} \mink{p_1}{k_2} - 
  2\mink{p_2}{k_2} \mink{p_1}{p_2} \mink{p_1}{k_2} +
  \mink{q}{k_2} \mink{p_1}{p_2} \mink{p_1}{k_2} - \notag\\[1mm]
  &&\mink{p_2}{k_1} \mink{q}{p_1} \mink{p_1}{k_2} + 
  \mink{p_2}{k_2} \mink{q}{p_1} \mink{p_1}{k_2} + 
  \mink{p_1}{k_1} \mink{q}{p_2} \mink{p_1}{k_2} - 
  \mink{p_2}{k_1} \mink{q}{p_2} \mink{p_1}{k_2} + \notag\\[1mm] 
  &&\mink{p_2}{k_2} \mink{q}{p_2} \mink{p_1}{k_2} +
  m_{\tilde{\nu}}^2 \mink{p_1}{p_2} \mink{p_1}{p_2} -  
  \mink{k_1}{k_2} \mink{p_1}{p_2} \mink{p_1}{p_2} + 
  \mink{p_1}{k_1} \mink{p_2}{k_2} \mink{p_1}{p_2} -\notag\\[1mm] 
  &&\mink{q}{k_1} \mink{p_2}{k_2} \mink{p_1}{p_2} - 
  \mink{p_1}{k_1} \mink{q}{k_2} \mink{p_1}{p_2} +  
  \mink{p_2}{k_2} \mink{q}{k_2} \mink{p_1}{p_2} -
  \mink{p_2}{k_2} \mink{p_2}{k_2} \mink{q}{p_1} + \notag\\[1mm] 
  &&\mink{p_2}{k_1} \mink{p_2}{k_2} \mink{q}{p_1} - 
  m_{\tilde{\nu}}^2 \mink{p_1}{p_2} \mink{q}{p_1} + 
  \mink{k_1}{k_2} \mink{p_1}{p_2} \mink{q}{p_1} - 
  m_{\tilde{\nu}}^2 \mink{p_1}{p_2} \mink{q}{p_2} +  \notag\\[1mm]
  &&\mink{k_1}{k_2} \mink{p_1}{p_2} \mink{q}{p_2}\big] 
  \\[2mm]
  T_{5 6} &=& -\frac{e^6 C_L a^4 |V_{21}|^4}{\mink{q}{p_2}}\Delta_{\chi_2^+}^2(p_1,k_1) 
  \Delta_{\chi_2^+}(p_2,k_2)\notag\\[1mm]
  &&\big[m_{\chi^+_2}^2 \mink{q}{k_1} \mink{p_1}{p_2} - 
  m_{\chi^+_2}^2 \mink{p_2}{k_1} \mink{q}{p_1} +   
  m_{\chi^+_2}^2 \mink{p_1}{k_1} \mink{q}{p_2} -
  4\mink{p_1}{k_1} \mink{p_2}{k_1} \mink{p_2}{k_2} +  \notag\\[1mm]
  &&4\mink{p_1}{k_1} \mink{q}{k_1} \mink{p_2}{k_2} + 
  2m_{\tilde{\nu}}^2 \mink{p_2}{k_2} \mink{p_1}{p_2} - 
  2m_{\tilde{\nu}}^2 \mink{p_2}{k_2} \mink{q}{p_1}\big] \\[2mm]
  T_{5 7} &=& \frac{e^6 C_L a^2 c f |V_{21}|^2}{2\mink{q}{p_1} \mink{q}{p_2}}
  \Delta_{\chi_2^+}(p_1,k_1) \real \{\Delta_Z(k_1,k_2)\}\notag\\[1mm] 
  &&\big[\mink{q}{p_2} \mink{p_1}{k_1} \mink{p_1}{k_1} +  
  2\mink{p_2}{k_1} \mink{p_1}{p_2} \mink{p_1}{k_1} - 
  \mink{q}{k_1} \mink{p_1}{p_2} \mink{p_1}{k_1} - 
  \mink{p_2}{k_2} \mink{p_1}{p_2} \mink{p_1}{k_1} +\notag\\[1mm] 
  &&\mink{q}{k_2} \mink{p_1}{p_2} \mink{p_1}{k_1} -  
  \mink{p_2}{k_1} \mink{q}{p_1} \mink{p_1}{k_1} - 
  \mink{p_2}{k_1} \mink{q}{p_2} \mink{p_1}{k_1} -  
  \mink{p_1}{k_2} \mink{q}{p_2} \mink{p_1}{k_1} - \notag\\[1mm]
  &&m_{\tilde{\nu}}^2 \mink{p_1}{p_2} \mink{p_1}{p_2} +  
  \mink{k_1}{k_2} \mink{p_1}{p_2} \mink{p_1}{p_2} - 
  \mink{p_2}{k_1} \mink{q}{k_1} \mink{p_1}{p_2} -  
  \mink{p_2}{k_1} \mink{p_1}{k_2} \mink{p_1}{p_2} + \notag\\[1mm]
  &&\mink{q}{k_1} \mink{p_2}{k_2} \mink{p_1}{p_2} +  
  \mink{p_2}{k_1} \mink{p_2}{k_1} \mink{q}{p_1} + 
  \mink{p_2}{k_1} \mink{p_1}{k_2} \mink{q}{p_1} -  
  \mink{p_2}{k_1} \mink{p_2}{k_2} \mink{q}{p_1} + \notag\\[1mm]
  &&m_{\tilde{\nu}}^2 \mink{p_1}{p_2} \mink{q}{p_1} -  
  \mink{k_1}{k_2} \mink{p_1}{p_2} \mink{q}{p_1} + 
  \mink{p_2}{k_1} \mink{q}{p_2} \mink{p_1}{k_2} +  
  m_{\tilde{\nu}}^2 \mink{p_1}{p_2} \mink{q}{p_2} - \notag\\[1mm]
  &&\mink{k_1}{k_2} \mink{p_1}{p_2} \mink{q}{p_2}\big] 
  \\[2mm]
  T_{5 8} &=& \frac{e^6 C_L a^2 c f |V_{21}|^2}{2 \mink{q}{p_2}} \Delta_{\chi_2^+}(p_1,k_1) 
  \real \{\Delta_Z(k_1,k_2)\}\notag\\[1mm]
  &&\big[2\mink{q}{k_1} \mink{p_1}{k_1} - \mink{q}{k_1} \mink{p_1}{k_2} - 
  \mink{p_1}{k_1} \mink{q}{k_2} - m_{\tilde{\nu}}^2 \mink{q}{p_1} +  
  \mink{k_1}{k_2} \mink{q}{p_1}\big] 
  \\[2mm] 
  T_{6 7} &=&  \frac{e^6 C_L a^2 c f |V_{21}|^2}{2 \mink{q}{p_1}}
  \Delta_{\chi_2^+}(p_1,k_1) \Delta_{\chi_2^+}(p_2,k_2) \real \{\Delta_Z(k_1,k_2)\}\notag\\[1mm]
  &&\big[m_{\chi^+_2}^2 (\mink{q}{k_1} \mink{p_1}{p_2} - 
  \mink{q}{k_2} \mink{p_1}{p_2} +  
  \mink{p_2}{k_1} \mink{q}{p_1} - 
  \mink{p_2}{k_2} \mink{q}{p_1} - 
  \mink{p_1}{k_1} \mink{q}{p_2} + \notag\\[1mm] 
  &&\mink{p_1}{k_2} \mink{q}{p_2}) - 
  2\mink{p_1}{k_1} \mink{p_2}{k_1} \mink{p_1}{k_2} -
  2\mink{p_1}{k_1} \mink{p_1}{k_1} \mink{p_2}{k_2} +  
  2\mink{p_1}{k_1} \mink{p_2}{k_2} \mink{q}{k_1} + \notag\\[1mm] 
  &&4\mink{p_1}{k_1} \mink{p_2}{k_2} \mink{p_1}{k_2} +
  2\mink{p_1}{k_1} \mink{p_2}{k_1} \mink{q}{k_2} - 
  4\mink{p_1}{k_1} \mink{p_2}{k_2} \mink{q}{k_2} -  
  2m_{\tilde{\nu}}^2 \mink{p_1}{p_2} \mink{p_1}{k_1} +\notag\\[1mm] 
  &&2\mink{k_1}{k_2} \mink{p_1}{k_1} \mink{p_1}{p_2} + 
  2m_{\tilde{\nu}}^2 \mink{p_1}{k_1} \mink{q}{p_2} -  
  2\mink{k_1}{k_2} \mink{p_1}{k_1} \mink{q}{p_2}\big] 
  \\[2mm]
  T_{6 8} &=&  -\frac{e^6 C_L a^2 c f |V_{21}|^2}{2 \mink{q}{p_2}} \Delta_{\chi_2^+}(p_1,k_1) 
  \Delta_{\chi_2^+}(p_2,k_2) \real \{\Delta_Z(k_1,k_2)\}\notag\\[1mm]
  &&\big[m_{\chi^+_2}^2 (\mink{q}{k_1} \mink{p_1}{p_2} - 
  \mink{q}{k_2} \mink{p_1}{p_2} -  
  \mink{p_2}{k_1} \mink{q}{p_1} + 
  \mink{p_2}{k_2} \mink{q}{p_1} + 
  \mink{p_1}{k_1} \mink{q}{p_2} - \notag\\[1mm] 
  &&\mink{p_1}{k_2} \mink{q}{p_2}) + 
  2\mink{p_1}{k_1} \mink{p_2}{k_2} \mink{p_2}{k_2} -
  4\mink{p_1}{k_1} \mink{p_2}{k_1} \mink{p_2}{k_2} +  
  4\mink{p_1}{k_1} \mink{p_2}{k_2} \mink{q}{k_1} + \notag\\[1mm]  
  &&2\mink{p_2}{k_1} \mink{p_1}{k_2} \mink{p_2}{k_2} - 
  2\mink{p_1}{k_2} \mink{p_2}{k_2} \mink{q}{k_1} - 
  2\mink{p_1}{k_1} \mink{p_2}{k_2} \mink{q}{k_2} +  
  2m_{\tilde{\nu}}^2 \mink{p_1}{p_2} \mink{p_2}{k_2} -\notag\\[1mm] 
  &&2\mink{k_1}{k_2} \mink{p_2}{k_2} \mink{p_1}{p_2} -
  2m_{\tilde{\nu}}^2 \mink{p_2}{k_2} \mink{q}{p_1} + 
  2\mink{k_1}{k_2} \mink{p_2}{k_2} \mink{q}{p_1}\big]  
  \\[2mm]
  T_{7 8} &=& 3\frac{e^6 f^2 (C_L c^2 + C_R d^2)}{4\mink{q}{p_1} \mink{q}{p_2}} |\Delta_Z(k_1,k_2)|^2\notag\\[1mm]
  &&\big[\mink{p_1}{k_1} \big(\mink{p_1}{k_1} \mink{q}{p_2} + 
                        2\mink{p_2}{k_1} \mink{p_1}{p_2} -  
                         \mink{q}{k_1} \mink{p_1}{p_2} -
                        2\mink{p_2}{k_2} \mink{p_1}{p_2} +  
                        \mink{q}{k_2} \mink{p_1}{p_2} - \notag\\[1mm]
                        &&\hspace*{20mm}\mink{p_2}{k_1} \mink{q}{p_1} +
                        \mink{p_2}{k_2} \mink{q}{p_1} - 
                        \mink{p_2}{k_1} \mink{q}{p_2} - 
                        2\mink{p_1}{k_2} \mink{q}{p_2} +
                        \mink{p_2}{k_2} \mink{q}{p_2}\big) +\notag\\[1mm]
  &&\mink{p_1}{p_2} \big(-2m_{\tilde{\nu}}^2 \mink{p_1}{p_2} + 
                   2\mink{k_1}{k_2} \mink{p_1}{p_2} -  
                   \mink{p_2}{k_1} \mink{q}{k_1} -
                   2\mink{p_2}{k_1} \mink{p_1}{k_2} + 
                   \mink{q}{k_1} \mink{p_1}{k_2} + \notag\\[1mm]
                   &&\hspace*{20mm}\mink{q}{k_1} \mink{p_2}{k_2} +
                   2\mink{p_1}{k_2} \mink{p_2}{k_2} + 
                   \mink{p_2}{k_1} \mink{q}{k_2} - 
                   \mink{p_1}{k_2} \mink{q}{k_2} -
                   \mink{p_2}{k_2} \mink{q}{k_2}\big) +\notag\\[1mm]
  &&\mink{q}{p_1} \big(\mink{p_2}{k_1} \mink{p_2}{k_1} + 
                 \mink{p_2}{k_2} \mink{p_2}{k_2} +  
                 \mink{p_2}{k_1} \mink{p_1}{k_2} -
                 2\mink{p_2}{k_1} \mink{p_2}{k_2} - \notag\\[1mm]
                 &&\hspace*{20mm}\mink{p_1}{k_2} \mink{p_2}{k_2} + 
                 2m_{\tilde{\nu}}^2 \mink{p_1}{p_2} -
                 2\mink{k_1}{k_2} \mink{p_1}{p_2}\big) +\notag\\[1mm]
  &&\mink{q}{p_2} \big(\mink{p_1}{k_2} \mink{p_1}{k_2} + 
                 \mink{p_2}{k_1} \mink{p_1}{k_2} - 
                 \mink{p_1}{k_2} \mink{p_2}{k_2} + 
                 2m_{\tilde{\nu}}^2 \mink{p_1}{p_2} -
                 2\mink{k_1}{k_2} \mink{p_1}{p_2}\big)\big] 
\end{eqnarray}
Formulas for the squared amplitudes for radiative sneutrino production
can also be found in Refs.~\cite{Franke:thesis,Franke:1994ph} for
longitudinal and transverse beam polarisations. Here, I give however
my calculated amplitudes for completeness.
I have calculated the squared amplitudes with \texttt{FeynCalc}~\cite{Kublbeck:1992mt}.
I neglect terms proportional to 
$\eps \imag \{\Delta_Z\}$,
see the discussion at the end of Appendix~\ref{sec:app:chifore}.
\chapter[Differential Cross section]{Definition of the Differential Cross Section and Phase Space}
\label{sec:phasespace}
I present some details of the phase space calculation for radiative neutralino production
\begin{equation}
\label{eq:scatter}
e^-(p_1) + e^+(p_2) \rightarrow \tilde{\chi}^0_1(k_1) +  
\tilde{\chi}^0_1(k_2) + \gamma(q). 
\end{equation} 
The differential cross section for (\ref{eq:scatter}) is given
by~\cite{Eidelman:2004wy}
\begin{eqnarray}
\dif \sigma &=& \half \frac{(2\pi)^4}{2 s}
\prod_f \frac{\dif^3 \mathbf{p}_f}{(2\pi)^3 2E_f}\delta^{(4)}(p_1 + 
p_2 - k_1 - k_2 - q)|\M|^2,
\label{phasespace}
\end{eqnarray}
where $\mathbf{p}_f$ and $E_f$ denote the final three-momenta
and the final energies
of the neutralinos and the photon.  The squared matrix element $|\M|^2$ is given in
Appendix~\ref{sec:app:chifore}.

I parametrise the four-momenta in the center-of-mass (cms) system of the
incoming particles, which I call the laboratory (lab) system. The beam
momenta  are then parametrised as 
\begin{eqnarray}
p_1 &={\displaystyle \half} 
\begin{pmatrix}
\sqrt{s}, & 0, & 0, & \sqrt{s}
\end{pmatrix},\qquad
p_2 &={\displaystyle \half}
\begin{pmatrix}
\sqrt{s}, & 0, & 0, & -\sqrt{s}
\end{pmatrix}.
\end{eqnarray}
For the outgoing neutralinos and the photon I consider in a first
step the local center-of-mass system of the two neutralinos.  The
photon shall escape along this $x_3$-axis. I start with general
momentum-vectors for the two neutralinos, boost them along their
$x_3$-axis and rotate them around the $x_1$-axis to reach the lab
system.  Note that the three-momenta of the outgoing particles lie in
a plane whose normal vector is inclined by an angle $\theta$ towards
the beam axis.  I parametrise the neutralino momenta in their cms
frame~\cite{Grassie:1983kq}
\begin{eqnarray}
\label{eq:momenta}
k_1^\ast &=&
 \begin{pmatrix}
\frac{1}{2}\sqrt{s^\ast}\\
\phantom{-}k^\ast \sin\theta^\ast\cos\phi^\ast \\
\phantom{-}k^\ast \sin\theta^\ast\sin\phi^\ast \\
\phantom{-}k^\ast \cos\theta^\ast 
\end{pmatrix},\qquad 
k_2^\ast = 
\begin{pmatrix}
\frac{1}{2}\sqrt{s^\ast}\\
-k^\ast \sin\theta^\ast\cos\phi^\ast \\
-k^\ast \sin\theta^\ast\sin\phi^\ast \\
-k^\ast \cos\theta^\ast
\end{pmatrix}, 
\end{eqnarray}
with the local cms energy $s^\ast$ of the two neutralinos
\begin{equation}
s^\ast = (k_1+k_2)^2 = 2 m_{\chi_1^0}^2 + 2\mink{k_1\,}{\,k_2},
\end{equation}
the polar angle $\theta^\ast$, the azimuthal angle $\phi^\ast$ and the
absolute value of the neutralino three-momenta $k^\ast$ in their cms
frame. These momenta are boosted to the lab system with the Lorentz
transformation
\begin{eqnarray}
\label{eq:lorentz}
L(\beta) = 
\begin{pmatrix}
\gamma & 0 & 0 & \gamma \beta \\
0 & 1 & 0 & 0 \\
0 & 0 & 1 & 0 \\
 \gamma \beta & 0 & 0 & \gamma
\end{pmatrix} ,
\end{eqnarray}
with $\gamma = \frac{1}{\sqrt{1-\beta^2}}$ and 
$\beta = \frac{|{\bf k_1 + k_2}|}{(k_1)^0 + (k_2)^0}_{|\mathrm{cms\,\, beam}}$ 
the boost velocity from the cms to the lab system
\begin{eqnarray}
\beta = \frac{|\mathbf{q}|}{\sqrt{s} -E_\gamma} = \frac{s - s^\ast}{s + s^\ast}.
\end{eqnarray}
Boosting the momenta $k_1^\ast$ and $k_2^\ast$, see
Eq.~(\ref{eq:momenta}), at first with the Lorentz transformation
Eq.~(\ref{eq:lorentz}) and then rotating with $\theta$ yields the
neutralino and photon momenta in the lab system~\cite{Grassie:1983kq}
\begin{eqnarray}
k_1 &=& 
\begin{pmatrix}
\gamma E^\ast + \beta\gamma k^\ast \cos\theta^\ast\\[1mm]
k^\ast\sin \theta^\ast \cos \phi^\ast\\[1mm]
k^\ast\sin \theta^\ast \sin \phi^\ast \cos\theta + 
(\beta \gamma E^\ast + \gamma k^\ast \cos\theta^\ast) \sin\theta\\[1mm]
 -k^\ast\sin \theta^\ast \sin \phi^\ast \sin\theta + (\beta \gamma 
E^\ast + \gamma k^\ast \cos\theta^\ast) \cos\theta
\end{pmatrix},\\[3mm]
k_2 &=&
\begin{pmatrix}
\gamma E^\ast - \beta\gamma k^\ast \cos\theta^\ast\\[1mm]
-k^\ast\sin \theta^\ast \cos \phi^\ast\\[1mm]
-k^\ast\sin \theta^\ast \sin \phi^\ast \cos\theta + 
(\beta\gamma E^\ast - \gamma k^\ast \cos\theta^\ast) \sin\theta\\[1mm]
 k^\ast\sin \theta^\ast \sin \phi^\ast \sin\theta + (\beta\gamma E^\ast 
- \gamma k^\ast \cos\theta^\ast) \cos\theta
\end{pmatrix},\\[3mm]
q &=&
\begin{pmatrix}
\frac{s-s^\ast}{2\sqrt{s}}\\[1mm]
0\\[1mm]
-\frac{s-s^\ast}{2\sqrt{s}}\sin\theta\\[1mm]
-\frac{s-s^\ast}{2\sqrt{s}}\cos\theta
\end{pmatrix},
\end{eqnarray} 
with
\begin{eqnarray}
k^\ast &=& \frac{1}{2}\sqrt{s^\ast-4 m_{\chi_1^0}^2},\\
E^\ast &= &\frac{\sqrt{s^\ast}}{2},\\ 
\beta\gamma &=&  \frac{s - s^\ast}{2 \sqrt{ss^\ast}} \enspace.
\end{eqnarray} 
The differential cross section for \signal~now reads~\cite{Grassie:1983kq}
\begin{eqnarray}
\dif\sigma &=& \frac{1}{4096 \pi^4 s}\left(1 - \frac{s^\ast}{s}\right)
\sqrt{1 - \frac{4 m_{\chi_1^0}^2}{s^\ast}}|\M|^2\,\dif\!\cos\theta \, \dif\!\cos\theta^\ast \, 
\dif\phi^\ast \, \dif s^\ast,
\end{eqnarray}
where the integration variables run over
\begin{equation}
\begin{array}{rcccl}
 0 & \le & \phi^\ast &\le & 2 \pi, \\[1mm]
  -1 & \le &\cos\theta^\ast &\le &1,\\[1mm]
  4 m_{\chi_1^0}^2 &\le &    s^\ast &\le &(1-x)s,\quad x = \frac{E_
\gamma}{E_{\mathrm{beam}}},\\[1mm]
   -0.99 & \le &\cos\theta & \le& 0.99.         
\end{array}     
\end{equation}

\chapter[Helicity amplitudes]{How to calculate helicity amplitudes for longitudinal polarisation states}

\section{Introduction}
Bouchiat and Michel presented \cite{Bouchiat:1958aa} formulae to perform helicity spin sums for
Dirac fermions and antifermions. Haber collected in
\cite{Haber:1994pe} mathematical tools to deal with such sums and presented example calculations.
Choi et al.\cite{Choi:1989yf} extended these formulae to spin 1 and spin $\frac{3}{2}$ fields.
In this paper I present a proof of the Bouchiat-Michel-formulae and I extend
these formulae to Majorana particles. All formulae are written in a covariant manner. 
%
\section{Spinor calculus}
\label{sec:spinor}
The Dirac spinors $u(p,\lambda)$ and $v(p,\lambda)$ obey the Dirac equation:
\begin{eqnarray}
(\ssl{p}-m)u = 0, \hspace*{15mm} (\ssl{p}+m)v = 0. 
\end{eqnarray}
The charge-conjugation-operator $C$ converts the spinor $u$ with positive energy into the spinor $v$ with negative
energy and vice versa:
\begin{eqnarray}
\label{eq:Cu}
u = C\vv^T,\enspace v = C \uu^T .
\end{eqnarray} 

There are two solutions of the Dirac equations for a given 4-momentum $p$, so there exists another good quantum
number to label these states. This is the helicity $\lambda$.
The helicity operator $\Lambda = \Sigma \mathbf{\hat{p}}$
commutes with the Dirac operator $(\ssl{p}\pm m)$, so the eigenvalues $\lambda = \pm\frac{1}{2}$ of $\Lambda$ are 
good quantum numbers.

The spinors $u$ and $v$ are normalized to (\cite{Peskin:1995qf})
\begin{eqnarray}
\uu(p,\lambda)u(p,\lambda') &=& \phantom{-}2m \delta_{\lambda\lambda'},\\[1mm] 
\vv(p,\lambda)v(p,\lambda') &=& -2m \delta_{\lambda\lambda'}.
\end{eqnarray}

When calculating cross sections or decay widths of fermions and antifermions, I sum over all spins states 
and average over initial spins, using the completeness relation \cite{Peskin:1995qf}:
\begin{eqnarray}
\label{eq:comu}
\sum_{\lambda} u(p,\lambda)\uu(p,\lambda) &=& \ssl{p} + m,\\[2mm]
\label{eq:comv}
\sum_{\lambda} v(p,\lambda)\vv(p,\lambda) &=& \ssl{p} - m.
\end{eqnarray}

When describing spin-polarized fermion ensemble one introduces spin vectors.
The longitudinal spin vector for a particle with mass $m$ is defined by
\begin{eqnarray}
\label{eq:spin3}
s = \frac{1}{m} 
\begin{pmatrix}
|\vec{p}|,&E\frac{\vec{p}}{|\vec{p}|}
\end{pmatrix}.
\end{eqnarray}
$s$ is normalized to $s\cdot s = -1$, and is orthogonal to the momentum vector $p$: $s\cdot p  = 0$.

I have to do distinguish between massive and massless particles. The spinvector for massless particles is 
obtained in the limit $m \rightarrow 0$.

\subsection{The massive case}

The operator $\gamma^5 \ssl{s}$ commutes with $\ssl{p}$, so both operators can be diagonalized
simultaneously. Their eigenvectors are known, 
and the eigenvalues are obtained by \cite{Schwabl:2000qm}  :
\begin{eqnarray}
\label{eq:case3}
&&\gamma^5\ssl{s}\ssl{p}u(p,\lambda) \notag\\[1mm]  
\hspace{-10mm}
& = & \gamma^5\Big(s\cdot p - \ie s_\mu p_\nu\sigma^{\mu\nu} \Big) u(p,\lambda)\notag \\[1mm]
& = &  -\ie\gamma^5\Big(\sigma^{0j}s_0 p_j + \sigma^{j0}s_j p_0\Big) u(p,\lambda)\notag  \\[1mm] 
& = &  \frac{-\ie}{m}\gamma^5 \sigma^{0j}\Big(|\vec{p}|p_j  
       - E\frac{p_j}{|\vec{p}|}E\Big)u(p,\lambda)\notag  \\[1mm] 
& = &  2 m \Sigma_j\frac{p_j}{|\vec{p}|}  u(p,\lambda)\notag  \\[1mm] 
& = &  2 \lambda  m u(p,\lambda)\notag \\[1mm] 
& = &  2 \lambda \ssl{p}u(p,\lambda).  
\end{eqnarray}   
I have used the relation $\gamma^5\gamma^0\gamma^j = \ie \gamma^0\gamma^1\gamma^2\gamma^3\gamma^0\gamma^j = 
-\ie\gamma^1\gamma^2\gamma^3\gamma^j = \ie\epsilon_{jkl}\gamma^k\gamma^l$ to realize that $\gamma^5 \ssl{s}$ 
is the helicity operator.
 
From the above calculation it follows immediately:
\begin{eqnarray}
\gamma^5 \ssl{s} u = 2 \lambda u.
\end{eqnarray}

\subsection{The massless case}
The Dirac equation for a massless spin \half-fermion is:
\begin{eqnarray}
\label{eq:weyl}
\ssl{p}u(p,\lambda) = 0
\end{eqnarray}
and multiply eq. (\ref{eq:weyl}) with $\gamma^5\gamma^0$ \cite{Schwabl:2000qm}: 
\begin{eqnarray}
0 &=& \gamma^5\gamma^0 \ssl{p}u(p,\lambda) = (\gamma^5 p_0 - \vec{\Sigma}\vec{p})u(p,\lambda);\\[1mm]
\label{eq:massless}  
  &\Rightarrow& \vec{\Sigma} \hat{p} u(p,\lambda) = \gamma^5 u(p,\lambda)
\end{eqnarray}
with $p^0 = |\vec{p}|$.
The chirality-operator $\gamma^5$ commutes with the helicity-operator, hence they have common eigenvectors,
a similar equation holds for $v$ spinors:
\begin{eqnarray}
\gamma^5 u(p,\lambda) &=& \pm u(p,\lambda) = 2\lambda u(p,\lambda),\\[1mm]
\gamma^5 v(p,\lambda) &=& \mp v(p,\lambda) = -2\lambda v(p,\lambda).
\end{eqnarray}
Since $(\gamma^5)^2 = 1, \mathrm{Trace}(\gamma^5)=0$, the eigenvalues of the chirality-operator are $\pm 1$.
   
\section{The Bouchiat-Michel-Formula}
The Bouchiat-Michel-formulae (BMF) tell us how to contract spinors with different polarisations. The BMF is interesting if the
initial state fermions have the same mass, but also in using density matrix techniques \cite{Haber:1994pe}.  

\subsection{Spin vectors}
I enlarge the set $s$, $p$ defined in sec. \ref{sec:spinor} with two other four-vectors $s^1$ and $s^2$ to 
a orthonormal basis in space-time:
\begin{eqnarray}
p \cdot s^a &=& 0\enspace \\[1mm]
s^a\cdot s^b &= &-\delta^{ab} \enspace \\[0mm]
\ssl{s}^a \ssl{s}^b &=& -\delta^{ab}+\frac{\ie \epsilon_{abc}\gamma^5 \ssl{p}\ssl{s}^c}{m}\label{eq:spinortho}
\end{eqnarray}
with $a= 1\ldots3$, $s^3 =s$.

The spinors $u(p,\lambda)$, $v(p,\lambda)$ satisfy:
\begin{eqnarray}
\gamma^5 \ssl{s}^a u(p,\lambda') = \sigma_{\lambda \lambda'}^a u(p,\lambda) \label{eq:upol}\\
\gamma^5 \ssl{s}^a v(p,\lambda') = \sigma_{\lambda' \lambda}^a v(p,\lambda) \label{eq:vpol}.
\end{eqnarray}
The proof of this equation can be found in/follows \cite{Bailin:1977ew}:
Define the Pauli-Lubanski-vector as
\begin{eqnarray}
W_\mu &=& \half \varepsilon_{\mu\alpha\beta\gamma}M^{\alpha\beta}P^\gamma\\
&=& 
\begin{pmatrix}
\vec{\Sigma} \vec{p},&E \vec{\Sigma} + \vec{K}\times \vec{p}
\end{pmatrix},
\end{eqnarray} 
where $M^{\alpha\beta}$, $K^i = M^{0i}$, $P^\gamma$ are the generators of the Poincare-group.
In the rest frame the spin vectors take a simple form: $s^i = \begin{pmatrix}0,&\hat{e}_i\end{pmatrix}$
where $\hat{e}_i$ is the ith unit vector. I build the scalar operator $W\cdot s^i$, which has the eigenvalues 
$\lambda m$ and evaluate it in the rest frame:
\begin{eqnarray}
W\cdot s^i = -\half m \vec{\Sigma} \hat{e}_i = -\half m \sigma^i
\end{eqnarray}
Then, I apply $W\cdot s^i$ on a spinor $u(p,\lambda)$:
\begin{eqnarray}
W\cdot s^i u(p,\lambda) &=& -\half m \sigma^i_{\lambda'\lambda} u(p,\lambda').  
\end{eqnarray}
On the other hand $W\cdot s^i = \frac{1}{4}\big[\ssl{s}^i, \ssl{p}\big]\gamma^5$, so
\begin{eqnarray}
W\cdot s^i u(p,\lambda) &=& -\frac{1}{4}\big[\ssl{s}^i, \ssl{p}\big]\gamma^5 u(p,\lambda)\notag\\
&=& -\frac{1}{4}\gamma^5(\ssl{s}^i\ssl{p} - \ssl{p}\ssl{s}^i) u(p,\lambda)\notag\\
&=& -\half m \gamma^5 \ssl{s}^i u(p,\lambda)\notag\\
&=& -\half m \sigma^i_{\lambda'\lambda} u(p,\lambda'). 
\end{eqnarray}
Now eq. (\ref{eq:upol}) follows. The proof for the spinor $v$ is similar.  

\subsection{BMF for massive Dirac fermions}
Now I have all ingredients to formulate and prove the BMF:
\begin{eqnarray}
u(p,\lambda') \uu(p,\lambda) &=& \half\big[\delta_{\lambda'\lambda} + \gamma^5 \ssl{s}^a \sigma^a_{\lambda'\lambda}\big]
                                  \big(\ssl{p} + m\big)\\ 
v(p,\lambda') \vv(p,\lambda) &=& \half\big[\delta_{\lambda\lambda'} + \gamma^5 \ssl{s}^a \sigma^a_{\lambda\lambda'}\big]
                                  \big(\ssl{p} - m\big). \label{eq:bmvv}
\end{eqnarray} 
The sum must have the form 
\begin{eqnarray}
u(p,\lambda') \uu(p,\lambda) &=& A\delta_{\lambda'\lambda} + B^a \sigma^a_{\lambda'\lambda}.
\end{eqnarray}
To determine the unknown coefficients $A$, $B^a$ we
multiply both sides with $\delta^{\lambda'\lambda}$ and with $\sigma_a^{\lambda'\lambda}$:
\begin{eqnarray}
\half u(p,\lambda') \uu(p,\lambda) \delta^{\lambda'\lambda} &=& \half \big(\ssl{p} + m\big) = A,\\ 
\half u(p,\lambda') \uu(p,\lambda) (\sigma^a)^{\lambda'\lambda} &=& \half \gamma^5 \ssl{s}^a\big(\ssl{p} + m \big) = B^a .\notag\\[-2mm]
&&
\end{eqnarray}
A similar proof holds for eq. (\ref{eq:bmvv})

\subsection{BMF for massless Dirac fermions}
To perform the limit $m\rightarrow 0$ I use (\ref{eq:massless}), and get
\begin{eqnarray}
&u(p,\lambda') \uu(p,\lambda) =\hspace{50mm} \notag\\[1mm]
&\half\Big(\delta_{\lambda'\lambda} + \gamma^5 \sigma^3_{\lambda'\lambda}
+ \gamma^5 \ssl{s}^1\sigma^1_{\lambda'\lambda} + \gamma^5 \ssl{s}^2\sigma^2_{\lambda'\lambda}\Big)\ssl{p},\\[2mm]
&v(p,\lambda') \vv(p,\lambda) =\hspace{50mm} \notag\\[1mm]
&\half\Big(\delta_{\lambda\lambda'} + \gamma^5 \sigma^3_{\lambda\lambda'}
+ \gamma^5 \ssl{s}^1\sigma^1_{\lambda\lambda'} + \gamma^5 \ssl{s}^2\sigma^2_{\lambda\lambda'}\Big)\ssl{p}.
\end{eqnarray}
I shall make some remarks to the spin vectors for particles with $E \gg m$. I start
with eq. (\ref{eq:spin3}) and expand $E = |\vec{p}|\sqrt{1 + m^2/|\vec{p}|^2}\approx 
|\vec{p}|\Big(1 + \half \frac{m^2}{|\vec{p}|^2} + o\big(\frac{m^4}{|\vec{p}|^4}\big)\Big)$.
This expansion preserves the normalization $s^3\cdot s^3 = -1$. 

\subsection{Majorana-Fermions} 
In supersymmetric (SUSY) field theories there appear Majorana fermions, for example the neutralinos, which are the SUSY partners
of the neutral weak gauge and Higgs bosons. In Feynman diagrams with Majorana fermions I find often
clashing arrows. So it is useful to have formulae to handle this case.
I use the relations (\ref{eq:Cu}), and I get
\begin{eqnarray}
u(p,\lambda')v^T(p,\lambda) &=& \half\big[\delta_{\lambda'\lambda} + \gamma^5 \ssl{s}^a \sigma^a_{\lambda'\lambda}\big]
                                  \big(\ssl{p} + m\big)C^{T},\notag\\[-3mm]
&&\\
\vv^T (p,\lambda')\uu(p,\lambda) &=&\half C^{-1}\big[\delta_{\lambda'\lambda} + \gamma^5 \ssl{s}^a \sigma^a_{\lambda'\lambda}\big]
                                  \big(\ssl{p} + m\big),\notag\\[-2mm]
&&\\
v(p,\lambda')u^T(p,\lambda) &=& \half\big[\delta_{\lambda'\lambda} + \gamma^5 \ssl{s}^a \sigma^a_{\lambda'\lambda}\big]
                                  \big(\ssl{p} - m\big)C^{T},\notag\\[-3mm]
&&\\
\uu^T (p,\lambda')\vv(p,\lambda) &=&\half C^{-1}\big[\delta_{\lambda'\lambda} + \gamma^5 \ssl{s}^a \sigma^a_{\lambda'\lambda}\big]
                                  \big(\ssl{p} - m\big).\notag\\[-2mm]
&&
\end{eqnarray}

\section{Calculation of the density matrix}
The methods described above can be used to compute squared matrix elements in the helicity mechanism described in 
\cite{Haber:1994pe}.
The only change in the usual mechanism is the completeness relation. I consider longitudinal polarized
electrons. I treat the electrons as massless.
I call this matrix the reaction matrix $R_{\lambda\lambda'}$.
\begin{eqnarray}
R_{\lambda\lambda'} = u(p,\lambda)\uu(p,\lambda') =  
\half\big(\delta_{\lambda\lambda'} + \gamma^5\sigma^3_{\lambda\lambda'}\big)\ssl{p}
\end{eqnarray}

The helicity indices are contracted with the beam matrix $B_{\lambda\lambda'}$:
\begin{eqnarray}
B_{\lambda\lambda'} = \half\Big(\delta^{\lambda\lambda'} + P_-^3\sigma_3^{\lambda\lambda'}\Big)
= \half 
\begin{pmatrix}
1 + P^3_- & 0\\[1mm]
0 & 1 - P^3_-
\end{pmatrix}
\end{eqnarray}
Now I calculate:
\begin{eqnarray}
\label{eq:polcom}
B_{\lambda\lambda'}R^{\lambda\lambda'}&=& \half\Big(\delta^{\lambda\lambda'} + P_-^3\sigma_3^{\lambda\lambda'}\Big)
                        \half\big(\delta_{\lambda\lambda'} + \gamma^5\sigma^3_{\lambda\lambda'}\big)\ssl{p}\notag \\[1mm]
                                      &=& \half\big(1 + P_-^3 \gamma^5\big)\ssl{p}\notag\\[1mm]
                                      &=& \left(\frac{1+P_-}{2}P_R + \frac{1 - P_-}{2}P_L\right)\ssl{p}          
\end{eqnarray}
I can use eq (\ref{eq:polcom}) instead of the usual completeness relations Eqs (\ref{eq:comu}), (\ref{eq:comv}) to do the spin sums.

\backmatter

\end{document}